\documentclass[twocolumn,runningheads]{svjour2}
\smartqed  % flush right qed marks, e.g. at end of proof
\usepackage{graphicx}
\usepackage{mathptmx}   % use Times fonts if available on your TeX system
\usepackage{amsmath}
\usepackage{amsfonts}
\usepackage{amssymb}
\usepackage{latexsym}
\usepackage[authoryear]{natbib}%\usepackage[numbers]{natbib}
\usepackage{journals}

\journalname{Astrophysics and Space Science (CoRoT/ESTA Volume)}
\newcommand{\msol}{\mbox{$\,{M}_{\odot}$}}

\newcommand{\corot}{{\small CoRoT}}

\newcommand{\esta}{{\small ESTA}}

\newcommand{\task}{{\small TASK}}

\newcommand{\ASTEC}{{\small\bf ASTEC}}
\newcommand{\CESAM}{{\small\bf CESAM}}
\newcommand{\CLES}{{\small\bf CL\'ES}}
\newcommand{\GARSTEC}{{\small\bf GARSTEC}}
\newcommand{\STAROX}{{\small\bf STAROX}}

\newcommand{\ADIPLS}{{\small\bf ADIPLS}}
\newcommand{\LOSC}{{\small\bf LOSC}}

\newcommand{\astec}{{\small ASTEC}}
\newcommand{\cesam}{{\small CESAM}}
\newcommand{\cesamMP}{{\small CESAM-MP}}
\newcommand{\cesamB}{{\small CESAM-B69}}
\newcommand{\cles}{{\small CL\'ES}}
\newcommand{\garstec}{{\small GARSTEC}}
\newcommand{\starox}{{\small STAROX}}

\newcommand{\MP}{{\small MP93}}
\newcommand{\BURG}{{\small B69}}

\newcommand{\PMS}{{\small PMS}}
\newcommand{\MS}{{\small MS}}
\newcommand{\ZAMS}{{\small ZAMS}}
\newcommand{\TAMS}{{\small TAMS}}
\newcommand{\SGB}{{\small SGB}}
\def\note #1]{{\bf #1]}}

\newcommand{\adipls}{{\small ADIPLS}}
\newcommand{\thisapss}{Astrophys. Space Sci. (CoRoT/ESTA Volume)}

\begin{document}

\title{CoRoT/ESTA--TASK~1 and TASK~3 comparison of the internal structure and seismic properties of representative stellar models}

\subtitle{Comparisons between the {\Large ASTEC}, {\Large CESAM}, {\Large CLES}, {\Large GARSTEC} and {\Large STAROX} codes}

\titlerunning{CoRoT/ESTA--TASK~1 and TASK~3 model comparisons for internal structure and seismic properties}

\author{Yveline Lebreton \and Josefina Montalb\'an   \and  J{\o}rgen Christensen-Dalsgaard \and Ian~W.~Roxburgh \and Achim Weiss
}

%\authorrunning{Short form of author list} % if too long for running head
\institute{Y. Lebreton \at
               GEPI, Observatoire de Paris, CNRS, Universit\'e Paris Diderot, 5 Place Janssen, 92195 Meudon, France \\
%              Tel.: +33-299236386\\              Fax: +33-299236957\\
              \email{Yveline.Lebreton@obspm.fr}           %  \\
%             \emph{Present address:} of F. Author  %  if needed
           \and
	   J. Montalb\'an \at Institut d'Astrophysique et G\'eophysique, Universit\'e de Li\`ege, Belgium\\
	   \email{j.montalban@ulg.ac.be} 
	  \and
          J. Christensen-Dalsgaard \at Institut for Fysik og Astronomi, Aarhus Universitet, Denmark
	  \and
	  I.W. Roxburgh \at Queen Mary University of London, England \\ LESIA, Observatoire de Paris, France
	   \and
           A. Weiss \at Max-Planck-Institut f\"ur Astrophysik, Garching, Germany	   
}

\date{Received: date / Accepted: date}
% The correct dates will be entered by the editor
\maketitle
%*******************************************************************
\begin{abstract}
We compare stellar models produced by different stellar evolution codes
for the {\corot}/{\esta} project, comparing their global quantities, their
physical structure, and their oscillation properties. We discuss the
differences between models and identify the underlying reasons
for these differences. The stellar models are representative of  potential
{\corot} targets. Overall we find very good agreement between the five
different codes, but with some significant deviations. We find noticeable discrepancies (though still at the per cent level) that result from the handling of the equation of state, of the opacities and of the convective boundaries. The results of our work will be helpful in interpreting future asteroseismology results from {\corot}.

\end{abstract}

%-------------------------------------------------------------------

\section{Introduction}\label{sec:intro}

The goals of \esta-{\small TASK}s~1 and 3 are to test the numerical tools used in stellar modelling, with the objective to be ready to interpret safely the asteroseismic data that will come from the {\corot} mission. This consists in quantifying the effects of different numerical implementations of the stellar evolution equations and related input physics on the internal structure, evolution and seismic properties of stellar models. As a result, we aim at improving the stellar evolution codes to get a good agreement between models built with different codes and same input physics. For that purpose, several study cases have been defined that cover a large range of stellar masses and evolutionary stages and stellar models have been calculated without (in {\task}~1) or with (in {\task}~3) microscopic diffusion \citep[see][]{2006corm.conf..363M,yl1-apss}. 

In this paper, we present the results of the detailed comparisons of the internal structures and seismic properties of {\small TASK}s~1 and 3 target models. The comparisons of the global parameters and evolutionary tracks are discussed in \cite{mm2-apss}. In order to ensure that the differences found are mainly determined by the way each code calculates the evolution and the structure of the model and not by significant differences in the input physics, we decided to use and compare models whose global parameters (age, luminosity, and radius) are very similar. Therefore, we selected models computed by five codes among the ten  participating in \esta: {\astec} \citep{jcd1-apss}, {\cesam} \citep{pm-apss}, {\cles} \citep{rs1-apss}, {\garstec} \citep{aw-apss}, and {\starox} \citep{ir1-apss}.

In Section \ref{sec:tasks} we recall the specifications of  {\small TASK}s~1 and 3 and present the five codes used in the present paper. We then present the comparisons for {\task}~1 in Sect.~\ref{sec:task1} and for {\task}~3 in Sect.~\ref{sec:task3}. For each case in each task, we have computed the relative differences of the physical quantities between pairs of models. We display the variation of the differences between the more relevant quantities inside the star and we provide the average and extreme values of the variations. We then compare the location of the boundaries of the convective regions as well as their evolution with time. For models including microscopic diffusion we examine how helium is depleted at the surface as a function of time. Finally, we analyse the effect of internal structure differences on seismic properties of the model.

\section{Presentation of the ESTA-TASKs and tools}\label{sec:tasks}

In the following we briefly recall the specifications and tools of {\task}~1 and {\task}~3 that have been presented in detail by \citet{yl1-apss}.

\subsection{TASK~1: basic stellar models}

The specifications for the seven cases that have been considered in {\task}~1 are recalled in Table \ref{tab:task1}. For each case, evolutionary sequences have been calculated for the specified values of the stellar mass and initial chemical composition ($X, Y, Z$ where $X$, $Y$ and $Z$ are respectively the initial hydrogen, helium and metallicity in mass fraction) up to the evolutionary stage specified. The masses are in the range $0.9-5.0$ \msol. For the initial chemical composition, different $(Y,Z)$ couples have been considered by combining two different values of $Z$ ($0.01$ and $0.02$) and two values of $Y$ ($0.26$ and $0.28$). 
The evolutionary stages considered are either on the pre main sequence (\PMS), the main sequence (\MS) or the subgiant branch (\SGB). On the {\PMS} the central temperature of the model ($T_{\mathrm c}=1.9\times 10^7\ {\mathrm K}$) has been specified. On the {\MS}, the value of the central hydrogen content has been fixed: $X_{\mathrm c}=0.69$ for a model close to the zero age main sequence (\ZAMS), $X_{\mathrm c}=0.35$ for a model in the middle of the {\MS} and $X_{\mathrm c}=0.01$ for a model close to the terminal age main sequence (\TAMS). On the \SGB, a model is chosen by specifying the value of the mass $M_{\mathrm c}^{\mathrm{He}}$ of the central region of the star where the hydrogen abundance is such that $X\leq 0.01$. We chose  $M_{\mathrm c}^{\mathrm{He}}=0.10$ \msol. 

All models calculated for {\task}~1 are based on rather simple input physics, currently implemented in stellar evolution codes and  one model has been calculated with overshooting. Also, reference values of some astronomical and physical constants have been fixed as well as the mixture of heavy elements to be used.
These specifications are described in \cite{yl1-apss}. 

%==========================
\begin{table}[htbp!]
\caption{Target models for {\task}~1.
We have considered 7 cases corresponding to different initial masses, chemical compositions and evolutionary stages. One evolutionary sequence (denoted by ``OV'' in the $\rm{5^{th}}$ column has been calculated with core overshooting (see text).}
\centering
\label{tab:task1}
\begin{tabular}[h]{cccclr}
\hline\noalign{\smallskip}
{\bf Case} & \boldmath$M/M_\odot$ & \boldmath$Y_0$ & \boldmath$Z_0$ &
  {\bf Specification} & {\bf Type} \\[3pt]
\tableheadseprule\noalign{\smallskip}
{\bf 1.1} & 0.9 & 0.28 & 0.02 &
  $X_{\mbox{\scriptsize c}}{=}0.35$ & 
  \begin{small}MS\end{small}\\[2pt]
{\bf 1.2} & 1.2 & 0.28 & 0.02 &
  $X_{\mbox{\scriptsize c}}{=}0.69$ & 
  \begin{small}ZAMS\end{small}\\[2pt]
{\bf 1.3} & 1.2 & 0.26 & 0.01 &
  $M^{\mbox{\scriptsize He}}_{\mbox{\scriptsize c}}{=}0.10$ \msol & 
  \begin{small}SGB\end{small}\\[2pt]
{\bf 1.4} & 2.0 & 0.28 & 0.02 &
  $T_{\mbox{\scriptsize c}}{=}1.9{\times}10^7$~K &
  \begin{small}PMS\end{small}\\[2pt]
{\bf 1.5} & 2.0 & 0.26 & 0.02 &
  $X_{\mbox{\scriptsize c}}{=}0.01, {\mbox{OV}}$ & 
  \begin{small}TAMS\end{small}\\[2pt]
{\bf 1.6} & 3.0 & 0.28 & 0.01 &
  $X_{\mbox{\scriptsize c}}{=}0.69$ &
  \begin{small}ZAMS\end{small}\\[2pt]
{\bf 1.7} & 5.0 & 0.28 & 0.02 &
  $X_{\mbox{\scriptsize c}}{=}0.35$ &
  \begin{small}MS\end{small}\\[2pt]
\noalign{\smallskip}\hline
%\hline
\end{tabular}
\end{table}
%==========================
%==========================

%==========================
%==========================

\subsection{TASK~3: stellar models including microscopic diffusion}

{\task}~3 is dedicated to the comparisons of stellar models including microscopic diffusion of chemical elements resulting from pressure, temperature and concentration gradients \citep[see][]{tm07}. The other physical assumptions proposed as the reference for the comparisons of  {\task}~3 are the same as used for {\task}~1, and  no overshooting.

Three study cases have been considered for the models to be compared. Each case corresponds to a given value of the stellar mass (see Table \ref{tab:task3}) and to a chemical composition close to the standard solar one ($Z/X=0. 0243$). For each case, models at different evolutionary stages have been considered. We focused on three particular evolution stages : middle of the \MS, {\TAMS} and {\SGB} (respectively stage A, B and C).

%==========================
\begin{table}[htbp!]
\caption{Target models for {\task}~3. Left: The 3 cases with corresponding masses and initial chemical composition. Right: The 3 evolutionary stages examined for each case. Stages A and B are respectively in the middle and end of the MS stage. Stage C is on the SGB.}
\centering
\label{tab:task3}      
\begin{tabular}{cccc}
\hline\noalign{\smallskip}
\begin{small}{\bf Case}\end{small} &\boldmath${\frac{M}{M_\odot}}$ & \boldmath$Y_0$ & \boldmath$Z_0$ \\[3pt]
\tableheadseprule\noalign{\smallskip}
\bfseries{3.1}& $1.0$ & $0.27$ & $0.017$ \\
\bfseries{3.2}& $1.2$ & $0.27$ & $0.017$ \\
\bfseries{3.3}& $1.3$ & $0.27$ & $0.017$ \\
\noalign{\smallskip}\hline
\end{tabular}
\hspace{0.5cm}
\begin{tabular}{ccc}
\hline\noalign{\smallskip}
\begin{small}{\bf Stage}\end{small} & \boldmath$X_{\rm c}$ & \boldmath$M^{\mbox{\scriptsize He}}_{\mbox{\scriptsize c}}$ \\[3pt]
\tableheadseprule\noalign{\smallskip}
\bfseries{A}& $0.35$ & - \\
\bfseries{B}& $0.01$ & - \\
\bfseries{C}& $0.00$ & $0.05\ M_{\mathrm{star}}$ \\
\noalign{\smallskip}\hline
\end{tabular}
\end{table}

\subsection{Numerical tools}\label{sec:tools}

Among the stellar evolution codes considered in the comparisons presented by \cite{mm2-apss}, we have considered  5 codes -- listed below -- for further more detailed comparisons. These codes have shown a very good agreement in the comparison of the global parameters which ensures that they closely follow the specifications of the tasks in terms of input physics and physical and astronomical constants.

\begin{itemize}
\item \ASTEC\ -- {\em Aarhus STellar Evolution Code}, described in \citet{jcd1-apss}.
\item \CESAM\ -- {\em Code d'\'Evolution Stellaire Adaptatif et Modulaire}, see \citet{pm-apss}.  
\item \CLES\ -- {\em Code Li\'egeois d'\'Evolution Stellaire}, see \citet{rs1-apss}. 
\item \GARSTEC\ -- {\em Garching Evolution Code}, presented in \linebreak[4] \citet{aw-apss}.
\item \STAROX\ -- {\em Roxburgh's Evolution Code}, see \citet{ir1-apss}.
\end{itemize}

The oscillation frequencies presented in this paper have been calculated with the {\LOSC} adiabatic oscillation code \citep[{\em Li\`ege Oscillations Code}, see][]{rs2-apss}. 
Part of the comparisons between the models has been performed with programs included in the {\em Aarhus Adiabatic Pulsation Package} {\ADIPLS}\footnote{available at {\tt\small http://astro.phys.au.dk/$\sim$jcd/adipack.n} } \citep[see][]{jcd2-apss}.

\section{Comparisons for TASK~1}\label{sec:task1}

\subsection{Presentation of the comparisons and general results}\label{sec:task1.1}

The {\task}~1  models  span different masses and evolutionary phases. Cases~1.1, 1.2 and 1.3 illustrate the internal structure of solar-like, low-mass $0.9${\msol} and $1.2${\msol} stars, at the beginning of the main sequence  of hydrogen burning (Case~1.2), in the middle of the {\MS} when the hydrogen mass fraction in the centre has been reduced to the half of the initial one (Case~1.1) and in the post-main sequence when the star has already built a He core of $0.1${\msol} (Case~1.3). Cases~1.4 and 1.5 correspond to intermediate-mass models ($2.0$\msol), the first one,  in a phase prior to the {\MS} when the nuclear reactions have not yet begun to play a relevant role, and the second one, at the end of the {\MS}, when the matter in the centre contains only $1$\% of hydrogen. Finally, Cases~1.6 ($3.0${\msol}) and 1.7 ($5.0${\msol}) sample the internal structure of models corresponding to middle and late B-type stars. For these more massive models, the beginning and the middle of their {\MS} are examined.

The models provided correspond to a different number of mesh points: the number of mesh points is $1202$ in the {\astec} models; it is in the range $2300-3700$ in {\cesam} models, $2200-2400$ in {\cles} models, $1500-2100$ in models by {\garstec} and $1900-2000$ in {\starox} models. As explained in the papers devoted to the description of the participating codes  \citep{jcd1-apss,pm-apss,ir1-apss,rs1-apss,aw-apss}, the numerical methods used to solve the equations and to interpolate in the tables containing physical inputs are specific to each code and so are the possibilites to choose the number and repartition of the mesh points in a model or the time step of the evolution calculation and, more generally, the different levels of precision of the computation. The specifications for the tasks have concerned mainly the physical inputs and the constants to be used \citep{yl1-apss} and have let the modelers free to tune up the numerous numerical parameters involved in their calculation which explains why each code deals with different numbers (and repartition) of mesh points.

%==========================
\begin{table}[htbp!]
\caption{{\task}~1 models: Global parameter differences given in per cent, between each code and {\cesam}. For each parameter we give the mean and maximum difference of the complete series of {\task}~1 models}
\centering
\label{tab:task1-glob}
\begin{tabular}[h]{lcccccc}
\hline\noalign{\smallskip}
{\bf Code} &  \multicolumn{2}{c}{\boldmath$\delta R/R$}&  \multicolumn{2}{c}{\boldmath$\delta L/L$}& \multicolumn{2}{c}{\boldmath$\delta T_{\rm eff}/T_{\rm eff}$} \\ [3pt]
& {\rm mean} & {\rm max} &{\rm mean} & {\rm max} & {\rm mean} & {\rm max}  \\ [3pt]
\tableheadseprule\noalign{\smallskip}
{\astec}        &  0.27 &   0.83  &  0.49  &  1.57 &   0.02 &   0.03   \\
{\cles}         &  0.20 &   0.43  &  0.16 &   0.52  &  0.01  &  0.01 \\
{\garstec}     &  0.37  &  0.59  &  0.23  &  0.46  &  0.03  &  0.06    \\
{\starox}    &  0.75 &   3.29  &  0.31  &  0.89  &  0.03  &  0.13 \\
\noalign{\smallskip}\hline
\end{tabular}
\end{table}

%==========================
%==========================
\begin{table*}[htbp!]
\caption{{\task}~1 models: Mean quadratic difference in the physical variables between each code and {\cesam} calculated according to Eq. \ref{eq:quad_mean}. The differences are given in per cent (except for $\delta X$) and represent  an average over the whole star from centre to photospheric radius. The local differences were computed at
fixed relative mass.}
\centering
\label{tab:task1-quad}
\begin{tabular}[h]{lccccccccccc}
\hline\noalign{\smallskip}
\multicolumn{11}{c} {\bf Case~1.1}\\
%\hline\noalign{\smallskip}
\hline\noalign{\smallskip}
{\bf Code} & \boldmath$\delta\ln c$      &  \boldmath$\delta\ln P$  
           & \boldmath$\delta\ln \rho$   & \boldmath$\delta\ln T$ & \boldmath$\delta\ln r$ & \boldmath$\delta\ln \Gamma_1$ 
           & \boldmath$\delta\ln \nabla_{\rm ad}$ & \boldmath$\delta\ln C_p$   
           & \boldmath$\delta\ln \kappa$ & \boldmath$\delta X$ & \boldmath$\delta\ln L_r$ \\ [3pt]
\tableheadseprule\noalign{\smallskip}

 \astec  & 0.02 & 0.15 & 0.16 & 0.04 & 0.05 & 2.69$\times 10^{-4}$ & 5.84$\times 10^{-4}$ & 7.36$\times 10^{-3}$ & 0.44 & 1.2$\times 10^{-4}$ & 0.17 \\ 
 \cles  & 0.04 & 0.25 & 0.22 & 0.07 & 0.06 & 3.14$\times 10^{-4}$ & 4.06$\times 10^{-3}$ & 2.64$\times 10^{-2}$ & 0.20 & 3.4$\times 10^{-4}$ & 0.03 \\ 
 \garstec  & 0.06 & 0.40 & 0.44 & 0.07 & 0.09 & 1.46$\times 10^{-1}$ & 1.13$\times 10^{-1}$ & 1.11$\times 10^{-1}$ & 0.45 & 4.7$\times 10^{-4}$ & 0.33 \\ 
 \starox  & 0.04 & 0.28 & 0.25 & 0.08 & 0.09 & 6.47$\times 10^{-4}$ & --- & --- & --- & 3.1$\times 10^{-4}$ & 0.08 \\ 

\noalign{\smallskip}\hline
%==========================
%==========================
\hline\noalign{\smallskip}
\multicolumn{11}{c} {\bf Case~1.2}\\
%\hline\noalign{\smallskip}
\hline\noalign{\smallskip}
{\bf Code} & \boldmath$\delta\ln c$      &  \boldmath$\delta\ln P$ 
           & \boldmath$\delta\ln \rho$   & \boldmath$\delta\ln T$ & \boldmath$\delta\ln r$ & \boldmath$\delta\ln \Gamma_1$ 
           & \boldmath$\delta\ln \nabla_{\rm ad}$ & \boldmath$\delta\ln C_p$   
           & \boldmath$\delta\ln \kappa$ & \boldmath$\delta X$ & \boldmath$\delta\ln L_r$ \\ [3pt]
\tableheadseprule\noalign{\smallskip}
 \astec  & 0.11 & 0.72 & 0.55 & 0.20 & 0.20 & 1.00$\times 10^{-3}$ & 2.21$\times 10^{-3}$ & 1.25$\times 10^{-2}$ & 0.58 & 2.0$\times 10^{-4}$ & 0.94 \\ 
 \cles  & 0.02 & 0.17 & 0.14 & 0.04 & 0.05 & 2.60$\times 10^{-4}$ & 3.60$\times 10^{-3}$ & 5.67$\times 10^{-3}$ & 0.19 & 9.3$\times 10^{-5}$ & 0.08 \\ 
 \garstec  & 0.06 & 0.23 & 0.23 & 0.04 & 0.07 & 1.55$\times 10^{-1}$ & 1.32$\times 10^{-1}$ & 1.36$\times 10^{-1}$ & 0.39 & 7.8$\times 10^{-5}$ & 0.24 \\ 
 \starox  & 0.02 & 0.09 & 0.09 & 0.05 & 0.03 & 8.63$\times 10^{-4}$ & --- & --- & --- & 3.7$\times 10^{-5}$ & 0.24 \\ 

 \noalign{\smallskip}\hline
%==========================
%==========================
\hline\noalign{\smallskip}
\multicolumn{11}{c} {\bf Case~1.3}\\
\hline\noalign{\smallskip}
{\bf Code} & \boldmath$\delta\ln c$      &  \boldmath$\delta\ln P$ 
           & \boldmath$\delta\ln \rho$   & \boldmath$\delta\ln T$ & \boldmath$\delta\ln r$  & \boldmath$\delta\ln \Gamma_1$ 
           & \boldmath$\delta\ln \nabla_{\rm ad}$ & \boldmath$\delta\ln C_p$   
           & \boldmath$\delta\ln \kappa$ & \boldmath$\delta X$ & \boldmath$\delta\ln L_r$ \\ [3pt]
\tableheadseprule\noalign{\smallskip}
 \astec  & 0.30 & 2.50 & 1.96 & 0.68 & 0.61 & 3.62$\times 10^{-3}$ & 8.16$\times 10^{-3}$ & 2.00$\times 10^{-1}$ & 1.32 & 2.1$\times 10^{-3}$ & 1.45 \\ 
 \cles  & 0.08 & 0.51 & 0.58 & 0.17 & 0.20 & 2.35$\times 10^{-3}$ & 4.88$\times 10^{-3}$ & 2.27$\times 10^{-1}$ & 0.55 & 2.0$\times 10^{-3}$ & 5.04 \\ 
 \garstec  & 0.12 & 0.76 & 0.69 & 0.16 & 0.21 & 1.81$\times 10^{-1}$ & 1.81$\times 10^{-1}$ & 3.01$\times 10^{-1 }$& 1.01 & 1.4$\times 10^{-3}$ & 0.91 \\ 

\noalign{\smallskip}\hline
%==========================
%==========================
\hline\noalign{\smallskip}
\multicolumn{11}{c} {\bf Case~1.4}\\
\hline\noalign{\smallskip}
{\bf Code} & \boldmath$\delta\ln c$      &  \boldmath$\delta\ln P$ 
           & \boldmath$\delta\ln \rho$   & \boldmath$\delta\ln T$ & \boldmath$\delta\ln r$ & \boldmath$\delta\ln \Gamma_1$ 
           & \boldmath$\delta\ln \nabla_{\rm ad}$ & \boldmath$\delta\ln C_p$   
           & \boldmath$\delta\ln \kappa$ & \boldmath$\delta X$ & \boldmath$\delta\ln L_r$ \\ [3pt]
\tableheadseprule\noalign{\smallskip}
 \cles  & 0.03 & 0.25 & 0.22 & 0.07 & 0.09 & 7.40$\times 10^{-4}$ & 4.80$\times 10^{-3}$ & 5.49$\times 10^{-3}$ & 0.27 & 6.9$\times 10^{-5}$ & 1.04 \\ 
 \garstec  & 0.20 & 0.89 & 0.66 & 0.27 & 0.23 & 1.71$\times 10^{-1}$ & 1.71$\times 10^{-1}$ & 2.18$\times 10^{-1}$ & 0.67 & 1.8$\times 10^{-5}$ & 0.62 \\ 
 \starox  & 0.12 & 0.89 & 0.84 & 0.24 & 0.33 & 3.07$\times 10^{-3}$ & --- & --- & --- & 4.3$\times 10^{-5}$ & 3.75 \\ 

\noalign{\smallskip}\hline
%==========================
%==========================
\hline\noalign{\smallskip}
\multicolumn{11}{c} {\bf Case~1.5}\\
\hline\noalign{\smallskip}
{\bf Code} & \boldmath$\delta\ln c$      &  \boldmath$\delta\ln P$ 
           & \boldmath$\delta\ln \rho$   & \boldmath$\delta\ln T$ & \boldmath$\delta\ln r$ & \boldmath$\delta\ln \Gamma_1$ 
           & \boldmath$\delta\ln \nabla_{\rm ad}$ & \boldmath$\delta\ln C_p$   
           & \boldmath$\delta\ln \kappa$ & \boldmath$\delta X$ & \boldmath$\delta\ln L_r$ \\ [3pt]
\tableheadseprule\noalign{\smallskip}
 \astec  & 0.33 & 1.33 & 1.15 & 0.44 & 0.34 & 1.93$\times 10^{-3}$ & 5.12$\times 10^{-3}$ & 5.93$\times 10^{-1}$ & 0.76 & 7.1$\times 10^{-3 }$& 0.99 \\ 
 \cles  & 0.14 & 1.03 & 0.85 & 0.20 & 0.23 & 1.56$\times 10^{-3}$ & 6.35$\times 10^{-3}$ & 2.34$\times 10^{-1}$ & 0.43 & 2.2$\times 10^{-3}$ & 0.85 \\ 
 \starox  & 0.78 & 7.03 & 5.90 & 1.22 & 1.53 & 8.94$\times 10^{-3}$ & --- & --- & --- & 1.0$\times 10^{-2}$ & 1.96 \\ 

 \noalign{\smallskip}\hline
%==========================
%==========================
\hline\noalign{\smallskip}
\multicolumn{11}{c} {\bf Case~1.6}\\
\hline\noalign{\smallskip}
{\bf Code} & \boldmath$\delta\ln c$      &  \boldmath$\delta\ln P$ 
           & \boldmath$\delta\ln \rho$   & \boldmath$\delta\ln T$ & \boldmath$\delta\ln r$ & \boldmath$\delta\ln \Gamma_1$ 
           & \boldmath$\delta\ln \nabla_{\rm ad}$ & \boldmath$\delta\ln C_p$   
           & \boldmath$\delta\ln \kappa$ & \boldmath$\delta X$ & \boldmath$\delta\ln L_r$ \\ [3pt]
\tableheadseprule\noalign{\smallskip}
 \astec  & 0.06 & 0.48 & 0.39 & 0.10 & 0.11 & 6.29$\times 10^{-4}$ & 1.66$\times 10^{-3}$ & 5.53$\times 10^{-2}$ & 0.25 & 7.2$\times 10^{-4}$ & 0.45 \\ 
 \cles  & 0.02 & 0.19 & 0.20 & 0.03 & 0.04 & 1.38$\times 10^{-3}$ & 4.20$\times 10^{-3}$ & 1.81$\times 10^{-2}$ & 0.26 & 2.1$\times 10^{-4}$ & 0.20 \\ 
 \garstec  & 0.23 & 2.58 & 1.95 & 0.60 & 0.64 & 1.96$\times 10^{-1}$ & 2.21$\times 10^{-1}$ & 3.10$\times 10^{-1}$ & 0.92 & 5.9$\times 10^{-4}$ & 1.17 \\ 
 \starox  & 0.04 & 0.25 & 0.19 & 0.07 & 0.08 & 8.51$\times 10^{-4}$ & --- & --- & --- & 1.2$\times 10^{-4}$ & 0.27 \\ 

\noalign{\smallskip}\hline
%==========================
%==========================
\hline\noalign{\smallskip}
\multicolumn{11}{c} {\bf Case~1.7}\\
\hline\noalign{\smallskip}
{\bf Code} & \boldmath$\delta\ln c$      &  \boldmath$\delta\ln P$ 
           & \boldmath$\delta\ln \rho$   & \boldmath$\delta\ln T$ & \boldmath$\delta\ln r$ & \boldmath$\delta\ln \Gamma_1$ 
           & \boldmath$\delta\ln \nabla_{\rm ad}$ & \boldmath$\delta\ln C_p$   
           & \boldmath$\delta\ln \kappa$ & \boldmath$\delta X$ & \boldmath$\delta\ln L_r$ \\ [3pt]
\tableheadseprule\noalign{\smallskip}
 \astec  & 0.27 & 1.77 & 1.53 & 0.26 & 0.38 & 8.85$\times 10^{-3}$ & 2.72$\times 10^{-2}$ & 4.27$\times 10^{-1}$ & 0.43 & 4.9$\times 10^{-3}$ & 1.21 \\ 
 \cles  & 0.05 & 0.16 & 0.19 & 0.03 & 0.04 & 2.79$\times 10^{-3}$ & 8.51$\times 10^{-3}$ & 1.06$\times 10^{-1}$ & 0.15 & 1.2$\times 10^{-3}$ & 0.24 \\ 
 \garstec  & 0.19 & 0.73 & 0.68 & 0.09 & 0.17 & 1.88$\times 10^{-1}$ & 2.62$\times 10^{-1}$ & 6.02$\times 10^{-1}$ & 0.69 & 2.4$\times 10^{-3}$ & 1.18 \\ 
 \starox  & 0.14 & 0.50 & 0.57 & 0.08 & 0.11 & 6.03$\times 10^{-3}$ & --- & --- & --- & 3.3$\times 10^{-3}$ & 0.77 \\ 

\noalign{\smallskip}\hline
%\hline
\end{tabular}
\label{tabla:d1.1}
\end{table*}
%==========================

%==========================
\begin{table*}[htbp!]
\caption{{\task}~1 models: Maximum variations given in per cent (except for $\delta X$) of the physical variables between each code and {\cesam} and value of the relative radius $(r/R)$ where they happen.
%JC-D, 070907%YL, 100907
%The local differences were computed at fixed relative mass.
The local differences were computed both at fixed relative mass and fixed relative radius and the maximum of the two values was searched (see  Sect.\ref{sec:task1.1}).
}
\centering
\label{tab:task1-maxdiff}
\begin{tabular}[h]{lcccccccccccc}
\hline\noalign{\smallskip}
\multicolumn{13}{c} {\bf Case~1.1}\\
\hline\noalign{\smallskip}
{\bf Code} & \boldmath$\delta\ln c$ & \boldmath$r/R$ & \boldmath$\delta\ln P$ & \boldmath$r/R$ 
          & \boldmath$\delta\ln \rho$ & \boldmath$r/R$ & \boldmath$\delta\ln \Gamma_1$ & \boldmath$r/R$          
          & \boldmath$\delta X$ & \boldmath$r/R$ 
          & \boldmath$\delta\ln L_r$ & \boldmath$r/R$\\[3pt]
\tableheadseprule\noalign{\smallskip}
 \astec  & 0.08 & 0.97540 & 1.03 & 0.98493 & 0.91 & 0.99114 & 0.14 & 0.99986 &  0.00082 & 0.10762 & 0.78 & 0.02275 \\ 
 \starox  & 0.16 & 0.69724 & 1.45 & 0.98064 & 1.31 & 0.98964 & 0.12 & 0.99984 & 0.00148 & 0.10886 & 0.55 & 0.07129 \\ 
 \garstec  & 0.28 & 0.36880 & 3.12 & 0.92362 & 2.72 & 0.94568 & 0.55 & 0.99987 & 0.00232 & 0.11734 & 3.60 & 0.00319 \\ 
 \cles  & 0.14 & 0.69663 & 0.97 & 0.79942 & 0.89 & 0.78455 & 0.09 & 0.99987 & 0.00116 & 0.11140 & 0.35 & 0.00441 \\ 
\noalign{\smallskip}\hline
%==========================
%==========================
\hline\noalign{\smallskip}
\multicolumn{13}{c} {\bf Case~1.2}\\
\hline\noalign{\smallskip}
{\bf Code} & \boldmath$\delta\ln c$ & \boldmath$r/R$ & \boldmath$\delta\ln P$ & \boldmath$r/R$ 
          & \boldmath$\delta\ln \rho$ & \boldmath$r/R$ & \boldmath$\delta\ln \Gamma_1$ & \boldmath$r/R$          
          & \boldmath$\delta X$ & \boldmath$r/R$ 
          & \boldmath$\delta\ln L_r$ & \boldmath$r/R$\\[3pt]
\tableheadseprule\noalign{\smallskip}
 \astec  & 0.61 & 0.83067 & 4.50 & 0.84074 & 4.19 & 0.86702 & 0.33 & 0.99167 & 0.00093 & 0.05118 & 9.76 & 0.00185 \\ 
 \starox  & 0.34 & 0.82990 & 2.52 & 0.85099 & 2.34 & 0.87925 & 0.18 & 0.99223 & 0.00095 & 0.04868 & 0.93 & 0.04651 \\ 
 \garstec  & 0.26 & 0.77098 & 2.22 & 0.86256 & 1.97 & 0.91370 & 0.43 & 0.98926 & 0.00102 & 0.05360 & 2.26 & 0.00236 \\ 
 \cles  & 0.08 & 0.99612 & 1.04 & 0.55596 & 0.94 & 0.61552 & 0.08 & 0.99990  & 0.00045 & 0.05124 & 1.20 & 0.00455 \\ 
\noalign{\smallskip}\hline
%==========================
%==========================
\hline\noalign{\smallskip}
\multicolumn{13}{c} {\bf Case~1.3}\\
\hline\noalign{\smallskip}
{\bf Code} & \boldmath$\delta\ln c$ & \boldmath$r/R$ & \boldmath$\delta\ln P$ & \boldmath$r/R$ 
          & \boldmath$\delta\ln \rho$ & \boldmath$r/R$ & \boldmath$\delta\ln \Gamma_1$ & \boldmath$r/R$          
          & \boldmath$\delta X$ & \boldmath$r/R$ 
          & \boldmath$\delta\ln L_r$ & \boldmath$r/R$\\[3pt]
\tableheadseprule\noalign{\smallskip}
 \astec  & 1.23 & 0.78937 & 4.86 & 0.78160 & 4.86 & 0.80785 & 0.73 & 0.98620 & 0.00740 & 0.04161 & 189.80 & 0.00042 \\ 
 \garstec  & 0.24 & 0.02940 & 1.63 & 0.00031 & 1.68 & 0.02892 & 0.50 & 0.99990  & 0.01018 & 0.02913 & 12.21 & 0.00091 \\ 
 \cles  & 0.49 & 0.02964 & 2.91 & 0.94624 & 2.62 & 0.97962 & 0.18 & 0.99989 & 0.00928 & 0.03105 & 25.42 & 0.00284 \\ 
\noalign{\smallskip}\hline
%==========================
%==========================
\hline\noalign{\smallskip}
\multicolumn{13}{c} {\bf Case~1.4}\\
\hline\noalign{\smallskip}
{\bf Code} & \boldmath$\delta\ln c$ & \boldmath$r/R$ & \boldmath$\delta\ln P$ & \boldmath$r/R$ 
          & \boldmath$\delta\ln \rho$ & \boldmath$r/R$ & \boldmath$\delta\ln \Gamma_1$ & \boldmath$r/R$          
          & \boldmath$\delta X$ & \boldmath$r/R$ 
          & \boldmath$\delta\ln L_r$ & \boldmath$r/R$\\[3pt]
\tableheadseprule\noalign{\smallskip}
 \starox  & 0.82 & 0.99902 & 5.04 & 0.99974 & 5.82 & 0.99985 & 1.05 & 0.99974 & 0.00034 & 0.13025 & 12.16 & 0.01056 \\ 
 \garstec  & 1.03 & 0.99988 & 2.68 & 0.99986 & 4.37 & 0.99989 & 0.45 & 0.99987 & 0.00024 & 0.13009 & 2.04 & 0.00158 \\ 
 \cles  & 0.24 & 0.99988 & 0.98 & 0.38180 & 1.20 & 0.99989 & 0.22 & 0.99834 & 0.00032 & 0.13011 & 3.31 & 0.01191 \\ 
\noalign{\smallskip}\hline
%==========================
%==========================
\hline\noalign{\smallskip}
\multicolumn{13}{c} {\bf Case~1.5}\\
\hline\noalign{\smallskip}
{\bf Code} & \boldmath$\delta\ln c$ & \boldmath$r/R$ & \boldmath$\delta\ln P$ & \boldmath$r/R$ 
          & \boldmath$\delta\ln \rho$ & \boldmath$r/R$ & \boldmath$\delta\ln \Gamma_1$ & \boldmath$r/R$          
          & \boldmath$\delta X$ & \boldmath$r/R$ 
          & \boldmath$\delta\ln L_r$ & \boldmath$r/R$\\[3pt]
\tableheadseprule\noalign{\smallskip}
 \astec  & 3.18 & 0.99689 & 7.17 & 0.98780 & 9.13 & 0.99602 & 3.13 & 0.99705 & 0.07693 & 0.06230 & 4.09 & 0.00382 \\ 
 \starox  & 10.02 & 0.99677 & 15.17 & 0.23988 & 21.26 & 0.99580 & 9.41 & 0.99687 & 0.05763 & 0.06285 & 11.55 & 0.00070 \\ 
 \cles  & 0.51 & 0.06442 & 2.29 & 0.83852 & 2.40 & 0.06442 & 0.19 & 0.98734 & 0.01582 & 0.06442 & 1.99 & 0.00584 \\ 
\noalign{\smallskip}\hline
%==========================
%==========================
\hline\noalign{\smallskip}
\multicolumn{13}{c} {\bf Case~1.6}\\
\hline\noalign{\smallskip}
{\bf Code} & \boldmath$\delta\ln c$ & \boldmath$r/R$ & \boldmath$\delta\ln P$ & \boldmath$r/R$ 
          & \boldmath$\delta\ln \rho$ & \boldmath$r/R$ & \boldmath$\delta\ln \Gamma_1$ & \boldmath$r/R$          
          & \boldmath$\delta X$ & \boldmath$r/R$ 
          & \boldmath$\delta\ln L_r$ & \boldmath$r/R$\\[3pt]
\tableheadseprule\noalign{\smallskip}
 \astec  & 0.53 & 0.16440 & 0.99 & 0.41038 & 0.83 & 0.40804 & 0.18 & 0.99567 & 0.01189 & 0.16416 & 1.41 & 0.09659 \\ 
 \starox  & 0.35 & 0.99977 & 0.52 & 0.99987 & 0.47 & 0.99987 & 0.57 & 0.99977 & 0.00408 & 0.16359 & 0.93 & 0.00230 \\ 
 \garstec  & 0.68 & 0.99407 & 2.53 & 0.00171 & 2.89 & 0.16320 & 0.65 & 0.99846 & 0.01387 & 0.16320 & 5.89 & 0.00192 \\ 
 \cles  & 0.21 & 0.99910 & 0.93 & 0.40301 & 0.85 & 0.43433 & 0.24 & 0.99969 & 0.00684 & 0.16392 & 1.10 & 0.00476 \\ 
\noalign{\smallskip}\hline
%==========================
%==========================
\hline\noalign{\smallskip}
\multicolumn{13}{c} {\bf Case~1.7}\\
\hline\noalign{\smallskip}
{\bf Code} & \boldmath$\delta\ln c$ & \boldmath$r/R$ & \boldmath$\delta\ln P$ & \boldmath$r/R$ 
          & \boldmath$\delta\ln \rho$ & \boldmath$r/R$ & \boldmath$\delta\ln \Gamma_1$ & \boldmath$r/R$          
          & \boldmath$\delta X$ & \boldmath$r/R$ 
          & \boldmath$\delta\ln L_r$ & \boldmath$r/R$\\[3pt]
\tableheadseprule\noalign{\smallskip}
\astec  & 0.90 & 0.12212 & 4.52 & 0.82037 & 3.73 & 0.85521 & 0.63 & 0.99536 & 0.01909 & 0.12436 & 6.71 & 0.00184 \\ 
\starox  & 0.55 & 0.11242 & 2.29 & 0.85773 & 1.95 & 0.87951 & 0.61 & 0.99948 & 0.01517 & 0.13663 & 1.55 & 0.00170 \\ 
\garstec  & 0.69 & 0.13697 & 2.38 & 0.83987 & 2.65 & 0.13598 & 0.45 & 0.99849 & 0.02655 & 0.13647 & 6.50 & 0.00148 \\ 
\cles  & 0.44 & 0.99317 & 1.52 & 0.59622 & 1.40 & 0.62262 & 0.47 & 0.99309 & 0.00580 & 0.13739 & 0.65 & 0.00962 \\ 
\noalign{\smallskip}\hline
%\hline
\end{tabular}
\label{tabla:v1.7}
\end{table*}
%==========================

%\clearpage
%==========================
% % Figures withe differences in the interior computed at constant M, and plotted
% versus  relative radius, 
%==========================
\begin{figure*}[htbp!]
\centering
\resizebox{\hsize}{!}{\includegraphics{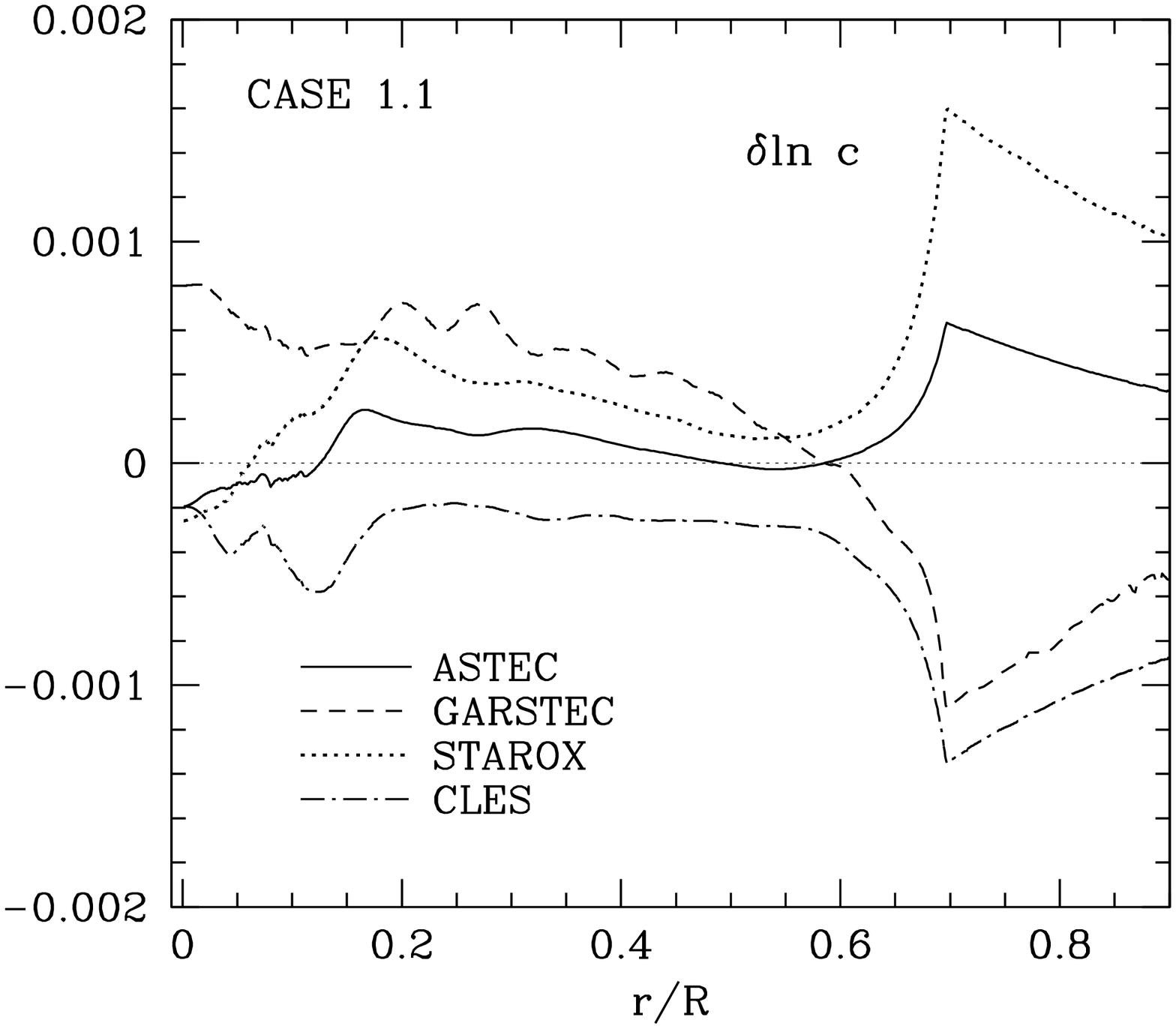}\includegraphics{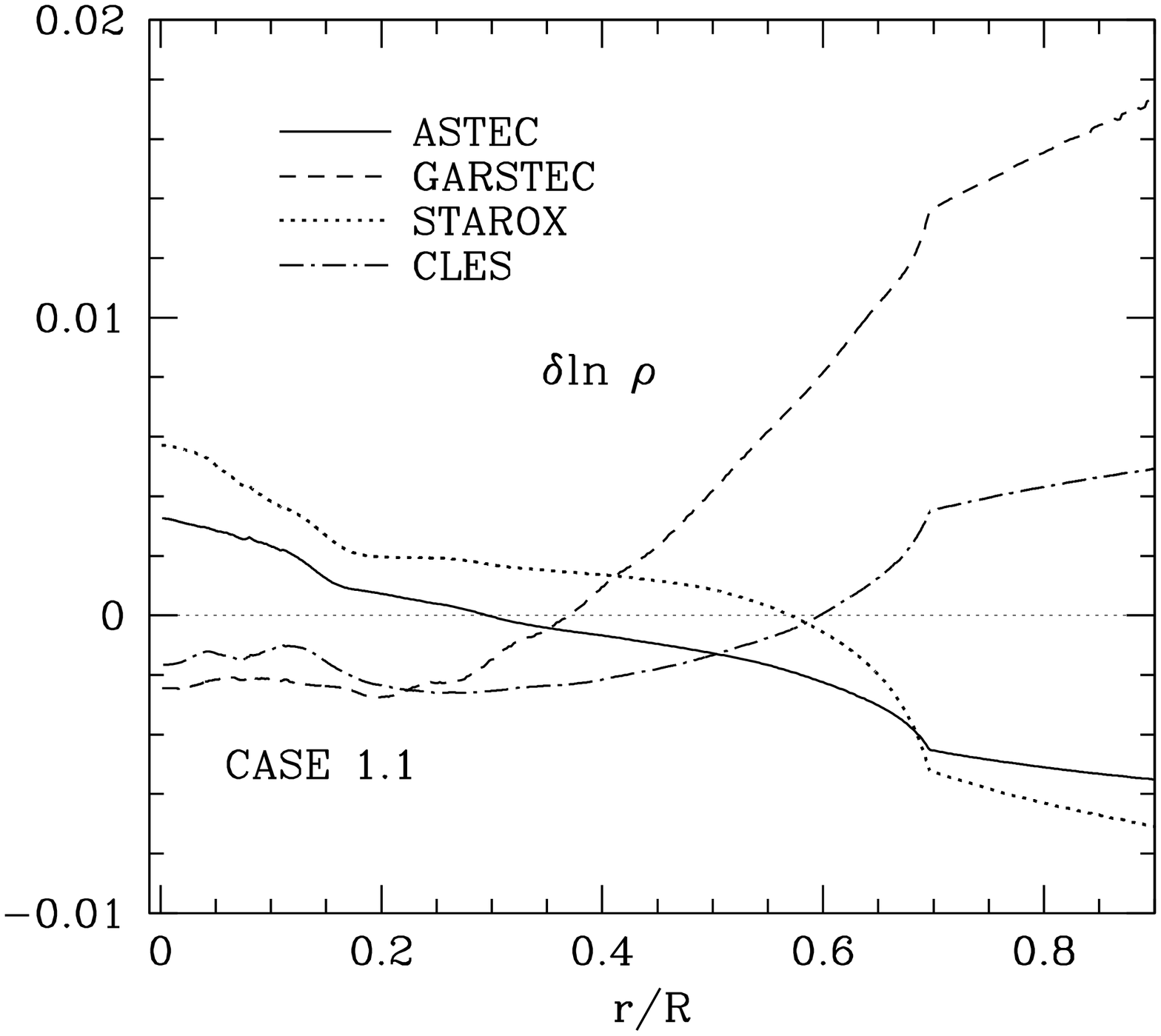}
\includegraphics{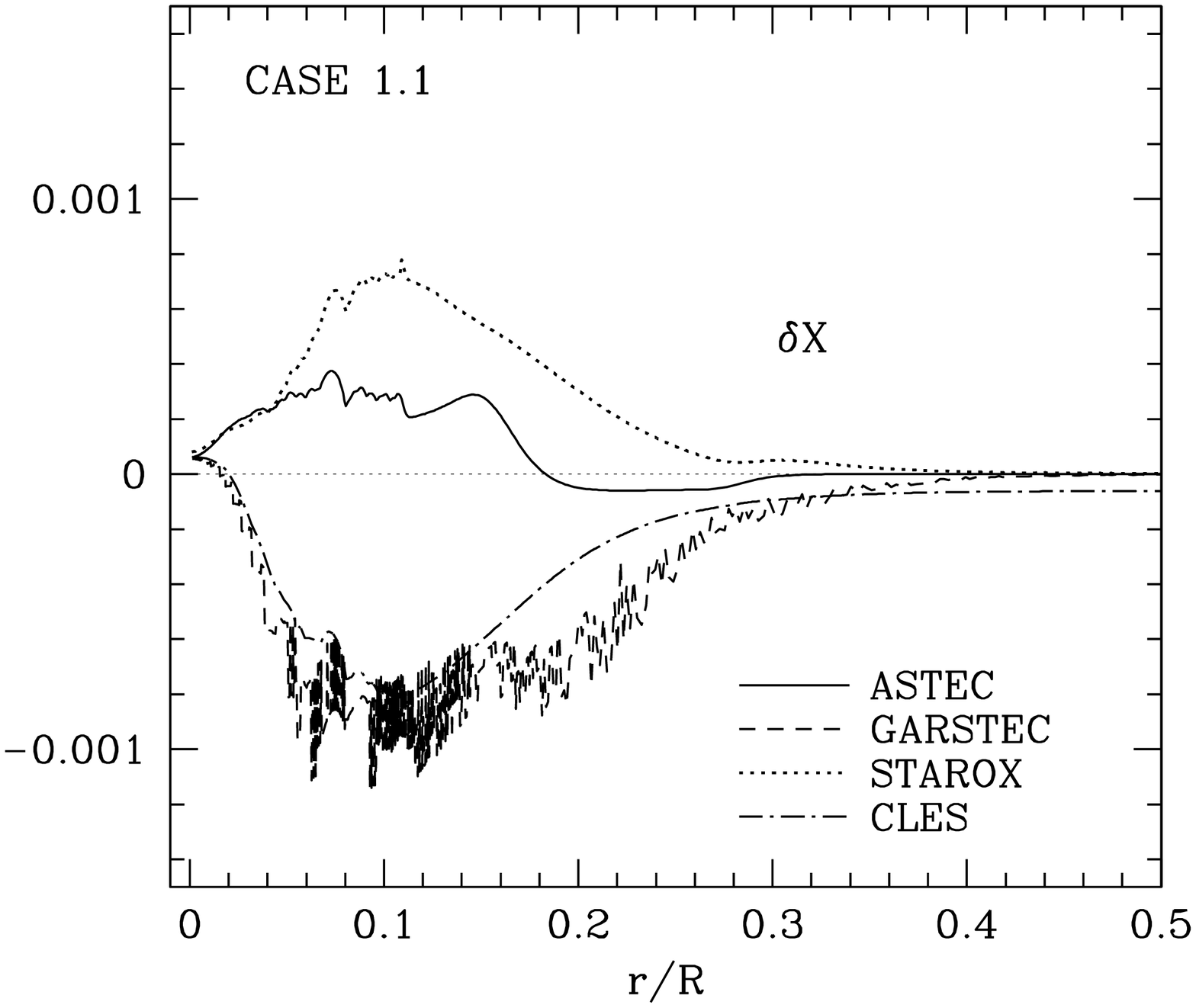}\includegraphics{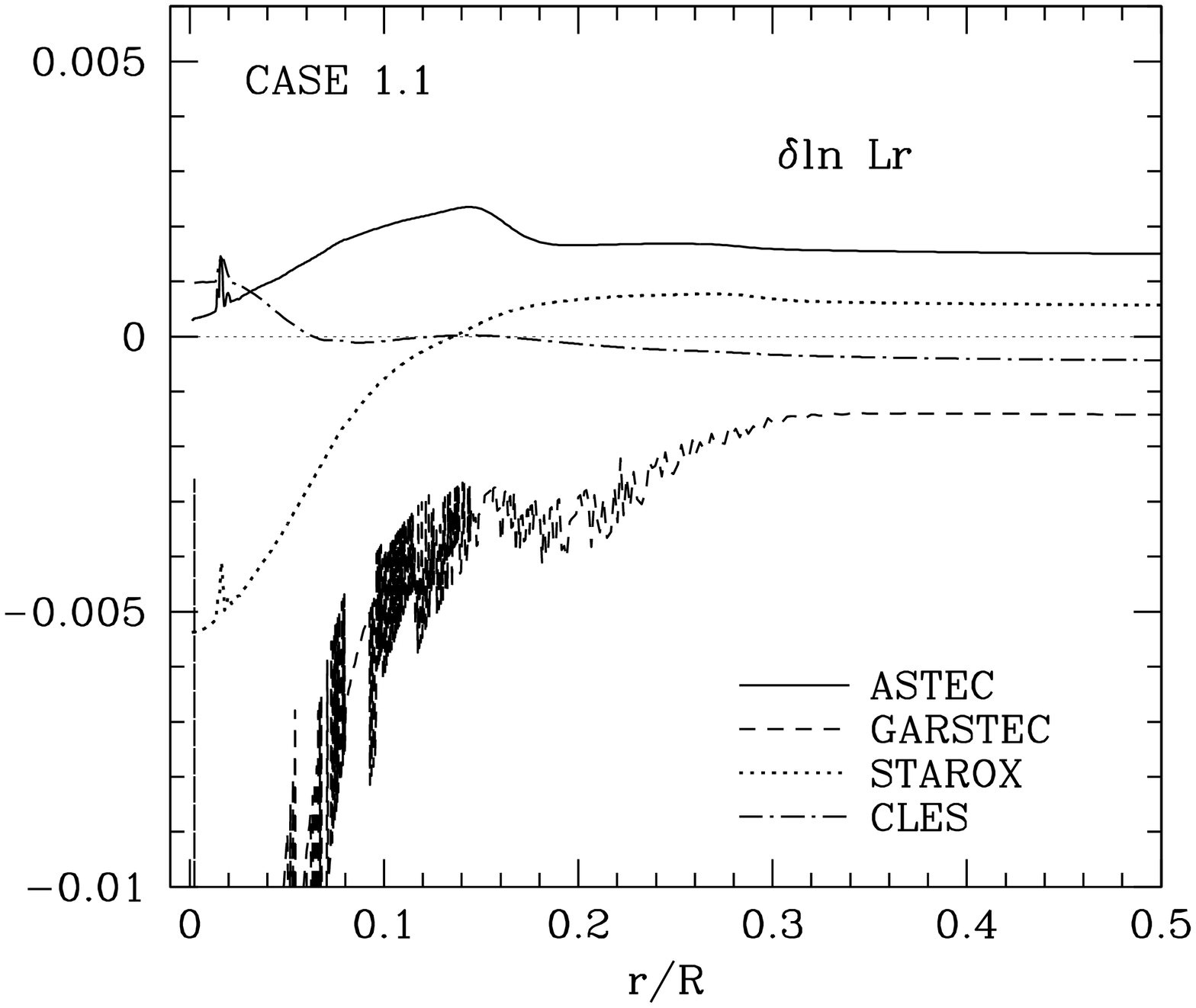}}
\resizebox{\hsize}{!}{\includegraphics{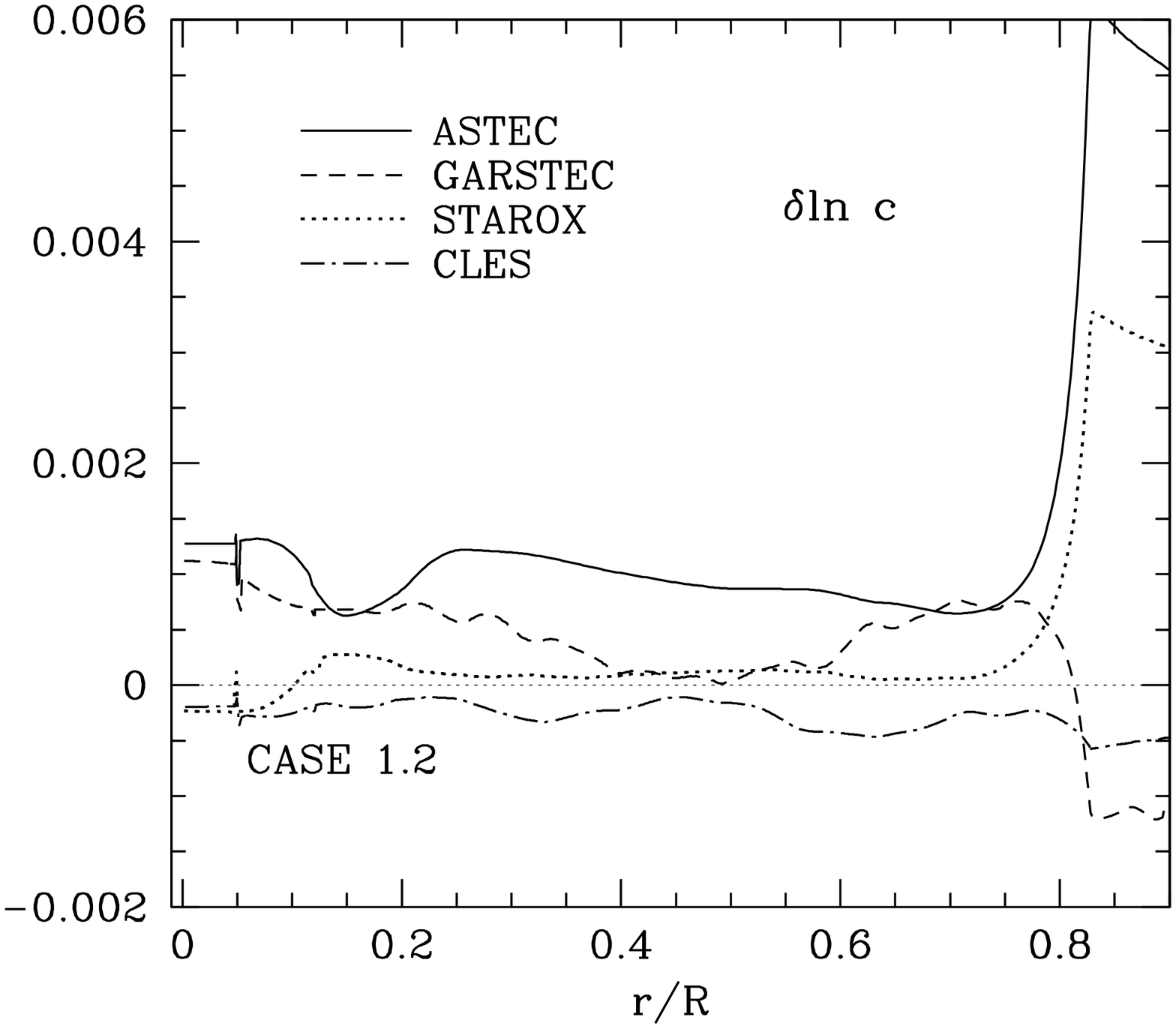}\includegraphics{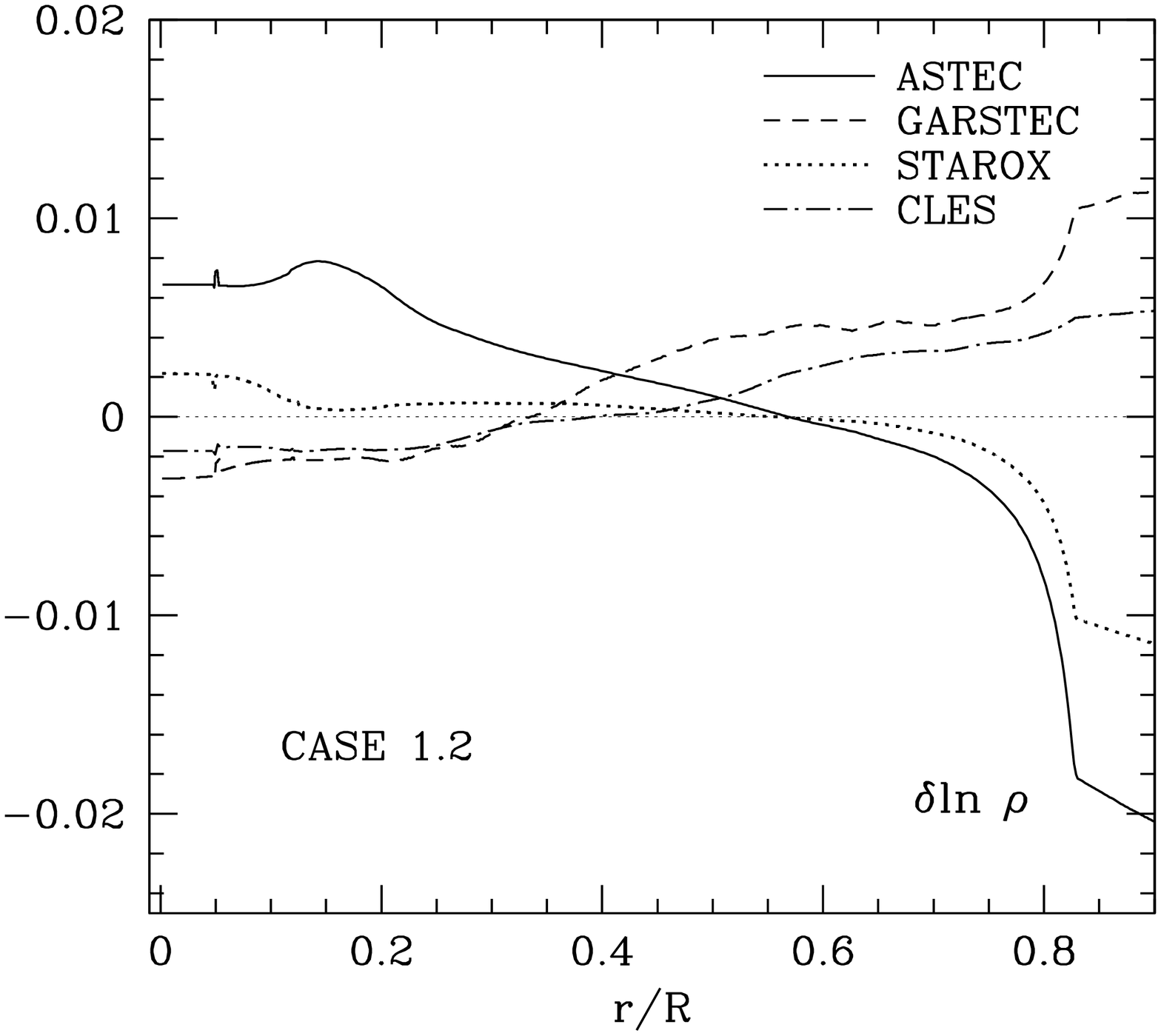}
\includegraphics{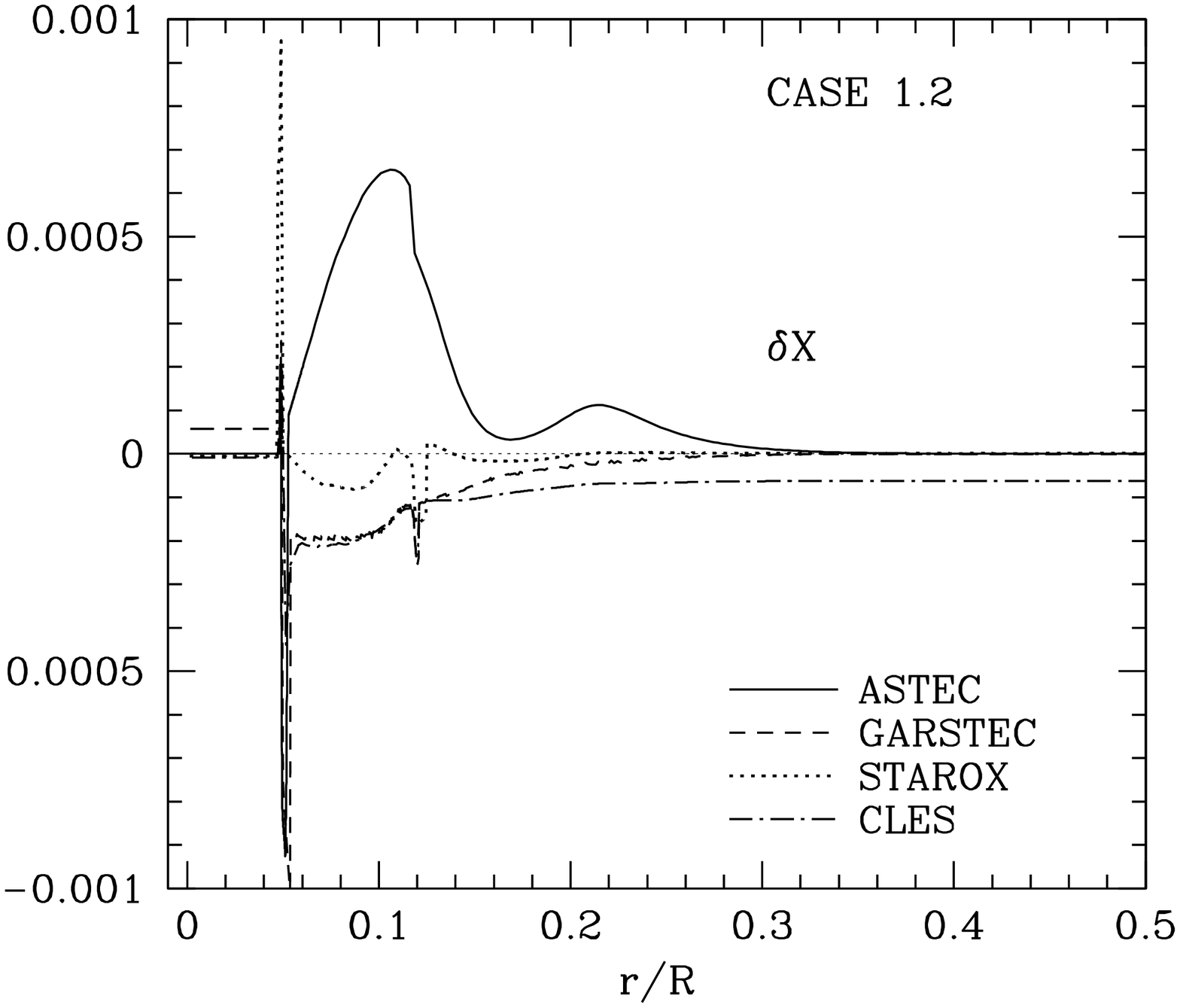}\includegraphics{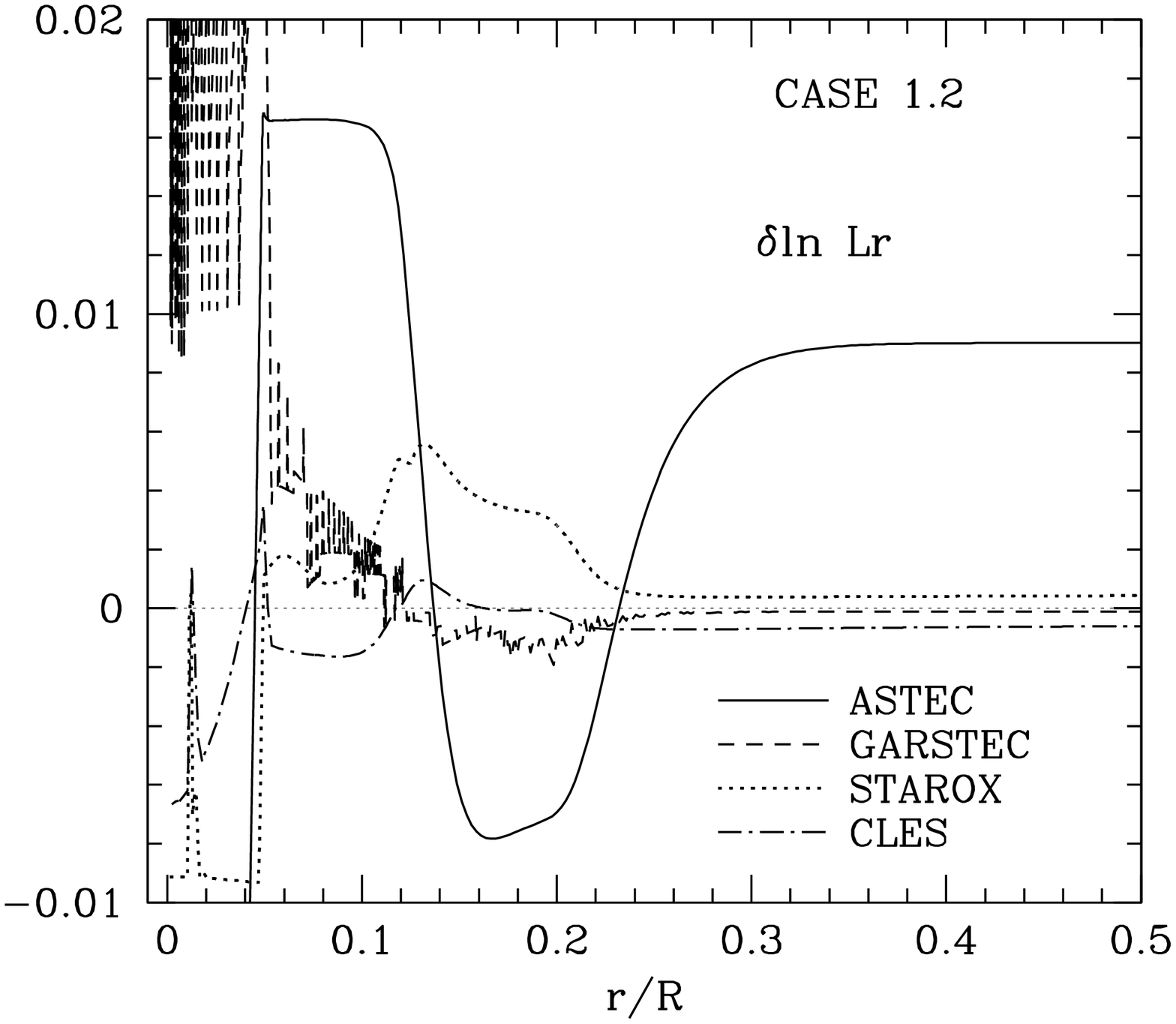}}
\resizebox{\hsize}{!}{\includegraphics{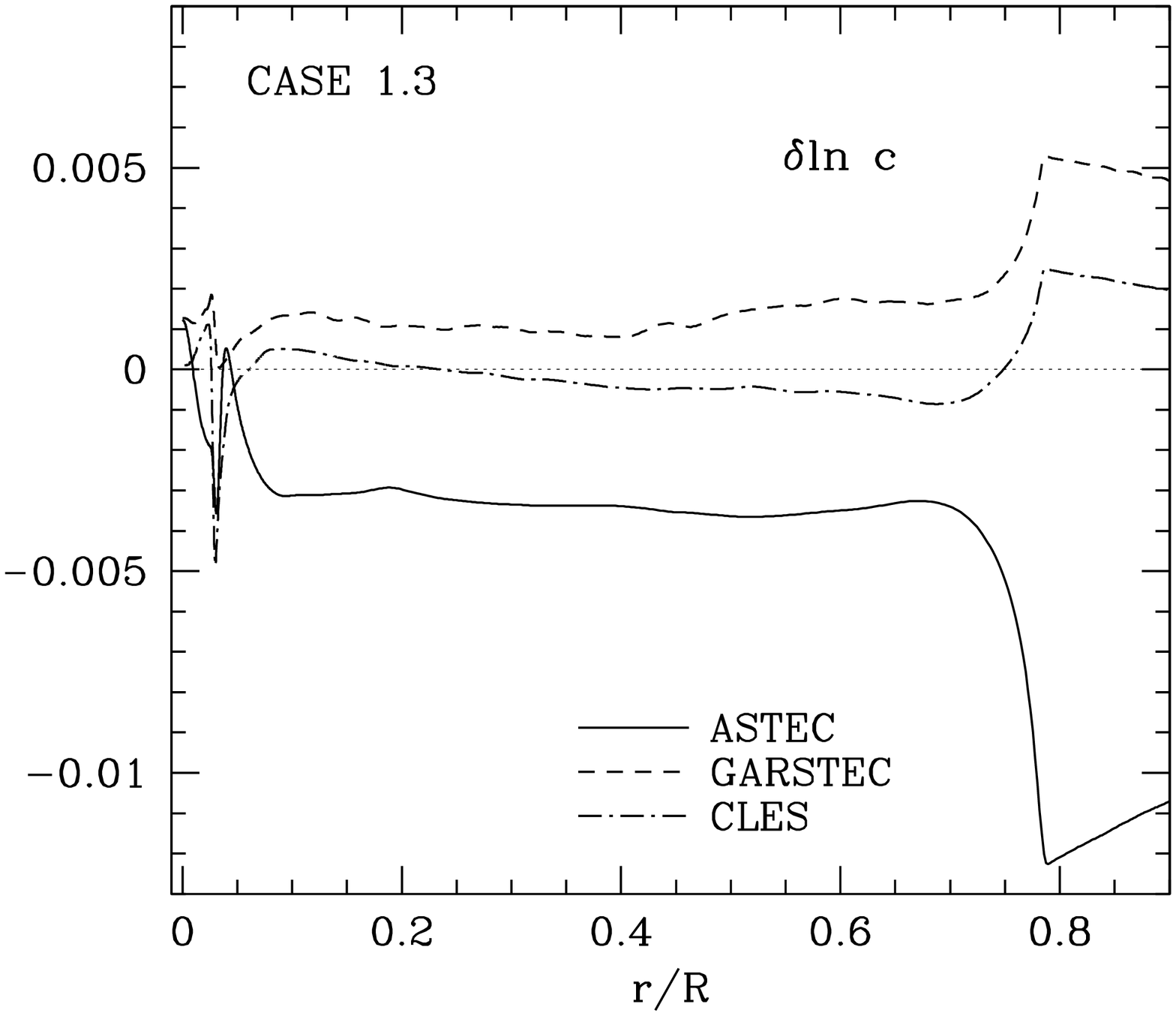}\includegraphics{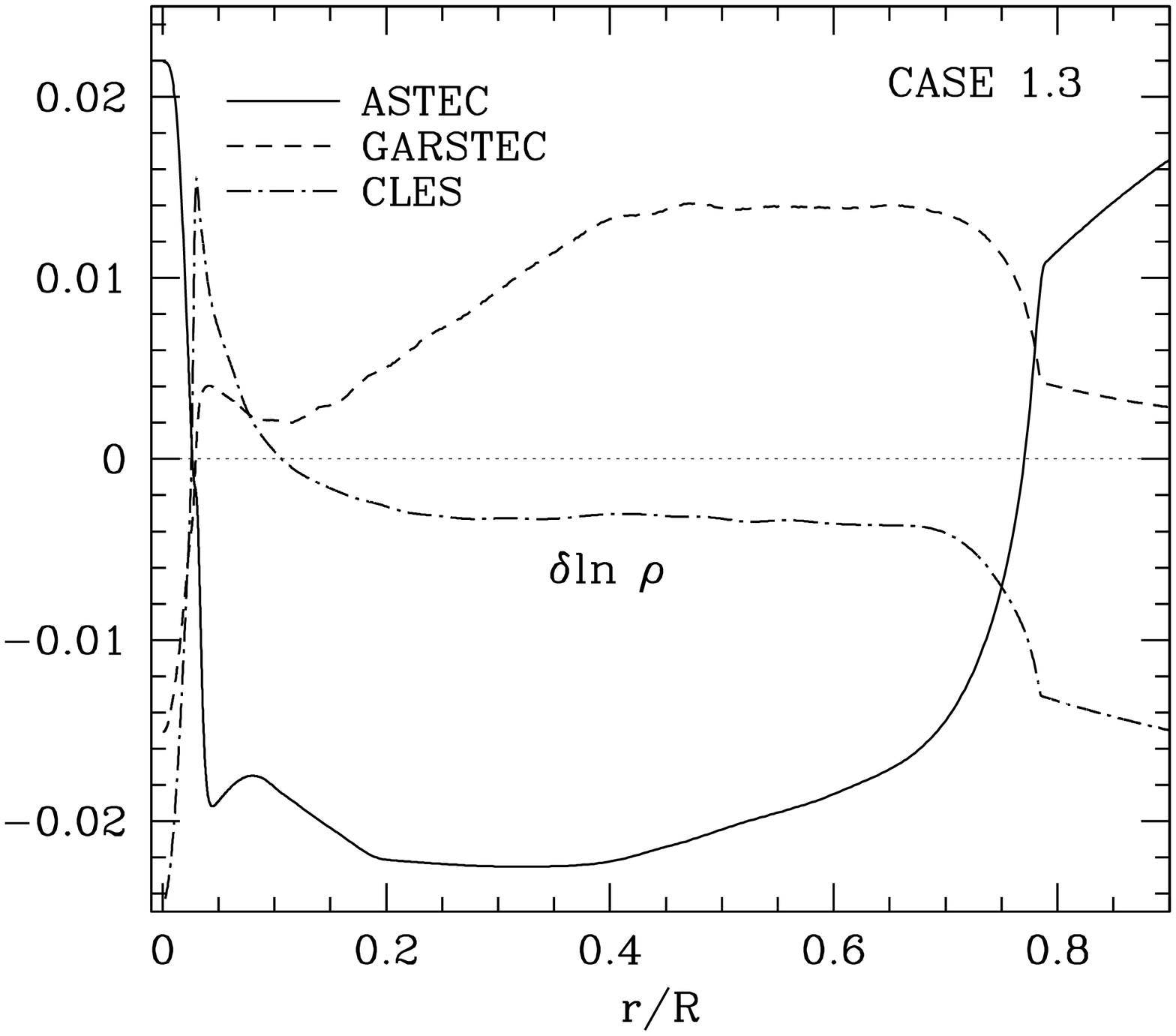}
\includegraphics{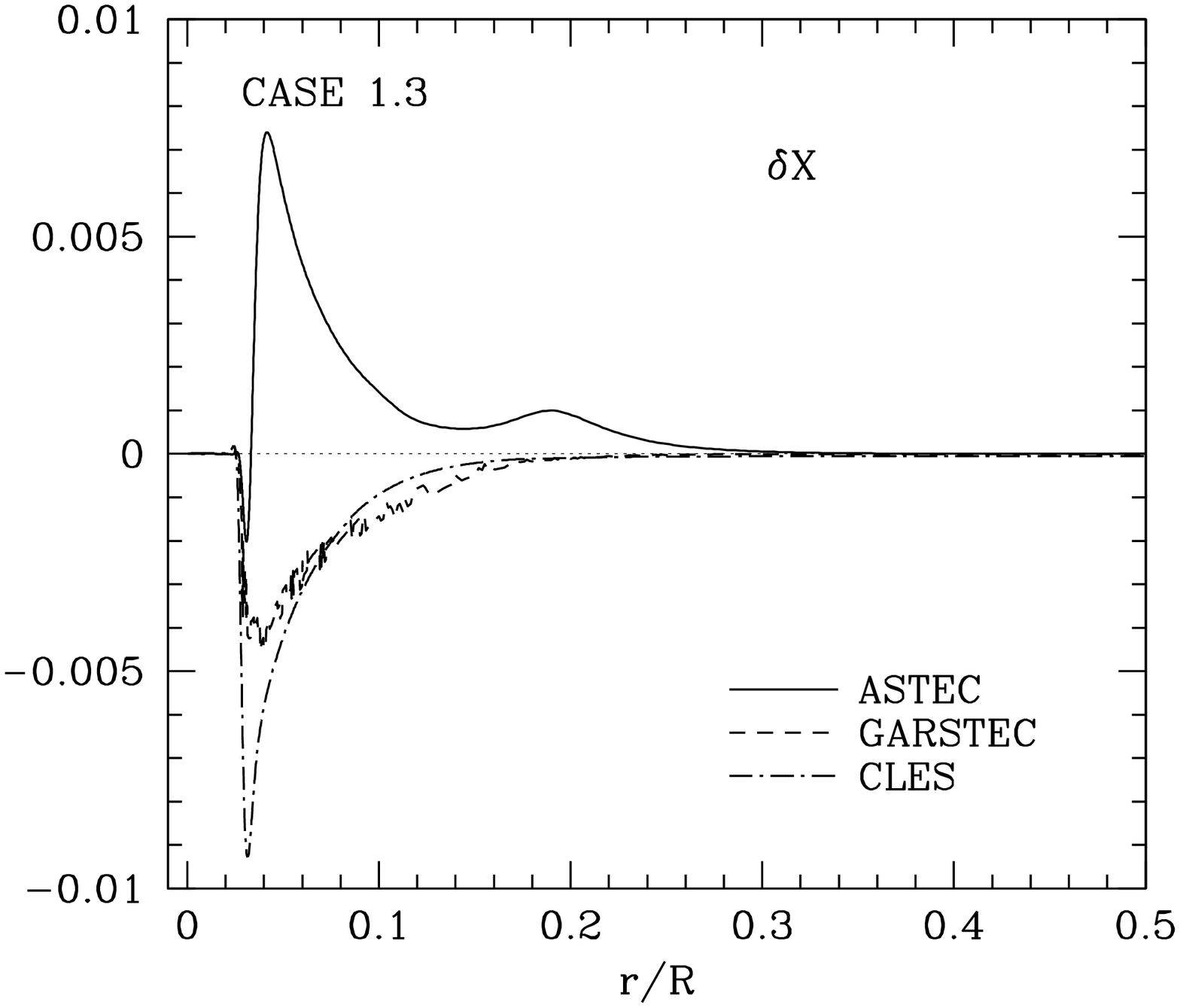}\includegraphics{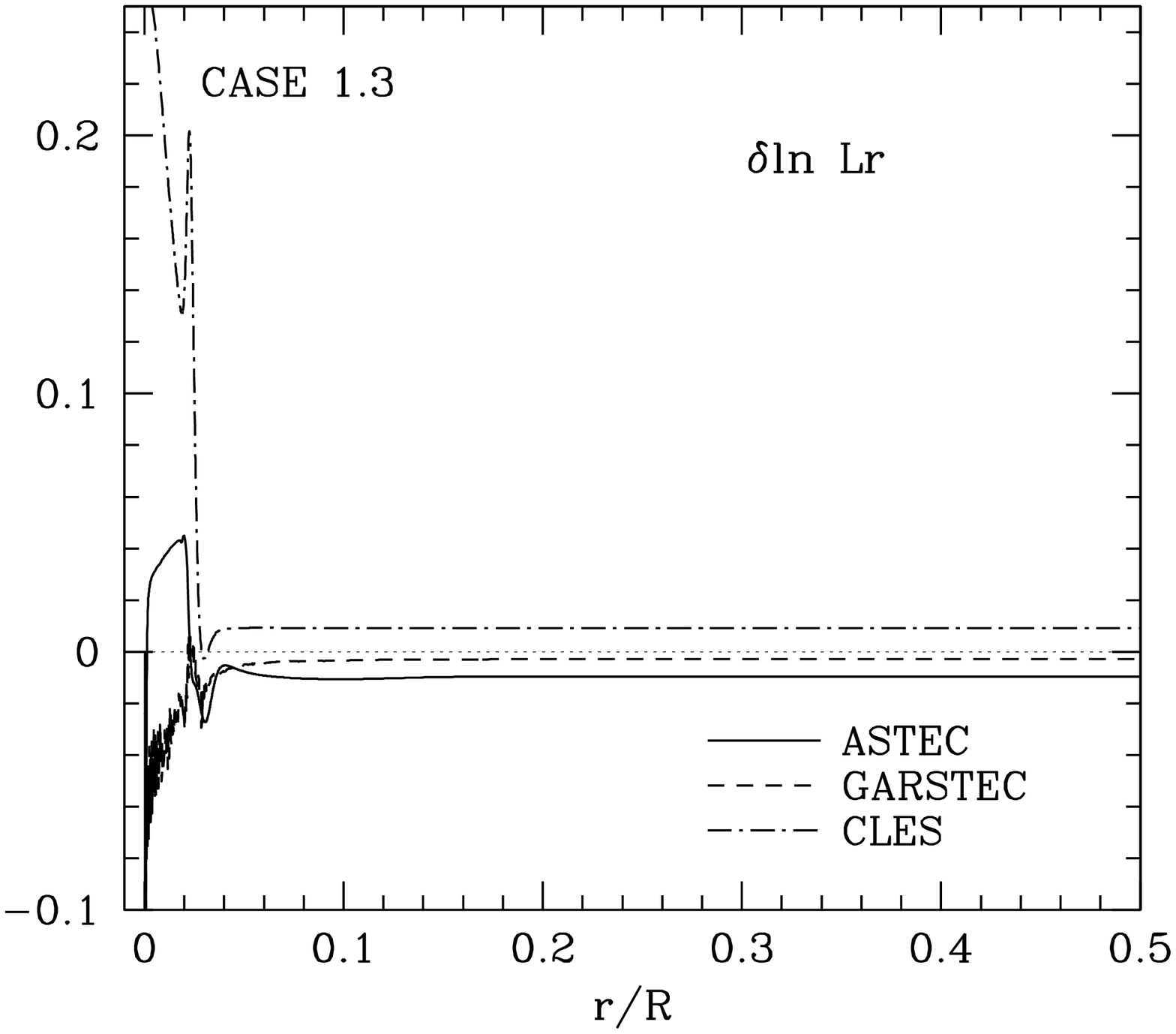}}
\caption{{\task}~1: Plots of the differences at fixed relative mass between pairs of models 
({\small CODE}-\cesam) corresponding to Cases~1.1, 1.2 and 1.3. {\it Left panel}: logarithmic sound speed differences.
{\it Centre left panel}: logarithmic density differences. {\it Centre right panel}: hydrogen mass fraction differences. {\it Right panel}: logarithmic luminosity differences.
Horizontal dotted line represents the reference model (\cesam).} 
\label{fig:task1-int123}
\end{figure*}

Table~\ref{tab:task1-glob} gives a brief summary of the differences in the global parameters of the models by providing the  mean and maximum differences in radius, luminosity and effective temperature obtained by each code with respect to {\cesam} models. The mean difference is obtained by averaging over all the cases calculated (not all cases have been calculated by each code). The differences are very small, i.e. below $0.5$ per cent for {\cles} and {\garstec}. They are a bit larger for two {\astec} models ($1$--$2$\% for Cases~1.2 and 1.7) and for two {\starox} models ($1$--$3$\% for Cases~1.2 and 1.5, but note that for the latter overshooting is treated differently than in other codes as explained in Sect.~\ref{sec:C14C15}). For a detailed discussion, see \cite{mm2-apss}.

For each model we have computed  the local differences in the physical variables with respect to the corresponding model built by \cesam. The physical variables we have considered are the following:
\begin{enumerate}
\item $P$: pressure
\item $\rho$: density
\item $L_{r}$: luminosity through the sphere with radius $r$
\item $X$: hydrogen mass fraction
\item $c$: sound speed
\item $\Gamma_1$: adiabatic exponent
\item $C_p$: specific heat at constant pressure
\item $\nabla_{\rm ad}$: adiabatic temperature gradient
\item $\kappa$: radiative opacity
\item $A=\frac{1}{\Gamma_1}\frac{\rm d \ln P}{\rm d \ln r}-\frac{\rm d \ln \rho}{\rm d \ln r}=N^2_{\rm BV} r/g$,
where $N_{\rm BV}$ is the Brunt-V\"ais\"al\"a frequency and $g$ the local gravity.
\end{enumerate}

To compute the differences we used the grid of a given model (either mass grid or radius grid, see below) and we interpolated the physical variables of the {\cesam} model on that grid. We used the so-called {\em diff-fgong.d} routine in the {\adipls} package. We performed both interpolations at fixed relative radius ($r/R$) and at fixed relative mass ($q=m/M$). In both cases, we have computed  the local logarithmic differences ($\delta \ln Q$) of each physical quantity $Q$ with respect to that of the corresponding {\cesam} model (except for X where we computed $\delta X$). The interpolation is cubic (either in $r/R$ or $q=m/M$) except for the innermost points where it is linear, either in $(r/R)^2$ or $(m/M)^{2/3}$, in order to improve the accuracy of the interpolation of $L$ and $m/M$.

Since not all the codes provide the atmosphere structure, we have calculated the differences inside the star up to the photospheric radius ($R$).
To provide an estimate of these differences  we have defined a kind of
 ``mean-quadratic error'':
\begin{equation}
\delta x =\left(\int _0^M (x_{\mbox{\scriptsize CODE}}-x_{\mbox{\scriptsize CESAM}})^2 \cdot 
\frac{{\rm d}m}{M}\right)^{1/2}
\label{eq:quad_mean}
\end{equation}
\noindent 

where the differences $x_{\mbox{\scriptsize CODE}}-x_{\mbox{\scriptsize CESAM}}$ are calculated at fixed mass.
The values of variations resulting from this computation are collected in Table~\ref{tabla:d1.1}. 
%We note that the ``mean-quadratic differences'' between the codes generally remain quite low except for a few particular cases: they are generally well below half a per cent but may sometimes reach 2 or 3 per cent for some variables ($P$, $\rho$, $L_r$ and $\kappa$). 
We note that the ``mean-quadratic differences'' between the codes generally remain quite low except for a few particular cases. For the unknowns of the stellar structure equations $P$, $T$, $L_r$ and for $r$, $\rho$ and $\kappa$, the differences range from $0.1$ to at most $7\%$. Concerning the variation in the thermodynamic quantities we note that while  the values of   $\delta \Gamma_1$, $\delta \nabla_{\rm ad}$, $\delta C_p$  for three of the codes are quite small, the differences are systematically larger than 0.1\% in the \garstec\ code. Some differences in the thermodynamic quantities might indeed be expected since each code has its own use of the {\small OPAL} equation of state package and variables \citep{2002ApJ...576.1064R}, see the discussion concerning {\cles} and {\cesam} in \citet{jm-apss}.

\begin{figure*}[htbp!]
\centering
\resizebox{0.75\hsize}{!}{\includegraphics{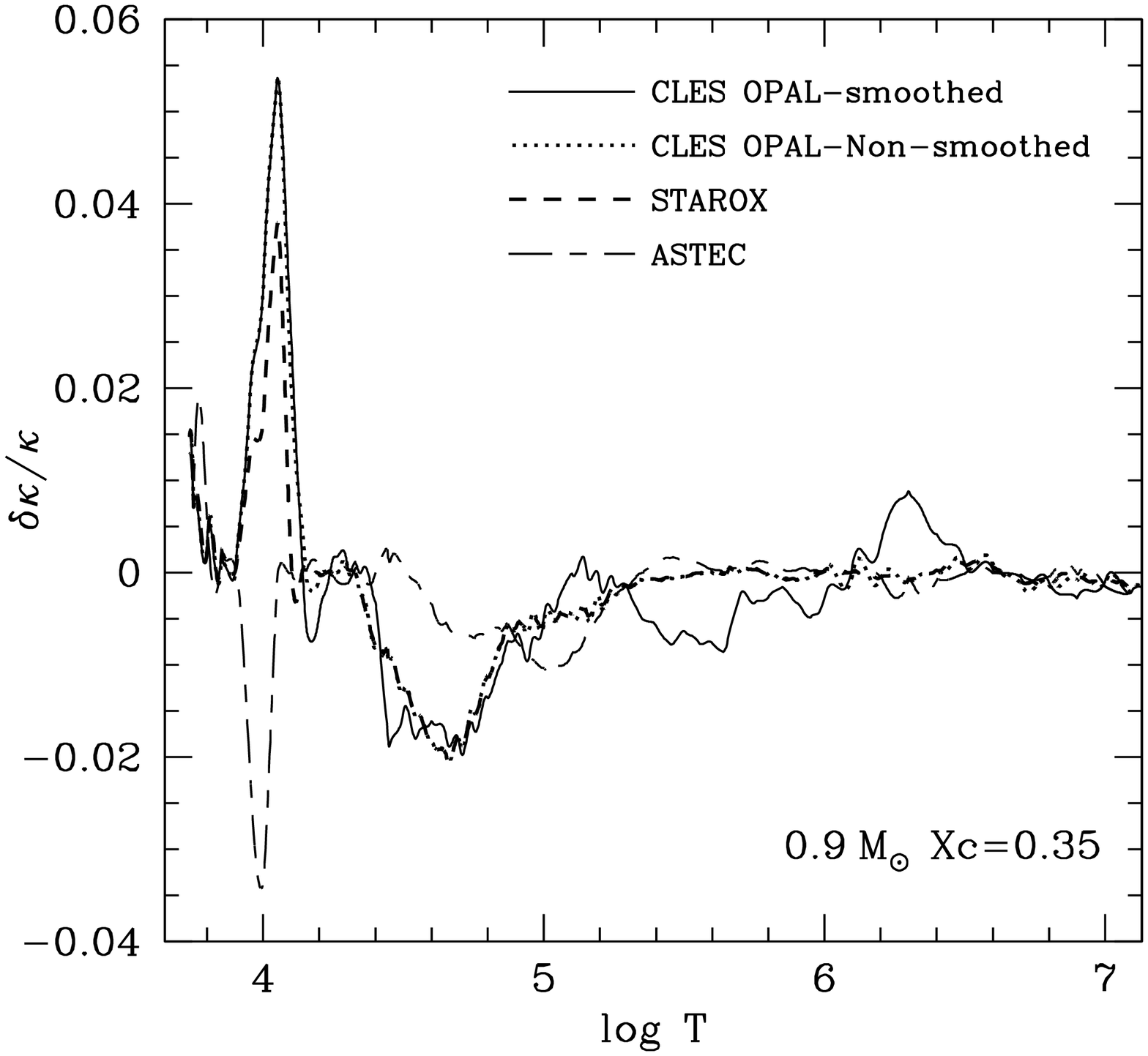}\includegraphics{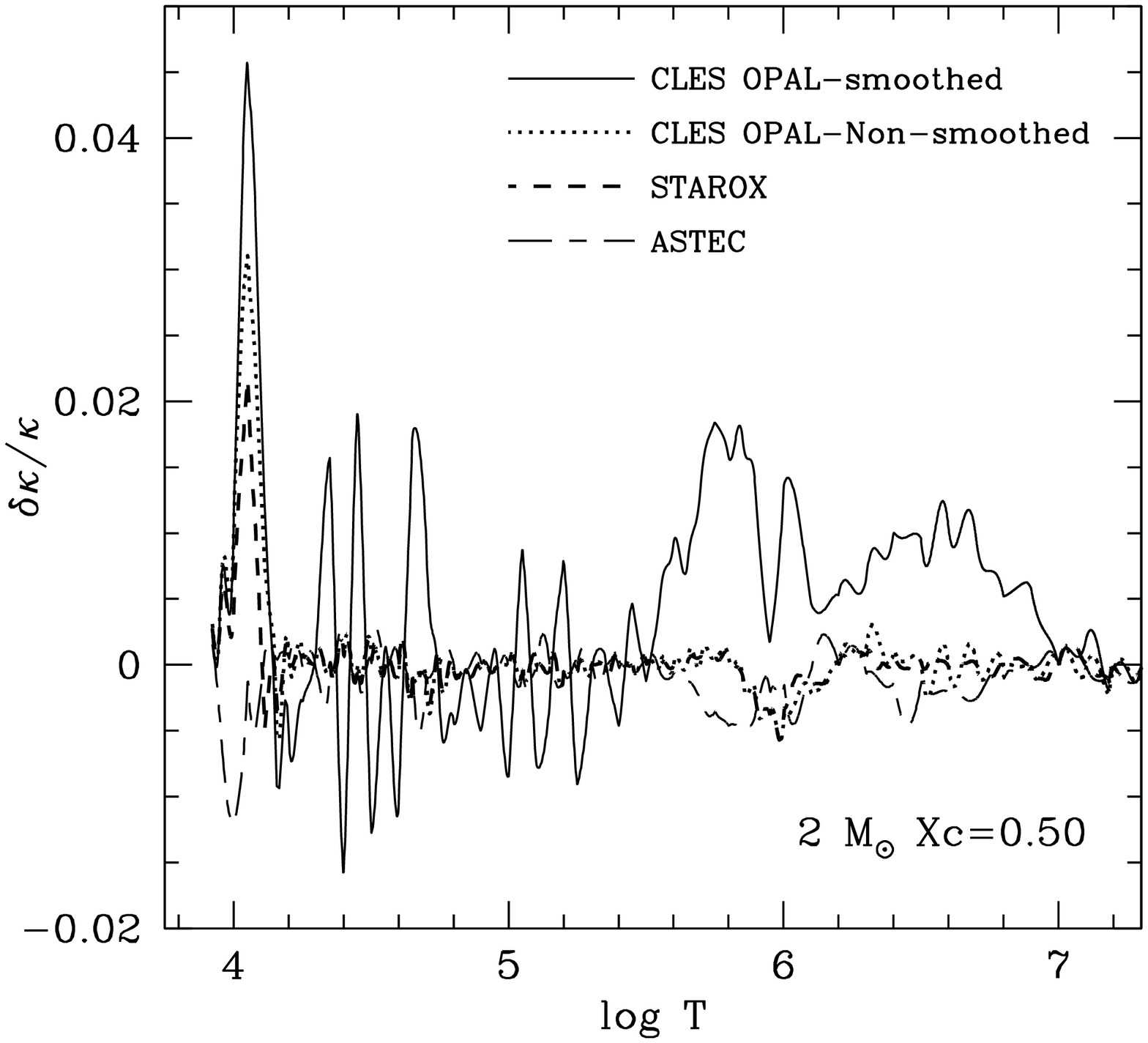}}
%\resizebox{\hsize}{!}{\includegraphics{diff_opal1.1.eps} \includegraphics{diff_opal1.5.eps}}
\caption{{\task}~1: Comparisons of opacities calculated by each code with respect to {\cesam} for fixed physical conditions corresponding to a model of 0.9{\msol} and $X_c=0.35$ (top panel) and a model of  2.0{\msol}, $X_c=0.50$ (bottom panel).}      
\label{fig:opa}
\end{figure*}

Similarly, we expect some differences in the opacities derived by the codes even though all codes use the {\small OPAL95} opacities \citep{ir96} and the {\small AF94} opacities \citep{af94} at low temperature. In Fig.~\ref{fig:opa} we provide the differences, with respect to {\cesam}, of the  opacities calculated by {\astec}, {\cles} and {\starox} for two  ($\rho, T, X, Z$) profiles extracted from {\cesam} models. The larger differences are in the range 2-6 \% and occur in a narrow zone around $\log T=4.0$. With {\garstec} differences are of the same order of magnitude. Those differences correspond to the joining of {\small OPAL95} and {\small AF94} opacity tables. Each code has its own method to merge the tables: {\cles}, {\garstec} and {\starox} interpolate between {\small OPAL95} and {\small AF94} values of $\log\kappa$ on a few temperature points of the domain where the tables overlap, {\cesam} looks for the temperature value where the difference in opacity is the smallest and  {\astec} merges the tables at $\log T=4.0$. However, in any case the differences obtained between the codes do not exceed the intrinsic differences between {\small OPAL95} and {\small AF94} tables in this zone. In the rest of the star differences in opacities are small and do not exceed 2\%. As shown by the detailed comparisons between {\cesam} and {\cles} codes performed by  \citet{jm-apss} differences in opacities at given physical conditions may amount to 2 percents due to the way the {\small OPAL95} data are generated and used (e.g. period when the {\small OPAL95} data were downloaded or obtained from the Livermore team, interpolation programme, mixture of heavy elements). As can be seen in Fig.~\ref{fig:opa}, a major source of difference (well noticeable for the 2.0{\msol}, $X_c=0.50$ model) is due to the fact that {\astec}, {\cesam} and {\starox} (and also {\garstec}) use early delivered {\small OPAL95} tables (hereafter unsmoothed) while {\cles} uses tables that were provided later on the {\small OPAL} web site together with a (recommanded) routine to smooth the data. Once this source of difference has been removed (see {\cles} curves with and without smoothed opacities) there remain differences which are probably due to interpolation schemes and to slight differences in the chemical mixture in the opacity tables \citep[see][]{jm-apss}. Finally,  when comparing models calculated by different codes (i.e. not simply comparing opacities), it is difficult to disentangle differences in the opacity computation from differences in the structure.  As an example, \citet{jm-apss} compared {\cles} and {\cesam} models based on harmonised opacity data and in some cases found a worsening of the agreement between the structures.

%=========================
%=========================

%=========================
%\begin{figure}[ht!]
%\centering
%\resizebox{0.75\hsize}{!}{\includegraphics{diff_opal1.1.eps}}
%\resizebox{0.75\hsize}{!}{ \includegraphics{diff_opal1.5.eps}}
%\resizebox{\hsize}{!}{\includegraphics{diff_opal1.1.eps} \includegraphics{diff_opal1.5.eps}}
%\caption{{\task}~1: Comparisons of opacities calculated by each code with respect to {\cesam} for fixed physical conditions corresponding to a model of 0.9{\msol} and $X_c=0.35$ (top panel) and a model of  2.0{\msol}, $X_c=0.50$ (bottom panel).}      
%\label{fig:opa}
%\end{figure}
%=========================

We have derived the maximal relative differences in $c$, $P$, $\rho$, $\Gamma_1$, $L_r$ and $X$ from the relative differences, considering the maximum of the differences calculated at fixed $r/R$ and of those obtained at fixed $m/M$. Note that for the latter estimate we removed the very external zones (i.e. located at $m>0.9999M$) where the differences may be very large. We report these maximal differences in Table~\ref{tabla:v1.7} together with the location ($r/R$) where they happen. We note that, except for $X$ and $L_r$, the largest differences are found in the most external layers and (or) at the boundary of the convection regions. This will be discussed in the following.

%=========================
\begin{figure*}[htbp!]
\centering
\resizebox{\hsize}{!}{\includegraphics{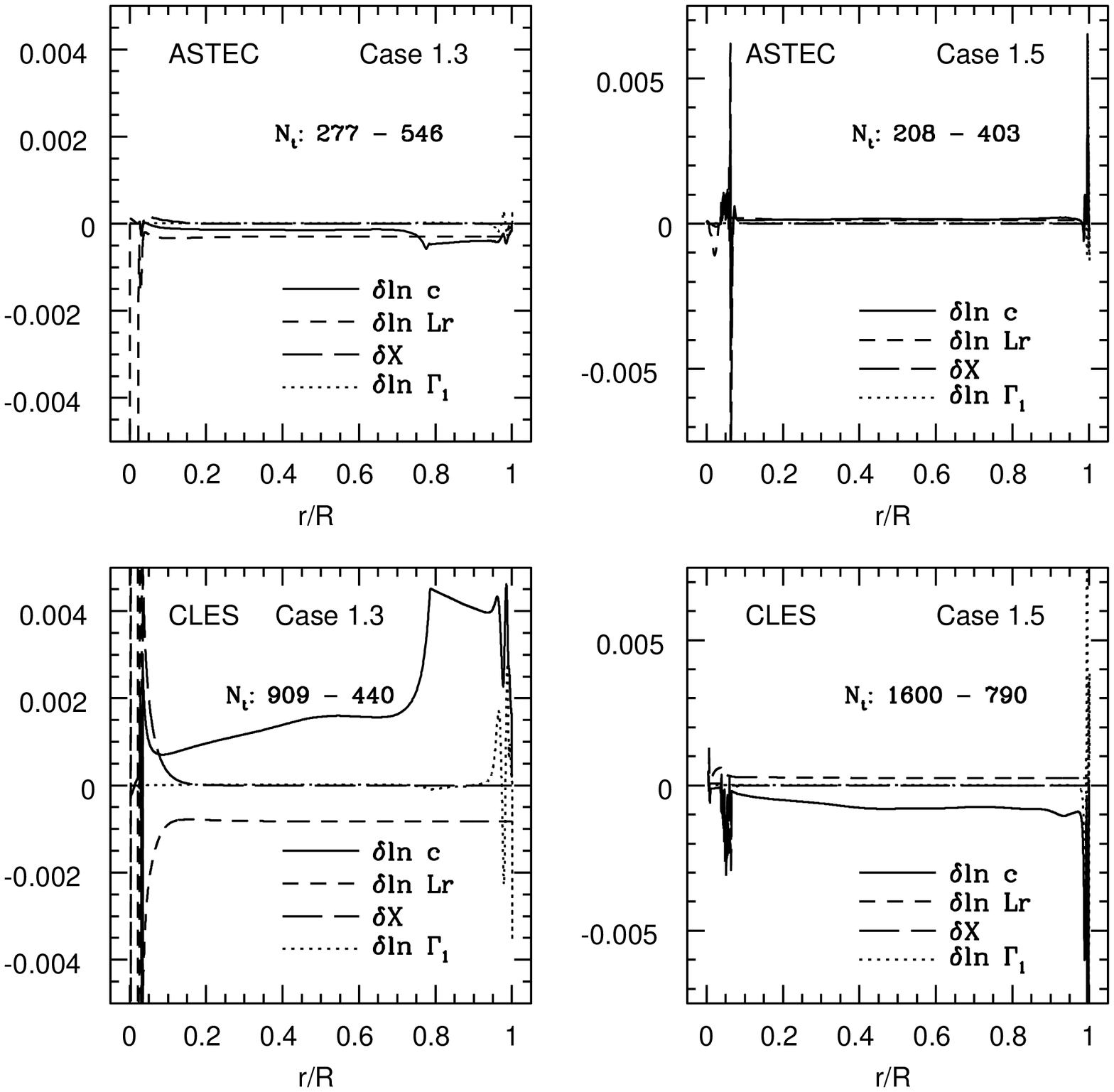}\includegraphics{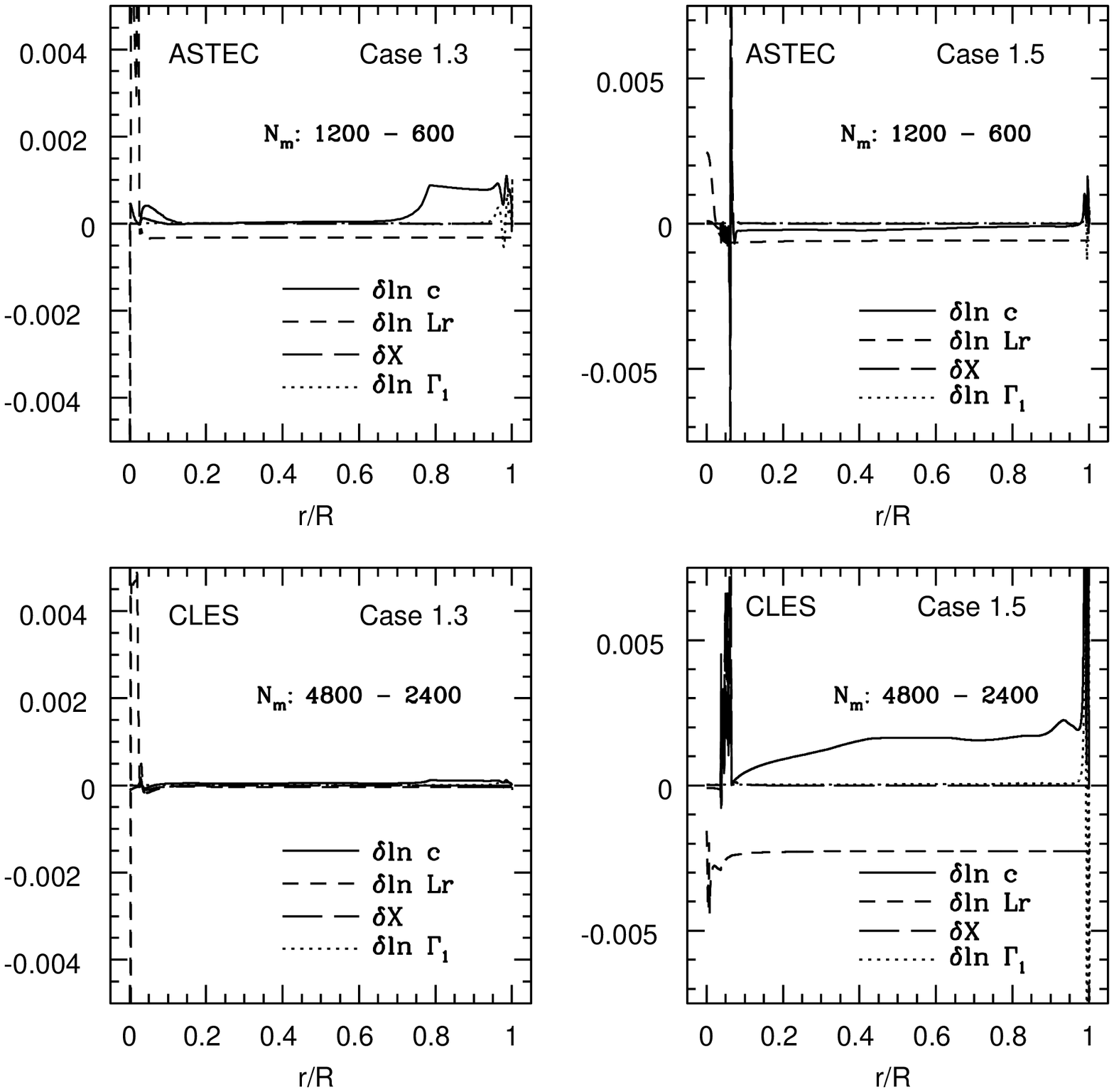}
}
%\resizebox{0.75\hsize}{!}{\includegraphics{fig1.3.dA.eps}}
\caption{{\task}~1: Effect of halving the time step in the evolution calculation (left and center left) and doubling the number of spatial  mesh points in a model (right and center right) for Cases 1.3 and 1.5 models obtained by {\astec} (top) and {\cles} (bottom). Differences between physical quantities have been calculated at fixed mass and plotted as a function of $r/R$.}      
\label{fig:mesh}
\end{figure*}
%=========================

The effects of doubling the number of (spatial) mesh points and of halving the time step for the computation of evolution have been examined in {\astec}  and {\cles} models for Cases~1.1, 1.3 and 1.5 \citep[][]{jcd-aarhus,am-aarhus,jm-apss}. Some results are displayed in Figs.~\ref{fig:mesh} for Cases~1.3 and 1.5. 
% For the Case 1.1 model of $0.9$ {\msol} on the \MS, the major differences are seen very close to the surface and are otherwise smaller than 0.15\%. 
For the Case 1.3 model of $1.2$ {\msol} on the \SGB, the main differences are obtained when the time step if halved and are seen at the very center (percent level), in the convective envelope and close to the surface (0.5\% for the sound speed). In the rest of the star they remain lower than 0.2\%. For the Case 1.5 model, which is a $2$ {\msol} model at the end of the {\MS} (with overshooting), differences at the percent level are noticeable in the region where the border of the convective core moved during the {\MS}. Differences may also be at the percent level close to the surface. For the Case 1.1 model of $0.9$ {\msol} on the \MS (not plotted), the differences are smaller by a factor 5 to 10 than those obtained for Cases 1.3 and 1.5. Further comparaisons of {\cesam} models with various {\cles} models (doubling either the number of mesh points or halving the time step) have shown different trends: doubling the number of mesh points did not change the results in Case~1.3 but improved the agreement in Case~1.5 for $L, \rho, c$ and the internal X-profile, halving the time step worsened the agreement in both cases.

\subsection{Internal structure}\label{sec:task1-str_int}

\subsubsection{Low-mass models: Cases~1.1, 1.2 and 1.3}

For these models, the differences in $c$, $\rho$, $X$, and  $L_r$ as a function of the relative radius are plotted in  Fig.~\ref{fig:task1-int123}.

Table~\ref{tabla:d1.1} and Fig.~\ref{fig:task1-int123} show that the five evolution codes provided quite similar stellar models for  Case~1.1 --
which has an internal structure and evolution stage quite similar to the
Sun -- and for Case~1.2. We note that the variations  
found in {\astec} model for Case~1.2 are larger than for Case~1.1 which is probably due to the lack of a {\PMS} evolution in present {\astec} computations. On the other hand, the systematic  difference in $X$  observed in {\cles} models, even in the outer layers, results from the detailed calculation of deuterium burning in the early {\PMS}.

%==========================
\begin{figure}[htbp!]
\centering
\resizebox{0.75\hsize}{!}{\includegraphics{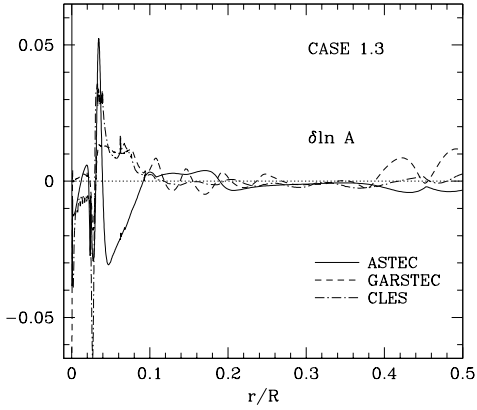}}
\caption{{\task}~1: Logarithmic differences of $A=N^2_{\rm BV} r/g$ calculated at fixed $m/M$ between pairs of models ({\small CODE}-{\cesam}) as a function of $r/R$ for Case~1.3}      
\label{fig:task1-A3}
\end{figure}
%=========================
%=========================
\begin{figure}[htbp!]
\centering
\resizebox{0.75\hsize}{!}{\includegraphics{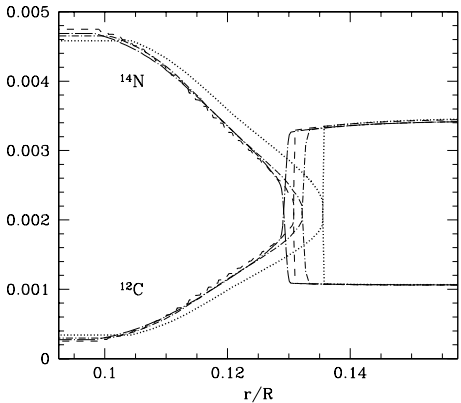}}
\caption{{\task}~1, Case~1.4: ${}^{12}{\rm C}$ and ${}^{14}{\rm N}$ abundances in the region where they become that of equilibrium for
models computed by {\cesam} (long-dash-dotted lines), {\cles} (dash-dotted lines), {\garstec} (dashed lines), and {\starox} (dotted lines).}
\label{fig:cno1.4}
\end{figure}
%==========================

For the most evolved model (Case~1.3) the differences increase drastically with respect to previous cases. It is worth to mention that at $m \sim 0.1M$ ($r\sim 0.03R$) the  variations of sound speed, Brunt-V\"ais\"al\"a frequency (Fig.~\ref{fig:task1-A3}) and hydrogen mass fraction are large. This can be understood as follows.  
Case~1.3 corresponds to a star of high central density which burns H in a shell.
The middle of the H-burning shell -- where the nuclear energy generation is maximum -- is located at $\sim 0.1M$. Moreover, before reaching that stage, i.e. during a large part of the {\MS}, the star had had a growing convective core (see Sect.~\ref{sec:task1-conv} below) which reached a maximum size of $m\sim0.05M$, when the central H content was $X\sim0.2$, before receding. The large differences seen at $m\sim0.1M$ can therefore be linked to the size reached by the convective core during the {\MS} as well as to the features of the composition gradients outside this core. 

For Case~1.3 we have looked at the values of the gravitational $\epsilon_{\rm grav}$ and nuclear energy $\epsilon_{\rm nuc}$ in the very central regions, i.e. in the He core (from the centre to $r/R\sim0.02$) and in the inner part of the H-burning shell ($r/R\in[0.02,0.03]$). We find that in the He core, from the border to the centre, {\cles} values of $\epsilon_{\rm grav}$ are larger by $0-30$ \% than the values obtained by {\garstec} and {\cesam}.
This probably explains the large differences in $L_r$, i.e. around $20-25$ per cent, seen for {\cles} model in the central regions (see Fig.~\ref{fig:task1-int123}). We also find differences in $\epsilon_{\rm grav}$ of a factor of $2$ ({\cles} vs. {\cesam}) and $3$ ({\garstec} vs. {\cesam}) in a very narrow region in the middle of the H-shell, but these differences appear in a region where $\epsilon_{\rm nuc}$ is large and therefore are less visible in the luminosity differences.

The differences in $X$ seen in the {\astec} model for Case 1.3 in the region where $r/R\in[0.1,0.3]$ are probably due to the nuclear reaction network it uses. {\astec} models have no carbon in their mixture since they assume that the CN part of the CNO cycle is in nuclear equilibrium at all times and include the original ${}^{12}{\rm C}$  abundance into that of ${}^{14}{\rm N}$. That means that the nuclear reactions of the CNO cycle that should take place at $r\sim 0.1R$ do not occur and hence the hydrogen in that region is less depleted than in models built by the other codes.

%==========================
% Variations for 2 Msum models
%==========================
\begin{figure*}[htbp!]
\centering
\resizebox{\hsize}{!}{\includegraphics{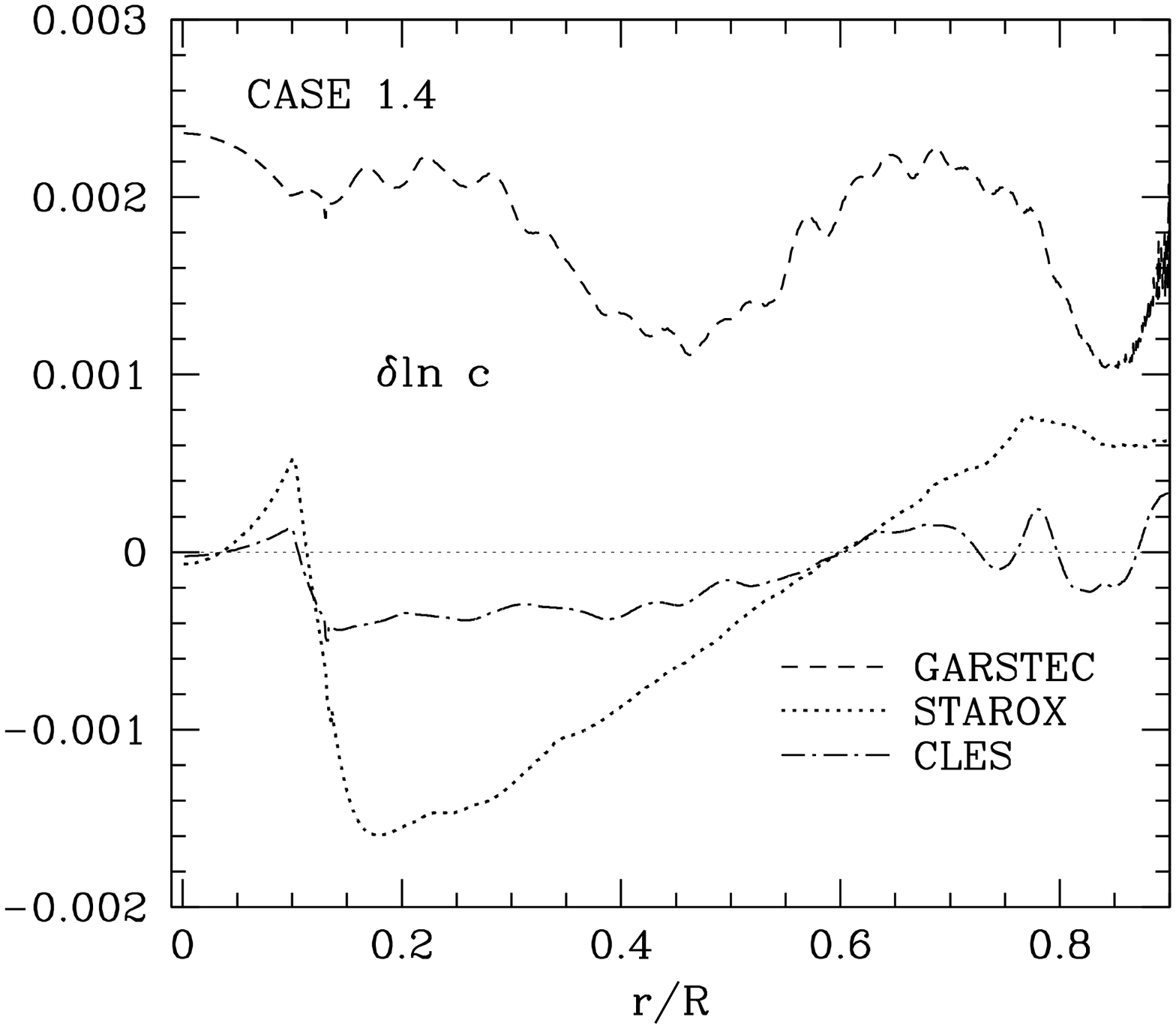}
                      \includegraphics{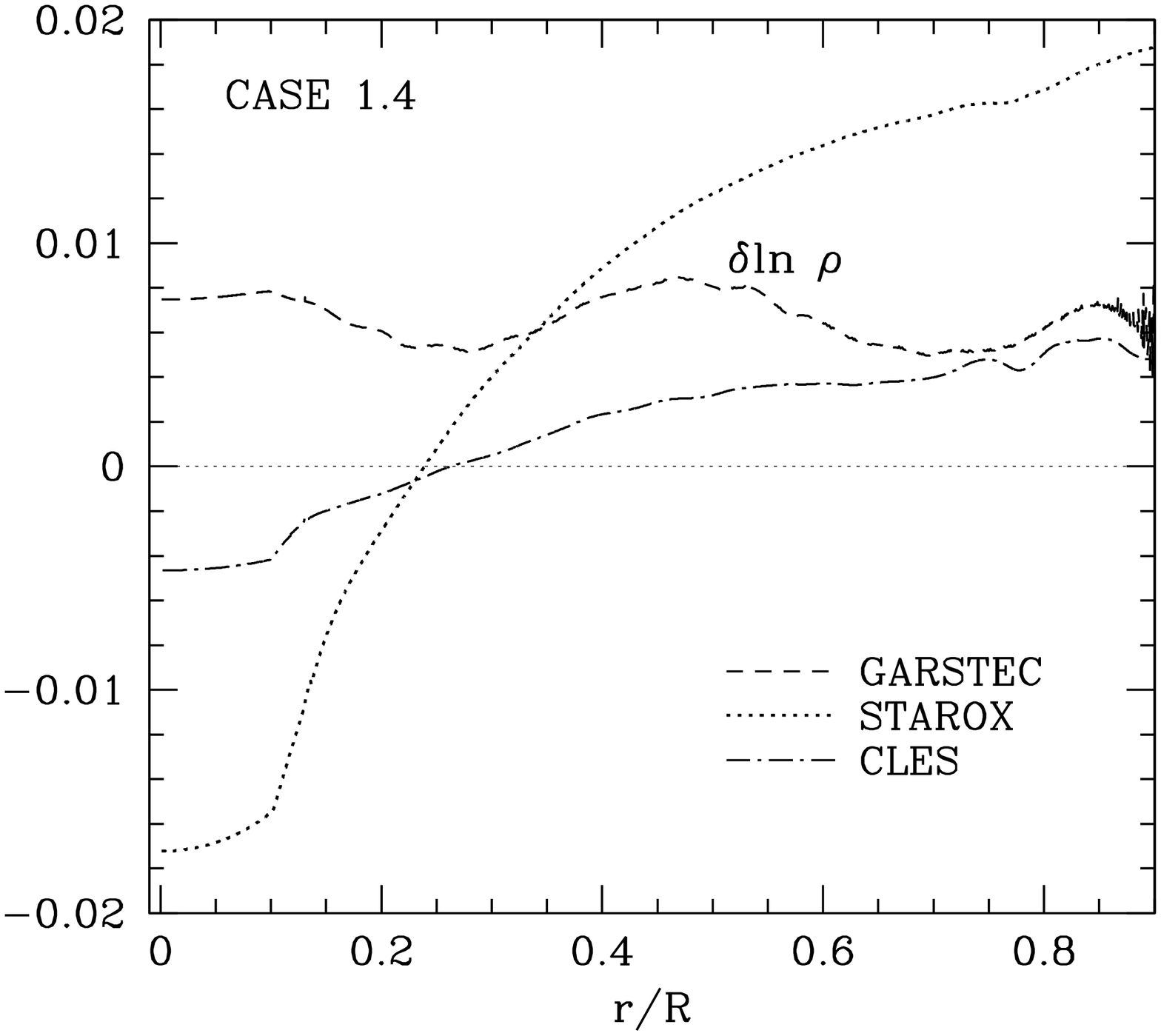}
                      \includegraphics{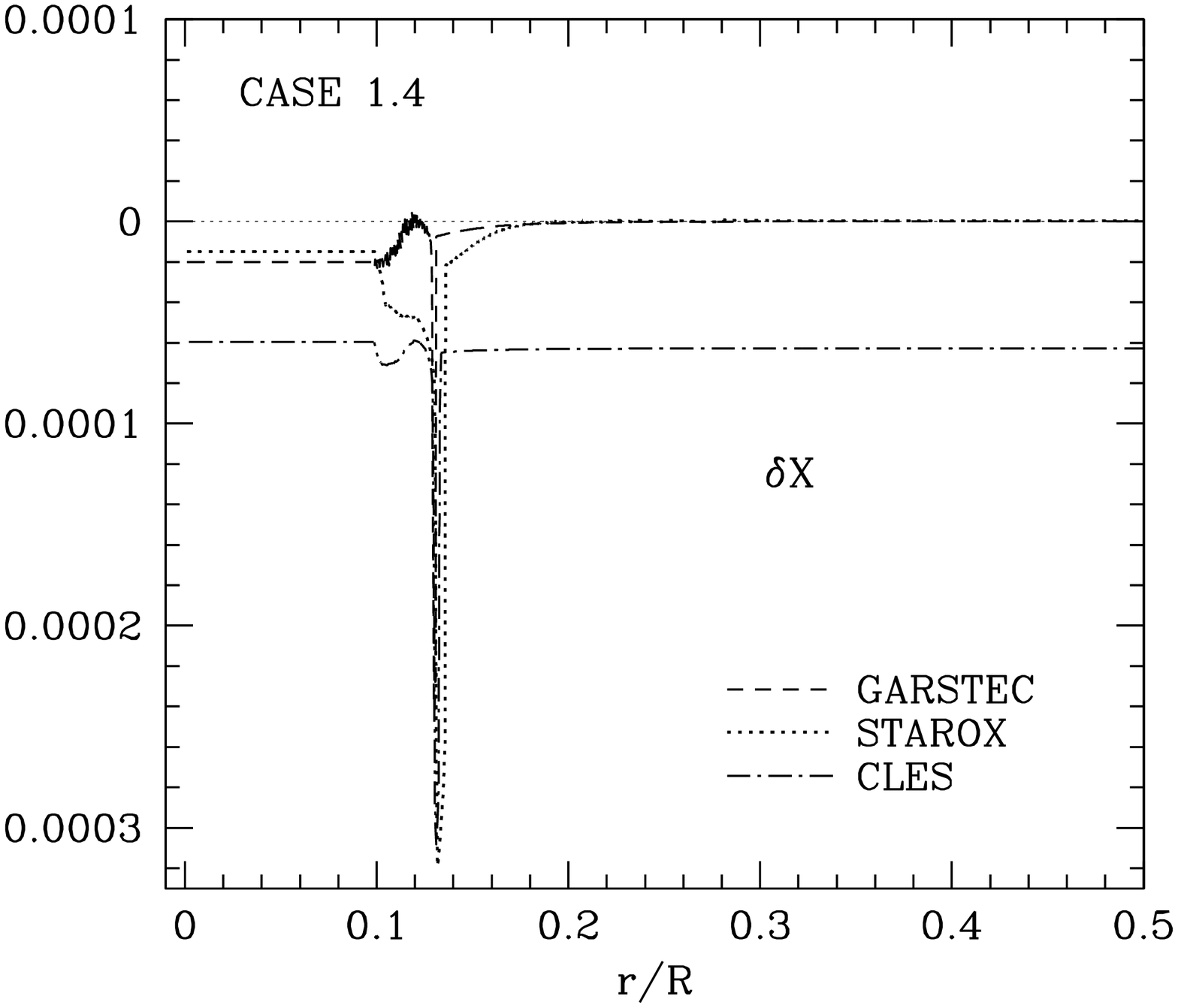}
		      \includegraphics{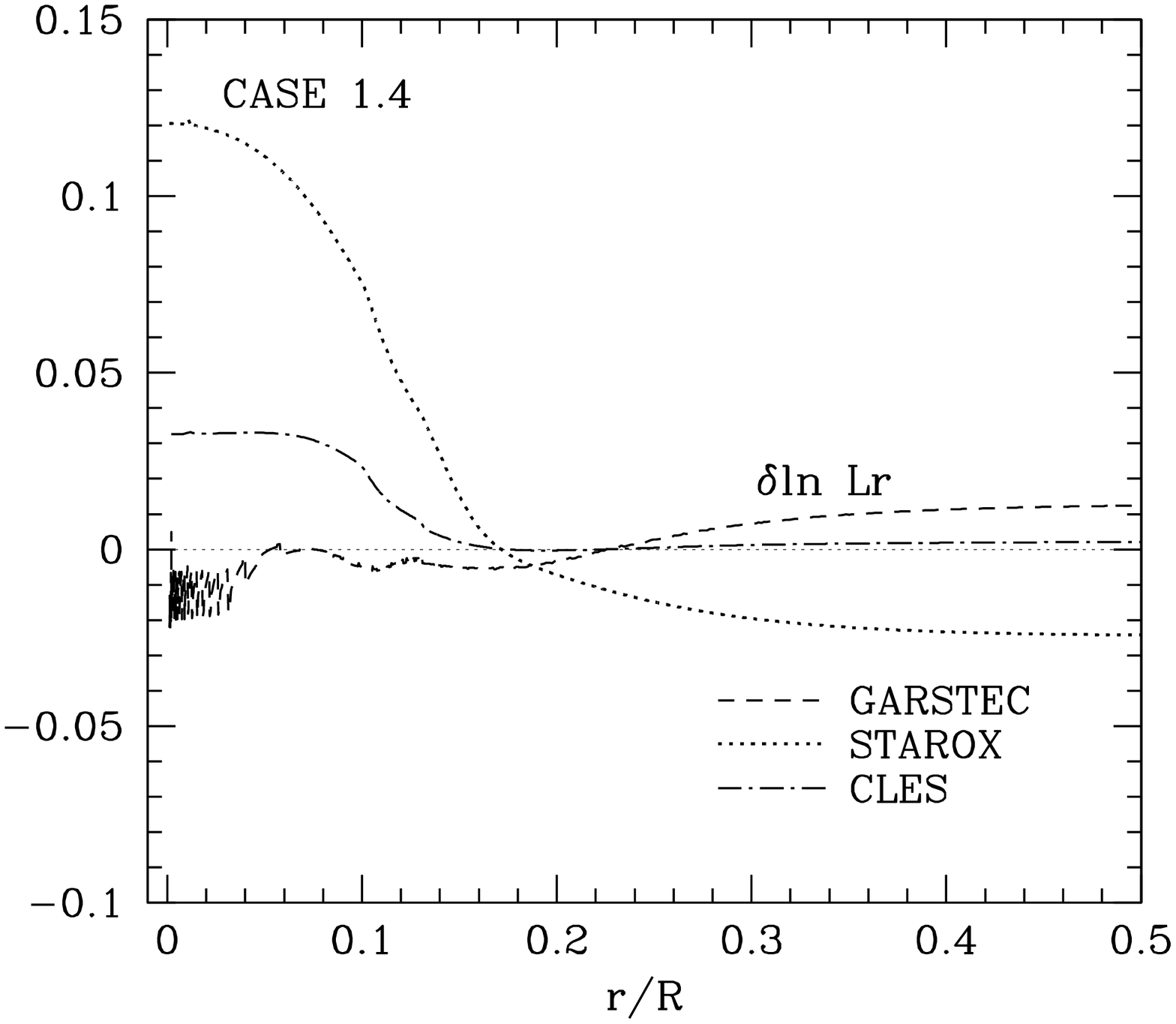}
}
\resizebox{\hsize}{!}{\includegraphics{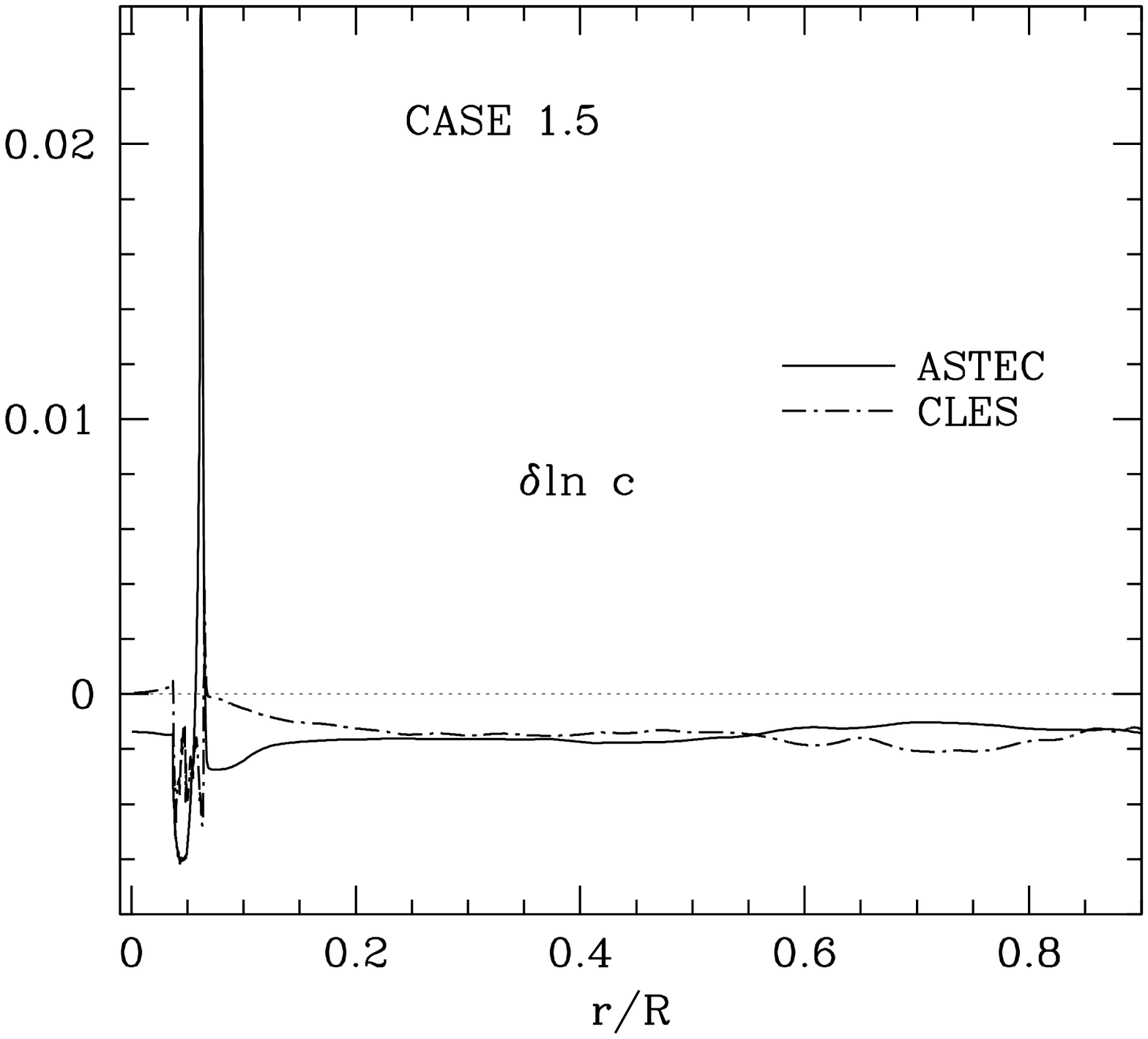}
                      \includegraphics{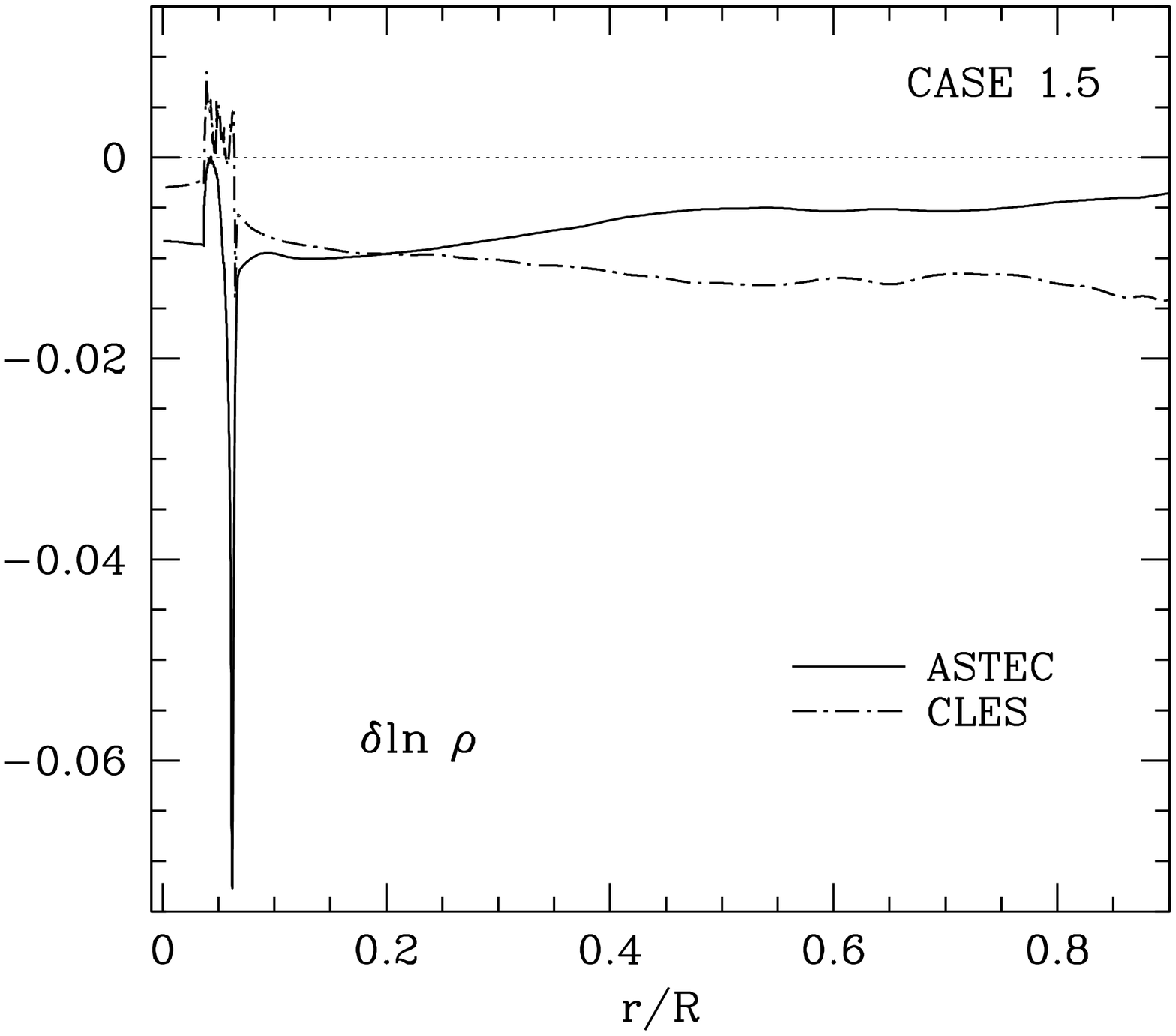}
                      \includegraphics{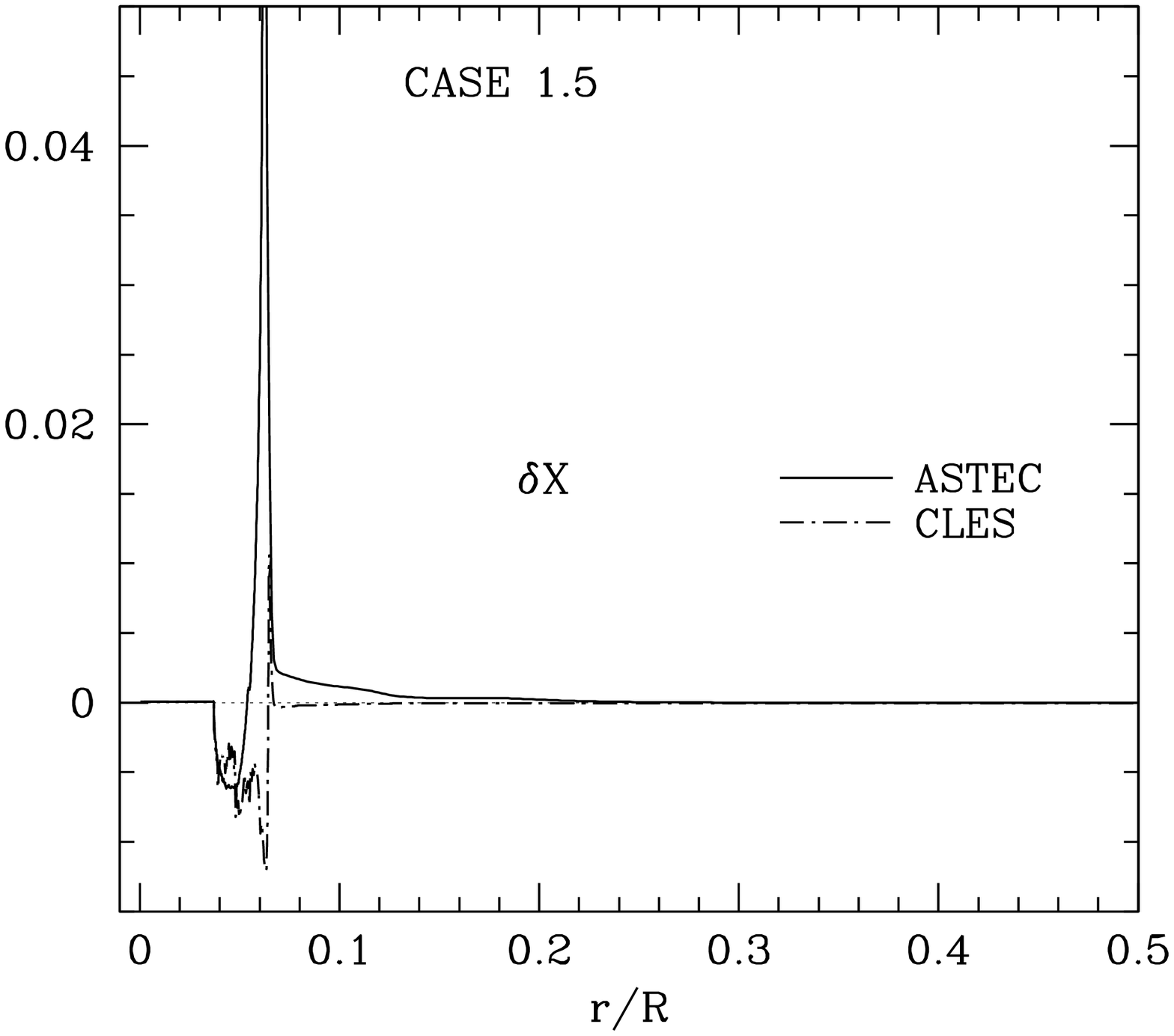}
		      \includegraphics{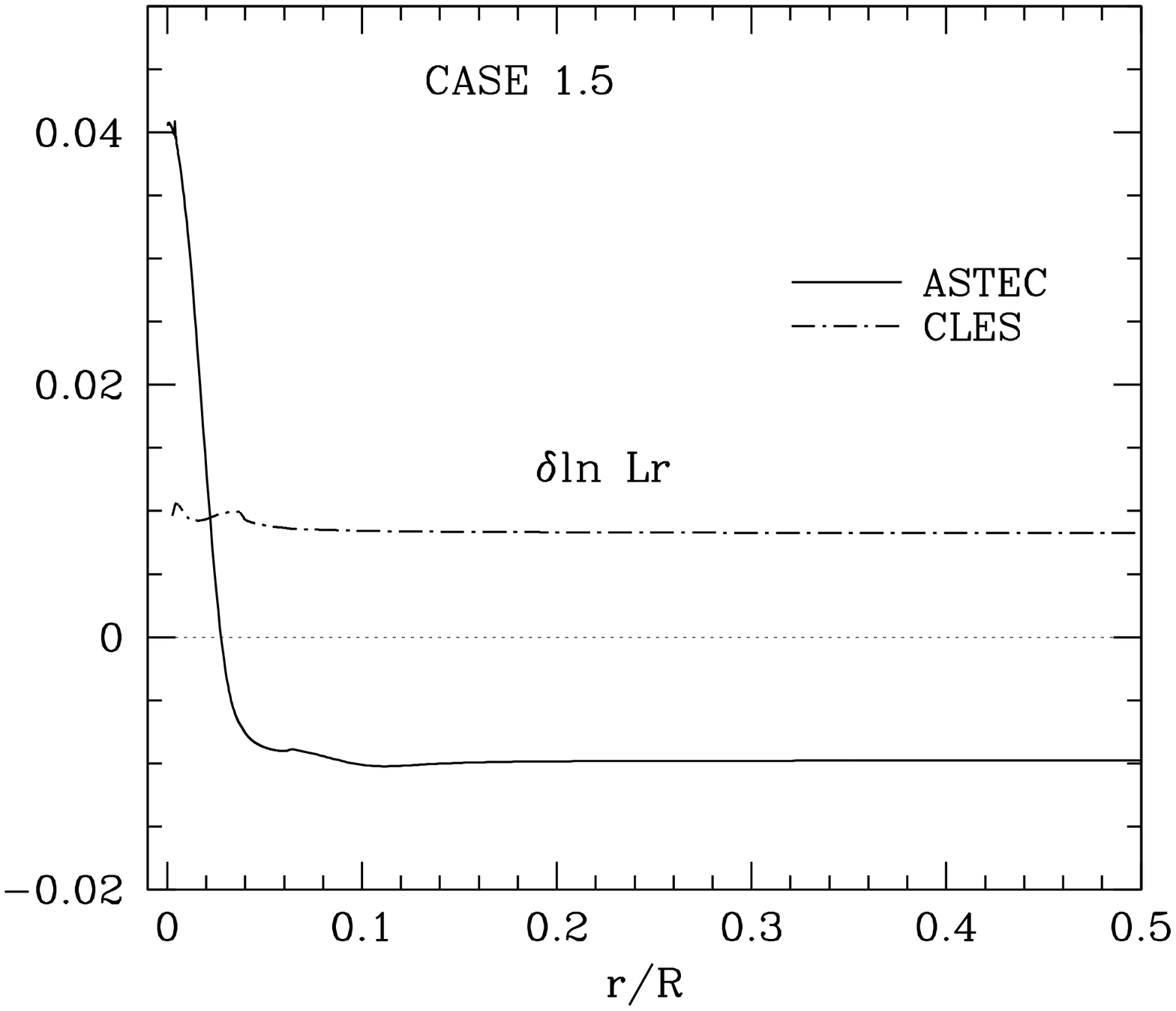}
}
\caption{{\task}~1: Plots of the differences at fixed relative mass between pairs of models 
({\small CODE}-\cesam) corresponding to Cases~1.4 and 1.5. {\it Left panel}: logarithmic sound speed differences.
{\it Centre left panel}: logarithmic density differences. {\it Centre right panel}: hydrogen mass fraction differences. {\it Right panel}: logarithmic luminosity differences.
Horizontal dotted line represents the reference model (\cesam).}      
\label{fig:task1-int45}
\end{figure*}

%==========================
% Variations for higer mass models
%==========================
%==========================

\begin{figure*}[htbp!]
\centering
\resizebox{\hsize}{!}{\includegraphics{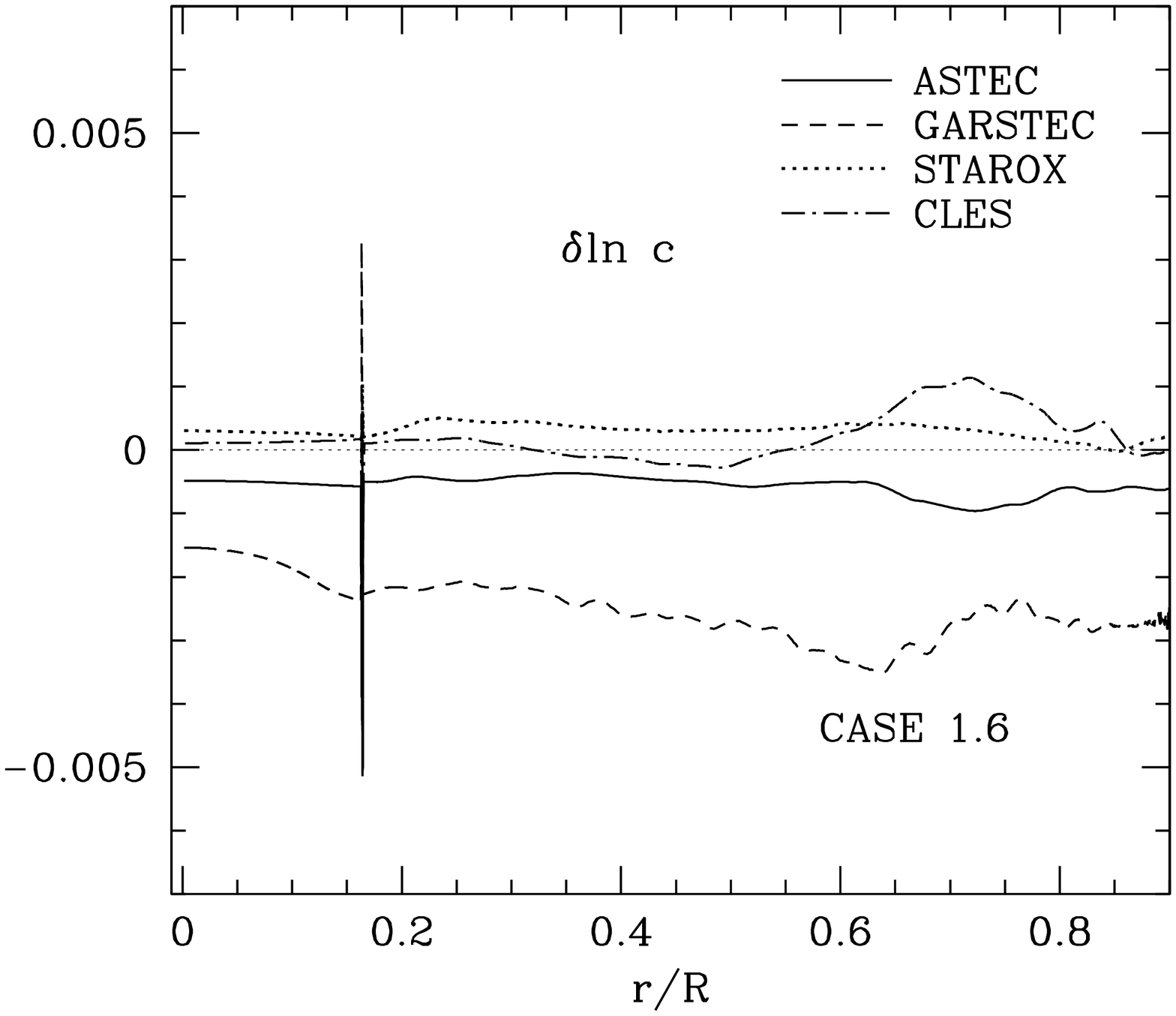}
                      \includegraphics{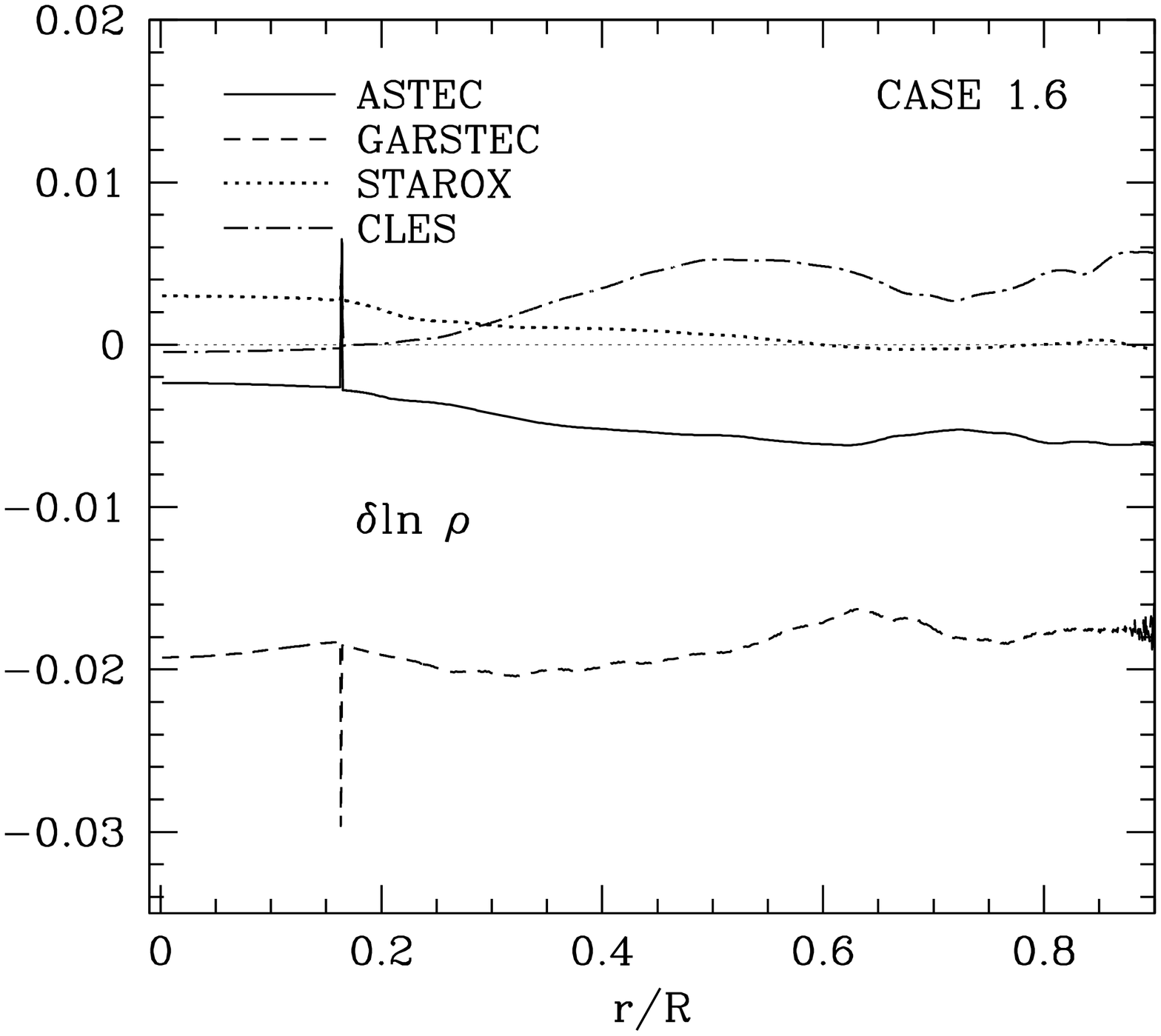}
                      \includegraphics{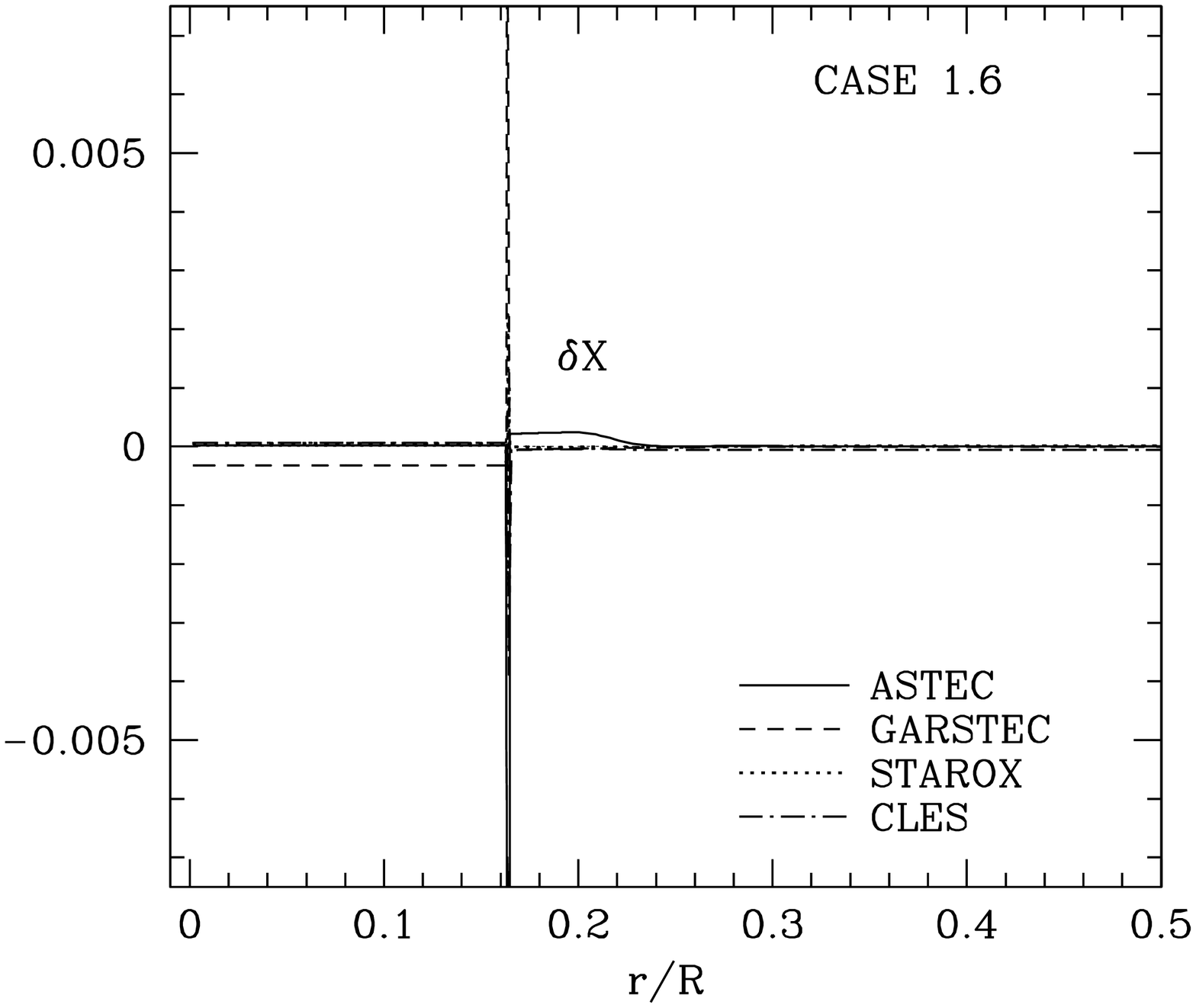}
                      \includegraphics{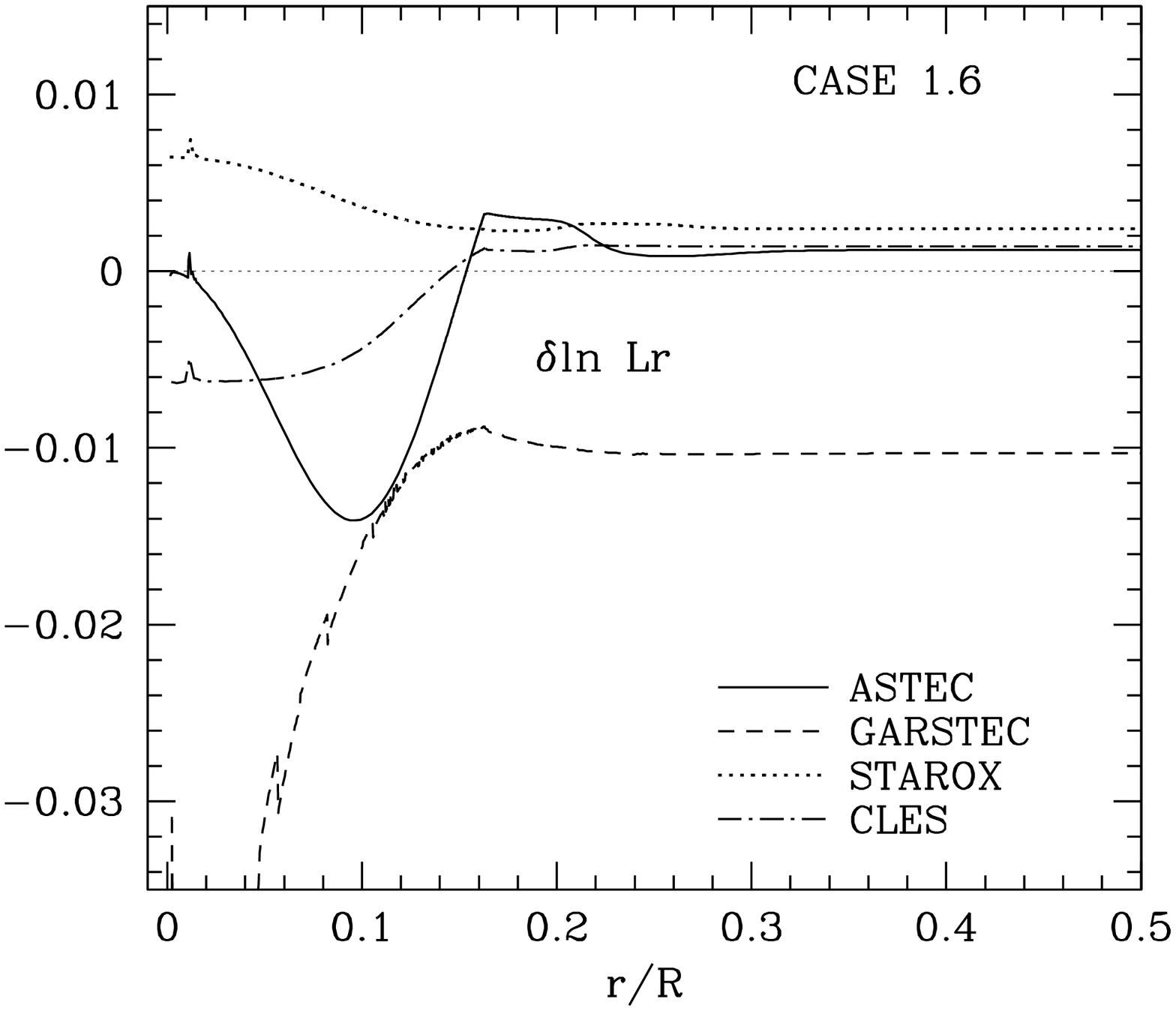}}

\resizebox{\hsize}{!}{\includegraphics{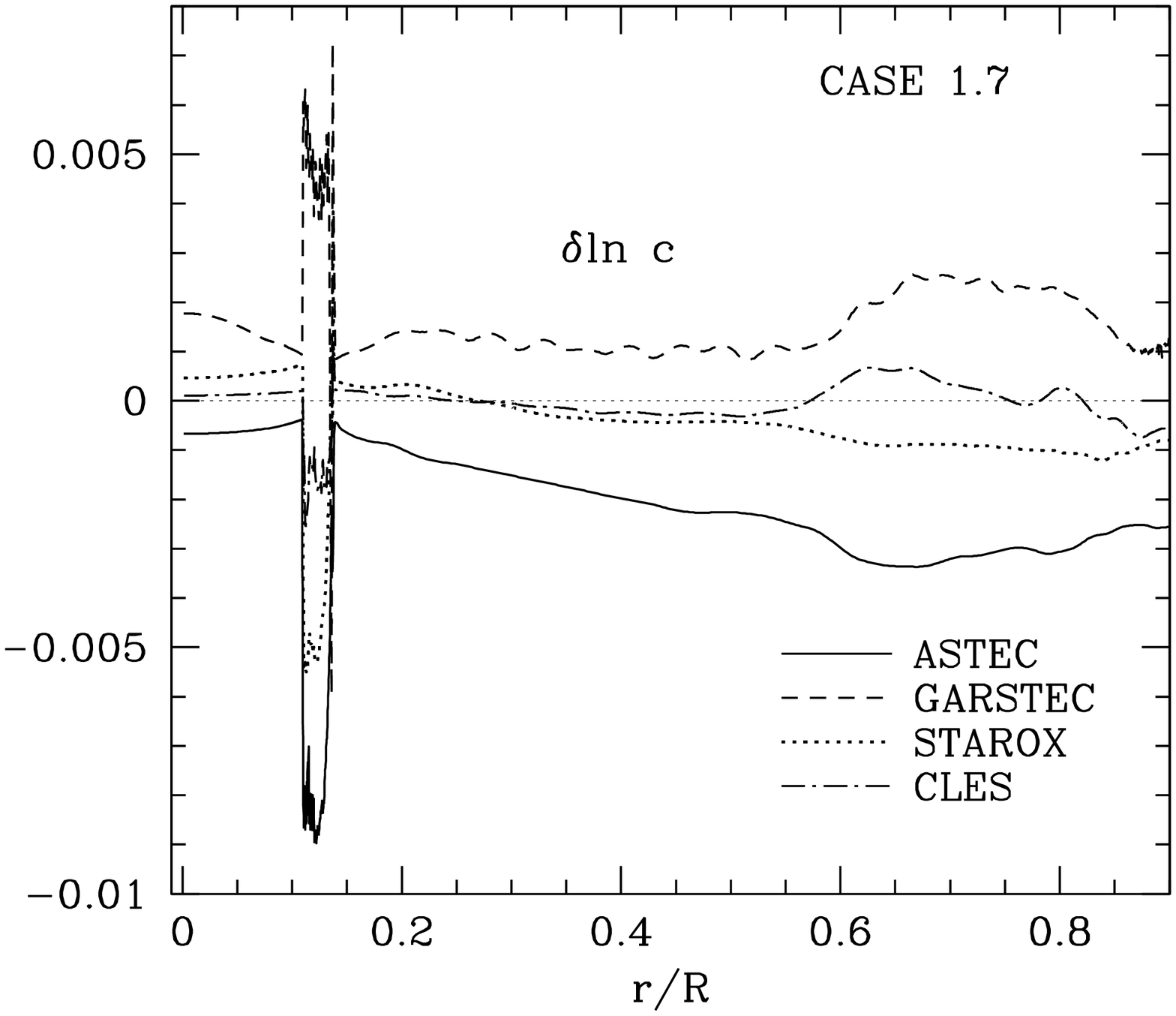}
                      \includegraphics{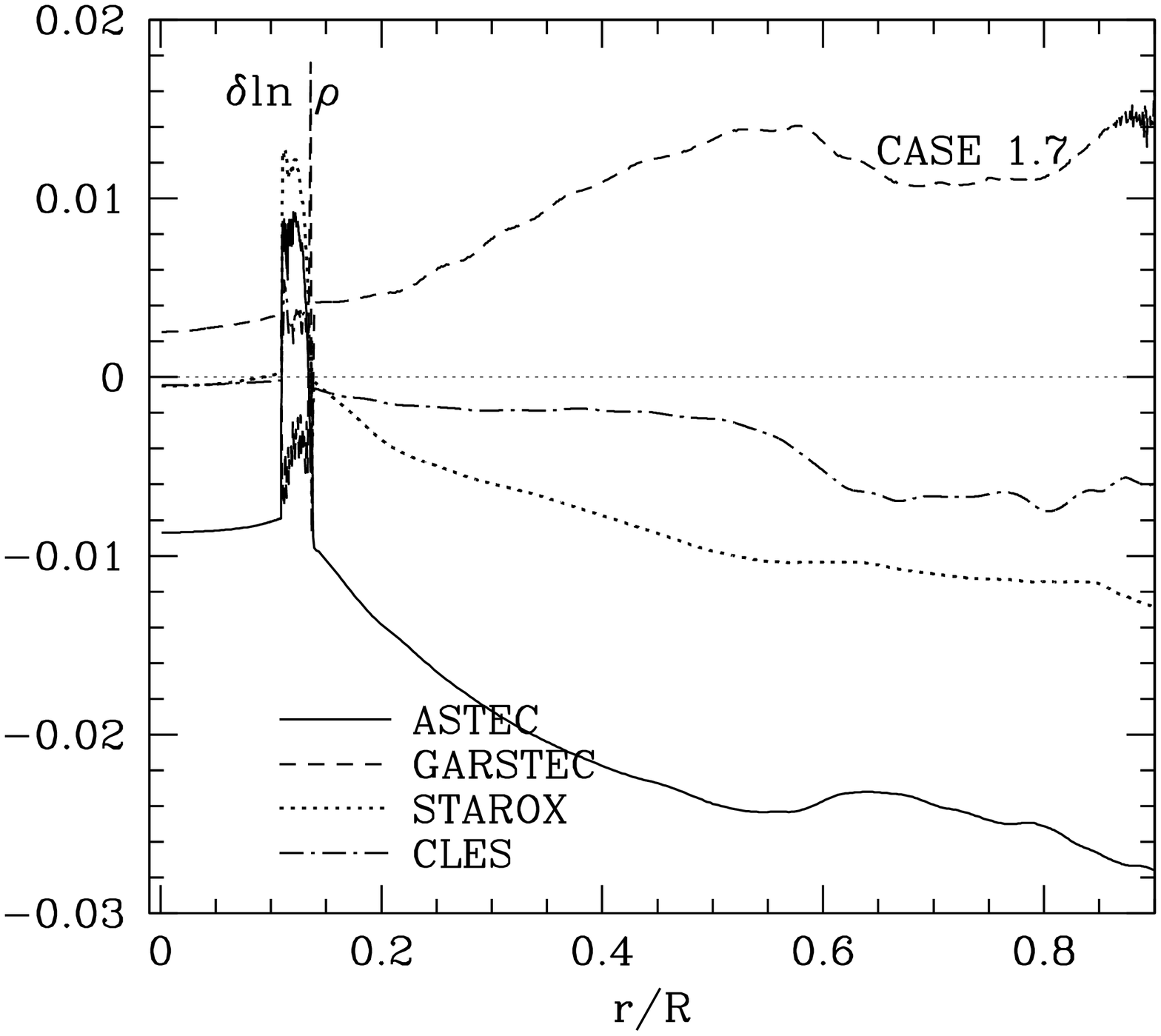}
                      \includegraphics{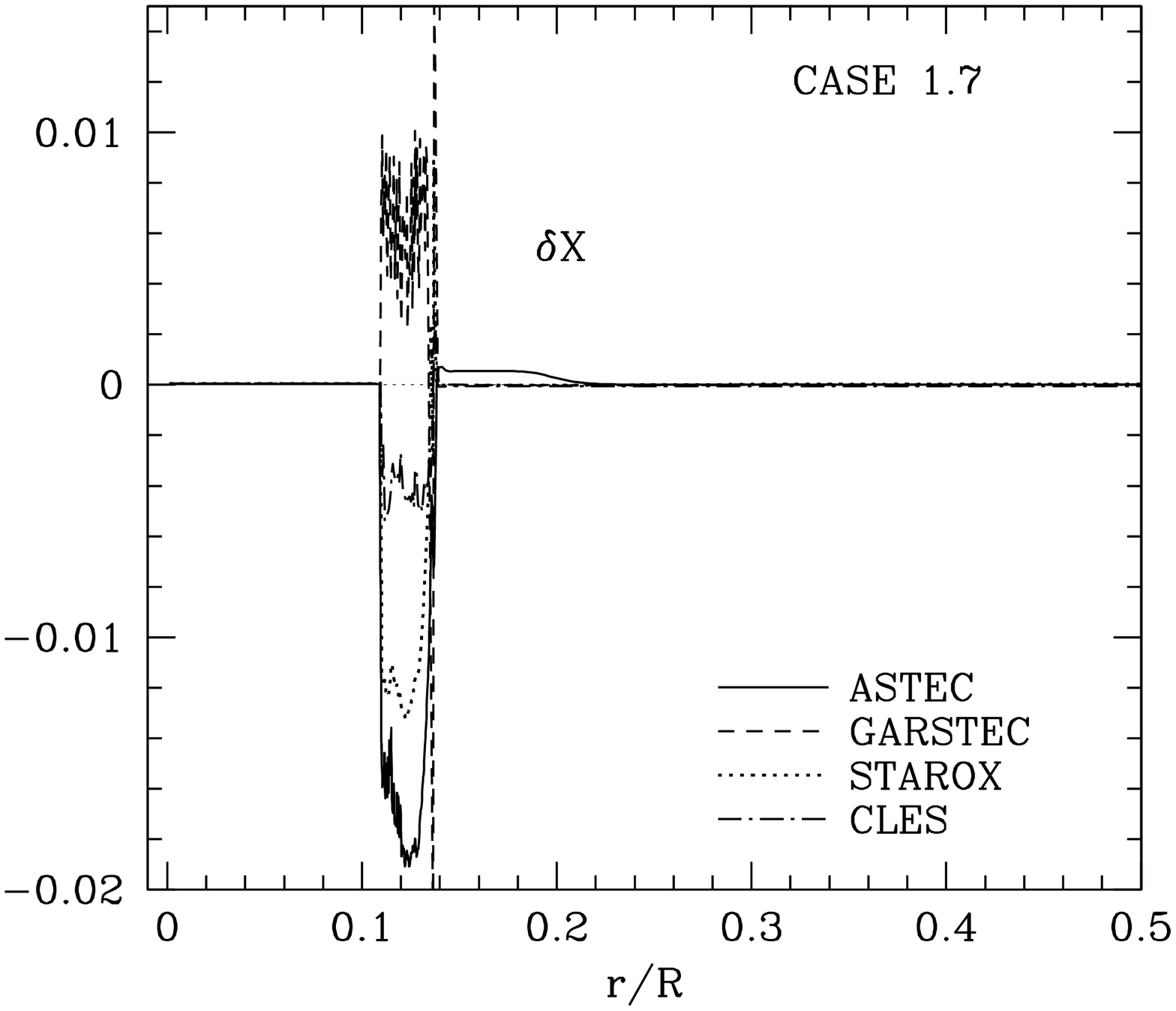}
		      \includegraphics{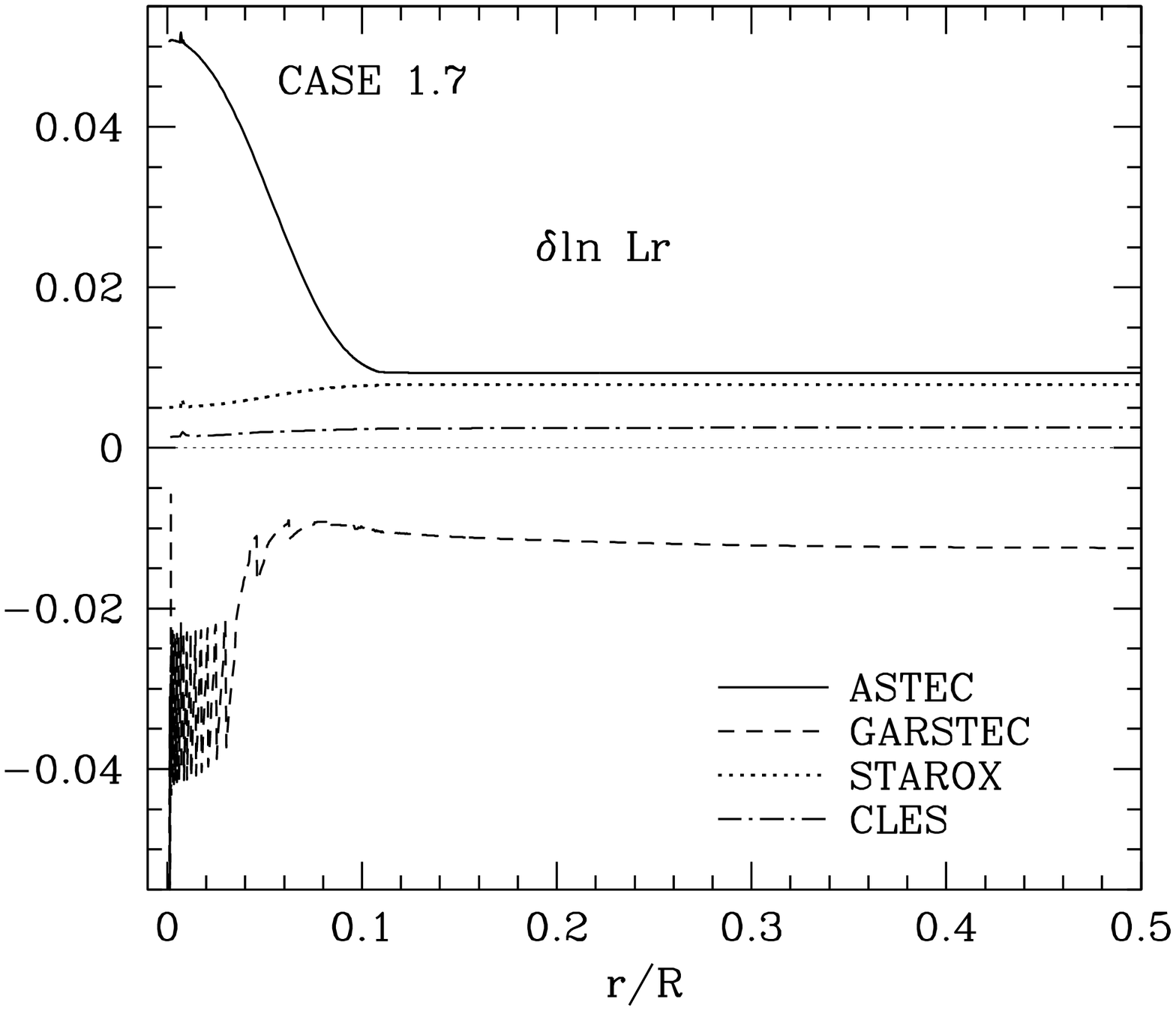}}
\caption{{\task}~1: Plots of the differences at fixed relative mass between pairs of models 
({\small CODE}-\cesam) corresponding to Cases~1.6 and 1.7. {\it Left panel}: logarithmic sound speed differences.
{\it Centre left panel}: logarithmic density differences. {\it Centre right panel}: hydrogen mass fraction differences. {\it Right panel}: logarithmic luminosity differences.
Horizontal dotted line represents the reference model (\cesam).}       
\label{fig:task1-int67}
\end{figure*}
%==========================

\subsubsection{Intermediate mass models: Cases~1.4 and 1.5}\label{sec:C14C15}

Case~1.4 illustrates the {\PMS} evolution phase  of a $2${\msol} star when the ${}^{12}{\rm C}$ and  ${}^{14}{\rm N}$ abundances become that of equilibrium. We can see in Fig.~\ref{fig:task1-int45} that the largest differences in $X$ are indeed found in the region where ${}^{12}{\rm C}$ is transformed into ${}^{14}{\rm N}$ (i.e. where $r\lesssim 0.14R$, see  Fig.~\ref{fig:cno1.4}), that is in the region in-between $m\sim0.1M$ (edge of the convective core) and  $m=0.2M$. 
 
In central regions, the contribution of the gravitational contraction to the total energy release is important: the ratio of the gravitational to the total energy $\epsilon_{\rm grav}/(\epsilon_{\rm grav}+\epsilon_{\rm nuc})$ varies from $\sim6$\% in the centre up to $\sim50$\% at $r/R\sim0.1$. Comparisons between {\cles}, {\cesam} and {\garstec} show differences in $\epsilon_{\rm grav}$ and $\epsilon_{\rm nuc}$ of a few per cent which eventually partially cancel. We note in Fig.~\ref{fig:task1-int45} a difference in $L_r$ of $\sim12\%$ for the {\starox} model but the data made available for this model do not allow to determine if the difference comes from the nuclear or the gravitational energy generation rate.
 
Case~1.5 deals with a 2~{\msol} model at the end of the {\MS} when the central H content is $X_{\rm c}=0.01$. In this model, the star was evolved with a central mixed  region increased by $0.15\,H_p$ ($H_p$ being the pressure scale height)
with respect to the size of the convective core determined by the Schwarzschild criterion. {\cesam}, {\astec} and {\cles} assume, as specified, an adiabatic stratification in the overshooting region while {\starox} generated the model assuming a radiative stratification in this zone. That smaller temperature gradient, even if it affects only a quite small region, works in practice like an increase in opacity. This leads to an evolution with a larger convective core, and therefore to a higher effective temperature and luminosity in the {\starox} model. Therefore in Fig.~\ref{fig:task1-int45} we only show the differences between the {\astec} and {\cles} models with respect to {\cesam}. 

The largest  differences in $c$ (as well as in $\rho$ and $X$) are found in the region in-between $r=0.03R$ and $0.06R$ ($m/M\in[0.07-0.2]$). They reflect the differences in the mean molecular weight gradient ($\nabla_\mu$) left  by the inwards displacement of the convective core during the {\MS} evolution. The strong peak  at $r \sim 0.06R$ ($m \sim 0.2M$) found in {\astec}-curves, as well as the plateau of $\delta X$ between $r\sim0.06R$ and $0.2R$ 
probably result from the treatment of chemical evolution in the {\astec} code, that assumes
the CN part of the CNO cycle to be in nuclear equilibrium at all times.

%=========================
\begin{figure*}[htbp!]
\centering
{\includegraphics[scale=0.25]{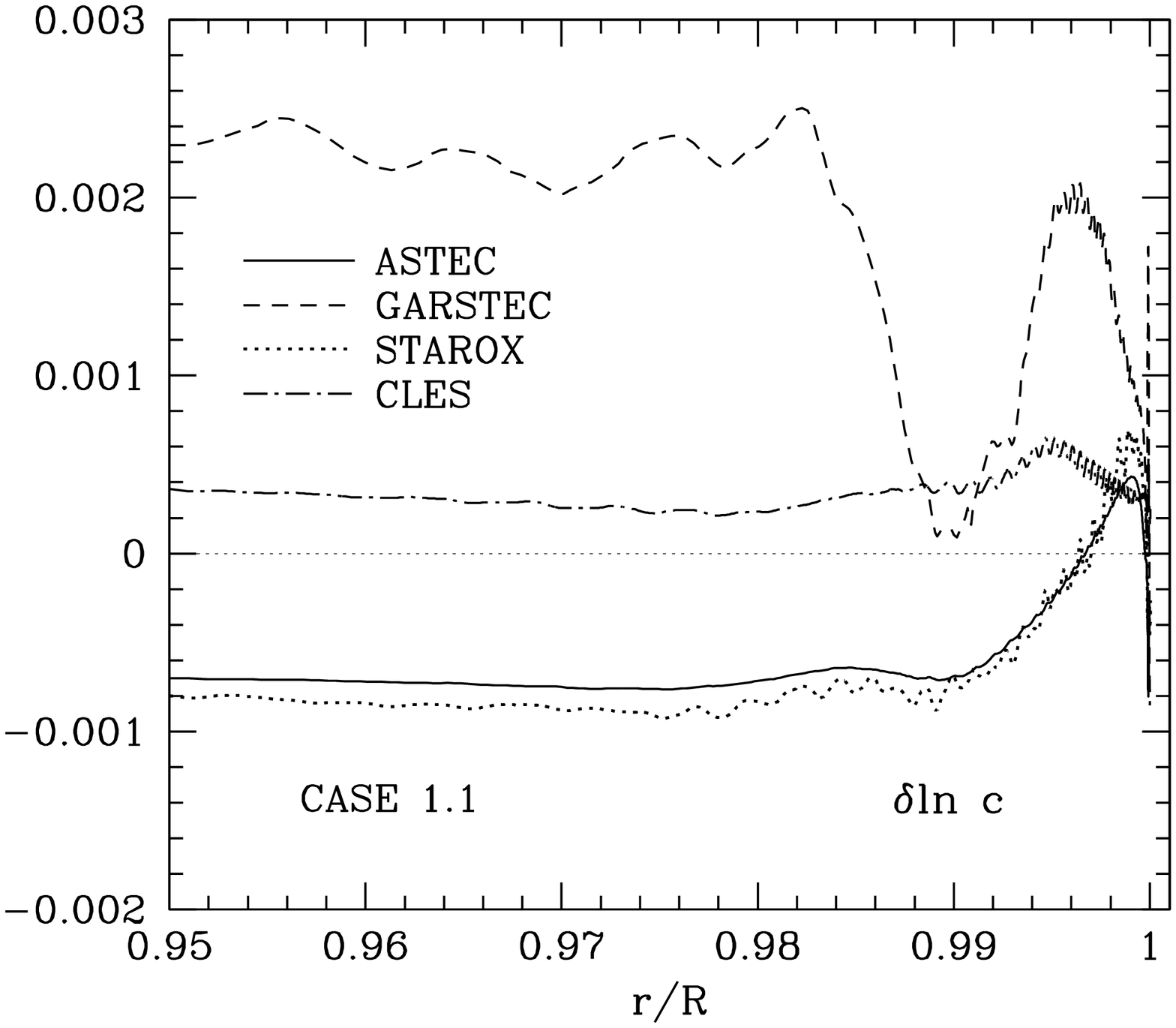}
 \includegraphics[scale=0.25]{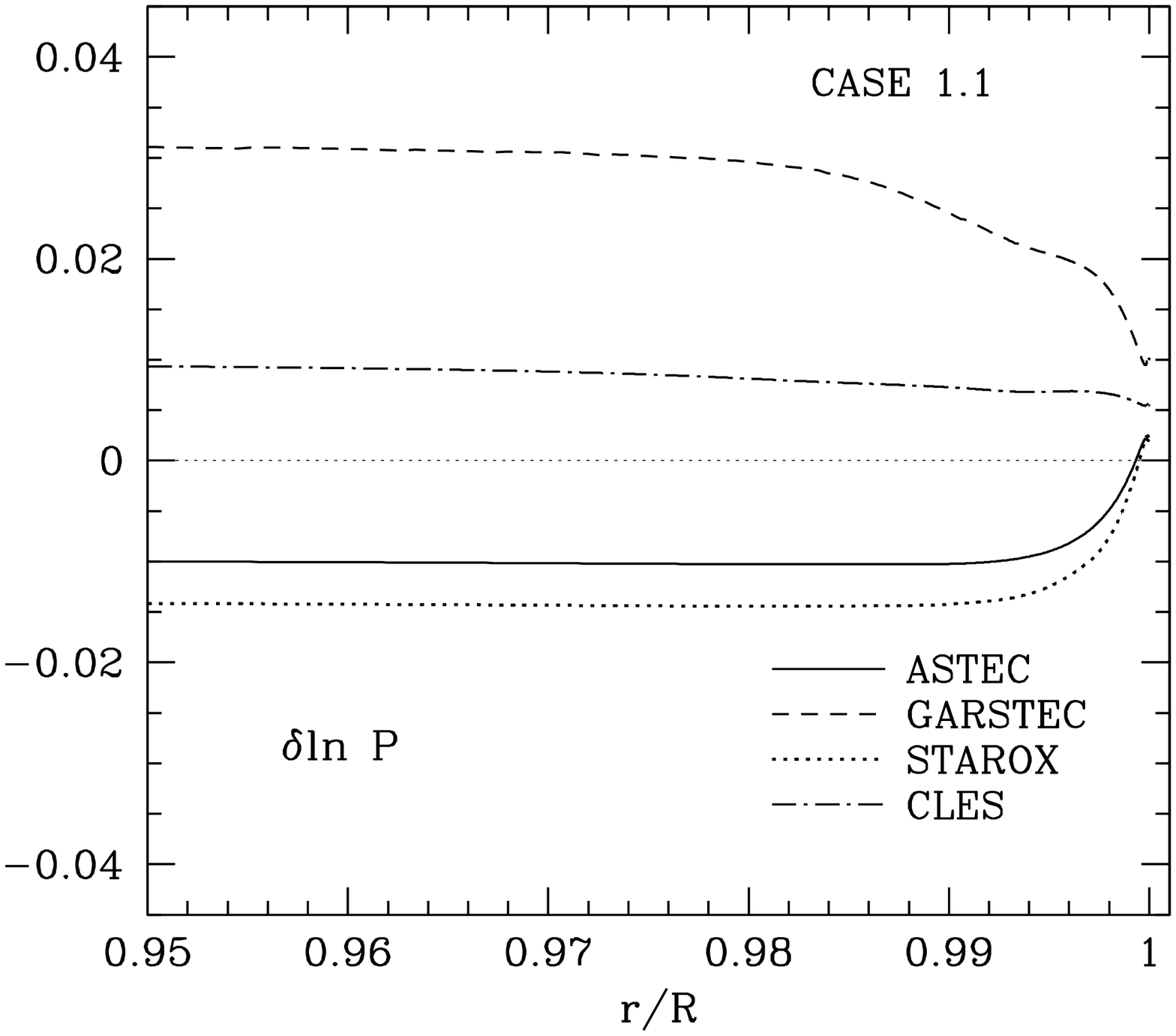}
 \includegraphics[scale=0.25]{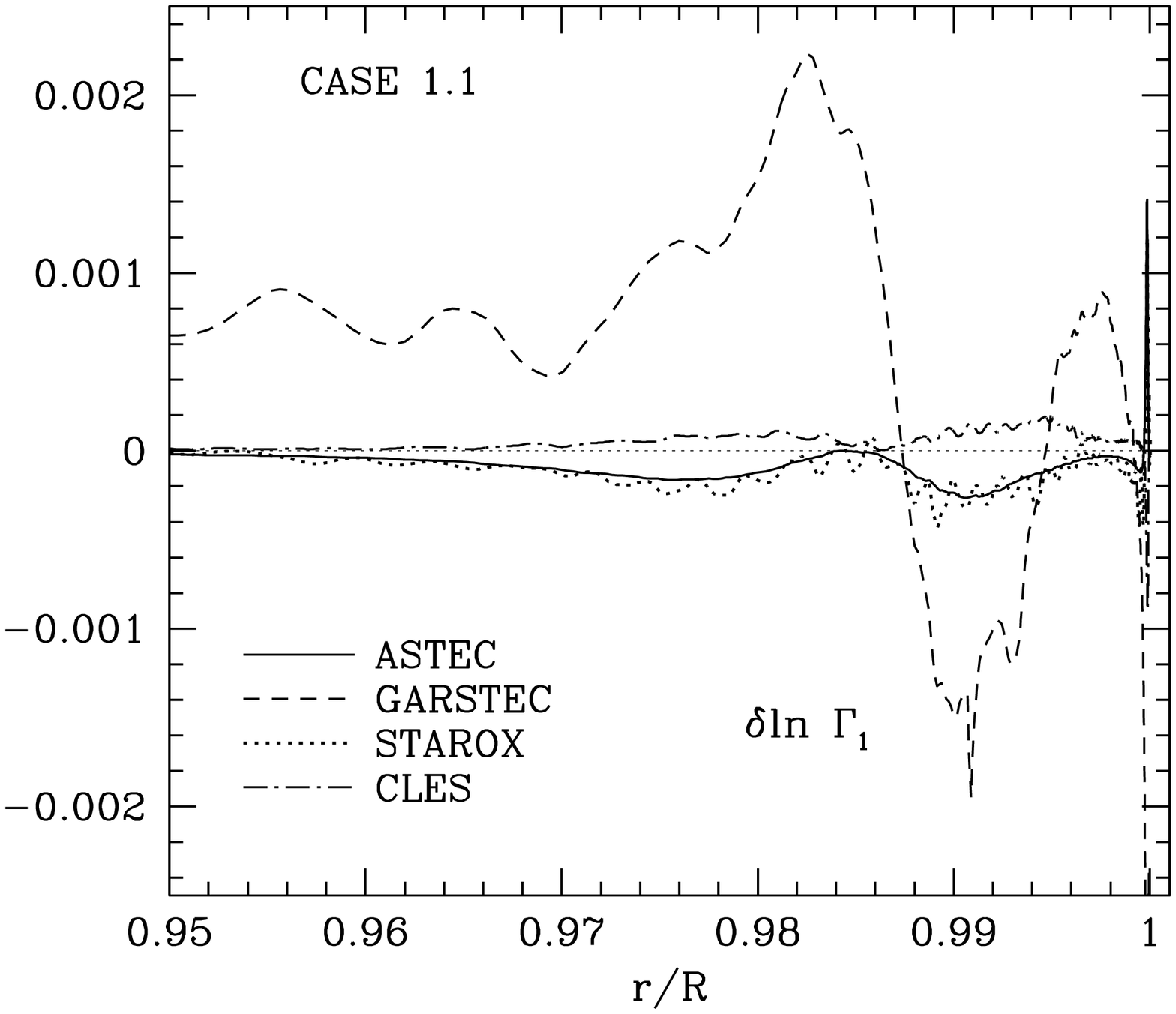}}
{\includegraphics[scale=0.25]{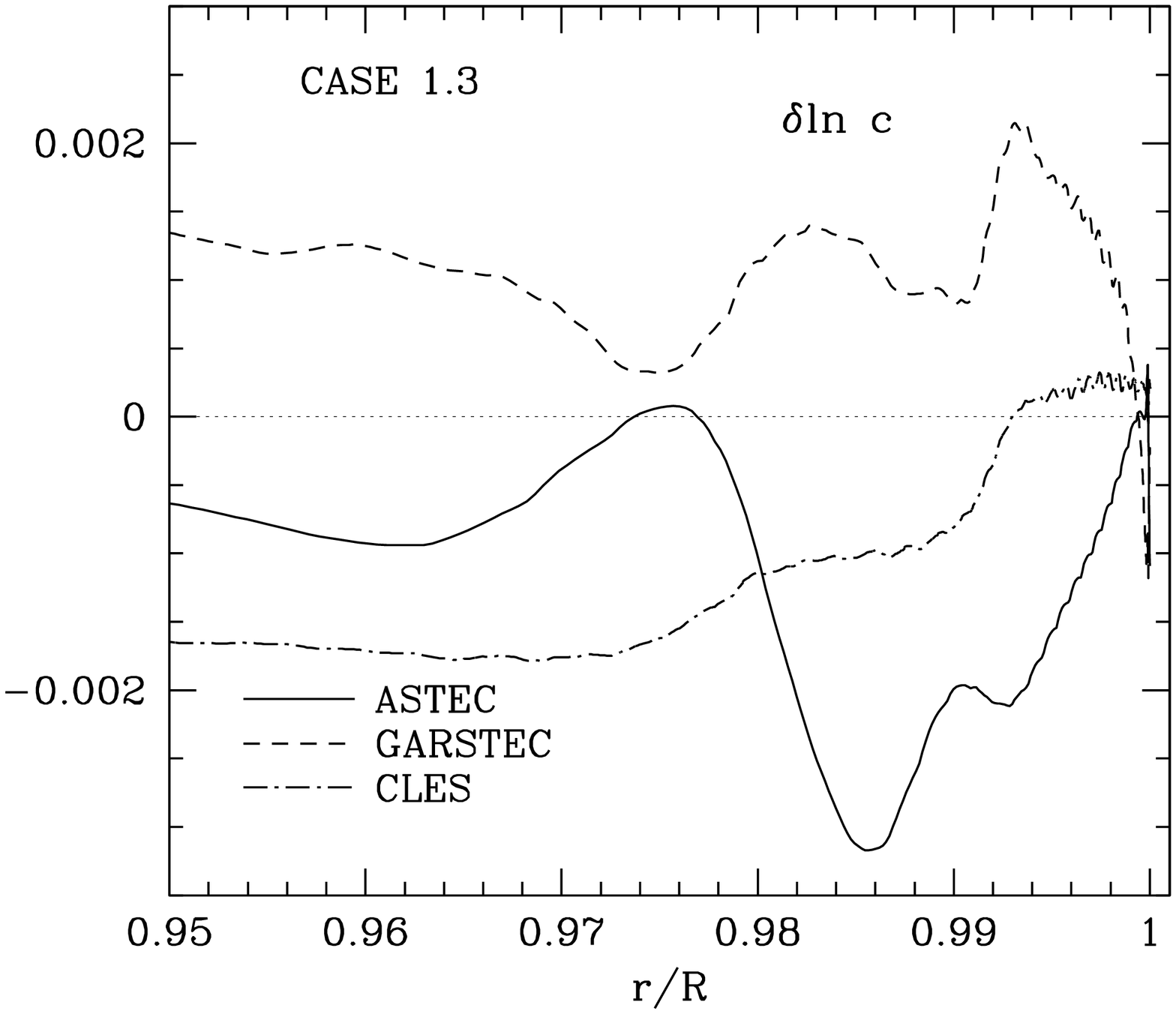}
 \includegraphics[scale=0.25]{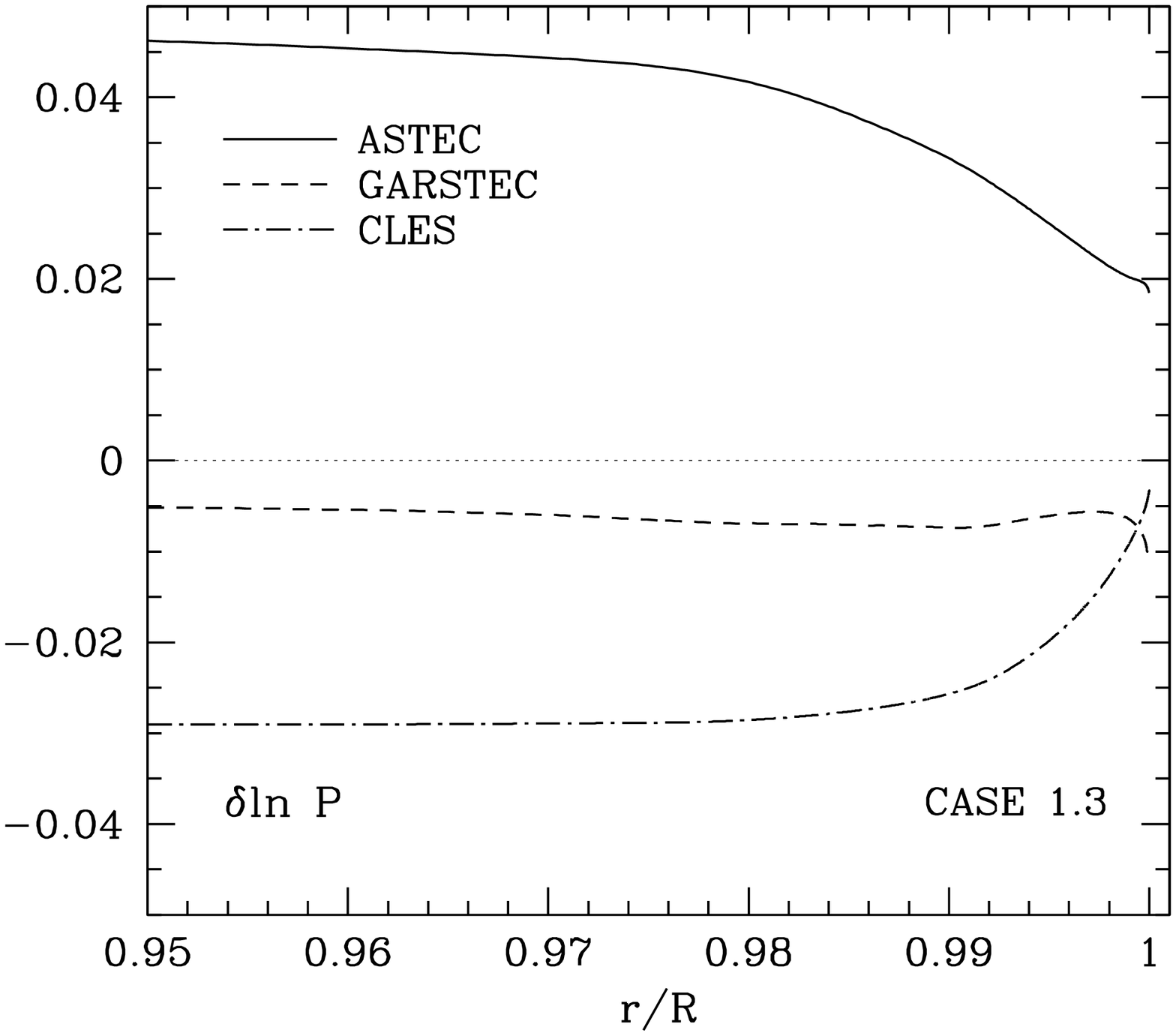}
 \includegraphics[scale=0.25]{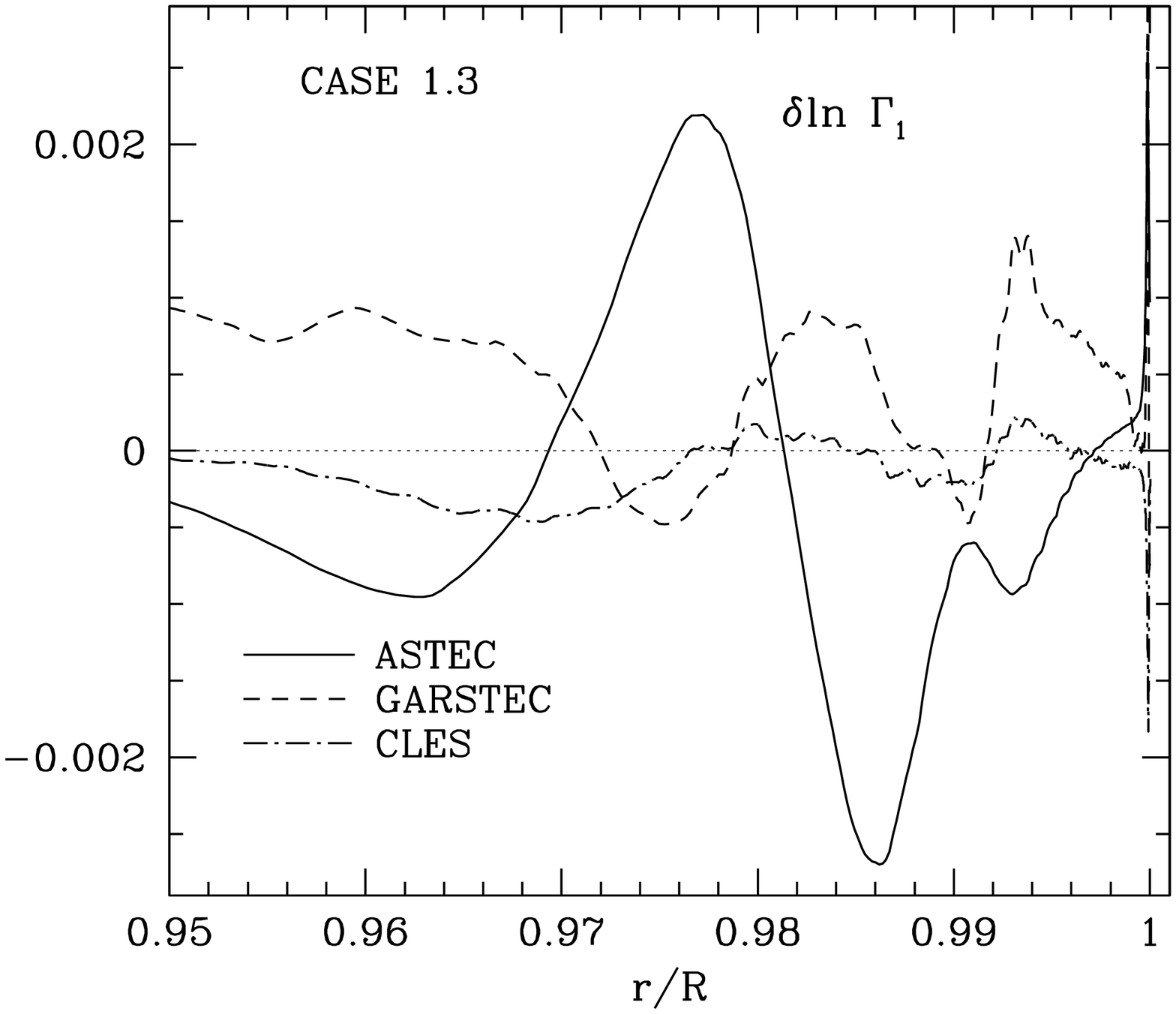}}
{\includegraphics[scale=0.25]{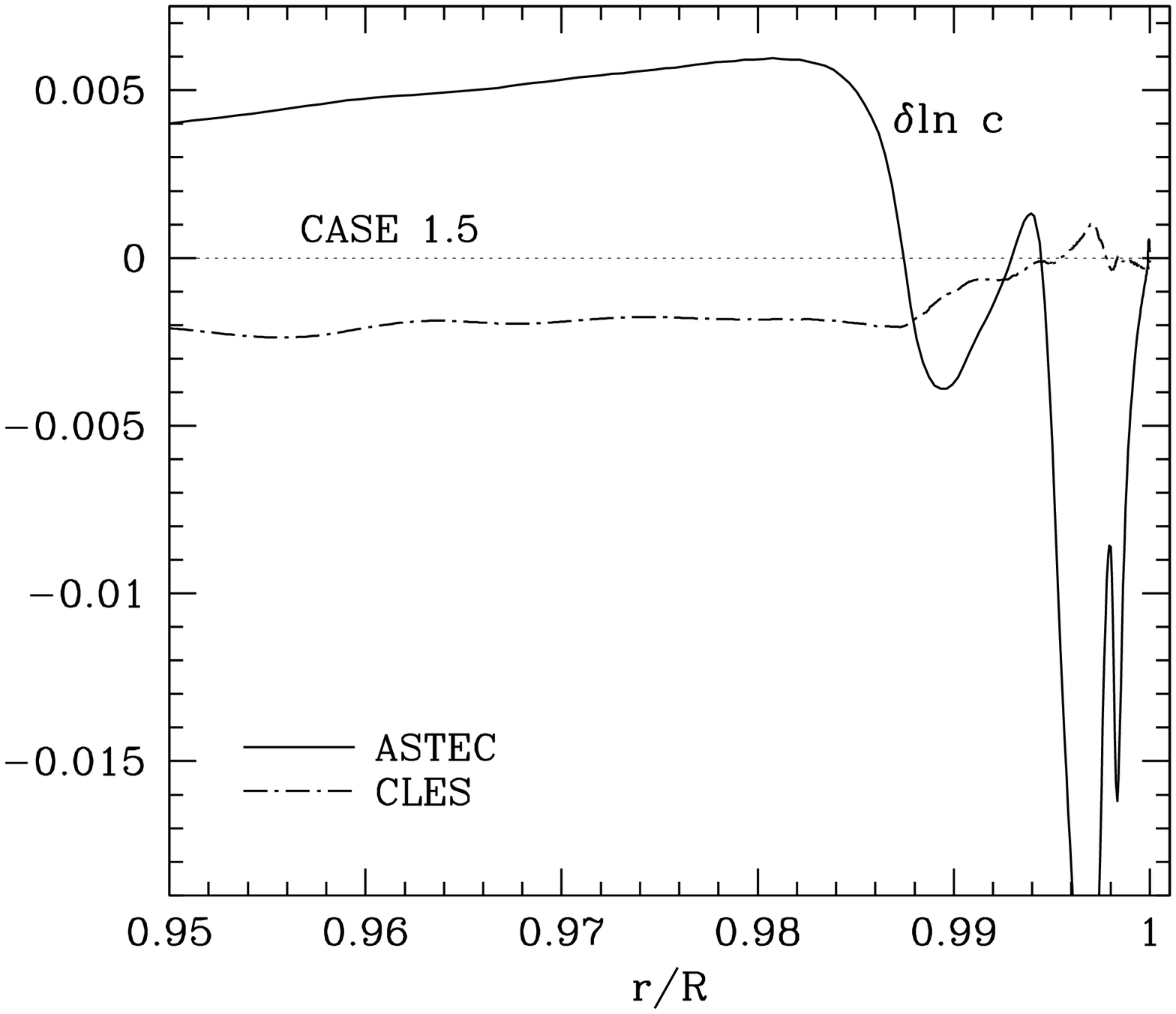}
 \includegraphics[scale=0.25]{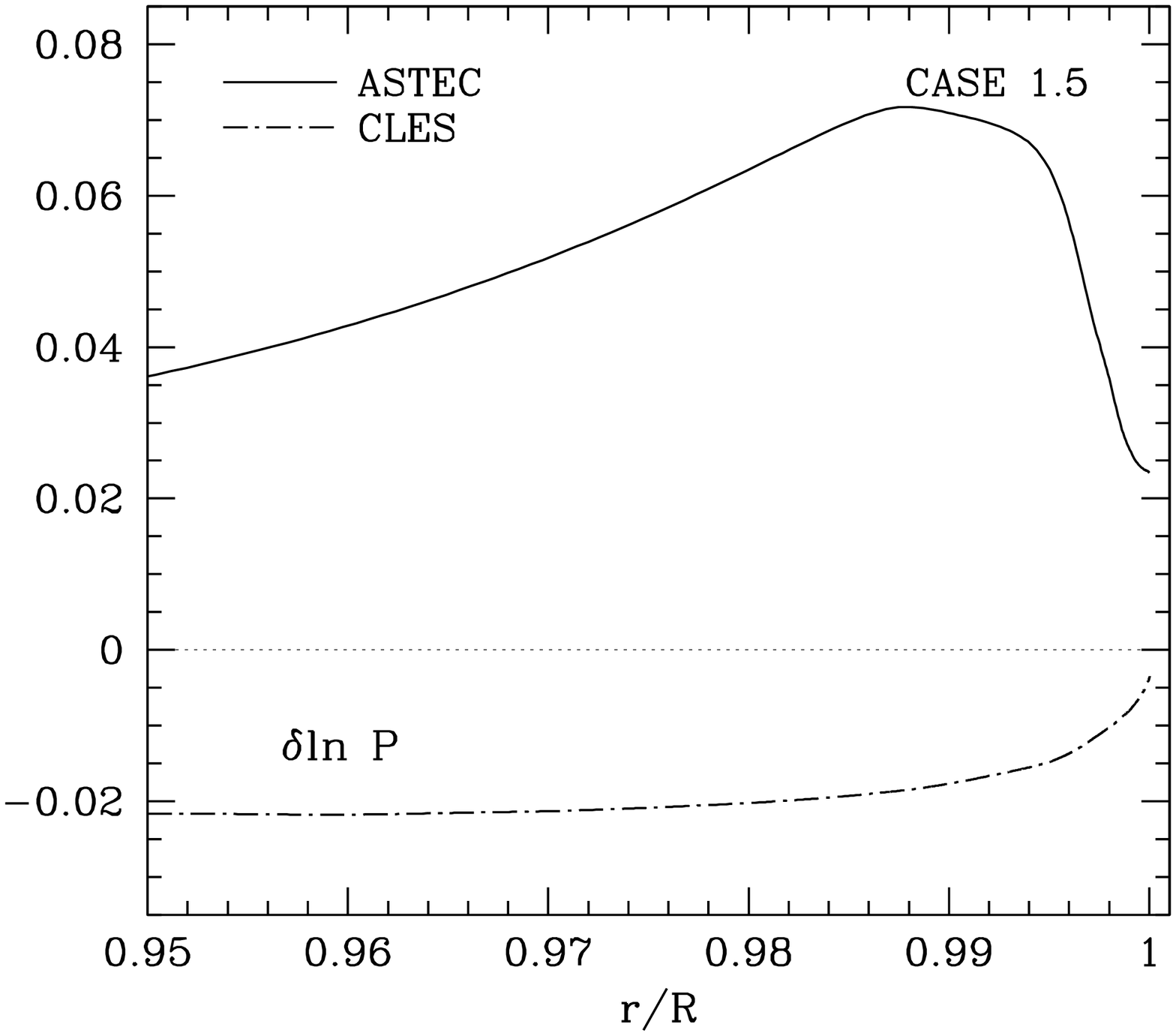}
 \includegraphics[scale=0.25]{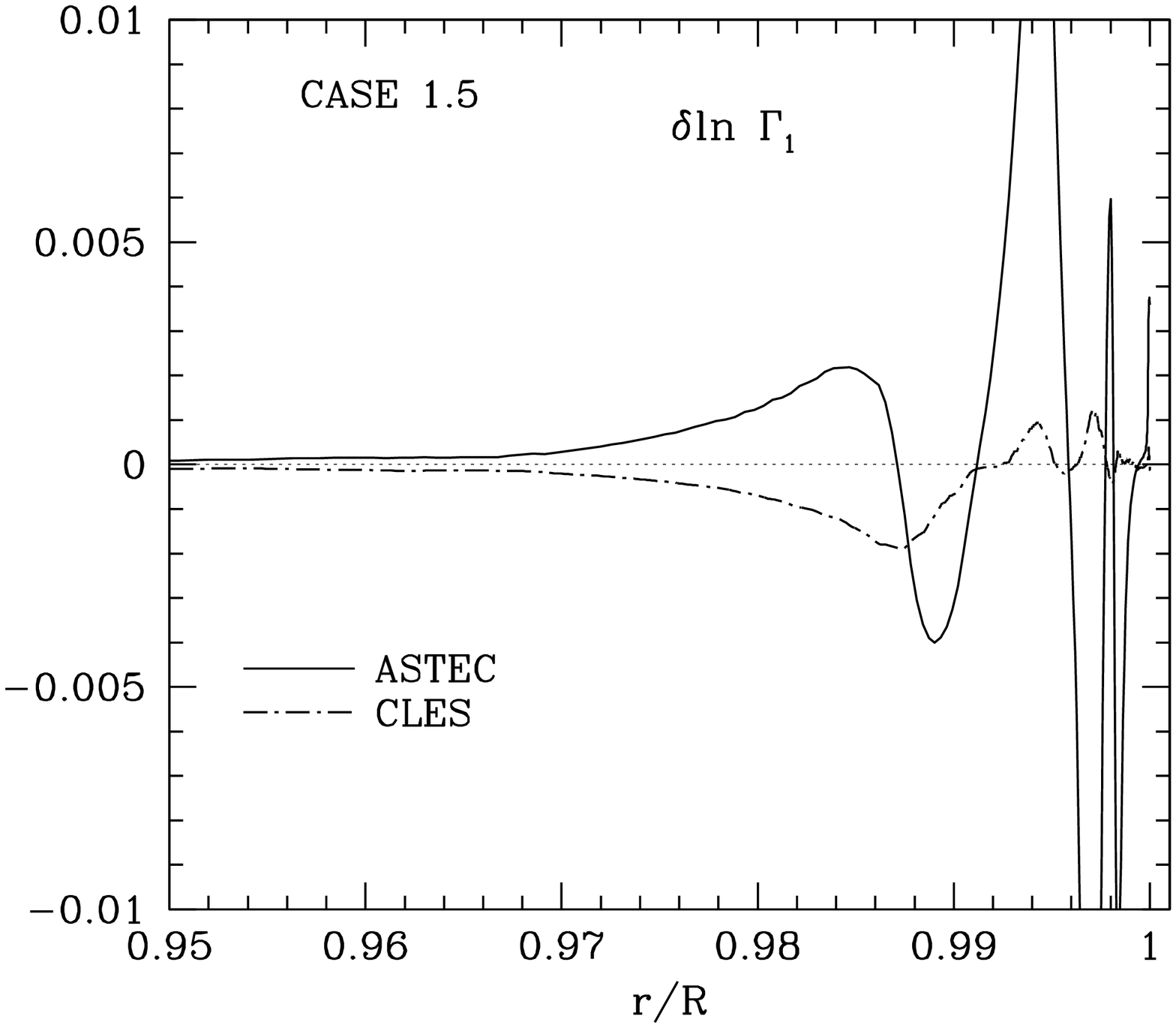}}
\caption{{\task}~1. Plots of the logarithmic differences calculated at fixed relative radius between pairs of models ({\small CODE}-\cesam) for the outer regions of  Case~1.1, 1.3 and 1.5 models. {\it Left panel}:  sound speed differences. {\it Central panel}:  pressure differences. {\it Right panel}: adiabatic exponent $\Gamma_1$ differences. Horizontal dotted line represents the reference model (\cesam).}
\label{fig:task1-extern135}
\end{figure*}
%==========================
%\clearpage
%==========================

\subsubsection{High mass models: Cases~1.6 and 1.7}

For the more massive models (Cases~1.6 and 1.7) the agreement between the 5 codes is generally quite good. In the {\ZAMS} model (Case~1.6, Fig.~\ref{fig:task1-int67}) only a spike  in $\delta X$ is found at the convective core boundary ($r\sim0.16R$). Again, we can see a plateau of $\delta X$ above the convective core boundary in the {\astec} models which probably results from the fact that it assumes 
the CN part of the CNO cycle to be in nuclear equilibrium at all times. We can also note that the model provided by {\garstec} corresponds to a model slightly more evolved than specified, with  $X_{\rm c}=0.6897$ instead of $0.69$. In the model in the middle of its {\MS} (Case~1.7), the features seen in the differences are similar to those in Case~1.5 models,  the largest differences being concentrated in the $\nabla_\mu$-region  left by the shrinking convective core.

\subsection{External layers}

The variations of $c$, $\rho$, and $\Gamma_1$ at fixed radius in the most external layers of the models are plotted for selected cases in Fig.~\ref{fig:task1-extern135}. As we shall show in Sect.~\ref{sec:task1-osc} these differences play an important role in the p-mode frequency variations.

\subsection{Convection regions and ionisation zones}\label{sec:task1-conv}

The location and evolution with time of the convective regions are essential elements in seismology. Rapid changes in the sound speed, like those arising at the boundary of a convective region, introduce a periodic signature in the oscillation frequencies of low-degree modes \citep{gough90} that in turn can be used to  derive the location of convective boundaries.  In addition, the location and displacement of the convective core edge leave a chemical composition gradient that affects the sound speed and the Brunt-V\"ais\"al\"a frequency and hence the frequencies of g and p-g mixed modes.

For each model considered we looked for the location of the borders of the convective regions by searching the zeros of the quantity $A=N^2_{\rm BV} r/g$. The results, expressed in relative radius and in acoustic depth, are collected in Table~\ref{tabla:conv}. Since variations of the adiabatic parameter $\Gamma_1$ can also introduce periodic signals in the oscillation frequencies, we also display in Table~\ref{tabla:conv} the values of the relative radius and acoustic depth of  the second He-ionisation region that were determined
by locating a minimum in  $\Gamma_1$.  We find a good agreement between the radii at the bottom of the convective envelope obtained  with the 5 codes. The dispersion in the values is smaller than 0.01\% for Cases~1.4, 1.6 and 1.7. It is of the order of 0.3\% for Cases~1.1, 1.2 and 1.5 while the largest dispersion (0.7\%) is found for Case~1.3. Concerning the mass of the convective core, the differences between codes increase as the stellar mass decrease: the differences are in the range 0.1--4\% for Case~1.7, 0.05--2\% for Case~1.6, 2.5--4\% for Case~1.5, 2.5 - 17\% for Case~1.4 and 3--30\% for Case~1.2. We point out that the convective core mass is larger in the Case~1.5 model provided by {\starox} which is due to the fact that {\starox} sets the temperature gradient to the radiative one in the overshooting region while the other codes take the adiabatic gradient.

\begin{figure*}[htbp!]
\centering
\resizebox{\hsize}{!}{\includegraphics[width=0.425\textwidth]{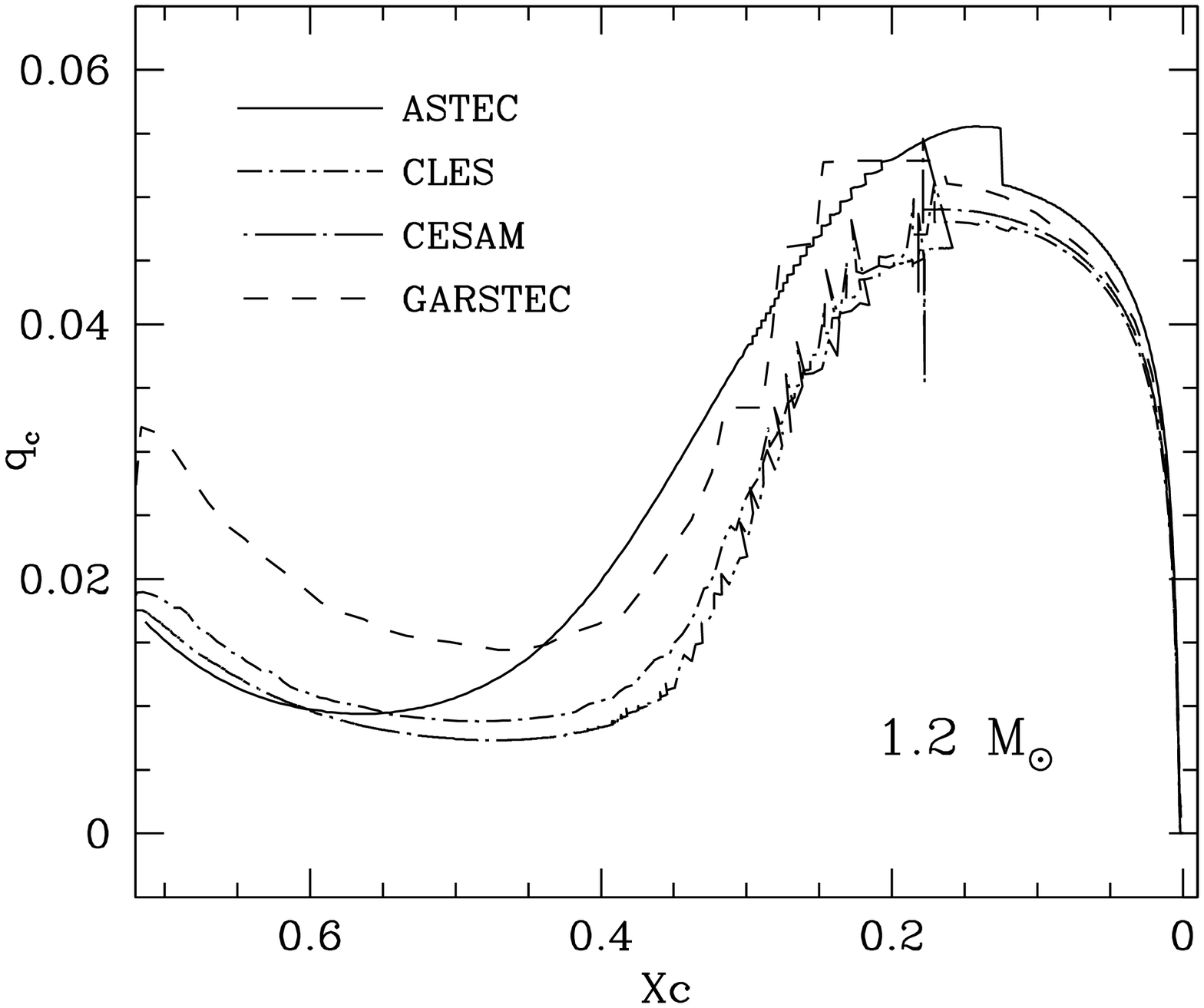}
                      \includegraphics[width=0.425\textwidth]{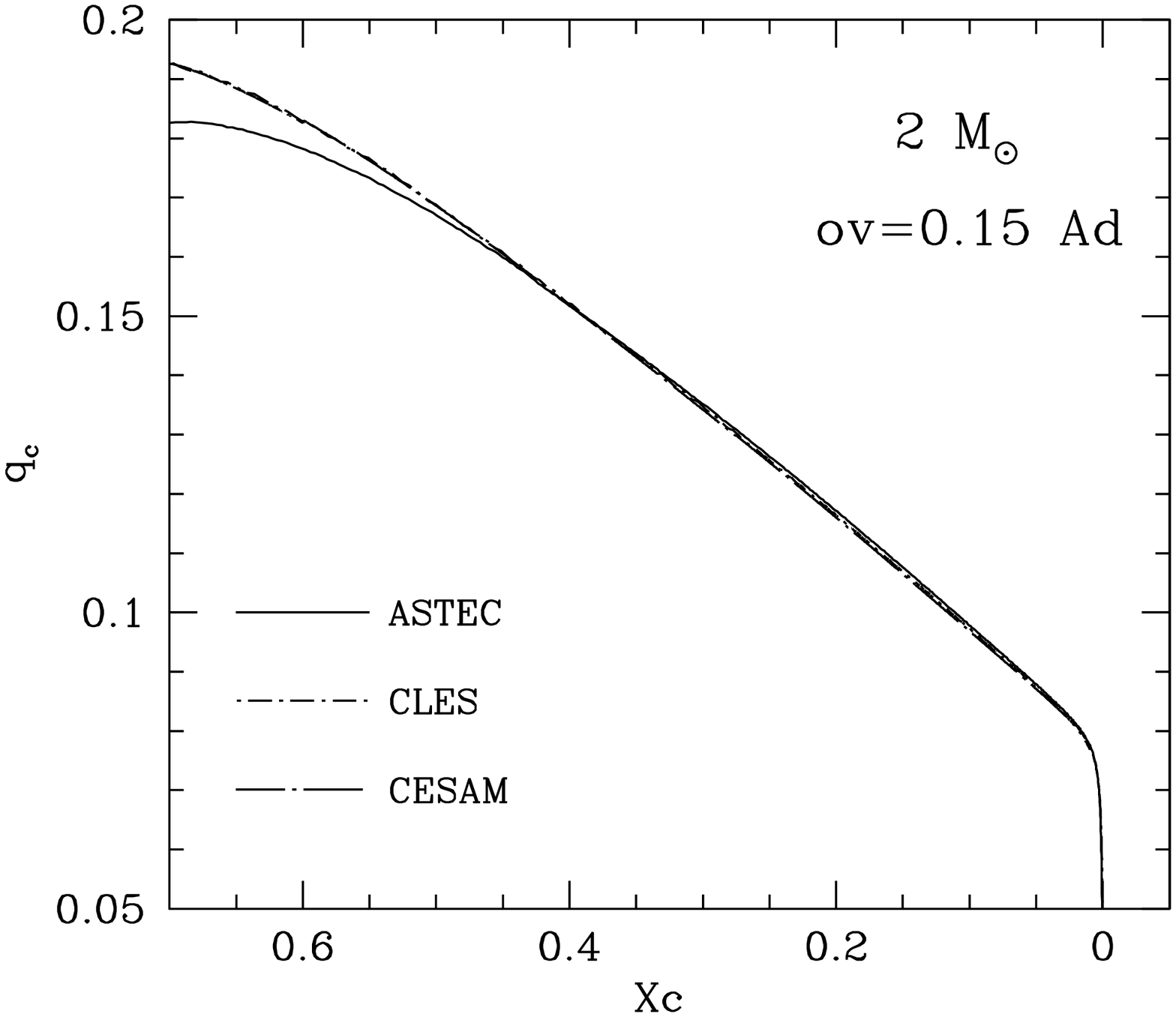}
                      \includegraphics[width=0.425\textwidth]{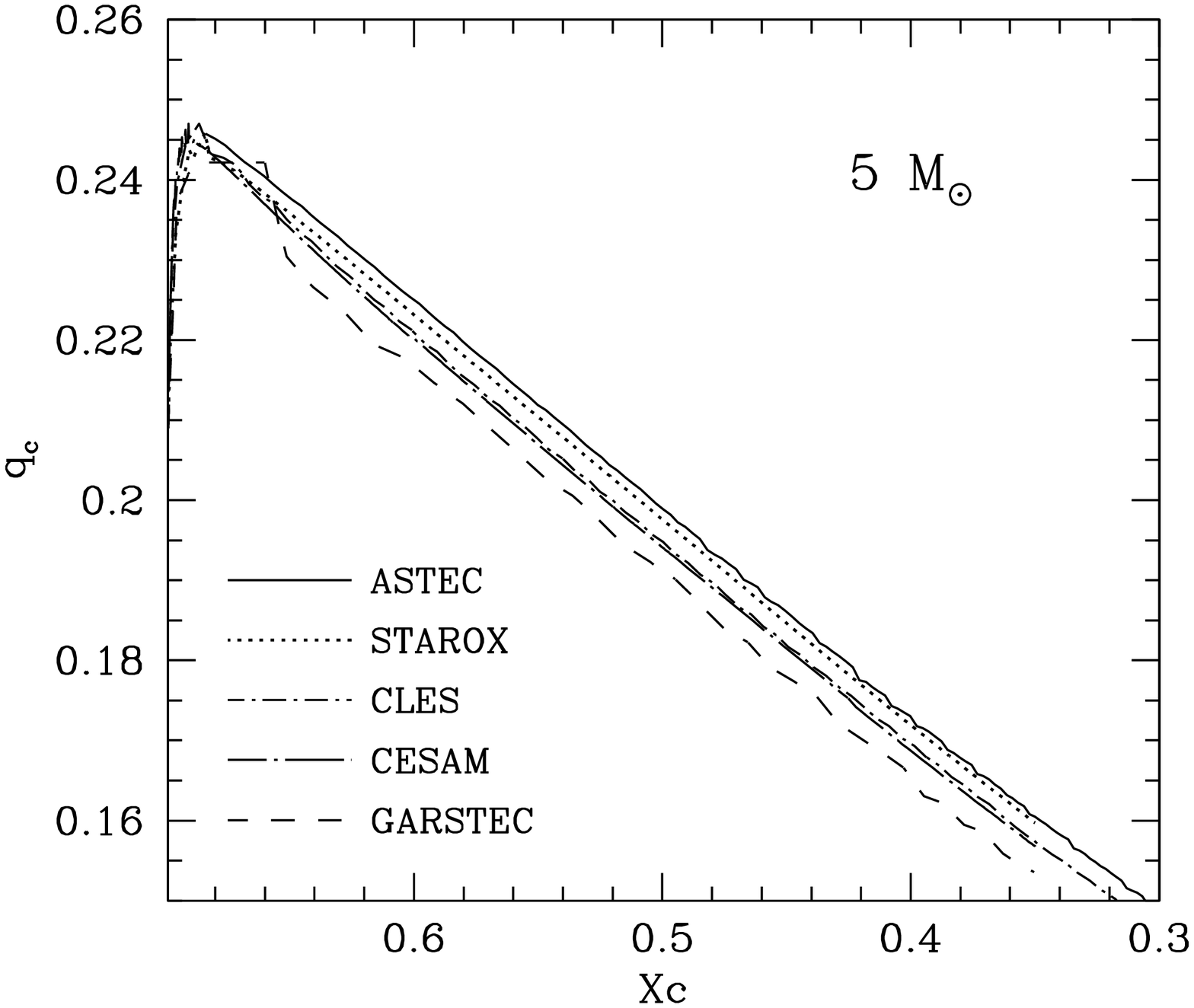}}
\caption{{\task}~1: Relative mass ($q_{\rm c}$) at the border of convective core as a function of the central hydrogen
mass fraction for Cases~1.3 (left), 1.5 (middle), and 1.7 (right).}    
\label{fig:cc}
\end{figure*}

We now focus on the models for Cases~1.3, 1.5 and 1.7. They illustrate the situations that can be found for the evolution of a convective core on the {\MS}. Case~1.3 deals with a $1.2${\msol} star which has a growing convective core during a large fraction of its {\MS}. Case~1.5 considers a $2${\msol} star for which the convective core is shrinking during the {\MS} and which undergoes nuclear reactions inside but also outside this core. Finally for the Case~1.7, which is for a 5{\msol} star, the convective core is shrinking on the {\MS} with nuclear reactions concentrated in the central region. In Fig.~\ref{fig:cc} we show, for the 3 cases,  the variation of the relative mass in the convective  core ($q_{\rm c}=m_{\rm cc}/M$)  as a function of the central H mass fraction (which decreases with evolution).

%==========================
\begin{table*}[htbp!]
\caption{{\task}~1: Features relevant for seismic analysis . Columns 1 and 2: acoustic radius $\tau_0$ of the model (at the photosphere) in seconds and acoustical cutoff frequency $\nu_{\rm ac}$ in $\mu$Hz. Columns 3 to 6: 
relative radius  $r_{\rm cz}/R$ at the bottom of the envelope convection zone(s) and corresponding acoustic depths $\tau_{\rm env}$ in seconds. Columns 7 to 9: relative mass $m_{\rm cc}/M$, radius $r_{\rm cc}/R$ and  acoustic depth $\tau_{\rm cc}$ of the convective core. Columns 10 and 11: relative radius of the second He-ionisation region and acoustic depth $\tau_{\rm HeII}$.}
\centering
\label{tab:task1-sismo}
\begin{tabular}[h]{lccccccccccc}
\hline\noalign{\smallskip}
&1&2&3&4&5&6&7&8&9&10&11\\
\noalign{\smallskip}\hline
\hline\noalign{\smallskip}
\multicolumn{12}{c} {\bf Case~1.1}\\
\hline\noalign{\smallskip}
{\bf Code} & \boldmath$\tau_0$ & \boldmath$\nu_{\rm ac}$ 
           & \boldmath$r_{\rm cz}/R$ & \boldmath$\tau_{\rm env}$ 
           & \boldmath$r_{\rm cz}/R$ & \boldmath$\tau_{\rm env}$ 
           & \boldmath$m_{\rm cc}/M$ & \boldmath$r_{\rm cc}/R$ & \boldmath$\tau_{\rm cc}$ & \boldmath$r_{\rm HeII}$
           & \boldmath$\tau_{\rm HeII}$\\[3pt]
\tableheadseprule\noalign{\smallskip}
 \astec  & 3134 & 5356 & ---- & ---- & 0.6985 & 1904 & ---- & ---- & ---- & 0.9808 &  531 \\ 
 \cesam  & 3128 & 5370 & ---- & ---- & 0.6959 & 1907 & ---- & ---- & ---- & 0.9807 &  531 \\ 
 \cles  & 3124 & 5379 & ---- & ---- & 0.6959 & 1905  & ---- & ---- & ---- & 0.9806 &  533 \\ 
 \garstec  & 3107 & 5401 & ---- & ----  & 0.6980 & 1889 & ---- & ---- & ---- & 0.9806 &  529 \\ 
 \starox  & 3135 & 5356 & ---- & ---- & 0.6972 & 1908 & ---- & ---- & ---- & 0.9806 &  533 \\ 
\noalign{\smallskip}\hline
%==========================
%==========================
\hline\noalign{\smallskip}
\multicolumn{12}{c} {\bf Case~1.2}\\
\hline\noalign{\smallskip}
{\bf Code} & \boldmath$\tau_0$ (s)& \boldmath$\nu_{\rm ac}$ (s)
           & \boldmath$r_{\rm cz}/R$ & \boldmath$\tau_{\rm env}$ (s)
           & \boldmath$r_{\rm cz}/R$ & \boldmath$\tau_{\rm env}$ (s)
           & \boldmath$m_{\rm cc}/M$ & \boldmath$r_{\rm cc}/R$ & \boldmath$\tau_{\rm cc}$ & \boldmath$r_{\rm HeII}$
           & \boldmath$\tau_{\rm HeII}$ (s)\\[3pt]
\tableheadseprule\noalign{\smallskip}
 \astec  & 3995 & 3993 & ---- & ---- & 0.8307 & 1832 & 1.0148$\times10^{-2}$ & 0.0512 & 3926 & 0.9839 &  577 \\ 
 \cesam  & 3976 & 4021 & ---- & ---- & 0.8281 & 1836 & 8.4785$\times10^{-3}$ & 0.0484 & 3910 & 0.9839 &  576 \\ 
 \cles  & 3969 & 4030 & ---- & ---- & 0.8285 & 1831  & 8.8180$\times10^{-3}$ & 0.0491 & 3902 & 0.9838 &  576 \\
 \garstec  & 3960 & 4028 & ---- & ---- & 0.8283 & 1829  & 1.1026$\times10^{-2}$ & 0.0531 & 3888 & 0.9840 &  572 \\
  \starox  & 3987 & 4009 & ---- & ---- & 0.8299 & 1832 & 7.6050$\times10^{-3}$ & 0.0465 & 3924 & 0.9839 &  576 \\ 
\noalign{\smallskip}\hline
%==========================
%==========================
\hline\noalign{\smallskip}
\multicolumn{12}{c} {\bf Case~1.3}\\
\hline\noalign{\smallskip}
{\bf Code} & \boldmath$\tau_0$ (s)& \boldmath$\nu_{\rm ac}$ (s)
           & \boldmath$r_{\rm cz}/R$ & \boldmath$\tau_{\rm env}$ (s)
           & \boldmath$r_{\rm cz}/R$ & \boldmath$\tau_{\rm env}$ (s)
           & \boldmath$m_{\rm cc}/M$ & \boldmath$r_{\rm cc}/R$ & \boldmath$\tau_{\rm cc}$ & \boldmath$r_{\rm HeII}$
           & \boldmath$\tau_{\rm HeII}$ (s)\\[3pt]
\tableheadseprule\noalign{\smallskip}
 \astec  & 9915 & 1136 & ---- & ---- & 0.7816 & 5244 & ---- & ---- & ---- & 0.9726 & 1868 \\ 
 \cesam  & 9922 & 1134 & ---- & ---- & 0.7844 & 5211  & ---- & ---- & ---- & 0.9726 & 1867 \\ 
 \cles  & 9971 & 1126 & ---- & ---- & 0.7860 & 5218 & ---- & ---- & ---- & 0.9725 & 1874 \\ 
  \garstec  & 9885 & 1139 & ---- & ---- & 0.7873 & 5159  & ---- & ---- & ---- & 0.9728 & 1850 \\ 
\noalign{\smallskip}\hline
%==========================
%==========================
\hline\noalign{\smallskip}
\multicolumn{12}{c} {\bf Case~1.4}\\
\hline\noalign{\smallskip}
{\bf Code} & \boldmath$\tau_0$ (s)& \boldmath$\nu_{\rm ac}$ (s)
           & \boldmath$r_{\rm cz}/R$ & \boldmath$\tau_{\rm env}$ (s)
           & \boldmath$r_{\rm cz}/R$ & \boldmath$\tau_{\rm env}$ (s)
           & \boldmath$m_{\rm cc}/M$ & \boldmath$r_{\rm cc}/R$ & \boldmath$\tau_{\rm cc}$ & \boldmath$r_{\rm HeII}$
           & \boldmath$\tau_{\rm HeII}$ (s)\\[3pt]
\tableheadseprule\noalign{\smallskip}
 \cesam  & 7012 & 1798 & 0.9946 &  329 & 0.9916 &  465 & 9.4057$\times10^{-2}$ & 0.0982 & 6809 & 0.9931 &  398 \\ 
 \cles  & 7000 & 1801 & 0.9946 &  327 & 0.9916 &  465 & 9.8622$\times10^{-2}$ & 0.1003 & 6793 & 0.9931 &  400 \\ 
 \garstec  & 6990 & 1788 & 0.9946 &  328 & 0.9916 &  463 & 9.1552$\times10^{-2}$ & 0.0972 & 6791 & 0.9698 & 400 \\ 
 \starox  & 6956 & 1826 & 0.9947 &  321 & 0.9917 &  457 & 1.0767$\times10^{-1}$ & 0.1044 & 6741 & 0.9932 &  389 \\ 
 \noalign{\smallskip}\hline
%==========================
%==========================
\hline\noalign{\smallskip}
\multicolumn{12}{c} {\bf Case~1.5}\\
\hline\noalign{\smallskip}
{\bf Code} & \boldmath$\tau_0$ (s)& \boldmath$\nu_{\rm ac}$ (s)
           & \boldmath$r_{\rm cz}/R$ & \boldmath$\tau_{\rm env}$ (s)
           & \boldmath$r_{\rm cz}/R$ & \boldmath$\tau_{\rm env}$ (s)
           & \boldmath$m_{\rm cc}/M$ & \boldmath$r_{\rm cc}/R$ & \boldmath$\tau_{\rm cc}$ & \boldmath$r_{\rm HeII}$
           & \boldmath$\tau_{\rm HeII}$ (s)\\[3pt]
\tableheadseprule\noalign{\smallskip}
 \astec  & 17059 &  645 & ---- & ---- & 0.9873 & 1359 &  7.7371$\times10^{-2}$ & 0.03711 & 16880 & 0.9919 & 1004 \\ 
 \cesam  & 17052 &  644 & ---- & ---- & 0.9879 & 1305  &  7.6814$\times10^{-2}$ & 0.03692 & 16874 & 0.9919 &  994 \\ 
 \cles  & 17159 &  639 & ---- & ---- & 0.9880 & 1309  &  7.5622$\times10^{-2}$  & 0.03656 & 16982 & 0.9918 & 1002 \\ 
 \starox  & 17805 &  611 & ---- & ---- & 0.9855 & 1575  &  7.9887$\times10^{-2}$ & 0.03635 & 17624 & 0.9911 & 1128 \\ 
\noalign{\smallskip}\hline
%==========================
%==========================
\hline\noalign{\smallskip}
\multicolumn{12}{c} {\bf Case~1.6}\\
\hline\noalign{\smallskip}
{\bf Code} & \boldmath$\tau_0$ (s) & \boldmath$\nu_{\rm ac}$ (s)
           & \boldmath$r_{\rm cz}/R$ & \boldmath$\tau_{\rm env}$ (s)
           & \boldmath$r_{\rm cz}/R$ & \boldmath$\tau_{\rm env}$ (s)
           & \boldmath$m_{\rm cc}/M$ & \boldmath$r_{\rm cc}/R$ & \boldmath$\tau_{\rm cc}$ & \boldmath$r_{\rm HeII}$
           & \boldmath$\tau_{\rm HeII}$ (s) \\[3pt]
\tableheadseprule\noalign{\smallskip}
 \astec  & 5848 & 1690 & 0.99897 &   82 & 0.99392 &  343 & 2.1263$\times10^{-1}$ & 0.1631 & 5548 & 0.9950 &  295 \\ 
 \cesam  & 5832 & 1696 & 0.99897 &   81 & 0.99393 &  342 & 2.0997$\times10^{-1}$ & 0.1624 & 5533 & 0.9950 &  297 \\ 
 \cles  & 5820 & 1700 & 0.99899 &   79 & 0.99392 &  341 & 2.1162$\times10^{-1}$ & 0.1632 & 5521 & 0.9950 &  296 \\ 
 \garstec  & 5878 & 1673 & 0.99896 &   80 & 0.99386 &  345 & 2.0774$\times10^{-1}$ & 0.1618 & 5579 & 0.9949 &  298 \\ 
 \starox  & 5831 & 1685 & 0.99898 &   81 & 0.99392 &  342 & 2.1177$\times10^{-1}$ & 0.1628 & 5532 & 0.9950 &  295 \\ 
\noalign{\smallskip}\hline
%==========================
%==========================
\hline\noalign{\smallskip}
\multicolumn{12}{c} {\bf Case~1.7}\\
\hline\noalign{\smallskip}
{\bf Code} & \boldmath$\tau_0$ (s) & \boldmath$\nu_{\rm ac}$ (s)
           & \boldmath$r_{\rm cz}/R$ & \boldmath$\tau_{\rm env}$ (s)
           & \boldmath$r_{\rm cz}/R$ & \boldmath$\tau_{\rm env}$ (s)
           & \boldmath$m_{\rm cc}/M$ & \boldmath$r_{\rm cc}/R$ & \boldmath$\tau_{\rm cc}$ & \boldmath$r_{\rm HeII}$
           & \boldmath$\tau_{\rm HeII}$ (s)\\[3pt]
\tableheadseprule\noalign{\smallskip}
 \astec  & 13546 &  556 & 0.99963 &   61 & 0.99291 &  807 & 1.5986$\times10^{-1}$ & 0.1098 & 13084 & 0.9944 &  668 \\ 
 \cesam  & 13383 &  565 & 0.99967 &   54 & 0.99297 &  794 & 1.5673$\times10^{-1}$ & 0.1096 & 12927 & 0.9945 &  659 \\ 
 \cles  & 13419 &  563 & 1.00000 &    0 & 0.99290 &  802 & 1.5642$\times10^{-1}$ & 0.1093 & 12964 & 0.9944 &  665 \\ 
 \garstec  & 13297 &  569 & 0.99967 &   51 & 0.99296 &  789 & 1.5286$\times10^{-1}$ & 0.1088 & 12847 & 0.9945 &  650 \\ 
 \starox  & 13454 &  562 & 0.99971 &   46 & 0.99294 &  799 & 1.5966$\times10^{-1}$ & 0.1100 & 12995 & 0.9945 &  654 \\ 
\noalign{\smallskip}\hline
%\hline
\end{tabular}
\label{tabla:conv}
\end{table*}
%==========================

%==========================
%==========================

For the most massive models (Cases~1.5 and 1.7), all the codes provide a similar evolution of the mass of the convective core, and the variations of $q_{\rm c}$ between them are in the range 0.5-5\% (corresponding to $\Delta m/M=2.10^{-4} - 7.10^{-3}$). We note that {\astec} behaves differently for the 2{\msol} model at the beginning of the {\MS} stage when $X_{\rm c} \gtrsim 0.5$. This is probably due to the fact that {\astec} does not include in the total energy the part coming from the nuclear reactions that transform ${}^{12}{\rm C}$ into ${}^{14}{\rm N}$.
%the reason is probably that {\astec} does not calculate the {\PMS} evolution.

Case~1.3 is the most problematic one. For a given chemical composition there is a range of stellar masses (typically between 1.1 and 1.6~{\msol}) where the convective core grows during a large part of the {\MS}. This generates a discontinuity in the chemical composition at its boundary and leads to the onset of semiconvection \citep[see e.g.][]{Gabriel77,Crowe82}. The crumple profiles of $q_{\rm c}$ in Fig.~\ref{fig:cc} (left) are the signature of  a semiconvection process that has not been adequately treated. In fact, none  of the codes participating in this comparison treat the semiconvection instability. The large difference between the {\astec} model curve and the {\cesam} and {\cles} ones results from the way the codes locate convective borders. While {\astec} searches these boundaries  downwards starting from the surface, {\cles} and {\cesam} search upwards beginning from the centre. We point out that semiconvection also appears below the convective envelope of these stars if  microscopic diffusion is included in the modelling \citep[see e.g.][ and Sec.~\ref{sec:task3} below]{Richard01}.

\subsection{Seismic properties}\label{sec:task1-osc}

\begin{figure*}[htbp!]
\centering
{\includegraphics[scale=0.33]{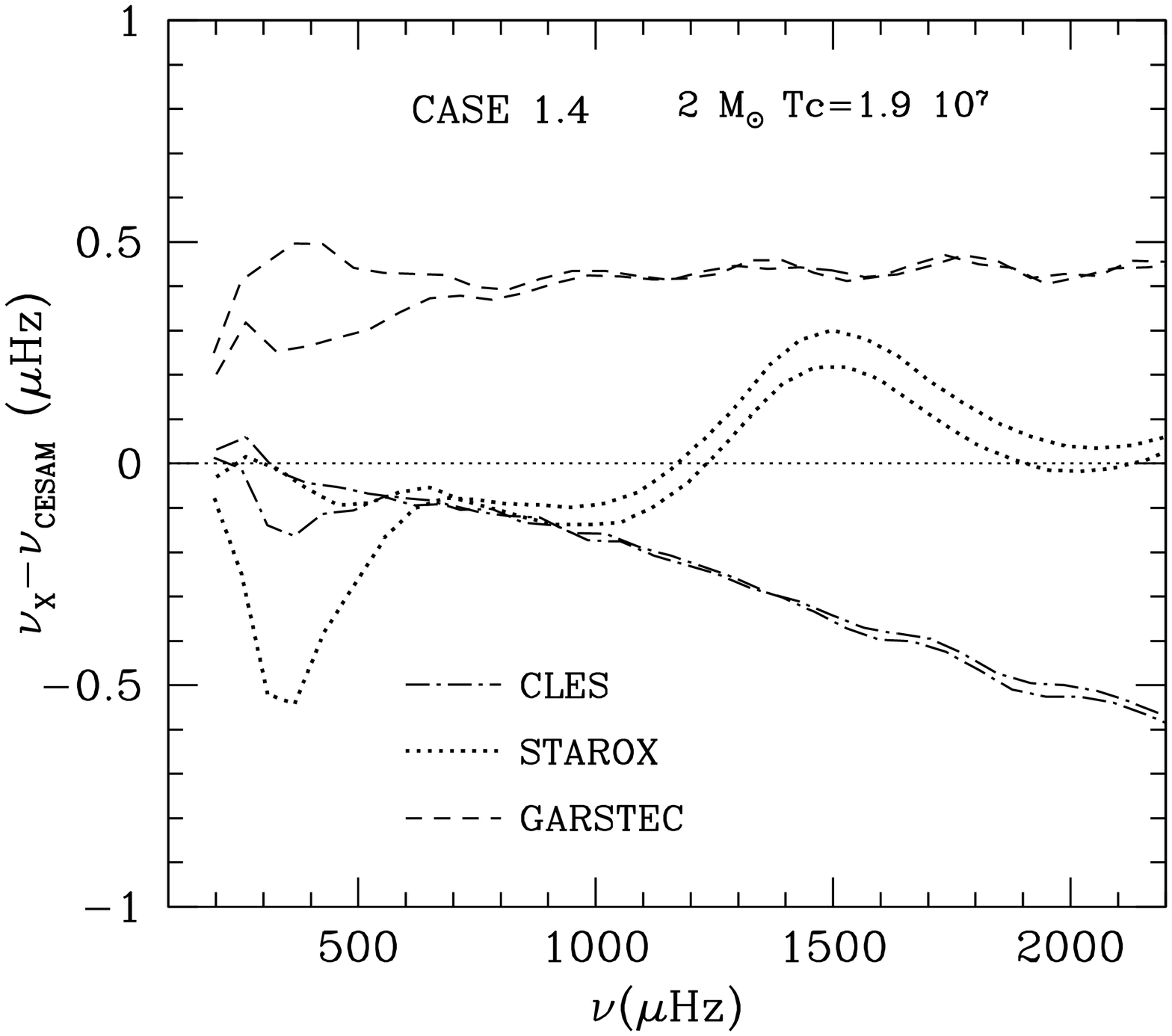}
 \includegraphics[scale=0.33]{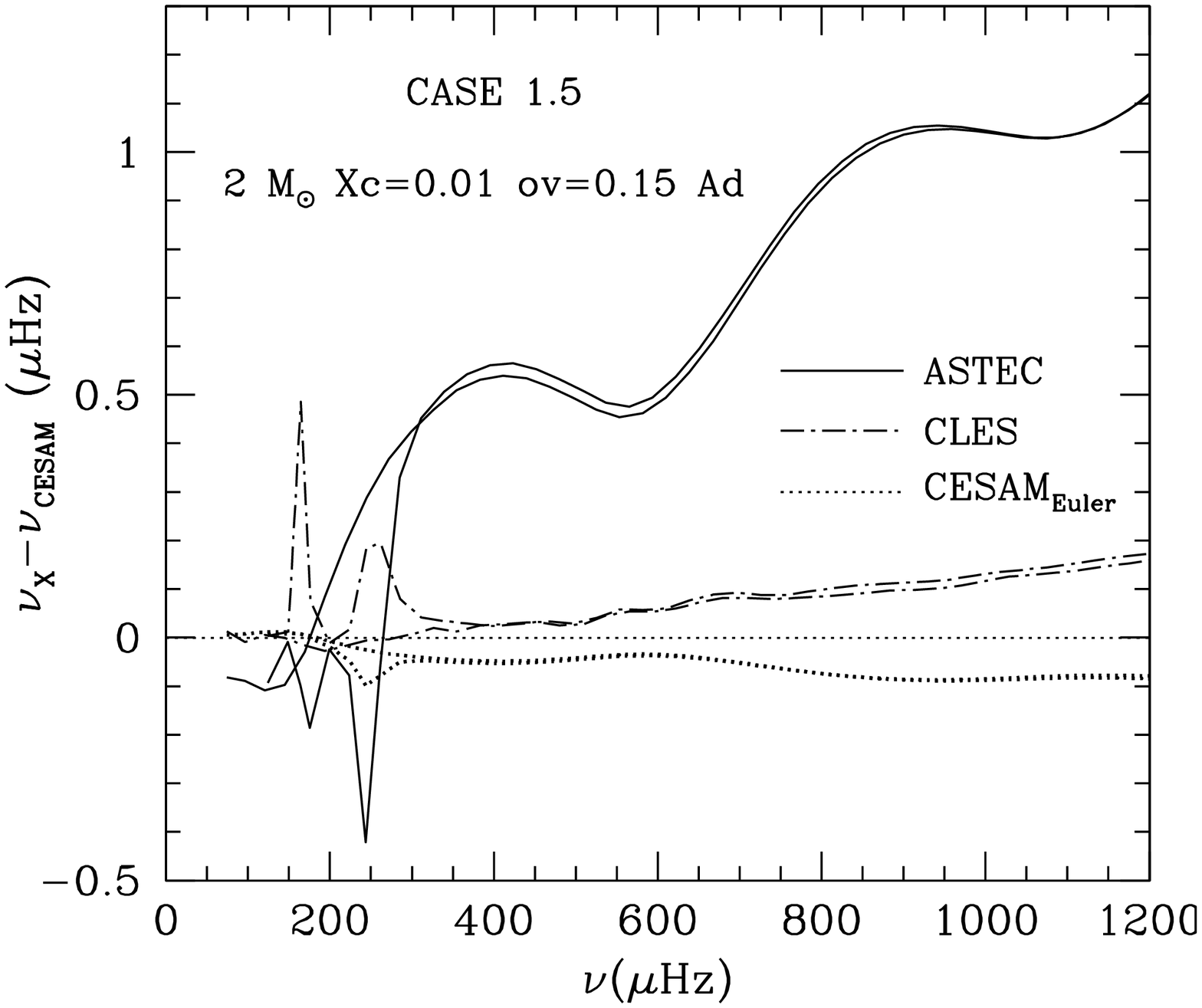}}
\caption{{\task}~1:  
p-mode frequency differences between models produced by different codes, for Case~1.4 (left) and 1.5 (right). {\cesam} model is taken as reference, and the frequencies have been scaled to remove the effect of different stellar radii. For each code, we plot two curves corresponding to modes with degrees $\ell=0$ and $\ell=1$. }       
\label{fig:freq45}
\end{figure*}

Using the adiabatic oscillation code {\small LOSC} we computed the oscillation frequencies of p and g modes with degree $\ell=0, 1, 2, 3$ and for frequencies in the range $\sigma=0.3-70/\tau_{\rm dyn}$ where $\sigma$ is the angular frequency and $\tau_{\rm dyn}=(R^3/GM)^{1/2}$ is the dynamical time. In these computations we used the standard option in \LOSC, that is, regularity of solution when $P=0$ at the surface ($\delta P/P+(4+\omega^2)\delta r/r=0$). The frequencies were computed on the basis of the model structure up to the photosphere (optical depth $\tau=2/3$).
When evaluating differences between different models they were scaled to correct for differences in stellar radius. The frequency differences $\nu_{\mbox{\scriptsize CODE}}-\nu_{\mbox{\scriptsize CESAM}}$ are displayed in Figs.~\ref{fig:freq12} to \ref{fig:freq67}. 

%==========================
\begin{table}[htbp!]
\caption{{\task}~1: Solar-like oscillations in Cases~1.1, 1.2 and 1.3: cutoff frequency at the photosphere $\nu_{\rm ac}$, frequency $\nu_{\rm max}$ expected at the maximum of the power spectrum and corresponding radial order $k_{\rm max}$,  and differences $\delta\nu(\ell=0)$ in the frequencies between the different codes}
\centering
\begin{tabular}[h]{lcccc}
\hline\noalign{\smallskip}
{\bf Case} & \boldmath$\nu_{\rm ac} (\mu$Hz) & \boldmath$\nu_{\rm max} (\mu$Hz) & \boldmath$k_{\rm max}$ & \boldmath$\delta\nu (\ell=0)\mu$Hz)\\[3pt]
\tableheadseprule\noalign{\smallskip}
{\bf 1.1} & 5400 & 3500 & 24 & 0.2--1 \\ 
{\bf 1.2} & 4000 & 2660 & 18 & 0.2--1 \\ 
{\bf 1.3} & 1100 & 770 & 8 & 0.05--0.2 \\ 
\noalign{\smallskip}\hline
%\hline
\end{tabular}
\label{tab:solarosc}
\end{table}

\subsubsection{Solar-like oscillations: Cases~1.1, 1.2 and 1.3}

On the basis of the \cite{1995A&A...293...87K} theory, we have estimated the frequency $\nu_{\rm max}$ at which we expect the maximum in the power spectrum. This value together with (1) the radial order corresponding to the  maximum ($k_{\rm max}$), (2) the acoustical cutoff frequency at the photosphere ($\nu_{\rm ac}=c/4\pi H_p$), and (3) the differences in the frequencies between different codes are collected in Table~\ref{tab:solarosc}.

%==========================
\begin{figure*}[htbp!]
\centering
\resizebox{\hsize}{!}{\includegraphics{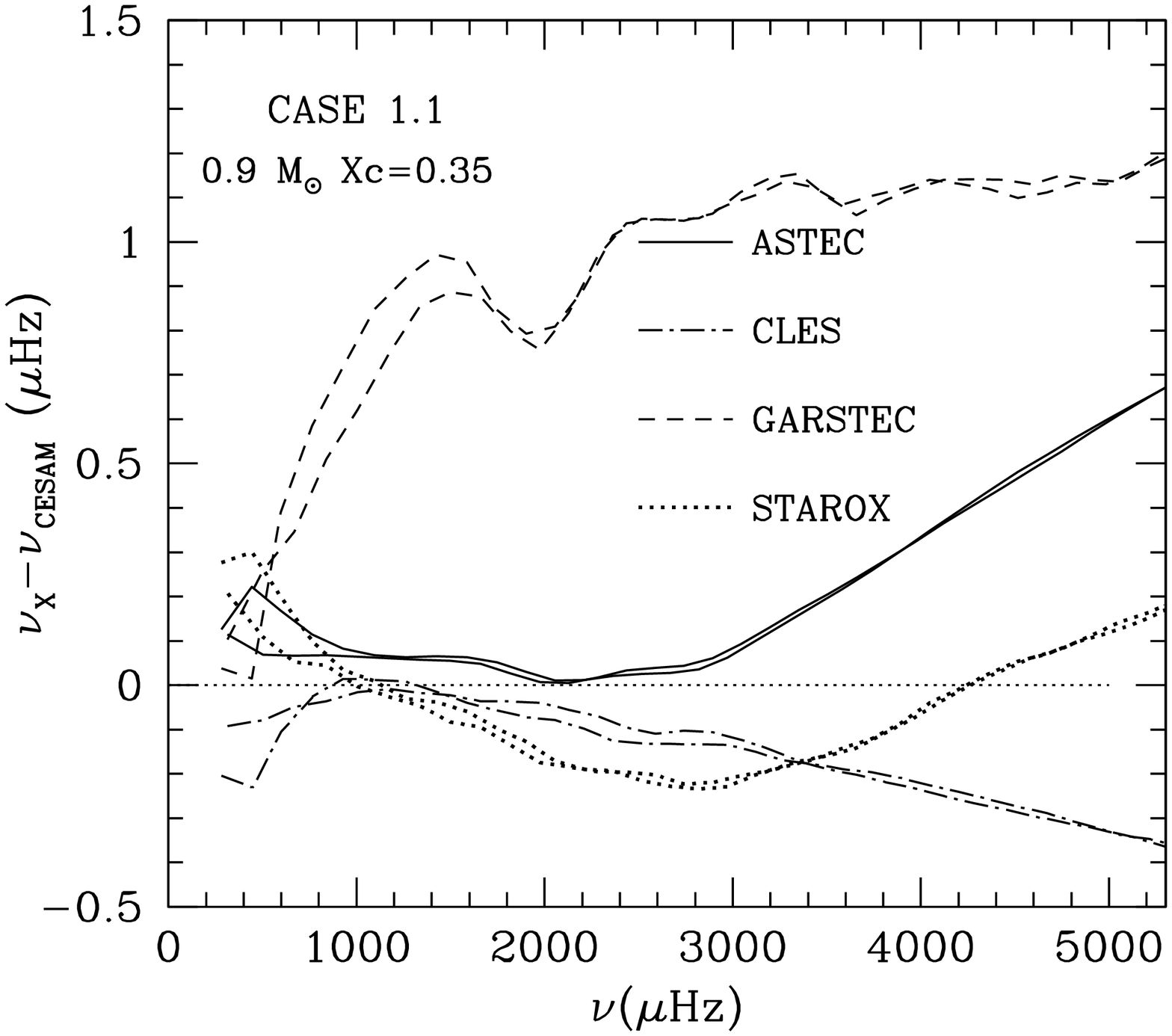}
                      \includegraphics{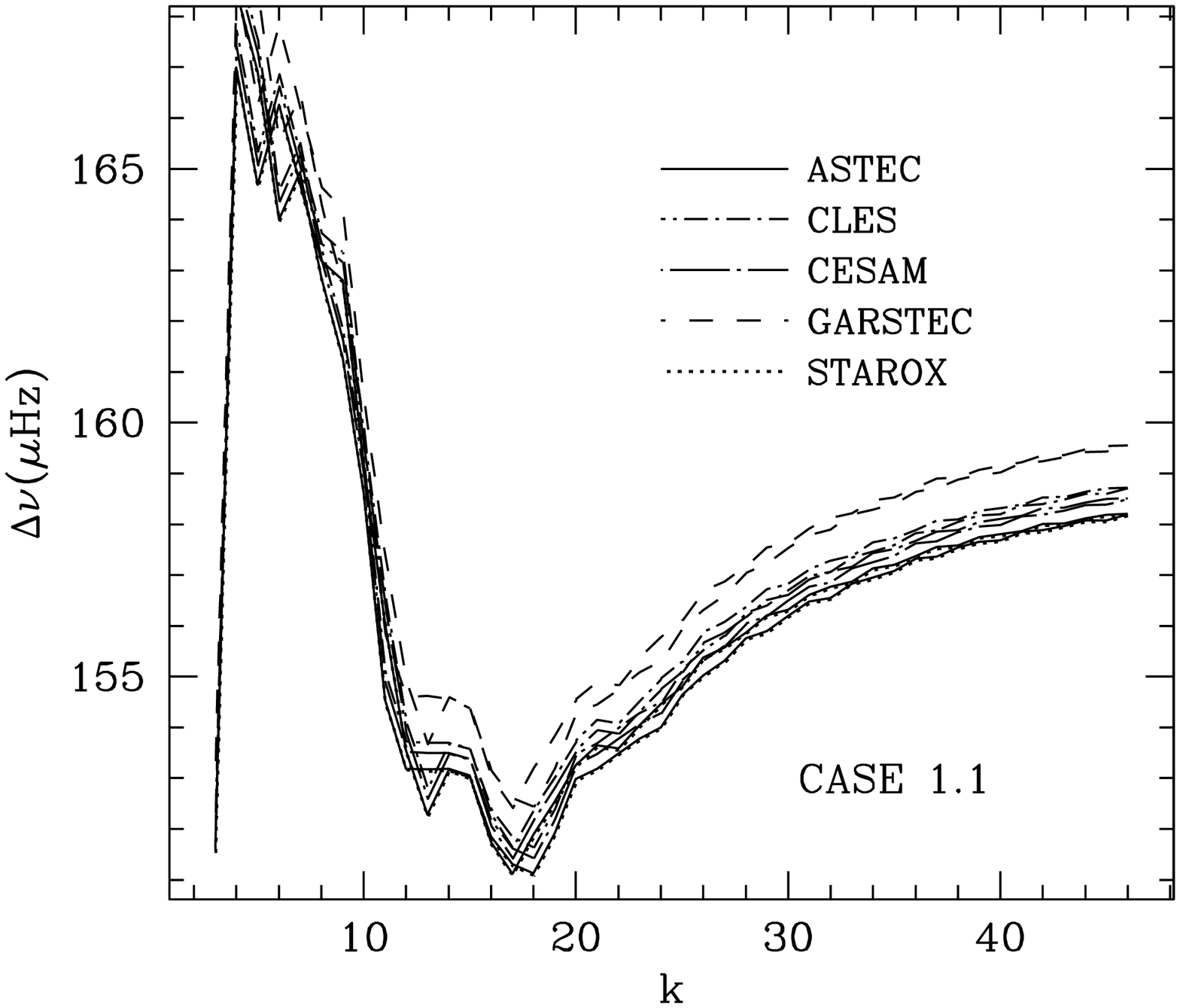}
                      \includegraphics{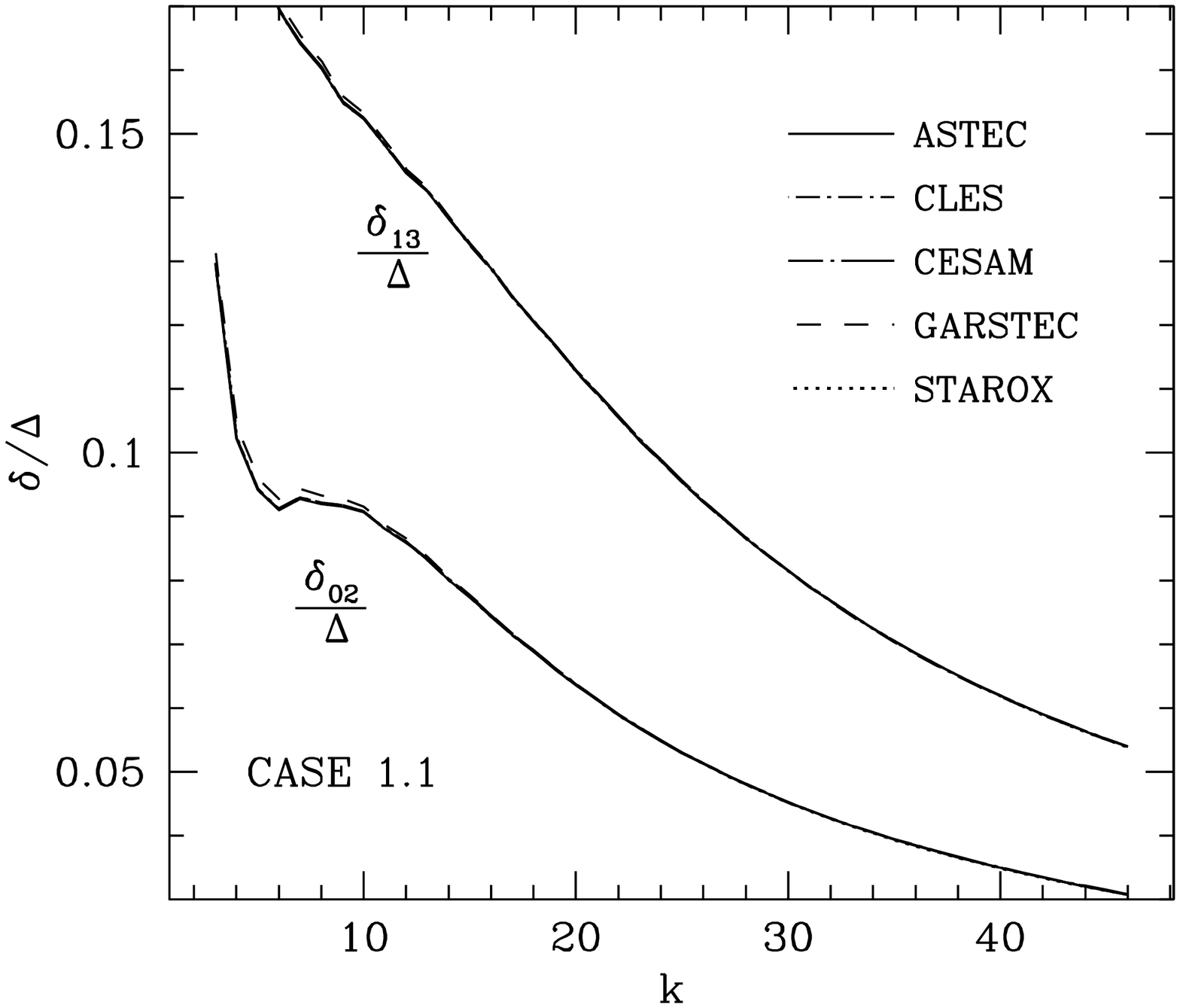}}
\resizebox{\hsize}{!}{\includegraphics{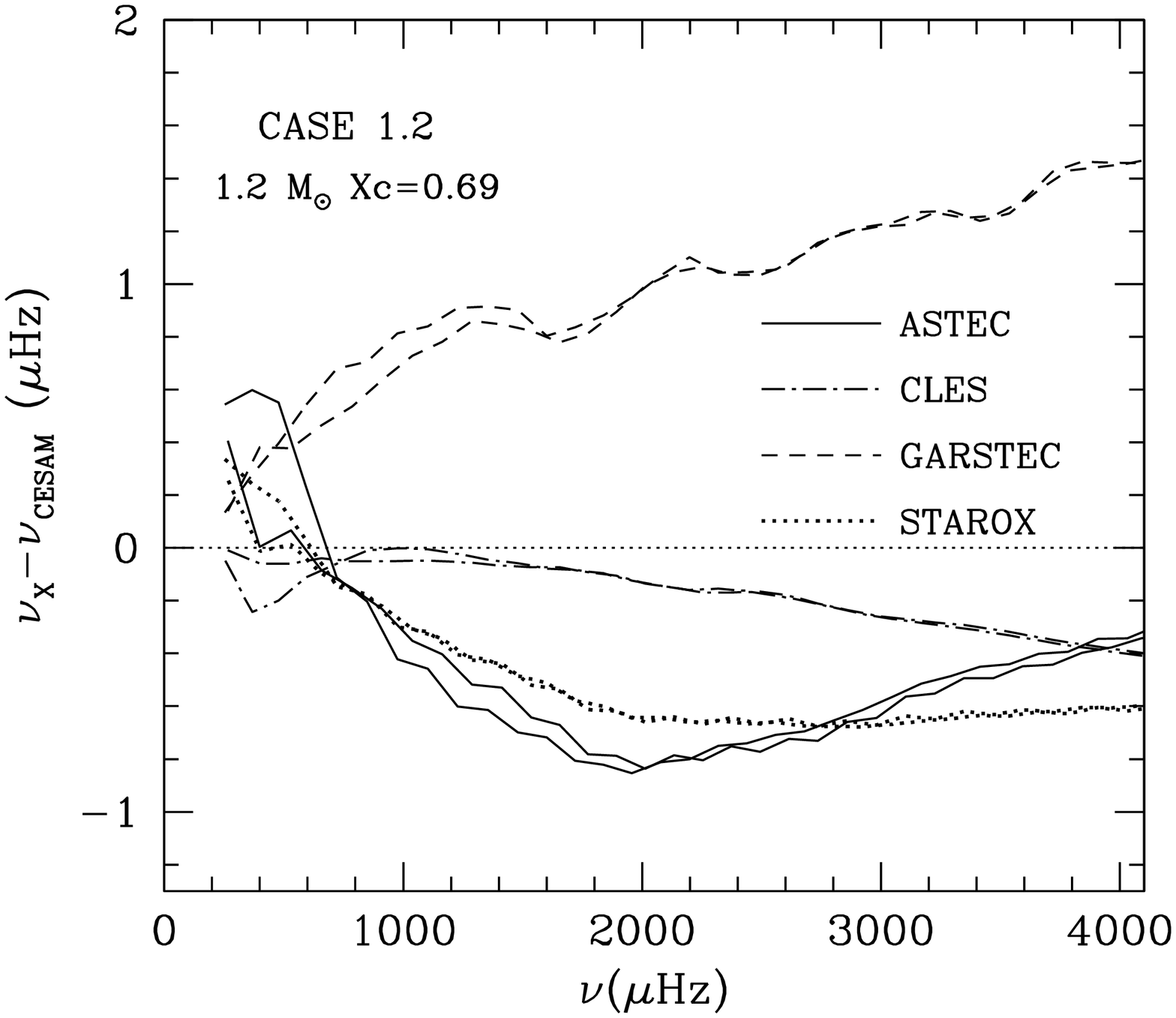}
                      \includegraphics{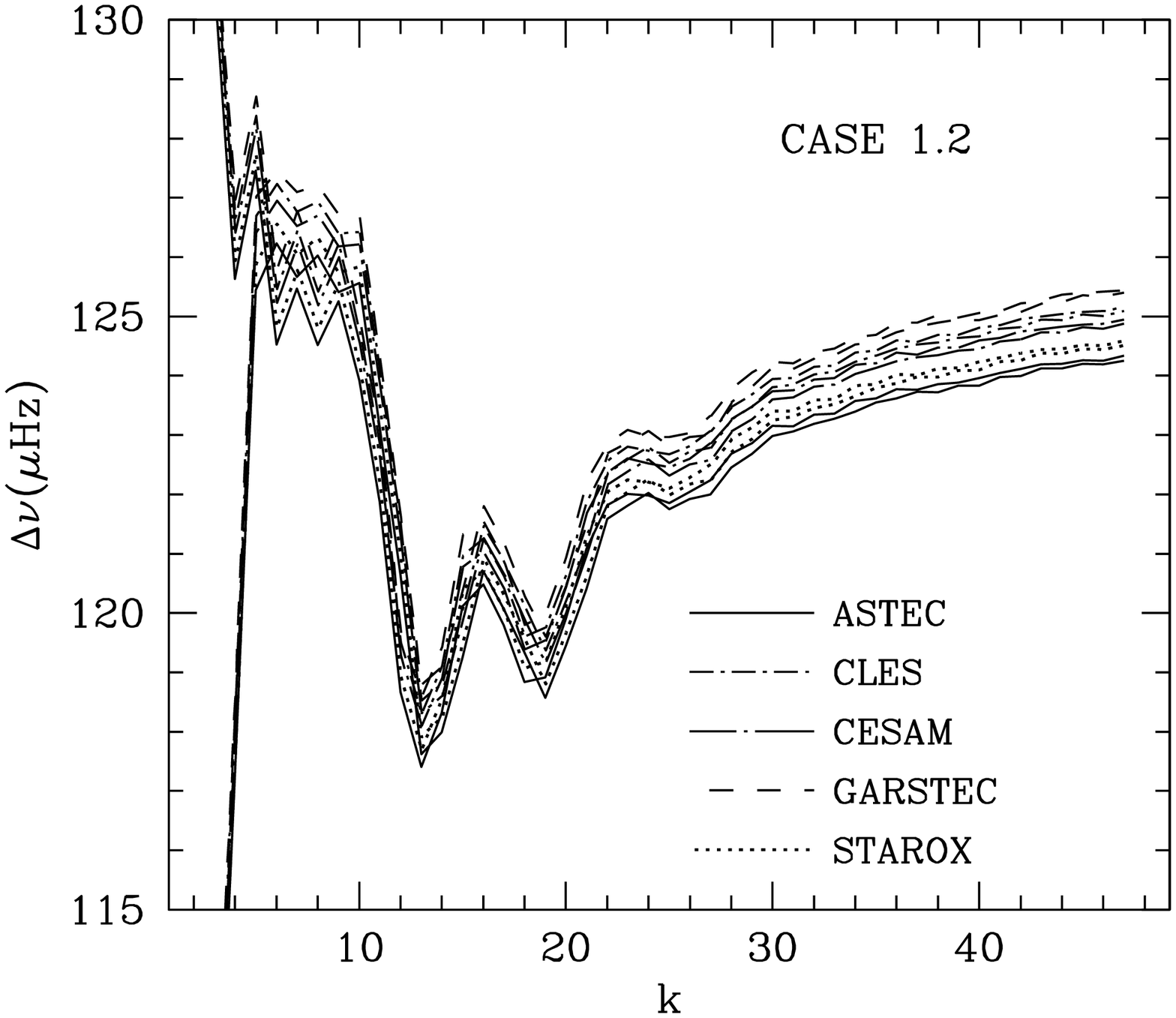}
                      \includegraphics{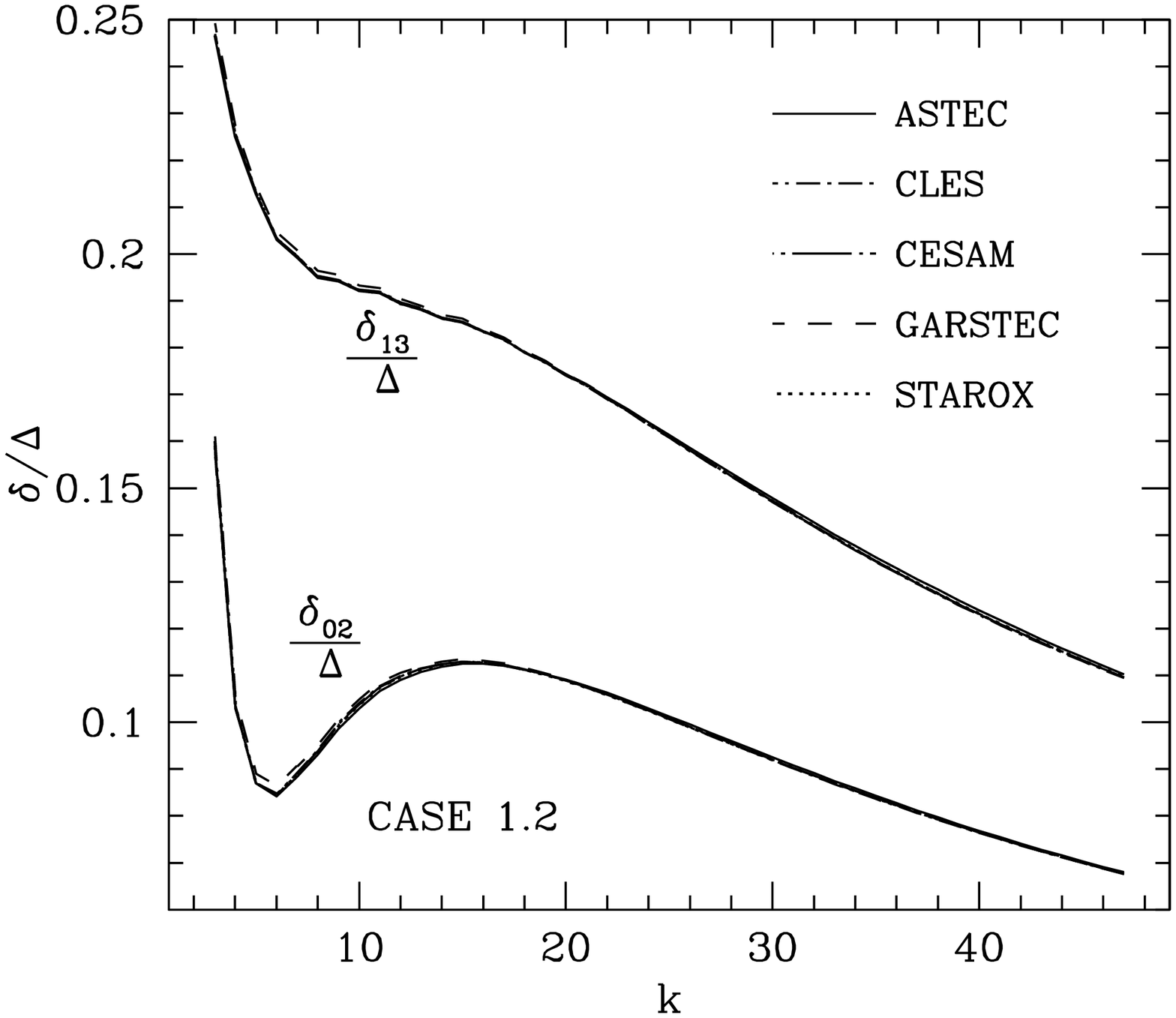}}
\caption{{\task}~1: {\it Left panel}: p-mode frequency differences between models produced by different codes, for Case~1.1 (top row) and 1.2 (bottom). {\cesam} model is taken as reference, and the frequencies have been scaled to remove the effect of different stellar radii. For each code, we plot two curves corresponding to modes with degrees $\ell=0$ and $\ell=1$. 
{\it Central panel}: Large frequency separations $\Delta\nu(\ell=0)$ and $\Delta\nu(\ell=1)$
 versus the radial order $k$ for Case~1.1 and 1.2 models;
 these are based on unscaled frequencies.
{\it Right panel}: Frequency separation ratios as a function of the radial order $k$.}
\label{fig:freq12}       
\end{figure*}

%==========================

\begin{figure*}[htbp!]
\centering
{\includegraphics[scale=0.33]{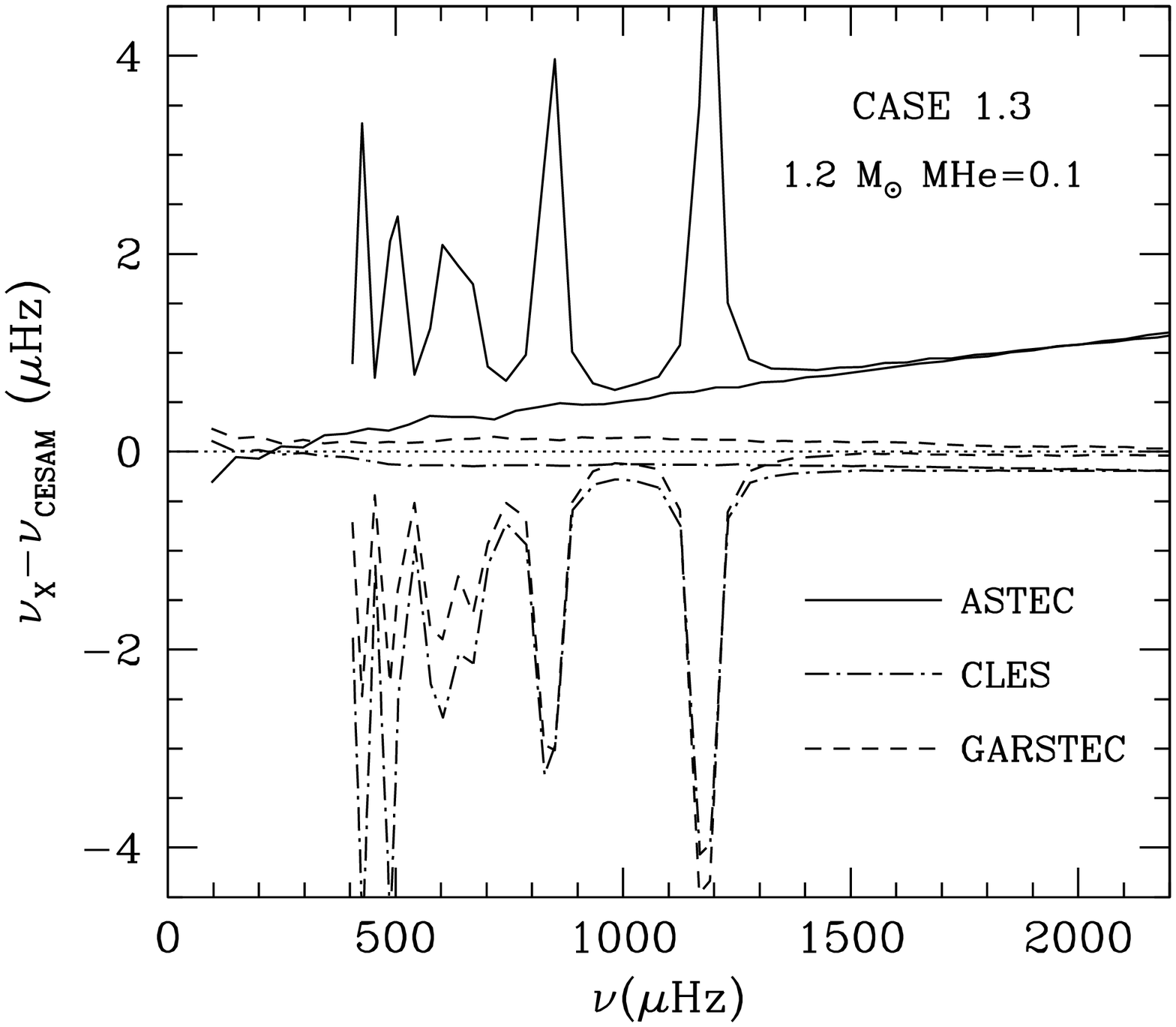}
 \includegraphics[scale=0.33]{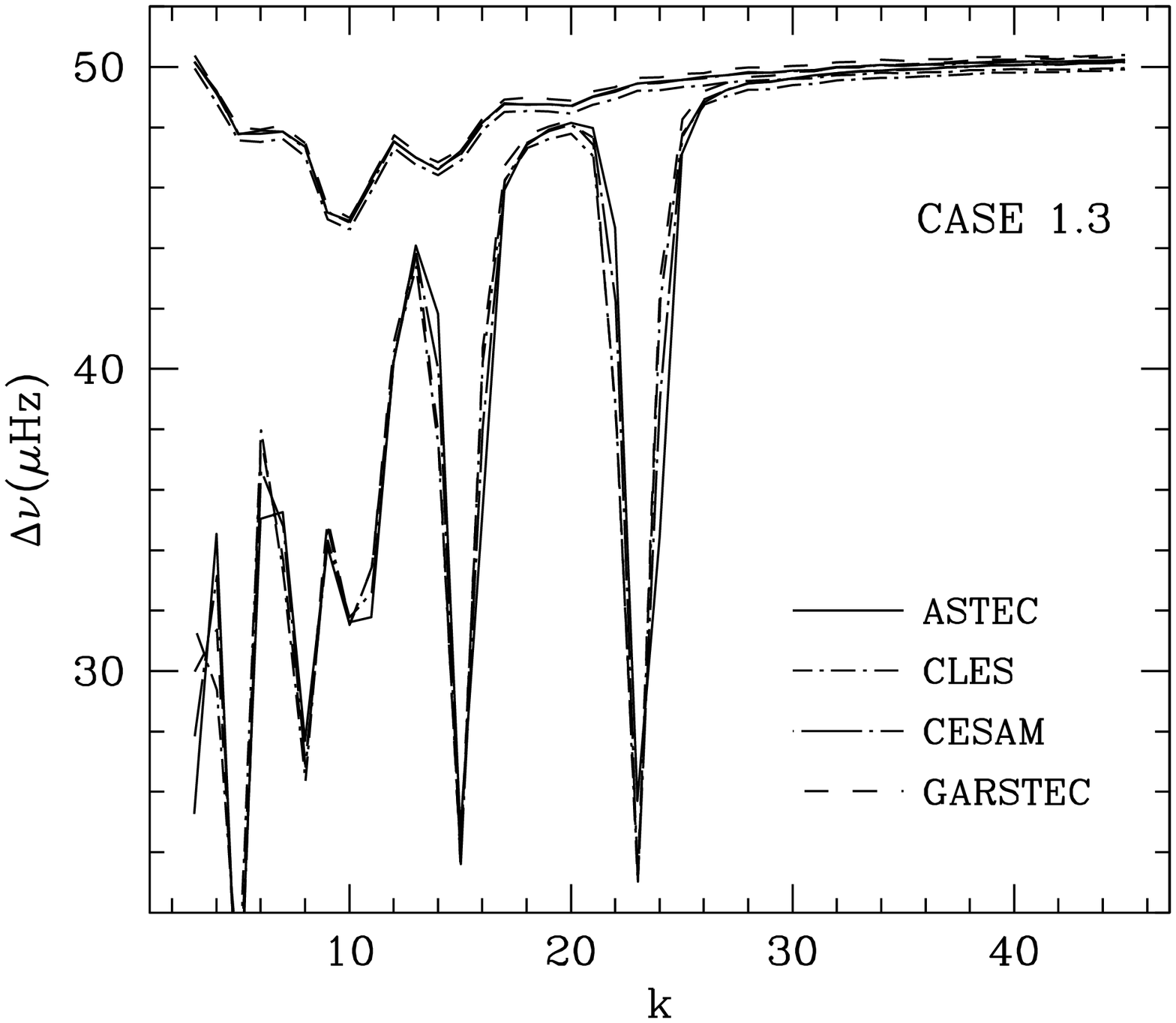}}
\caption{{\task}~1: {\it Left panel}: p-mode frequency differences between models produced by different codes, for Case~1.3. {\cesam} model is taken as reference, and the frequencies have been scaled to remove the effect of different stellar radii. For each code, we plot two curves corresponding to modes with degrees $\ell=0$ and $\ell=1$. 
{\it Right panel}: Large frequency separations $\Delta\nu(\ell=0)$ and $\Delta\nu(\ell=1)$
 versus the radial order $k$ for Case~1.3, based on unscaled frequencies.} 
\label{fig:freq3}       
\end{figure*}
%----------------------------------------------------------------------------

The frequency domains covered are in the range $\nu\sim200-5000\ \mu$Hz for Case~1.1, $200-4000\ \mu$Hz for Case~1.2 and $100-2000\ \mu$Hz for Case~1.3 models. The radial orders are in the range $k\sim0-50$.
To explore the effects of the model frequencies in the asymptotic p-mode region we have included
modes well above the acoustical cutoff frequency.
In addition to the differences $\delta \nu= \nu_{\mbox{\scriptsize CODE}}-\nu_{\mbox{\scriptsize CESAM}}$, we have computed the large frequency separation for $\ell=0$ and $1$  ($\Delta\nu_{\ell,k}=\nu_{\ell,k}-\nu_{\ell,k-1}$) (see Figs.~\ref{fig:freq12} and \ref{fig:freq3}), and for Cases~1.1 and 1.2 we derived the frequency-separation ratios defined in \cite{roxvoron03}.
For these quantities the original model frequencies, without corrections for
differences in radius, were used;
indeed a substantial part of the visible differences in $\Delta \nu$ are
caused by the radius differences.
As shown by Roxburgh and Vorontsov \citep[see also][]{2005MNRAS.356..671F} the frequency-separation ratios have the advantage to be independent of the physical properties of the outer layers. The almost perfect agreement between the value of these ratios for the models computed with all the codes indicates that the differences observed in the frequencies and in the large frequency separation are only determined by the differences in the surface layers (see also Fig.~\ref{fig:task1-extern135}).

For the highly condensed Case~1.3 model, the differences in $\ell=0$ mode frequencies come from surface differences.
On the other hand, the peaks observed in the $\ell=1$ mode frequency differences and in the large frequency separation come from variations of Brunt-V\"ais\"al\"a frequency and from the mixed character of the corresponding modes, see \citet{jcd95} and references therein. The frequencies of the modes trapped in the $\mu$-gradient region depend not only on the location of this gradient but also on its profile. Differences shown in Fig.~\ref{fig:task1-A3} reflect the different behaviour of the $\mu$ gradient in {\astec} with respect to {\cles}, {\garstec}, and {\cesam} which in turn can explain the different behaviour of the {\astec} frequencies seen in Fig.~\ref{fig:freq3}.

%----------------------------------------------------------------------------
%----------------------------------------------------------------------------

\subsubsection{Cases~1.4 and 1.5}

Figure \ref{fig:freq45} (left panel) displays the differences in the p-mode frequencies for the {\PMS} model of $2${\msol}. Two bumps appear in the differences between {\starox} and {\cesam}. The inner one at $\nu\sim300\mu$Hz can be attributed to differences in the sound speed close to the centre as seen in Fig.~\ref{fig:task1-int45}. The outer bump at $\nu\sim1500\mu$Hz results from differences in $\Gamma_1$ in the second He-ionisation region.

\begin{figure}[htbp!]
\centering
%\resizebox{0.8\hsize}{!}{\includegraphics{./c1.5/fig1.5.A.eps}}
\resizebox{0.8\hsize}{!}{\includegraphics{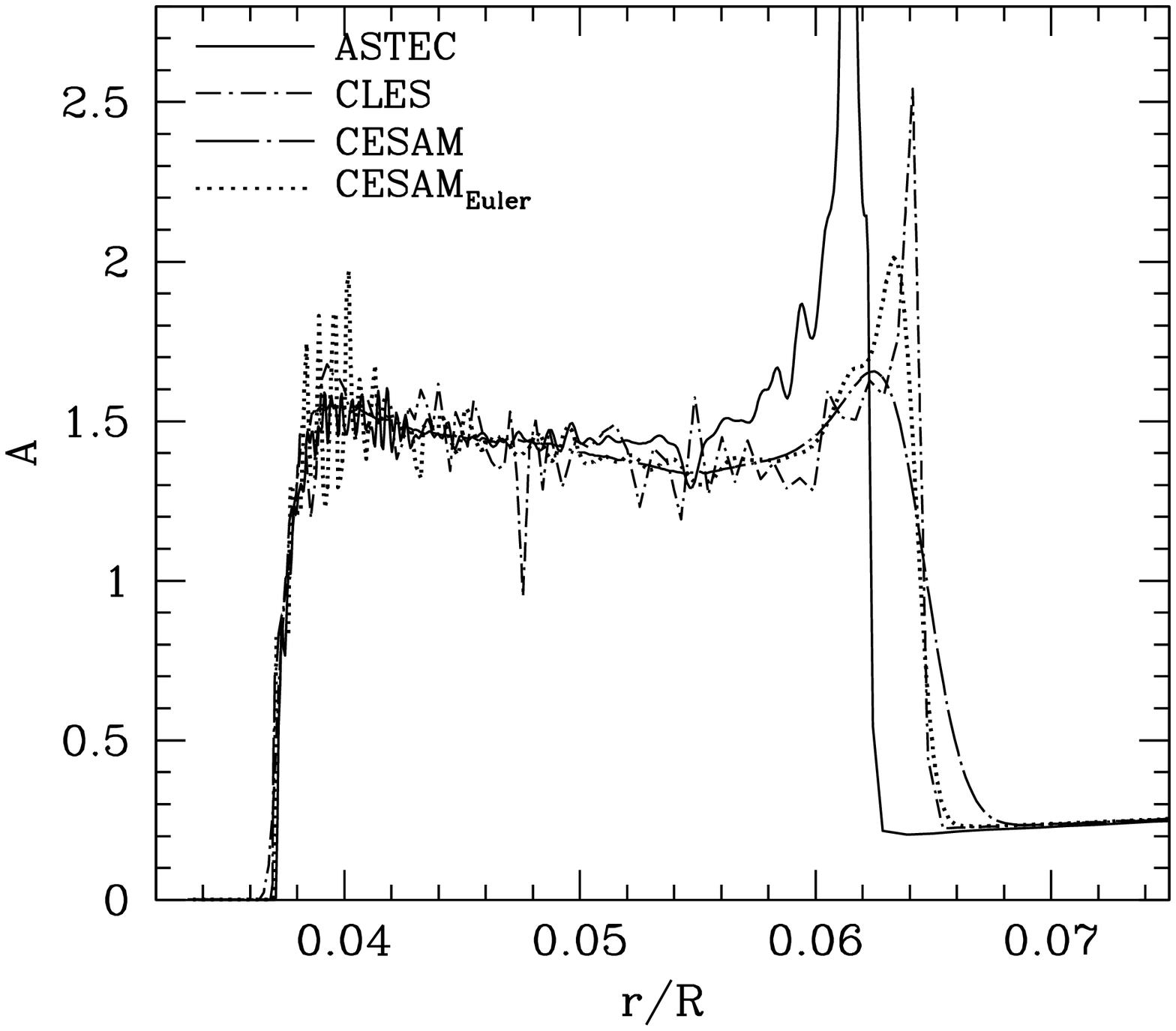}}
\caption{{\task}~1: Run of the quantity $A=N^2_{\rm BV} r/g$ in the deep interior of Case~1.5 models}
\label{fig:task1-A5}
\end{figure}

Figure \ref{fig:freq45} (right panel) shows the differences in the p-mode frequencies for the evolved Case~1.5 model. As in Case~1.3, the differences mainly result from differences in the surface layers (see Fig.~\ref{fig:task1-extern135}).  Also, this model is sufficiently evolved to present g-p mixed modes. In fact, the peaks observed at low frequency for $\ell=1$ modes correspond to mo\-des trapped in the $\mu$-gradient region. Figure \ref{fig:task1-A5} displays the profile of $A$ showing that even though the $\mu$ gradient is generated at the same depth  in the star, its slope is quite different, and therefore the mixed-mode frequencies also differ. 
We point out that the smoother decrease of $A$ observed in {\cesam} models with respect to others at $r\sim0.065R$ is due to the scheme used for the integration of the temporal evolution of the chemical composition, i.e. an L-stable implicit Runge-Kutta scheme of order 2 \citep[see][]{pm-apss}. We have checked that when the standard Euler backward scheme is used, the $A$ profile becomes quite similar to what is obtained by other codes (see Fig. \ref{fig:task1-A5}) and that, as can be seen in Fig.~\ref{fig:freq45} the frequencies of the mixed modes are also modified.

%----------------------------------------------------------------------------
%----------------------------------------------------------------------------

\subsubsection{Cases~1.6 and 1.7}

The frequency differences for p modes in Case~1.6 and 1.7 are smaller than  $0.2~\mu\rm{Hz}$ except for the {\garstec} models, for which the differences can be slightly larger than  $0.2~\mu\rm{Hz}$ for the more massive model, and reach $0.8~\mu\rm{Hz}$ for the {\ZAMS} one. We recall, however, that this latter 
has a central hydrogen content slightly smaller than specified, differing by $-3.4\times10^{-4}$ from the specified  $X_{\rm c}=0.69$. To investigate the effect of such a small difference in the central H content, the frequencies of two {\cles} models differing by $\delta X_{\rm c}=3.4\times10^{-4}$ have been calculated: they show differences in the range $-0.05$ to $\sim0.3\mu$Hz that only partially account for the differences found.

%----------------------------------------------------------------------------
\begin{figure*}[hbtp!]
\resizebox*{\hsize}{!}{\includegraphics*{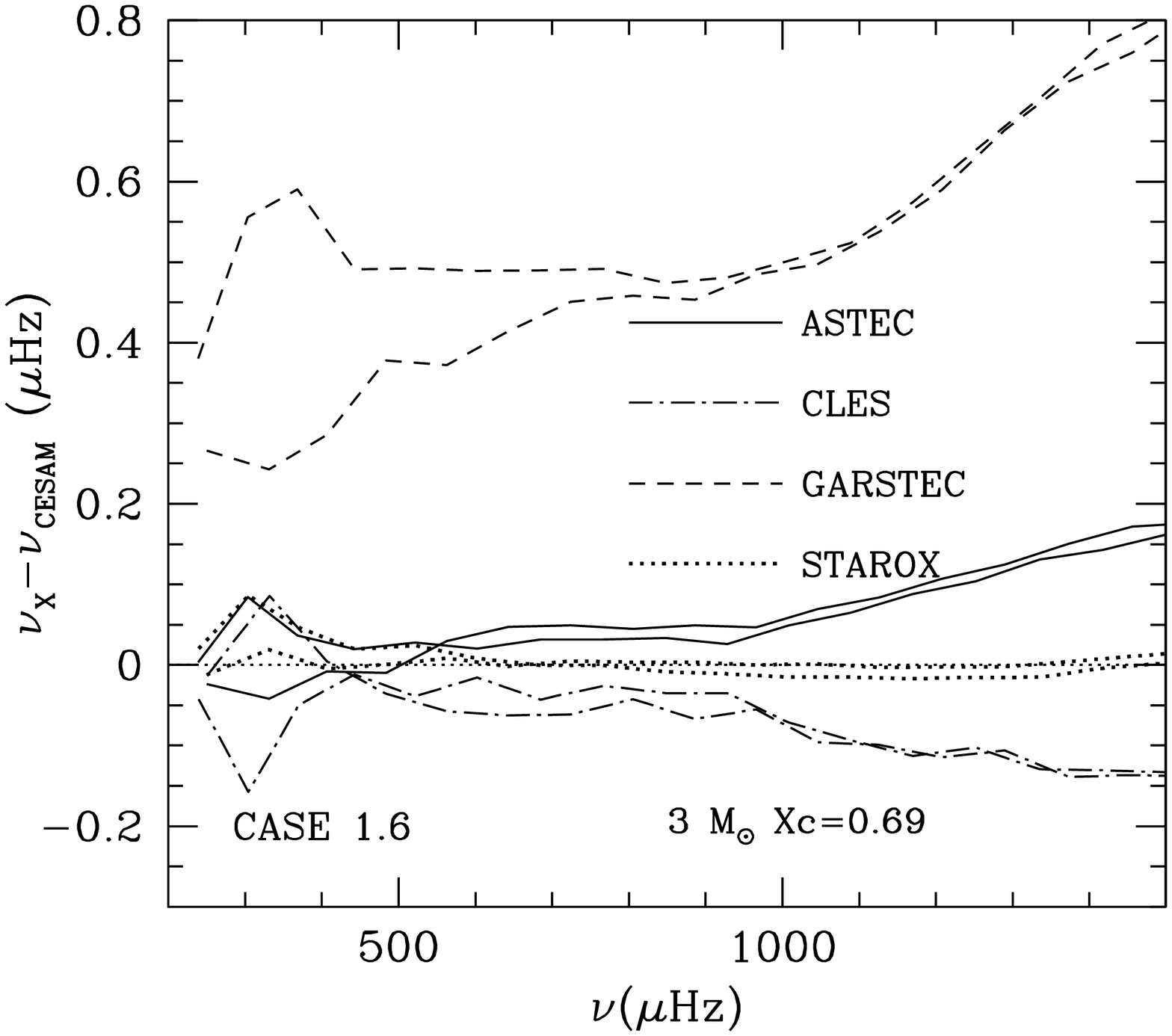}
                      \includegraphics*{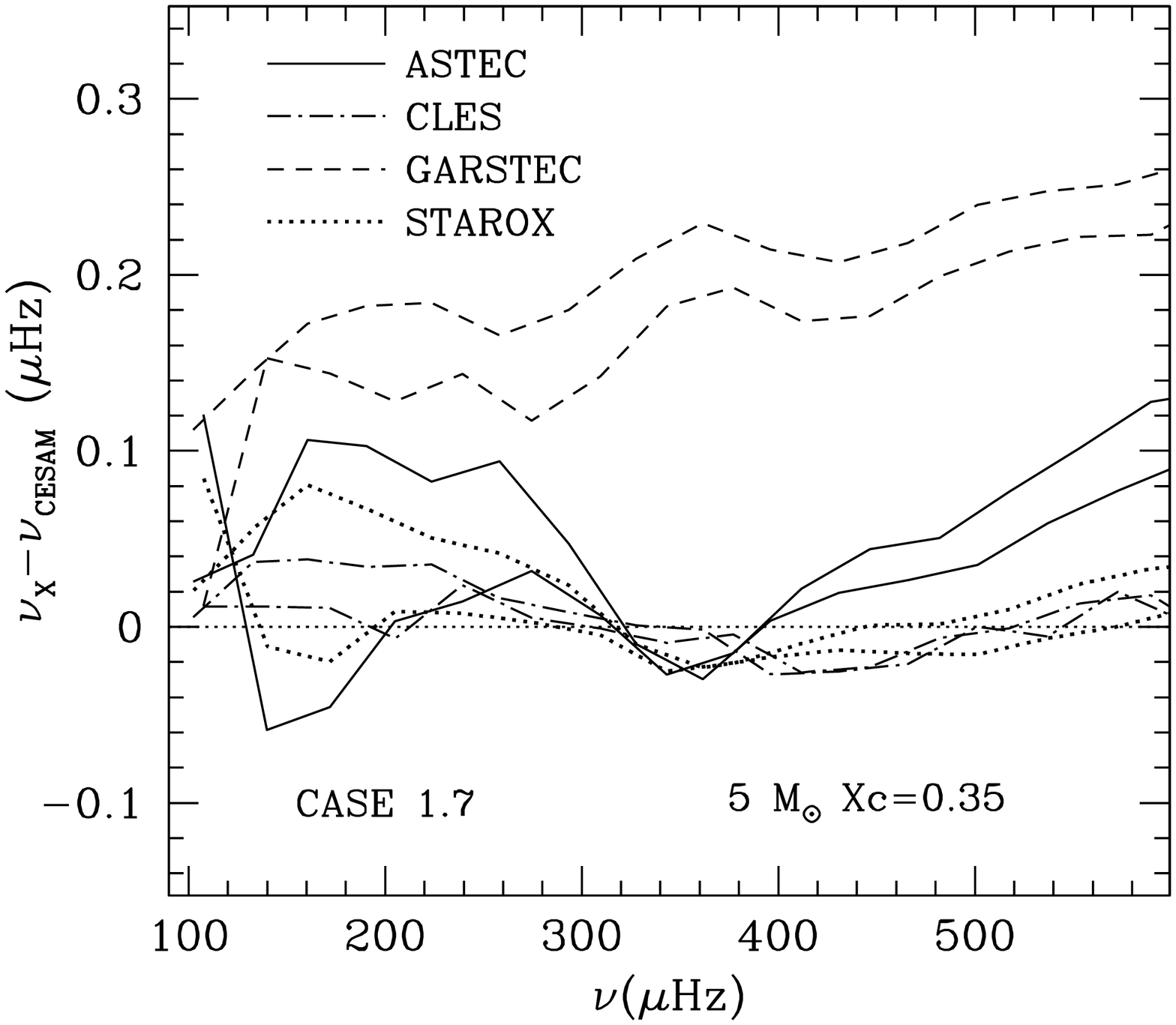}
                      \includegraphics*{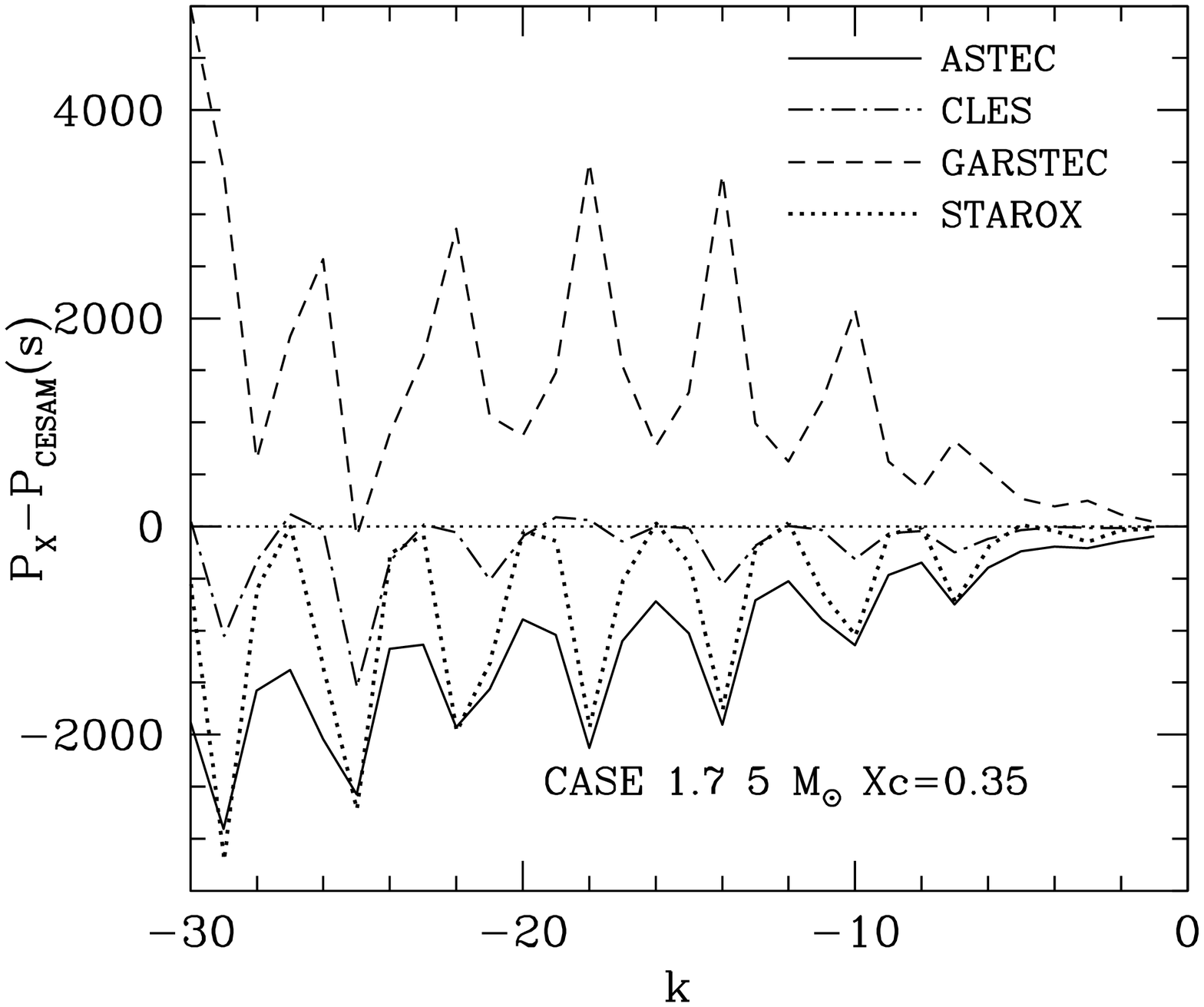}}
\caption{{\task}~1: {\it Left and central panels:} p-mode frequency differences between models produced by different codes, for Case~1.6 (left) and 1.7 (centre). {\cesam} model is taken as reference, and the frequencies have been scaled to remove the effect of different stellar radii. For each code, we plot two curves corresponding to modes with degrees $\ell=0$ and $\ell=1$. {\it Right panel:} Plots of the g-mode period differences, between models produced by different codes for Case~1.7 (the {\cesam} model -- horizontal dotted line -- is taken as reference).
}       
\label{fig:freq67}
\end{figure*}
%----------------------------------------------------------------------------

The stellar parameters of Case~1.7 models match quite well those of a typical SPB star (Slowly Pulsating B type star). This type of pulsators presents high-order g modes with periods ranging from $0.4$ to $3.5$ days for modes with low degree ($\ell=1$ and $2$) \citep{1993MNRAS.265..588D}. We have estimated for Case~1.7 models,  the period differences for g modes with radial order $k=-30$ to $-1$. As has been shown in \citet{migliomg} the periods of g modes can present also a oscillatory signal, whose periodicity depends on the location of the $\mu$ gradient, and  whose amplitude is determined by the  slope of the chemical composition gradient. The profile of the quantity $A$ for Case~1.7 is quite similar to that of Case~1.5 (Fig.~\ref{fig:task1-A5}), that is with the profile in {\cesam} model being smoother than in models obtained by the other codes. 

The effect on the variation of g-mode periods is shown in Fig.~\ref{fig:freq67} (right) where the periodicity of the signature is related to the location of the $\mu$ gradient and the amplitude of the difference is increasing with the steepness of the gradient.

%----------------------------------------------------------------------------
\begin{table*}[htbp!]
\caption{{\task}~3 models: Global parameter differences given in per cent, between each code and {\cesam}-MP. For each parameter we give the mean difference and the maximum difference of the complete series of {\task}~3 models (i.e. each case and each phase are included)}
\centering
\label{tab:task3-glob}
\begin{tabular}[h]{lcccccccccc}
\hline\noalign{\smallskip}
{\bf Code} & \multicolumn{2}{c}{\boldmath$\delta M/M$}& \multicolumn{2}{c}{\boldmath$\delta R/R$}&  \multicolumn{2}{c}{\boldmath$\delta L/L$}& \multicolumn{2}{c}{\boldmath$\delta T_{\rm eff}/T_{\rm eff}$} &\multicolumn{2}{c}{\boldmath$\delta\rm age/age$}\\ [3pt]
& {\rm mean} & {\rm max} &{\rm mean} & {\rm max} &{\rm mean} & {\rm max} & {\rm mean} & {\rm max} &{\rm mean} & {\rm max} \\ [3pt]
\tableheadseprule\noalign{\smallskip}
{\astec}      & 0.01  &  0.01  &  0.45 &   0.85  &  1.06 &   1.89  &  0.04  &  0.06 & --     &  --  \\
{\cesam}-B69  & 0.00  &  0.00  &  0.23 &   0.54  &  0.11 &   0.31  &  0.10  &  0.22 &  0.35  &  1.02\\
{\cles}       & 0.00  &  0.00  &  0.23 &   0.54  &  0.23 &   0.45  &  0.07  &  0.20 &  0.47  &  0.77\\
{\garstec}   & 0.05  &  0.08  &  3.21 &  26.52  &  4.30 &  37.02  &  0.54  &  3.92 &  --    &--     \\
\noalign{\smallskip}\hline
\end{tabular}
\end{table*}

%-------------------------------------------------------------------
\begin{table*}[htbp!]
\caption{{\task}~3 models: Mean quadratic difference in the physical variables between each code and {\cesam} calculated according to Eq. \ref{eq:quad_mean}. The differences are given in per cent (except for $\delta X$) and represent  an average over the whole star from centre to photospheric radius.  The local differences were calculated at fixed relative mass.}
\centering
\label{tab:task3-quad}
\begin{tabular}[h]{lccccccccccc}
\hline\noalign{\smallskip}
\multicolumn{12}{c} {\bf Case~3.1A - Case~3.1B - Case~3.1C}\\
\hline\noalign{\smallskip}
{\bf Code} & \boldmath$\delta\ln c$      &  \boldmath$\delta\ln P$ 
           & \boldmath$\delta\ln \rho$   &$\delta\ln T$&$\delta\ln r$& \boldmath$\delta\ln \Gamma_1$ 
           & \boldmath$\delta\ln \nabla_{\rm ad}$ & \boldmath$\delta\ln C_p$   
           & \boldmath$\delta\ln \kappa$ & \boldmath$\delta X$ & \boldmath$\delta\ln L_r$ \\ [3pt]
\tableheadseprule\noalign{\smallskip}
 {\astec}     &   0.03 &   0.14 &   0.14 &   0.06 &   0.06 &  4.07$\times10^{-4}$ &  7.42$\times10^{-4}$ &  3.51$\times10^{-2}$ &   1.37 & 0.00049 &   0.19 \\ 
  {\cesam}-B69 &   0.02 &   0.17 &   0.17 &   0.03 &   0.04 &  1.55$\times10^{-4}$ &  3.21$\times10^{-4}$ &  3.27$\times10^{-2}$ &   0.16 & 0.00042 &   0.04 \\ 
  {\cles}      &   0.02 &   0.18 &   0.18 &   0.04 &   0.04 &  2.41$\times10^{-3}$ &  5.93$\times10^{-3}$ &  3.84$\times10^{-2}$ &   0.24 & 0.00046 &   0.17 \\ 
  {\garstec}   &   0.04 &   0.46 &   0.43 &   0.07 &   0.11 &  1.51$\times10^{-1}$ &  1.23$\times10^{-1}$ &  1.40$\times10^{-1}$ &   0.59 & 0.00065 &   0.51 \\ 
%\noalign{\smallskip}\hline\hline\noalign{\smallskip}
%\hline\noalign{\smallskip}\hline\noalign{\smallskip}
\tableheadseprule\noalign{\smallskip}
   {\astec}     &   0.15 &   1.66 &   1.59 &   0.53 &   0.58 &  1.73$\times10^{-3}$ &  3.87$\times10^{-3}$ &  4.36$\times10^{-1}$ &   2.37 & 0.00406 &   1.91 \\ 
  {\cesam}-B69 &   0.04 &   0.50 &   0.48 &   0.09 &   0.12 &  5.10$\times10^{-4}$ &  1.02$\times10^{-3}$ &  7.77$\times10^{-2}$ &   0.29 & 0.00085 &   0.35 \\ 
  {\cles}      &   0.05 &   0.53 &   0.49 &   0.13 &   0.16 &  3.98$\times10^{-3}$ &  9.06$\times10^{-3}$ &  7.45$\times10^{-2}$ &   0.39 & 0.00091 &   0.76 \\ 
  {\garstec}   &   0.06 &   0.69 &   0.56 &   0.16 &   0.15 &  1.60$\times10^{-1}$ &  1.42$\times10^{-1}$ &  2.80$\times10^{-1}$ &   0.61 & 0.00186 &   0.67 \\ 
%\noalign{\smallskip}\hline\hline\noalign{\smallskip}
%\hline\noalign{\smallskip}\hline\noalign{\smallskip}
\tableheadseprule\noalign{\smallskip}
 {\astec}     &   0.15 &   1.21 &   1.00 &   0.41 &   0.35 &  1.32$\times10^{-3}$ &  2.55$\times10^{-3}$ &  2.32$\times10^{-1}$ &   2.68 & 0.00241 &   5.87 \\ 
  {\cesam}-B69 &   0.05 &   0.50 &   0.46 &   0.11 &   0.11 &  5.56$\times10^{-4}$ &  1.22$\times10^{-3}$ &  8.49$\times10^{-2}$ &   0.44 & 0.00106 &   0.75 \\ 
  {\cles}      &   0.06 &   0.40 &   0.41 &   0.09 &   0.11 &  4.53$\times10^{-3}$ &  1.02$\times10^{-2}$ &  1.34$\times10^{-1}$ &   0.39 & 0.00117 &   2.87 \\ 
  {\garstec}   &   0.10 &   0.87 &   0.75 &   0.23 &   0.19 &  1.72$\times10^{-1}$ &  1.62$\times10^{-1}$ &  3.81$\times10^{-1}$ &   0.64 & 0.00242 &   1.37 \\ 
 
\noalign{\smallskip}\hline
\hline\noalign{\smallskip}
\multicolumn{12}{c} {\bf Case~3.2A - Case~3.2B - Case~3.2C}\\
\hline\noalign{\smallskip}
{\bf Code} & \boldmath$\delta\ln c$      &  \boldmath$\delta\ln P$ 
           & \boldmath$\delta\ln \rho$ &$\delta\ln T$&$\delta\ln r$  & \boldmath$\delta\ln \Gamma_1$ 
           & \boldmath$\delta\ln \nabla_{\rm ad}$ & \boldmath$\delta\ln C_p$   
           & \boldmath$\delta\ln \kappa$ & \boldmath$\delta X$ & \boldmath$\delta\ln L_r$  \\ [3pt]
\tableheadseprule\noalign{\smallskip}
 {\cesam}-B69 &   0.02 &   0.15 &   0.15 &   0.02 &   0.03 &  2.26$\times10^{-4}$ &  4.75$\times10^{-4}$ &  3.76$\times10^{-2}$ &   0.12 & 0.00041 &   0.07 \\ 
  {\cles}      &   0.03 &   0.15 &   0.16 &   0.02 &   0.04 &  2.05$\times10^{-3}$ &  4.96$\times10^{-3}$ &  5.20$\times10^{-2}$ &   0.22 & 0.00054 &   0.43 \\ 
  {\garstec}   &   0.12 &   0.33 &   0.39 &   0.07 &   0.10 &  1.61$\times10^{-1}$ &  1.45$\times10^{-1}$ &  2.29$\times10^{-1}$ &   0.44 & 0.00200 &   0.80 \\ 
\tableheadseprule\noalign{\smallskip}
   {\cesam}-B69 &   0.21 &   1.64 &   1.38 &   0.30 &   0.35 &  1.55$\times10^{-3}$ &  4.05$\times10^{-3}$ &  3.23$\times10^{-1}$ &   0.56 & 0.00276 &   0.59 \\ 
  {\cles}      &   0.23 &   1.71 &   1.48 &   0.28 &   0.37 &  3.64$\times10^{-3}$ &  8.86$\times10^{-3}$ &  3.96$\times10^{-1}$ &   0.71 & 0.00339 &   0.80 \\ 
  {\garstec}   &   0.18 &   0.95 &   0.86 &   0.15 &   0.22 &  1.68$\times10^{-1}$ &  1.66$\times10^{-1}$ &  2.40$\times10^{-1}$ &   0.69 & 0.00145 &   0.77 \\ 
 
\tableheadseprule\noalign{\smallskip}
{\cesam}-B69 &   0.06 &   0.22 &   0.24 &   0.04 &   0.06 &  2.95$\times10^{-4}$ &  1.02$\times10^{-3}$ &  1.26$\times10^{-1}$ &   0.28 & 0.00117 &   2.90 \\ 
  {\cles}      &   0.20 &   1.47 &   1.25 &   0.32 &   0.34 &  4.30$\times10^{-3}$ &  1.02$\times10^{-2}$ &  3.35$\times10^{-1}$ &   0.29 & 0.00294 &   1.49 \\ 
  {\garstec}   &   0.17 &   1.47 &   1.26 &   0.30 &   0.34 &  1.74$\times10^{-1}$ &  1.71$\times10^{-1}$ &  5.12$\times10^{-1}$ &   0.46 & 0.00348 &   3.50 \\ 
    
\noalign{\smallskip}\hline
\hline\noalign{\smallskip}
\multicolumn{12}{c} {\bf Case~3.3A - Case~3.3B - Case~3.3C}\\
\hline\noalign{\smallskip}
{\bf Code} & \boldmath$\delta\ln c$      &  \boldmath$\delta\ln P$ 
           & \boldmath$\delta\ln \rho$   &$\delta\ln T$&$\delta\ln r$ & \boldmath$\delta\ln \Gamma_1$ 
           & \boldmath$\delta\ln \nabla_{\rm ad}$ & \boldmath$\delta\ln C_p$   
           & \boldmath$\delta\ln \kappa$ & \boldmath$\delta X$ & \boldmath$\delta\ln L_r$ \\ [3pt]
\tableheadseprule\noalign{\smallskip}
  {\cesam}-B69 &   0.18 &   0.39 &   0.50 &   0.06 &   0.09 &  1.29$\times10^{-3}$ &  3.42$\times10^{-3}$ &  3.61$\times10^{-1}$ &   0.32 & 0.00373 &   0.41 \\ 
  {\cles}      &   0.04 &   0.06 &   0.11 &   0.03 &   0.02 &  2.58$\times10^{-3}$ &  6.69$\times10^{-3}$ &  7.85$\times10^{-2}$ &   0.24 & 0.00088 &   0.84 \\ 
  {\garstec}   &   0.10 &   0.26 &   0.31 &   0.05 &   0.06 &  1.65$\times10^{-1}$ &  1.55$\times10^{-1}$ &  2.10$\times10^{-1}$ &   0.54 & 0.00122 &   0.49 \\ 

\tableheadseprule\noalign{\smallskip}
   {\cesam}-B69 &   0.18 &   1.09 &   0.92 &   0.21 &   0.25 &  1.07$\times10^{-3}$ &  3.08$\times10^{-3}$ &  2.90$\times10^{-1}$ &   0.30 & 0.00264 &   0.42 \\ 
  {\cles}      &   0.18 &   1.21 &   0.98 &   0.26 &   0.28 &  3.23$\times10^{-3}$ &  7.89$\times10^{-3}$ &  2.71$\times10^{-1}$ &   0.22 & 0.00248 &   0.42 \\ 
  {\garstec}   &   0.34 &   2.09 &   1.77 &   0.41 &   0.45 &  1.69$\times10^{-1}$ &  1.69$\times10^{-1}$ &  7.50$\times10^{-1}$ &   0.53 & 0.00634 &   0.36 \\ 
\tableheadseprule\noalign{\smallskip}
  {\cesam}-B69 &   0.09 &   0.48 &   0.43 &   0.11 &   0.11 &  7.23$\times10^{-4}$ &  1.64$\times10^{-3}$ &  1.45$\times10^{-1}$ &   0.27 & 0.00138 &   4.40 \\ 
  {\cles}      &   0.11 &   0.28 &   0.38 &   0.11 &   0.10 &  3.32$\times10^{-3}$ &  8.15$\times10^{-3}$ &  2.41$\times10^{-1}$ &   0.30 & 0.00211 &   1.28 \\ 
  {\garstec}   &   0.30 &   1.84 &   1.67 &   0.37 &   0.43 &  1.76$\times10^{-1}$ &  1.76$\times10^{-1}$ &  7.72$\times10^{-1}$ &   0.46 & 0.00601 &   3.57 \\ 
\noalign{\smallskip}\hline
\end{tabular}
\end{table*}
%-----------------------------------------------------------------------------------------------------------------------------------
%-----------------------------------------------------------------------------------------------------------------------------------
\begin{table*}[htbp!]
\caption{{\task}~3 models: Maximum variations given in per cent (except for $\delta X$) of the physical variables between each code and {\cesam} and value of the relative radius $(r/R)$ where they happen.
%JC-D, 070907%YL, 100907
%The local differences were computed at fixed relative mass.
The local differences were computed both at fixed relative mass and fixed relative radius and the maximum of the two values was searched (see footnote~\ref{calculo}).
}

\centering
\label{tab:task3-maxdiff}
\begin{tabular}[h]{lcccccccccccc}
\hline\noalign{\smallskip}
\multicolumn{13}{c} {\bf Case~3.1A - Case~3.1B - Case~3.1C}\\
\hline\noalign{\smallskip}
{\bf Code} & \boldmath$\delta\ln c$ & \boldmath$r/R$ & \boldmath$\delta\ln P$ & \boldmath$r/R$ 
          & \boldmath$\delta\ln \rho$ & \boldmath$r/R$ & \boldmath$\delta\ln \Gamma_1$ & \boldmath$r/R$          
          & \boldmath$\delta X$ & \boldmath$r/R$ 
          & \boldmath$\delta\ln L_r$ & \boldmath$r/R$\\[3pt]
\tableheadseprule\noalign{\smallskip}
  {\astec}     &    0.13 & 0.68223 &    0.94 & 0.96261 &    0.85 & 0.96261 &    0.01 & 0.96261 & 0.00271 & 0.69608 &    1.14 & 0.00155 \\ 
  {\cesam}{\small -B69} &    0.07 & 0.73259 &    0.64 & 0.42254 &    0.57 & 0.44886 &    0.00 & 0.96260 & 0.00192 & 0.68474 &    0.19 & 0.06786 \\ 
  {\cles}      &    0.12 & 0.68285 &    1.11 & 0.58835 &    1.12 & 0.67438 &    0.02 & 0.79926 & 0.00220 & 0.67861 &    0.63 & 0.06038 \\ 
  {\garstec}   &    0.23 & 0.25979 &    2.23 & 0.51009 &    2.07 & 0.66870 &    0.23 & 0.00000 & 0.00271 & 0.08724 &    4.03 & 0.00215 \\ 
 
 %\noalign{\smallskip}\hline
%==========================
%\hline\noalign{\smallskip}
%\multicolumn{13}{c} {\bf Case~3.1B}\\
%\hline\noalign{\smallskip}
%\hline\noalign{\smallskip}
%{\bf Code} & \boldmath$\delta c$ & \boldmath$r/R$ & \boldmath$\delta P$ & \boldmath$r/R$ 
 %         & \boldmath$\delta \rho$ & \boldmath$r/R$ & \boldmath$\delta \Gamma_1$ & \boldmath$r/R$
%          & \boldmath$\delta X$ & \boldmath$r/R$ 
%         & \boldmath$\delta L_r$ & \boldmath$r/R$\\[3pt]
\tableheadseprule\noalign{\smallskip}
  {\astec}     &    0.49 & 0.96005 &    4.96 & 0.42170 &    4.34 & 0.42410 &    0.04 & 0.96005 & 0.01364 & 0.05022 &   10.63 & 0.00644 \\ 
  {\cesam}{\small -B69} &    0.21 & 0.71495 &    1.59 & 0.35641 &    1.41 & 0.37767 &    0.02 & 0.96000 & 0.00343 & 0.70558 &    2.59 & 0.01962 \\ 
  {\cles}      &    0.19 & 0.75105 &    2.07 & 0.49726 &    1.91 & 0.63770 &    0.03 & 0.95661 & 0.00442 & 0.71375 &    5.95 & 0.01206 \\ 
  {\garstec}   &    0.35 & 0.01139 &    1.62 & 0.52826 &    1.65 & 0.63443 &    0.30 & 0.00000 & 0.00682 & 0.06100 &    7.81 & 0.00130 \\  
%\noalign{\smallskip}\hline
%==========================
%\hline\noalign{\smallskip}
%\multicolumn{13}{c} {\bf Case~3.1C}\\
%\hline\noalign{\smallskip}
%\hline\noalign{\smallskip}
%{\bf Code} & \boldmath$\delta c$ & \boldmath$r/R$ & \boldmath$\delta P$ & \boldmath$r/R$ 
%          & \boldmath$\delta \rho$ & \boldmath$r/R$ & \boldmath$\delta \Gamma_1$ & \boldmath$r/R$
%          & \boldmath$\delta X$ & \boldmath$r/R$ 
 %         & \boldmath$\delta L_r$ & \boldmath$r/R$\\[3pt]
\tableheadseprule\noalign{\smallskip}
  {\astec}     &    0.69 & 0.66252 &    2.55 & 0.40052 &    3.41 & 0.66153 &    0.02 & 0.96068 & 0.02401 & 0.66153 &  232.00 & 0.00216 \\ 
  {\cesam}{\small -B69} &    0.44 & 0.65858 &    1.37 & 0.29892 &    1.98 & 0.65858 &    0.03 & 0.96042 & 0.01742 & 0.65858 &    9.27 & 0.02439 \\ 
  {\cles}      &    0.41 & 0.03876 &    1.62 & 0.41953 &    2.05 & 0.65804 &    0.03 & 0.96067 & 0.01116 & 0.65804 &   16.81 & 0.02402 \\ 
  {\garstec}   &    0.36 & 0.65262 &    2.26 & 0.86451 &    2.60 & 0.65653 &    0.32 & 0.00000 & 0.01186 & 0.65996 &   15.42 & 0.00266 \\ 
 
 \noalign{\smallskip}\hline
%==========================
\hline\noalign{\smallskip}
\multicolumn{13}{c} {\bf Case~3.2A -- Case~3.2B -- Case~3.2C} \\
%\hline\noalign{\smallskip}
\hline\noalign{\smallskip}
{\bf Code} & \boldmath$\delta\ln c$ & \boldmath$r/R$ & \boldmath$\delta\ln P$ & \boldmath$r/R$ 
          & \boldmath$\delta\ln \rho$ & \boldmath$r/R$ & \boldmath$\delta\ln \Gamma_1$ & \boldmath$r/R$          
          & \boldmath$\delta X$ & \boldmath$r/R$ 
          & \boldmath$\delta\ln L_r$ & \boldmath$r/R$\\[3pt]
\tableheadseprule\noalign{\smallskip}

 % {\astec}     &    2.54 & 0.04666 &    5.84 & 0.76670 &    5.13 & 0.80432 &    0.06 & 0.97267 & 0.04314 & 0.04666 &   17.30 & 0.00846 \\ 
   {\cesam}{\small -B69} &    0.44 & 0.04631 &    1.43 & 0.88799 &    1.12 & 0.88799 &    0.01 & 0.88799 & 0.00895 & 0.04631 &    0.85 & 0.04674 \\ 
  {\cles}      &    0.68 & 0.04632 &    0.59 & 0.79763 &    1.49 & 0.04632 &    0.05 & 0.84435 & 0.01382 & 0.04632 &    7.17 & 0.00173 \\ 
  {\garstec}   &    2.42 & 0.04531 &    1.38 & 0.36304 &    5.22 & 0.04531 &    0.26 & 0.00000 & 0.04600 & 0.04556 &    7.25 & 0.00438 \\ 
 
%\noalign{\smallskip}\hline
%==========================
%\hline\noalign{\smallskip}
%\multicolumn{13}{c} {\bf Case~3.2B}\\
%\hline\noalign{\smallskip}
%\hline\noalign{\smallskip}
%{\bf Code} & \boldmath$\delta c$ & \boldmath$r/R$ & \boldmath$\delta P$ & \boldmath$r/R$ 
%          & \boldmath$\delta \rho$ & \boldmath$r/R$ & \boldmath$\delta \Gamma_1$ & \boldmath$r/R$
%          & \boldmath$\delta X$ & \boldmath$r/R$ 
%         & \boldmath$\delta L_r$ & \boldmath$r/R$\\[3pt]
\tableheadseprule\noalign{\smallskip}
   {\cesam}{\small -B69} &    1.22 & 0.04918 &    2.96 & 0.66237 &    3.08 & 0.74359 &    0.10 & 0.88773 & 0.02190 & 0.04968 &    2.98 & 0.00520 \\ 
  {\cles}      &    1.40 & 0.04903 &    3.15 & 0.65759 &    3.42 & 0.74393 &    0.05 & 0.94536 & 0.02501 & 0.04942 &    4.07 & 0.00378 \\ 
  {\garstec}   &    0.91 & 0.04599 &    2.19 & 0.68604 &    2.61 & 0.74353 &    0.33 & 0.00000 & 0.01171 & 0.04614 &    6.09 & 0.00065 \\ 

 %\noalign{\smallskip}\hline
%==========================
%\hline\noalign{\smallskip}
%\multicolumn{13}{c} {\bf Case~3.2C}\\
%\hline\noalign{\smallskip}
%\hline\noalign{\smallskip}
%{\bf Code} & \boldmath$\delta c$ & \boldmath$r/R$ & \boldmath$\delta P$ & \boldmath$r/R$ 
%          & \boldmath$\delta \rho$ & \boldmath$r/R$ & \boldmath$\delta \Gamma_1$ & \boldmath$r/R$
 %         & \boldmath$\delta X$ & \boldmath$r/R$           & \boldmath$\delta L_r$ & \boldmath$r/R$\\[3pt]
\tableheadseprule\noalign{\smallskip} 
   {\cesam}{\small -B69} &    0.40 & 0.03964 &    2.18 & 0.72949 &    1.93 & 0.72949 &    0.10 & 0.73083 & 0.00788 & 0.04007 &   22.20 & 0.02376 \\ 
  {\cles}      &    0.94 & 0.03764 &    2.07 & 0.41034 &    2.41 & 0.03764 &    0.06 & 0.79075 & 0.01781 & 0.03881 &   10.13 & 0.03093 \\ 
  {\garstec}   &    0.87 & 0.03799 &    1.93 & 0.28747 &    2.95 & 0.03781 &    0.32 & 0.01557 & 0.02111 & 0.03835 &   26.89 & 0.02166 \\ 
\noalign{\smallskip}\hline
%==========================
\hline\noalign{\smallskip}
\multicolumn{13}{c} {\bf Case~3.3A -- Case~3.3B -- Case~3.3C}\\
%\hline\noalign{\smallskip}
\hline\noalign{\smallskip}
{\bf Code} & \boldmath$\delta\ln c$ & \boldmath$r/R$ & \boldmath$\delta\ln P$ & \boldmath$r/R$ 
          & \boldmath$\delta\ln \rho$ & \boldmath$r/R$ & \boldmath$\delta\ln \Gamma_1$ & \boldmath$r/R$          
          & \boldmath$\delta X$ & \boldmath$r/R$ 
          & \boldmath$\delta\ln L_r$ & \boldmath$r/R$\\[3pt]
\tableheadseprule\noalign{\smallskip} 
  {\cesam}{\small -B69} &    3.00 & 0.06563 &    5.22 & 0.88988 &    6.15 & 0.06563 &    0.02 & 0.86685 & 0.06233 & 0.06563 &    1.75 & 0.00764 \\ 
  {\cles}      &    1.51 & 0.88954 &    7.43 & 0.88881 &    7.41 & 0.88954 &    0.07 & 0.89961 & 0.03196 & 0.88954 &    7.91 & 0.00159 \\ 
  {\garstec}   &    2.83 & 0.89137 &   14.12 & 0.88512 &   14.07 & 0.88583 &    0.27 & 0.00000 & 0.04040 & 0.88583 &    3.68 & 0.00092 \\ 

\noalign{\smallskip}\hline
%==========================
%\hline\noalign{\smallskip}
%\multicolumn{13}{c} {\bf Case~3.3B}\\
%\hline\noalign{\smallskip}
%\hline\noalign{\smallskip}
%{\bf Code} & \boldmath$\delta c$ & \boldmath$r/R$ & \boldmath$\delta P$ & \boldmath$r/R$ 
 %         & \boldmath$\delta \rho$ & \boldmath$r/R$ & \boldmath$\delta \Gamma_1$ & \boldmath$r/R$
    %      & \boldmath$\delta X$ & \boldmath$r/R$ 
       %   & \boldmath$\delta L_r$ & \boldmath$r/R$\\[3pt]
\tableheadseprule\noalign{\smallskip}
   {\cesam}{\small -B69} &    1.37 & 0.83342 &   11.12 & 0.83342 &    8.25 & 0.83342 &    0.12 & 0.83342 & 0.01953 & 0.05247 &    1.65 & 0.03892 \\ 
  {\cles}      &    1.02 & 0.05241 &    1.48 & 0.34861 &    1.68 & 0.05198 &    0.06 & 0.78878 & 0.01949 & 0.05241 &    1.51 & 0.03895 \\ 
  {\garstec}   &    2.41 & 0.05219 &    2.65 & 0.33358 &    4.82 & 0.05219 &    0.34 & 0.00000 & 0.05126 & 0.05258 &    1.84 & 0.00062 \\ 

\tableheadseprule\noalign{\smallskip}
%==========================
%\hline\noalign{\smallskip}
%\multicolumn{13}{c} {\bf Case~3.3C}\\
%\hline\noalign{\smallskip}
%\hline\noalign{\smallskip}
%{\bf Code} & \boldmath$\delta c$ & \boldmath$r/R$ & \boldmath$\delta P$ & \boldmath$r/R$ 
   %       & \boldmath$\delta \rho$ & \boldmath$r/R$ & \boldmath$\delta \Gamma_1$ & \boldmath$r/R$
      %    & \boldmath$\delta X$ & \boldmath$r/R$ 
         % & \boldmath$\delta L_r$ & \boldmath$r/R$\\[3pt]
   {\cesam}{\small -B69} &    0.68 & 0.03866 &    2.14 & 0.73005 &    2.08 & 0.03866 &    0.11 & 0.73005 & 0.01320 & 0.03907 &   24.23 & 0.01925 \\ 
  {\cles}      &    0.86 & 0.03902 &    1.52 & 0.85187 &    2.42 & 0.03902 &    0.06 & 0.76734 & 0.01759 & 0.03902 &   12.60 & 0.00123 \\ 
  {\garstec}   &    2.15 & 0.03863 &    2.44 & 0.30766 &    5.86 & 0.03845 &    0.32 & 0.01262 & 0.04691 & 0.03900 &   26.80 & 0.02018 \\ 
\noalign{\smallskip}\hline
%==========================
%\hline
\end{tabular}
\label{tab2:task3}
\end{table*}

\section{Comparisons for TASK~3}\label{sec:task3}

\subsection{Presentation of the comparisons and general results}

{\task}~3 deals with models that include microscopic diffusion of helium and metals due to pressure, temperature and concentration gradients. The codes examined here have adopted different treatments of the diffusion processes. The {\astec} code follows the simplified formalism of \citet{MP93} (hereafter {\MP}) while the {\cles} and {\garstec} codes compute the diffusion coefficients by solving Burgers' equations \citep[][hereafter {\BURG}]{B69} according to the formalism of \cite{TBL94}. On the other hand, {\cesam} provides two approaches to compute diffusion velocities: one, which will be denoted by {\cesamMP} is based on the {\MP} approximation, the other (hereafter {\cesamB}) is based on Burger's formalism, with collisions integrals derived from \citet{1986ApJS...61..177P}. We point out that after preliminary comparisons for {\task}~3 models presented by \citet{jm-eas} and \citet{yl-eas}, we fixed some numerical problems found in the {\cesam} calculations including diffusion with the {\BURG} approach. Therefore, all the {\cesam} models presented here (both {\cesamMP} and {\cesamB} ones) are new re\-calculated models.

Low stellar masses ($1.0, 1.2$ and $1.3${\msol}) corresponding to solar-type stars, for which diffusion resulting from radiative forces can be safely neglected, have been considered at 3 stages of evolution (middle of the {\MS} when $X_{\rm c}=0.35$, end of the {\MS} when $X_{\rm c}=0.01$ and on {\SGB} when the helium core mass represents 5 per cent of the total mass of the star).

The models provided again have a different number of mesh points: the number of mesh points is $1200$ in the {\astec} and {\cesam} models, $2300-2500$ in {\cles} models and $1700$-$2000$ in models by {\garstec}.  
Table~\ref{tab:task3-glob} gives the mean and maximum differences in the global parameters (mass, radius, luminosity, effective temperature and age) obtained by each code with respect to {\cesamMP} models. For each code, the mean difference has been obtained by averaging over the number of cases and phases calculated. The differences are generally very small, i.e. below $0.5$ per cent for {\cesamB}, {\cles} and {\garstec}. They are a bit larger for {\astec} evolved models and they are high (25--37\%) for one {\garstec} model (the Case~3.2C, subgiant model). We note that there are small differences in mass in {\astec} and {\garstec} models: {\astec} uses a value of the solar mass slightly smaller than the one specified for the comparisons while {\garstec} starts from the specified mass but takes into account the decrease of mass during the evolution which results from the energy lost by radiation.

As in {\task}~1 we have examined the differences in the physical variables computed by the codes. {\astec} results are only considered for Case~3.1 as further studies are under way for models including convective cores \citep[see][]{jcd-eas}. Table~\ref{tab:task3-quad} provides the ``mean quadratic differences'' (Eq.~\ref{eq:quad_mean}).  

%As in {\task}~1 we note that the differences are generally low -- often below $0.5$ per cent -- but may reach 2--4 per cent (sometimes a bit more) for some variables ($P$, $\rho$, $L_r$, $\kappa$). 
As in {\task}~1 we note that the ``mean-quadratic differences'' between the codes generally remain quite low.
The differences in $P$, $T$, $L_r$, $r$, $\rho$ and $\kappa$ range from $0.1$ to $6\%$. The differences in the thermodynamic quantities ($\Gamma_1$, $\nabla_{\rm ad}$, $C_p$)  are often well below $1$ per cent with, as in {\task}~1, larger differences in the {\garstec} code which are probably due to a different use of the {\small OPAL} equation of state package and variables.

The maximal differences\footnote{The method of calculation is the same as in Sect.~\ref{sec:task1.1}. \label{calculo}}  in $c$, $P$, $\rho$, $\Gamma_1$, $L_r$ and $X$ between codes, and the location $(r/R)$ where they happen are reported in Table~\ref{tab:task3-maxdiff}. Again we note that the largest differences are mainly found in the most external layers and (or) at the boundary of the convection regions.

%----------------------------------------------------------------------------
\begin{figure*}[htbp!]
\centering
\resizebox*{\hsize}{!}{\includegraphics*{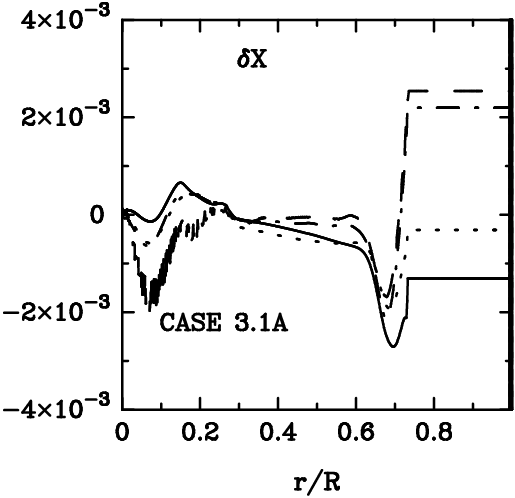}\includegraphics*{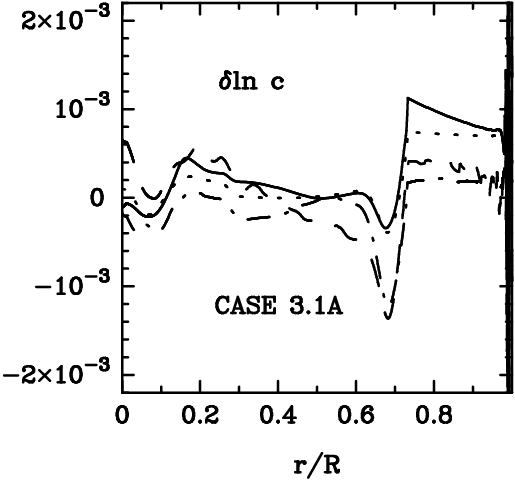}\includegraphics*{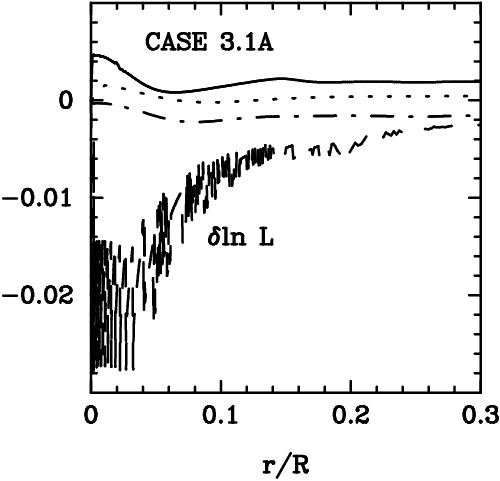}\includegraphics*{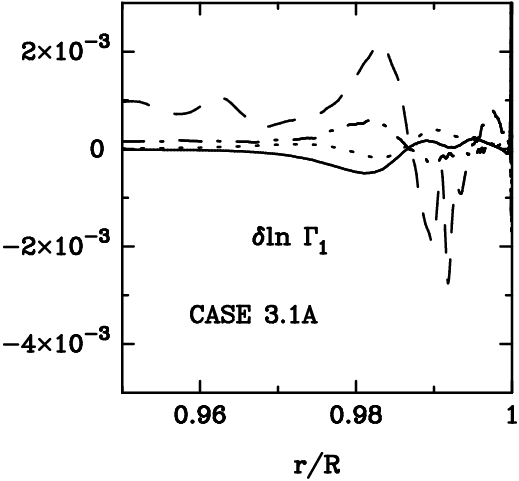}}
\resizebox*{\hsize}{!}{\includegraphics*{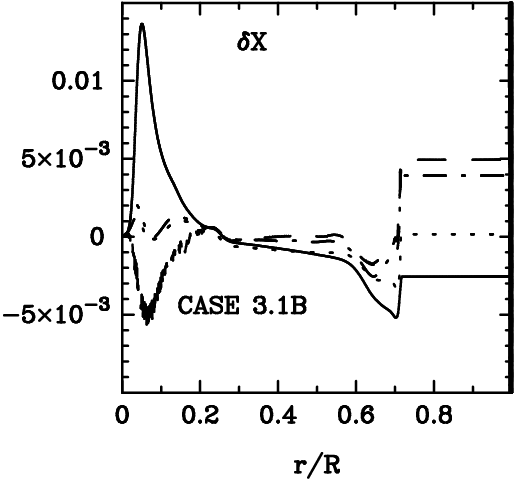}\includegraphics*{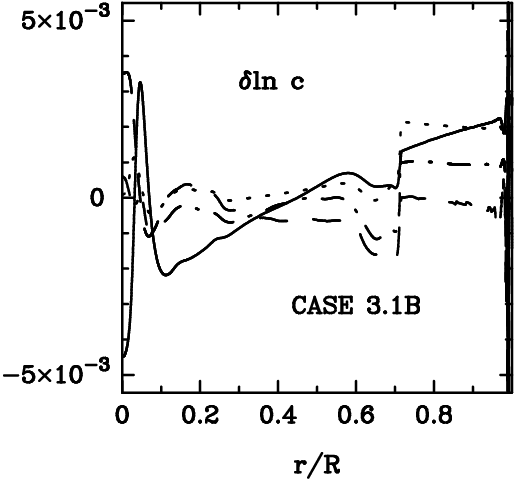}\includegraphics*{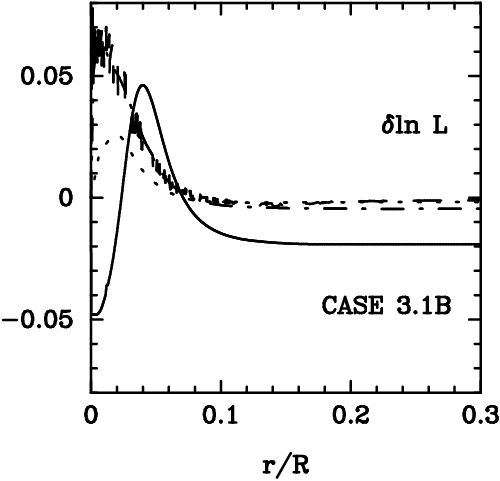}\includegraphics*{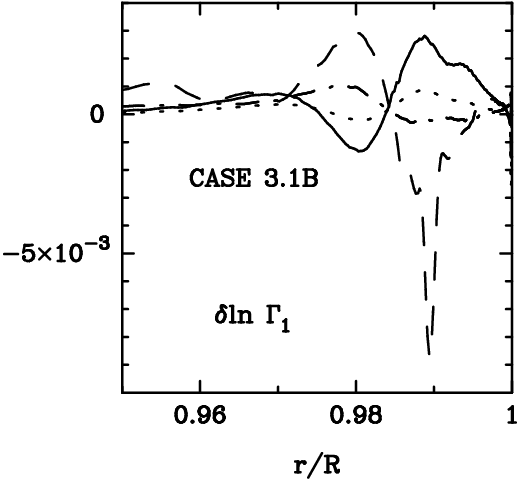}}
\resizebox*{\hsize}{!}{\includegraphics*{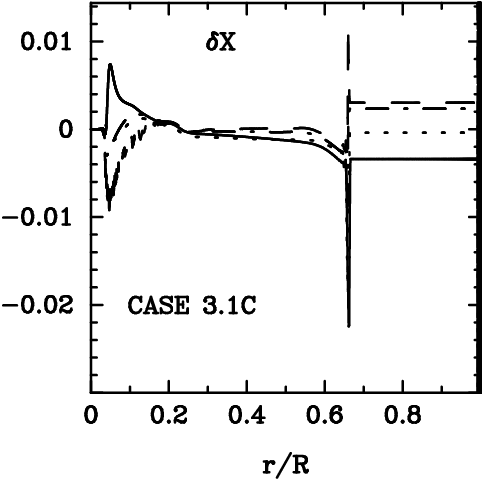}\includegraphics*{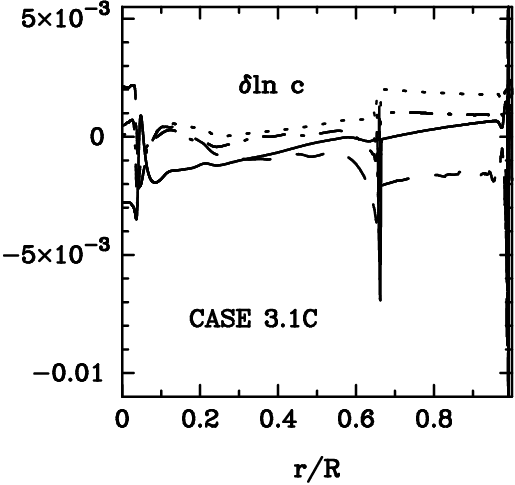}\includegraphics*{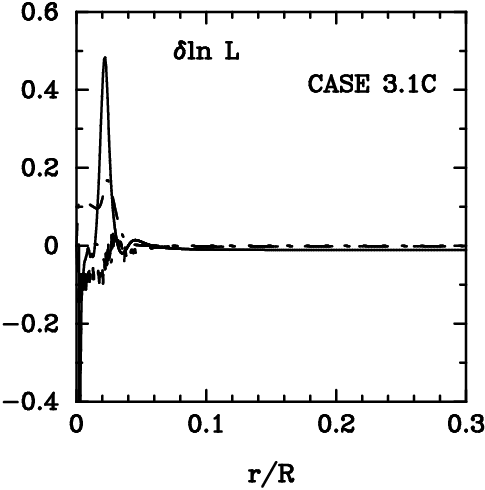}\includegraphics*{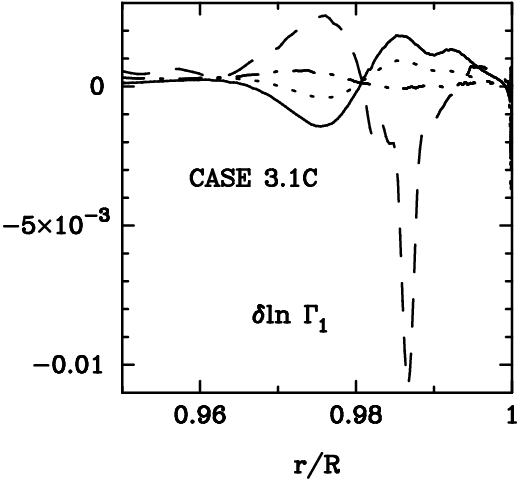}}
\caption{{\task}~3: Differences between pairs of models ({\small CODE}-\cesam) corresponding to Cases~3.1, phase A (top), B (middle) and C (bottom) plotted as a function of radius. From left to right:  for hydrogen mass fraction, logarithmic sound speed, logarithmic luminosity and adiabatic exponent $\Gamma_1$. Differences have been calculated at fixed relative mass (for $X$, $c$, $L_r$) or fixed relative radius (for $\Gamma_1$). Results are given for {\astec} (continuous line), {\cesamB} (dotted), {\cles} (dot-dash) and {\garstec} (dashed).
} 
\label{fig:task3-int31}
\end{figure*}
%----------------------------------------------------------------------------

%----------------------------------------------------------------------------
\begin{figure*}[htbp!]
\centering
\resizebox{\hsize}{!}{\includegraphics{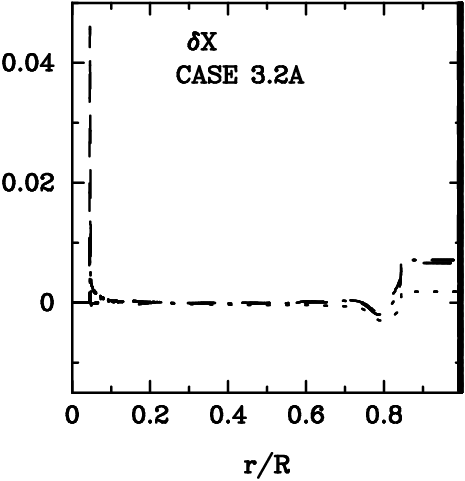}\includegraphics{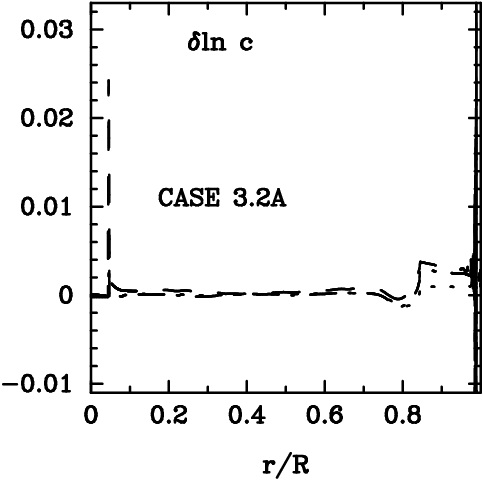}\includegraphics{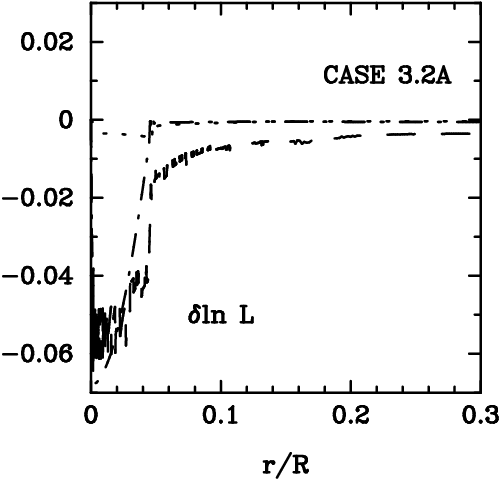}\includegraphics{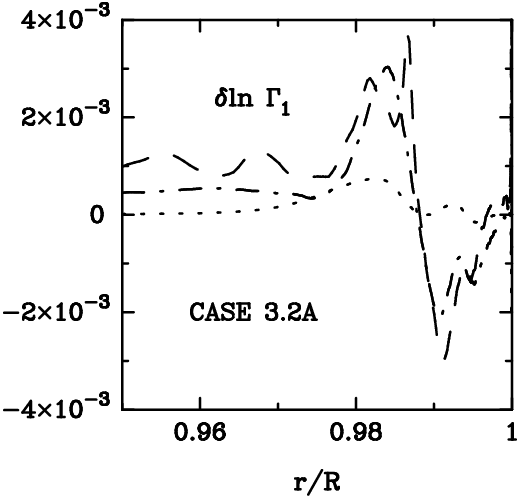}}
\resizebox{\hsize}{!}{\includegraphics{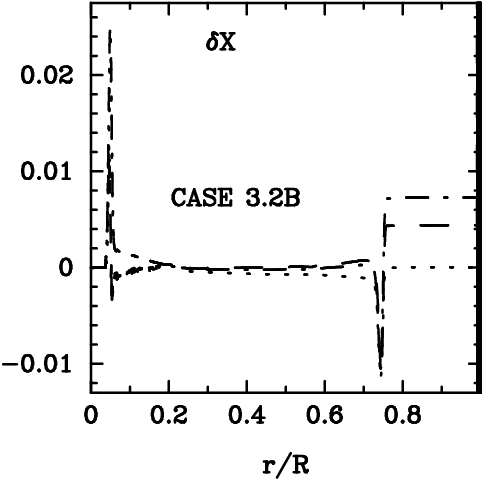}\includegraphics{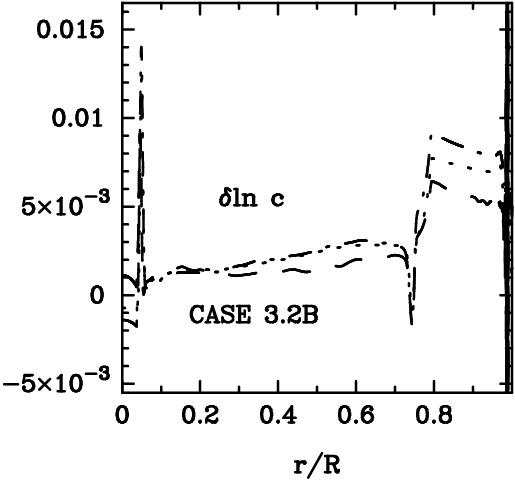}\includegraphics{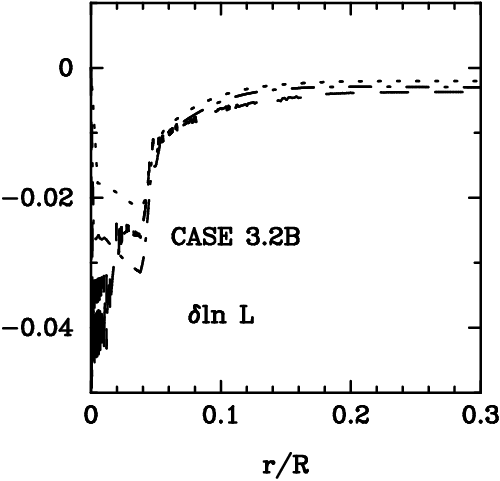}\includegraphics{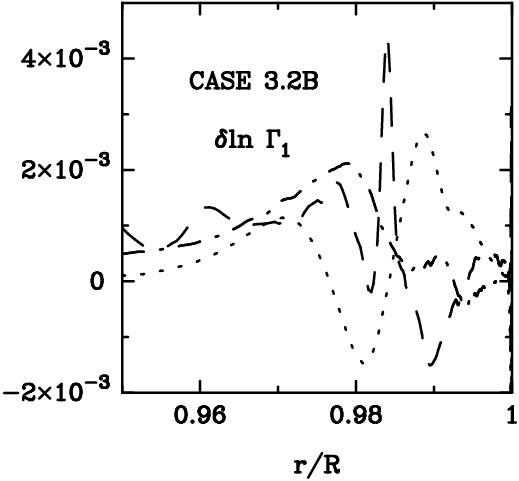}}
\resizebox{\hsize}{!}{\includegraphics{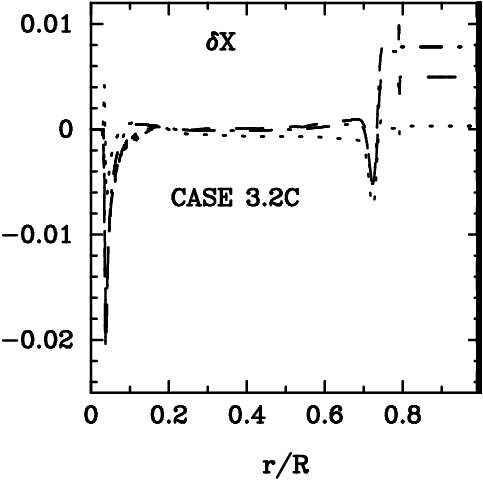}\includegraphics{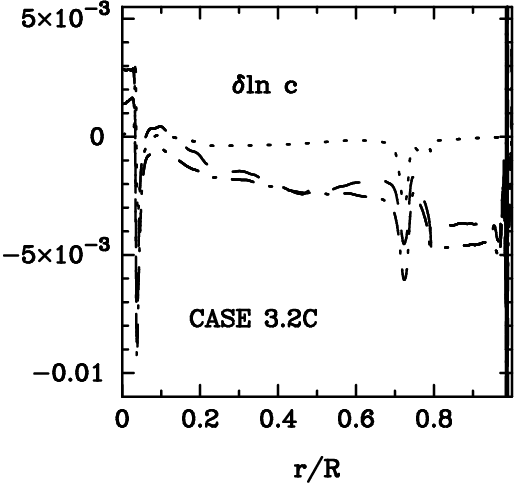}\includegraphics{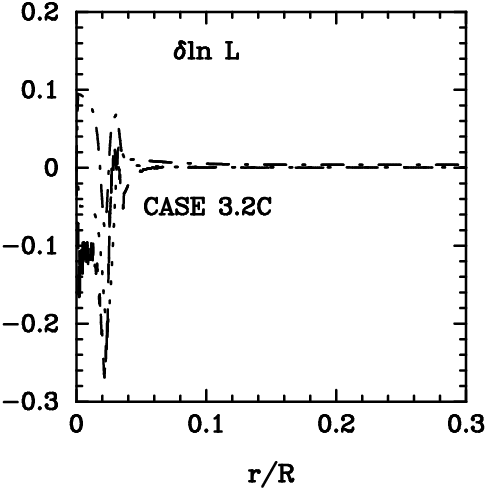}\includegraphics{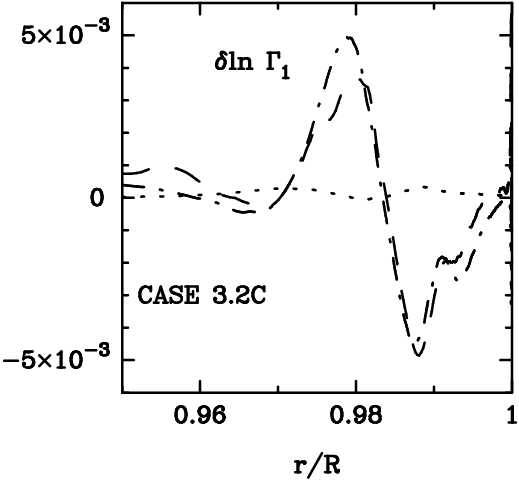}}
\caption{{\task}~3: Differences between pairs of models ({\small CODE}-\cesam) corresponding to Cases~3.2, phase A (top), B (middle) and C (bottom) plotted as a function of radius. From left to right:  for hydrogen mass fraction, logarithmic sound speed, logarithmic luminosity and adiabatic exponent $\Gamma_1$. Differences have been calculated at fixed relative mass (for $X$, $c$, $L_r$) or fixed relative radius (for $\Gamma_1$). Results are given for {\astec} (continuous line), {\cesamB} (dotted), {\cles} (dot-dash) and {\garstec} (dashed).
} 
\label{fig:task3-int32}
\end{figure*}
%----------------------------------------------------------------------------
%\begin{figure*}[ht!]
%\centering
%\resizebox*{0.9\hsize}{!}{\hspace*{-0.cm}\includegraphics*{./FIG-T3-ZC/ce_cles.eps}}
%\caption{{\task}~3: Convective envelope radius evolution for $1.0$ (left), $1.2$ (centre) and $1.3${\msol} (right) models computed with {\cles}. Black regions correspond to the convective envelope, and grey ones are convectively unstable regions outside the convective envelope.}
%\label{fig:task3-zce}       % Give a unique label
%\end{figure*}

%\begin{figure}[b!]
%\centering
%\resizebox{0.75\hsize}{!}{\includegraphics{./FIG-T3-ZC/mcc_cles.eps}}
%\caption{{\task}~3: Convective core mass evolution for $1.2${\msol} models computed with {\cles}. The black region correspond to the convective core, and grey ones are convectively unstable regions outside the convective core.}
%\label{fig:task3-cc}   
%\end{figure}

%\begin{figure*}[ht!]\centering
%\resizebox*{0.75\hsize}{!}{\hspace{-0.cm}\includegraphics*{./FIG-T3-ZC/RZC.eps}\hspace{2.cm}\includegraphics*{./FIG-T3-ZC/MCC.eps}}
%\caption{{\task}~3: Convective envelope radius evolution (right) and convective core mass evolution (left) for Cases 3.1, 3.2 and 3.3 (repectively $1.0$, $1.2$ and $1.3${\msol}) in models computed with {\cesam} with two different approaches to treat microscopic diffusion ({\MP} and {\BURG})}
%\label{fig:task3-zce-ces}       % Give a unique label
%\end{figure*}
%----------------------------------------------------------------------------
\begin{figure*}[htbp!]
\centering
\resizebox{\hsize}{!}{\includegraphics{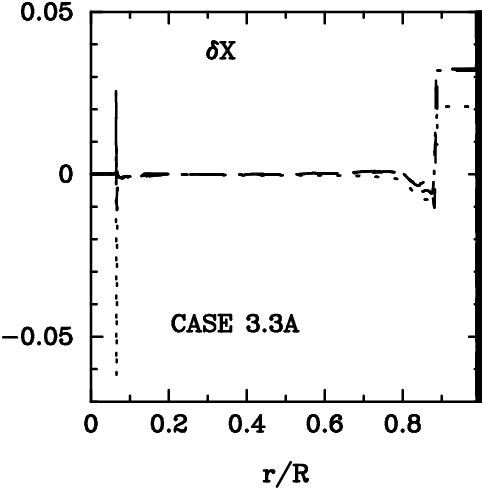}\includegraphics{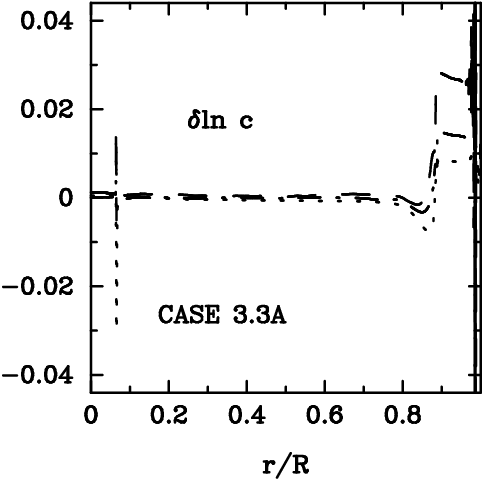}\includegraphics{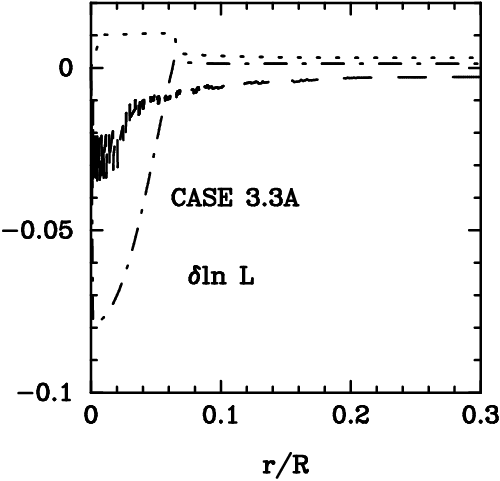}\includegraphics{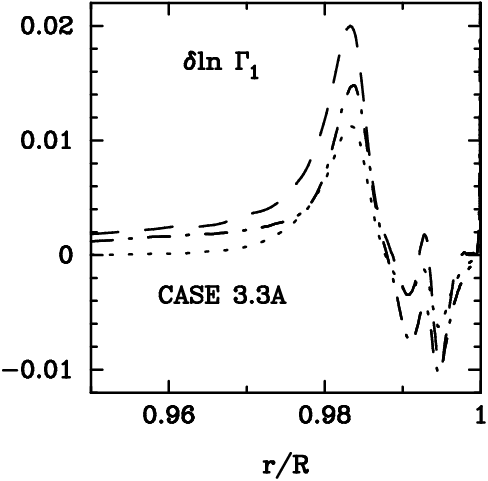}}
\resizebox{\hsize}{!}{\includegraphics{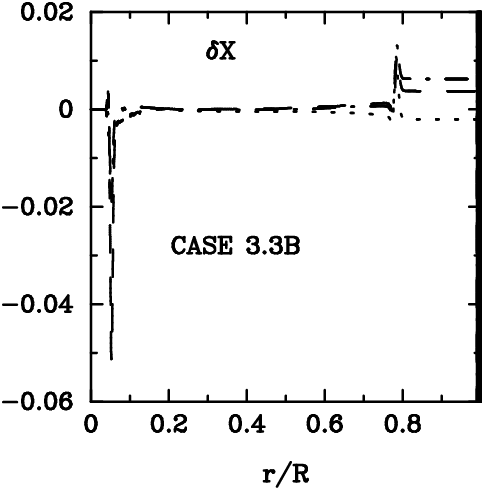}\includegraphics{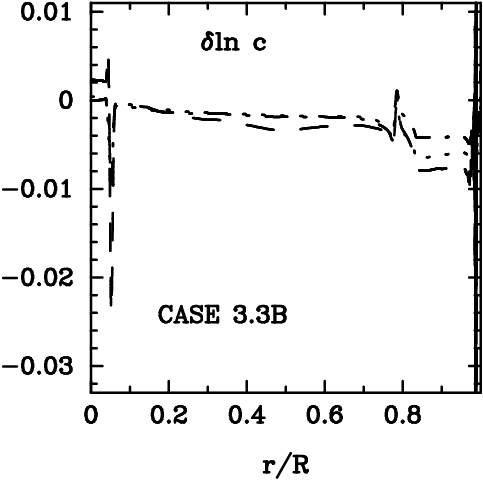}\includegraphics{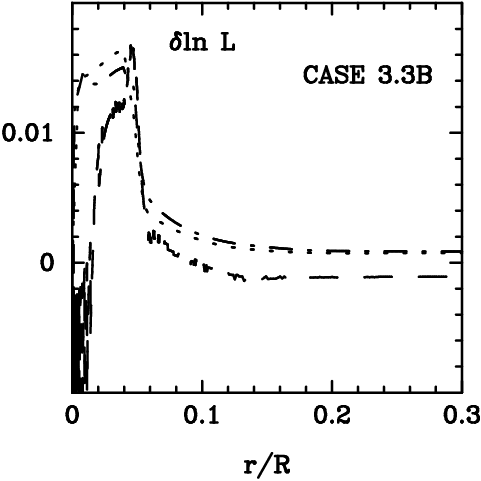}\includegraphics{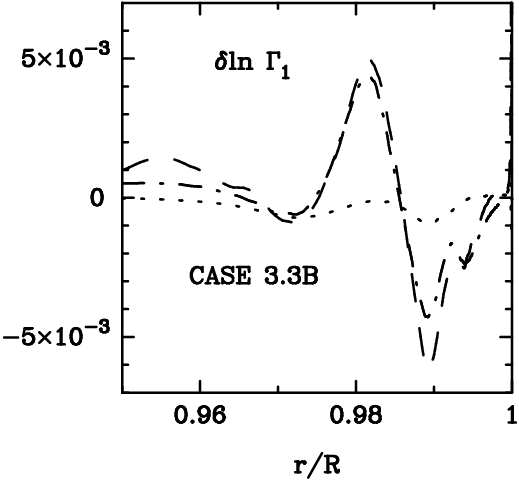}}
\resizebox{\hsize}{!}{\includegraphics{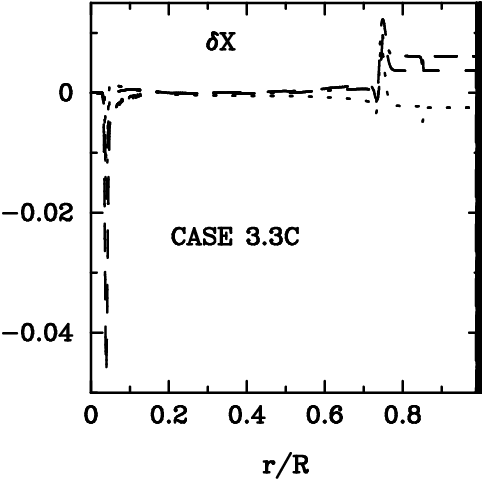}\includegraphics{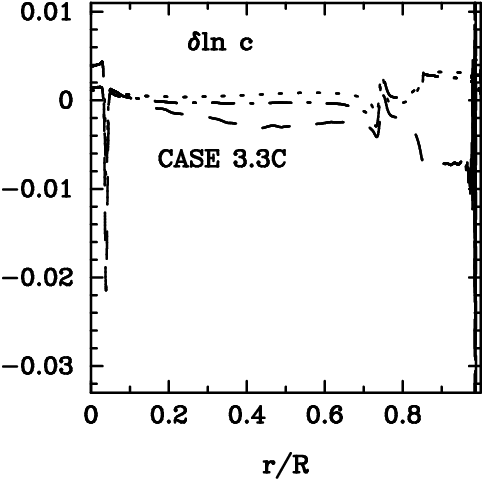}\includegraphics{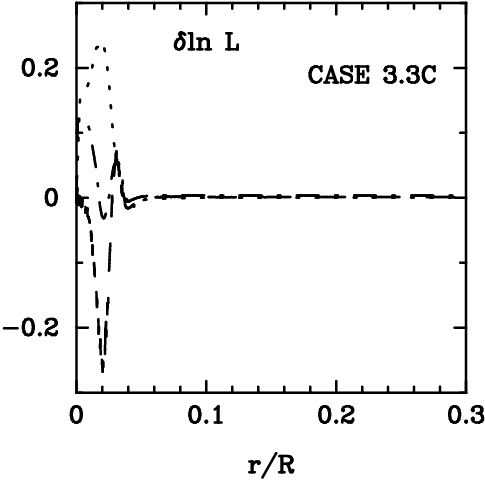}\includegraphics{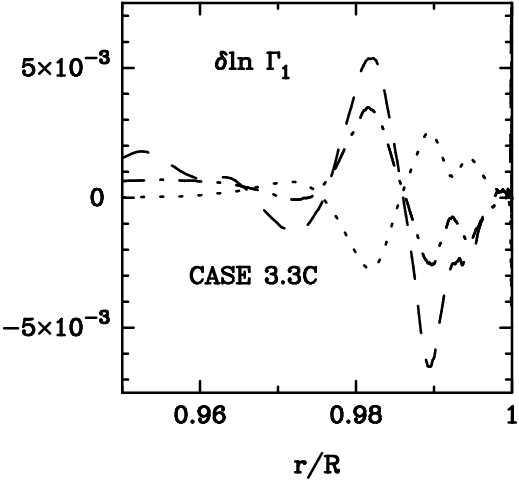}}
\caption{{\task}~3: Differences between pairs of models ({\small CODE}-\cesam) corresponding to Cases~3.3, phase A (top), B (middle) and C (bottom) plotted as a function of radius. From left to right:  for hydrogen mass fraction, logarithmic sound speed, logarithmic luminosity and adiabatic exponent $\Gamma_1$. Differences have been calculated at fixed relative mass (for $X$, $c$, $L_r$) or fixed relative radius (for $\Gamma_1$). Results are given for {\astec} (continuous line), {\cesamB} (dotted), {\cles} (dot-dash) and {\garstec} (dashed).
} 
\label{fig:task3-int33}
\end{figure*}
%----------------------------------------------------------------------------

\subsection{Internal structure}\label{sec:task3-int}

The variations in $X$, $c$, $L_r$ and $\Gamma_1$ are displayed in Figures~\ref{fig:task3-int31}, \ref{fig:task3-int32}, \ref{fig:task3-int33}, for Cases~3.1, 3.2 and 3.3 respectively. Here, to spare space, we do not plot differences in $\rho$ which are reflected in those in $c$.

%----------------------------------------------------------------------------
\begin{figure}[htbp!]
\centering
\resizebox*{0.85\hsize}{!}{\includegraphics*{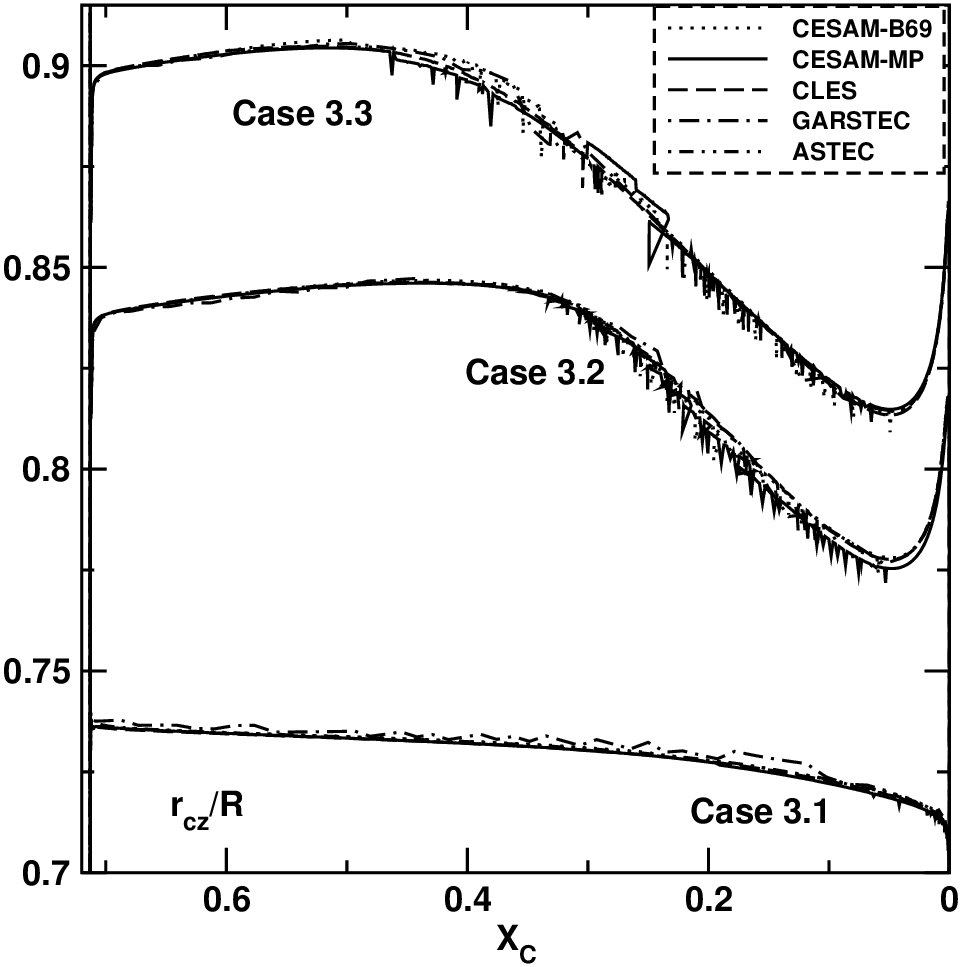}}
\caption{{\task}~3: Evolution of the radius of the convective envelope for Cases~3.1, 3.2 and 3.3 (respectively $1.0$, $1.2$ and $1.3${\msol}) in models computed by {\astec}, {\garstec}, {\cles} and {\cesam} with for the latter two different approaches to treat microscopic diffusion ({\MP} and {\BURG})}
\label{fig:task3-zce}       % Give a unique label
\end{figure}
%----------------------------------------------------------------------------
%----------------------------------------------------------------------------
\begin{figure}[htbp!]
\centering
\resizebox*{0.85\hsize}{!}{\includegraphics*{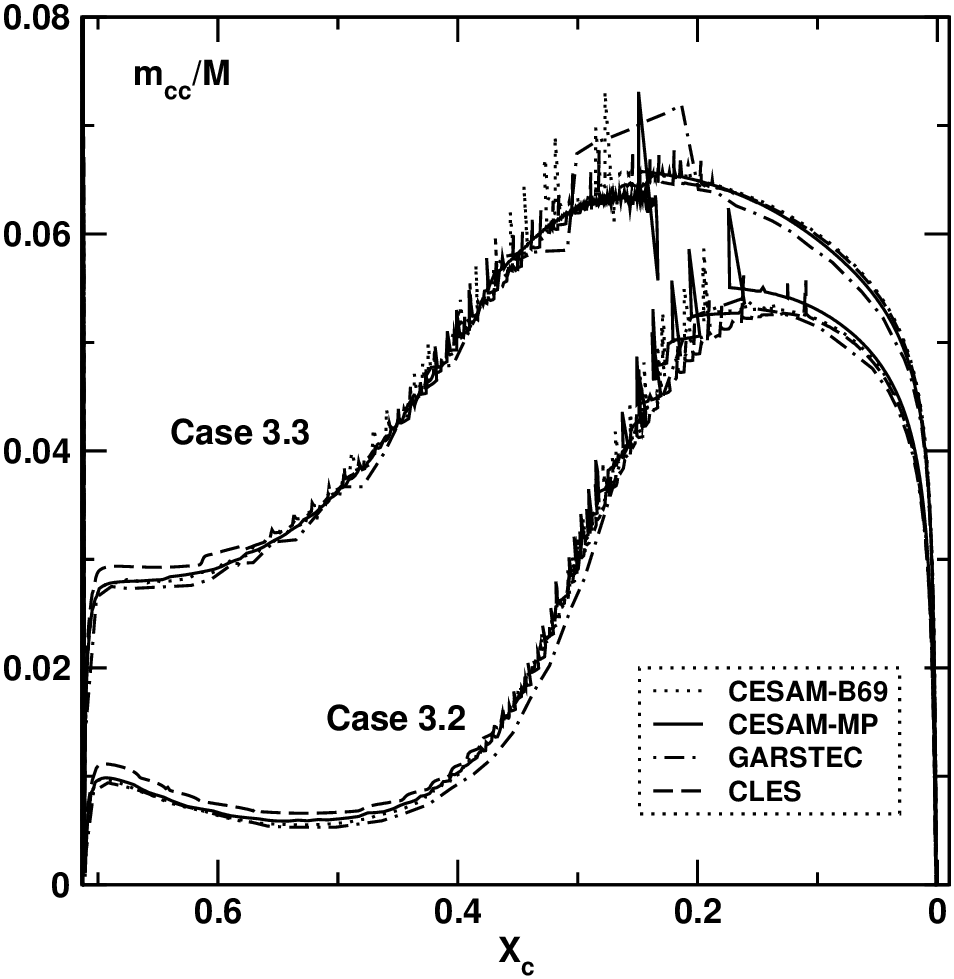}}
\caption{{\task}~3: Evolution of the mass of the convective core for Cases~3.2 and 3.3 (respectively $1.0$, $1.2$ and $1.3${\msol}) in models computed by {\cles}, {\garstec} and {\cesam} with for the latter two different approaches to treat microscopic diffusion ({\MP} and {\BURG})}
\label{fig:task3-cc}       % Give a unique label
\end{figure}
%----------------------------------------------------------------------------

\subsubsection{Solar models: Case~3.1}

The solar model is characterised by a radiative interior and a convective envelope which deepens as evolution proceeds. The differences in the hydrogen abundance $X$ seen in Fig.~\ref{fig:task3-int31} can be compared to those found in {\task}~1, Case~1.1 model (Fig.~\ref{fig:task1-int123}, top-right). In the centre, where there exists an H gradient built by nuclear reactions and where (in the present case) H is drawn outwards by diffusion, the differences are roughly of the same order of magnitude. In the middle-upper radiative zone, where the settling of He and metals leads to an H enrichment, and in the convection zone, much larger differences are found which reflect different diffusion velocities and also depend on the extension and downward progression of the convective envelope. We note that differences grow with evolution from phase A to C. The sound speed differences reflect differences (i) in the stellar radius, (ii) in the chemical composition gradients in the central regions and below the convective envelope (see the features in the region where $R\in [0.65,0.9]$)  and finally (iii) in the location of the convective regions boundaries. They remain quite modest except at the border of the convective envelope and in the zone close to the surface. The differences in $\Gamma_1$, seen in the external regions, reflect differences in the He abundance in the regions of second He ionisation. We also note that large differences in luminosity are found on the {\SGB} (phase C).

\subsubsection{Solar-type stars with convective cores: Cases~3.2 and 3.3}

Those stars of $1.2$ and $1.3${\msol} have, on the {\MS}, a convective envelope and a convective core. The differences in the hydrogen abundance $X$ seen in the centre in Fig.~\ref{fig:task3-int32} and \ref{fig:task3-int33} are rather similar to those found in {\task}~1, Case~1.2 and 1.3 models (Fig.~\ref{fig:task1-int123}, centre and bottom, right). As explained in Sect.~\ref{sec:task1-conv}, the core mass is growing during a large fraction of the {\MS}. Due to nuclear reactions a helium gradient builds up at the border of the core. In these regions, the diffusion due to the He concentration gradient competes with the He settling term and finally dominates which makes He move outwards from the core regions. As a consequence metals also diffuse outwards preventing the metal settling. Because of the metal enrichment which induces an opacity increase, the zone at the border of the convective core is the seat of semiconvection \citep{Richard01}.

Also, large differences appear at the bottom of and in the convective envelope in the presence of diffusion. In fact, in these models, diffusion makes the metals pile up beneath the convective envelope which induces an increase of opacity in this zone which in turn triggers convective instability in the form of semiconvection \citep[see][]{2001ApJ...555..990B}.

\subsection{Convection zones}

Fig.~\ref{fig:task3-zce} %taken from \cite{jm-eas} 
shows the evolution of the radius of the convective envelope in the models for Cases~3.1, 3.2 and 3.3 while Fig.~\ref{fig:task3-cc} displays the evolution of the mass of the convective core in Cases~3.2 and 3.3. The crumpled zones in the $r_{\rm cz}/R$ and $m_{\rm cc}/M$ profiles are the signature of the regions of semiconvection which we find for Cases~3.2 and 3.3 either in the regions above the convective core or beneath the convective envelope. As pointed out by \citet{jm-eas}, in the presence of metal diffusion, it is difficult to study the evolution of the boundaries of convective regions. The numerical treatment of those boundaries in the codes is crucial for the determination of the evolution of the unstable layers: it affects the outer convective zone depths and surface abundances as well as the masses of the convective cores and therefore the evolution of the star.

Rather small differences in the location of the convective boundaries are seen in Figs.~\ref{fig:task3-zce} and \ref{fig:task3-cc}. Table~\ref{tab:task3-conv} displays the properties of the convective zone boundaries. The radii found at the base of the convective envelope in all cases differ by $0.1$--$0.7$ per cent. On the other hand the mass in the convective cores differs by $2$--$4$ per cent except for Case~3.2A where the mass in {\garstec} model differs from the others by more than $10$ per cent. 

\subsection{Helium surface abundance}\label{sec:task3-ys}

\begin{figure*}[htbp!]
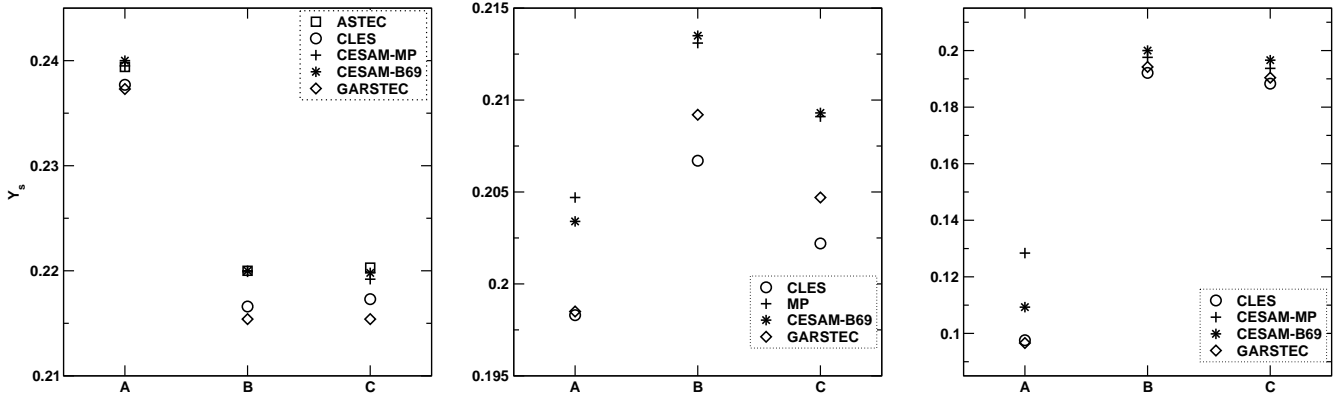

\centering
\resizebox*{\hsize}{!}{\hspace*{-0.2cm}\includegraphics*{Ys1.0.eps}\hspace*{1.5cm}\includegraphics*{Ys1.2.eps}\hspace*{1.8cm}\includegraphics*{Ys1.3.eps}}
%\resizebox{\hsize}{!}{ \includegraphics{./FIGS-TASK3/Ys1.0.eps}\hspace{-0.2cm}\includegraphics{./FIGS-TASK3/Ys1.2.eps}\hspace{0.2cm}\includegraphics{./FIGS-TASK3/Ys1.3.eps}}
%\resizebox{\hsize}{!}{\includegraphics{./FIGS-TASK3/Ys1.0.eps}}\resizebox{0.3\hsize}{!}{\includegraphics{./FIGS-TASK3/Ys1.2.eps}}\resizebox{0.3\hsize}{!}{\includegraphics{./FIGS-TASK3/Ys1.3.eps}}
\caption{{\task}~3: Helium content $Y_{\rm s}$ in the convection envelope for the different codes and cases considered (A and B are for middle and end of the {\MS} respectively while C is for {\SGB}, see Table~\ref{tab:task3}) for $1.0${\msol} (left),  $1.2${\msol} (centre) and  $1.3${\msol} (right).}
\label{fig:task3-ys}       % Give a unique label
\end{figure*}

Figure~\ref{fig:task3-ys} displays the helium abundance $Y_{\rm s}$ in the convective envelope for the different cases and phases considered. The evolution of $Y_{\rm s}$  is linked to the efficiency of microscopic diffusion inside the star and to the evolution with time of the internal border of the convective envelope. We note that the surface helium abundance differs by less than $2\%$ for Case~3.1 (any phase) and $3\%$ for Case~3.2 (any phase) and Case~3.3 (phases B and C) whatever the prescription for the diffusion treatment is. For Case~3.3A which is hotter with a thinner convection envelope, the differences between the {\MP} prescription for diffusion used in  {\cesamMP} models and the complete solution of Burger's equations {\BURG} are rather large. For {\cesamB} models this difference is of the order of 16\%, and a  maximum of about $30\%$ in the diffusion efficiency is found between {\garstec} and {\cesamMP}. Such differences can indeed be expected and result from the different approaches used to treat microscopic diffusion ({\MP} vs. {\BURG}) and from the approximations made to calculate the collision integrals. 
For the solar model, \citet{TBL94} found differences of about $15$ per cent between their results -- based on the solution of Burger's equations but with approximations made to estimate the collision integrals- and the {\MP} formalism of \cite{MP93} -- where the diffusion equations are simplified but in which the collision integrals are obtained according to \citet{1986ApJS...61..177P}. Further tests made by one of us (JM) have shown that for Case~3.2, the use of collision integrals of \citet{1986ApJS...61..177P} within the formalism by \citet{TBL94}  leads to differences in the surface He abundance that may amount to $\sim 2.5$ per cent. Since we can expect that the differences of the diffusion coefficients increase as the depth of the convective zone decreases, it is not surprising that in Case~3.3 models differences in the surface helium abundance are even larger. We also point out that the helium depletion is also sensitive to the numerical treatment of convective borders, in particular in the presence of semiconvection.

\subsection{Seismic properties}\label{sec:task3-seismo}

\begin{figure*}[htbp!]
\centering
\resizebox*{\hsize}{!}{\hspace*{-0.2cm}\includegraphics*{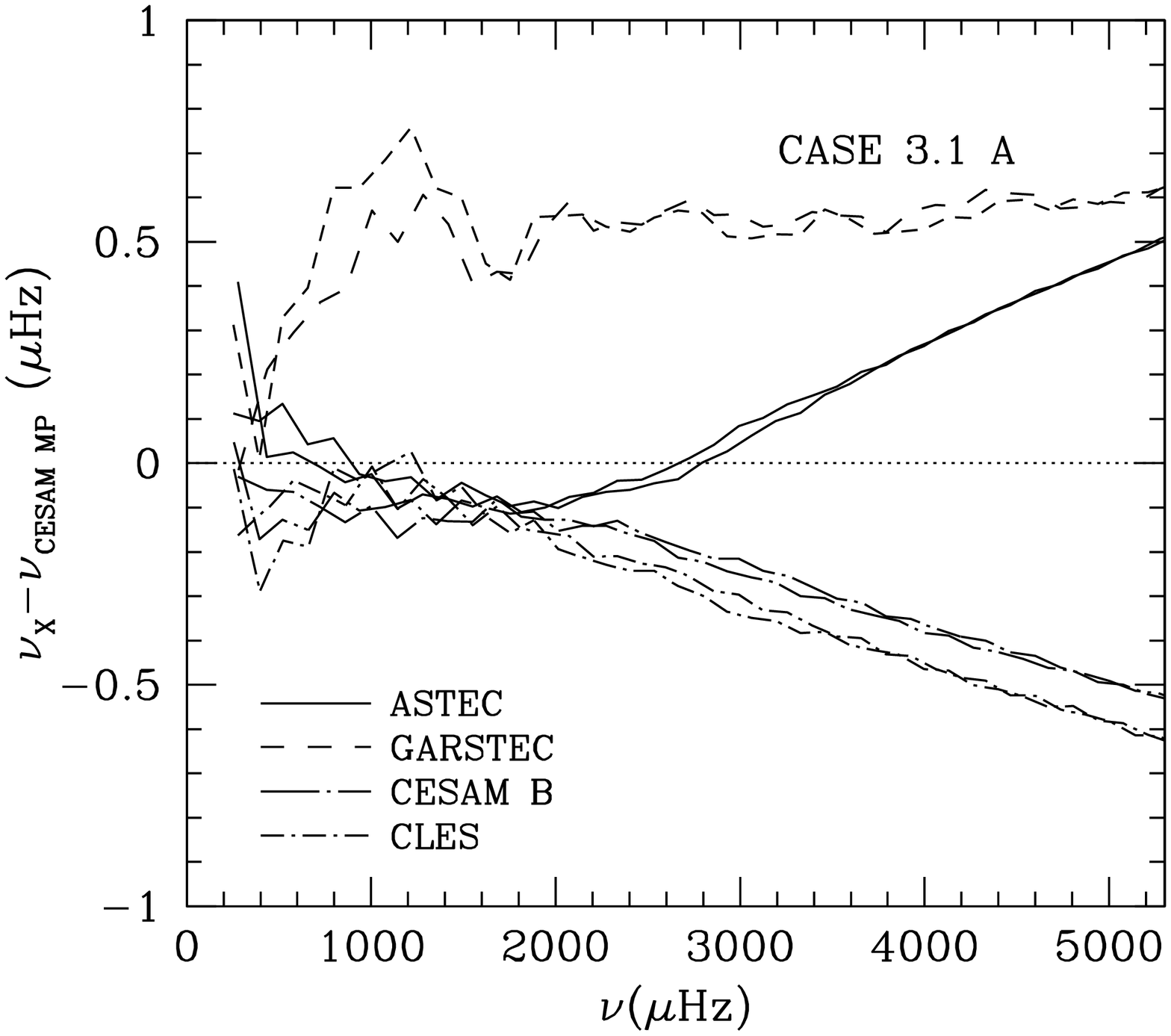}\hspace*{1.5cm}\includegraphics*{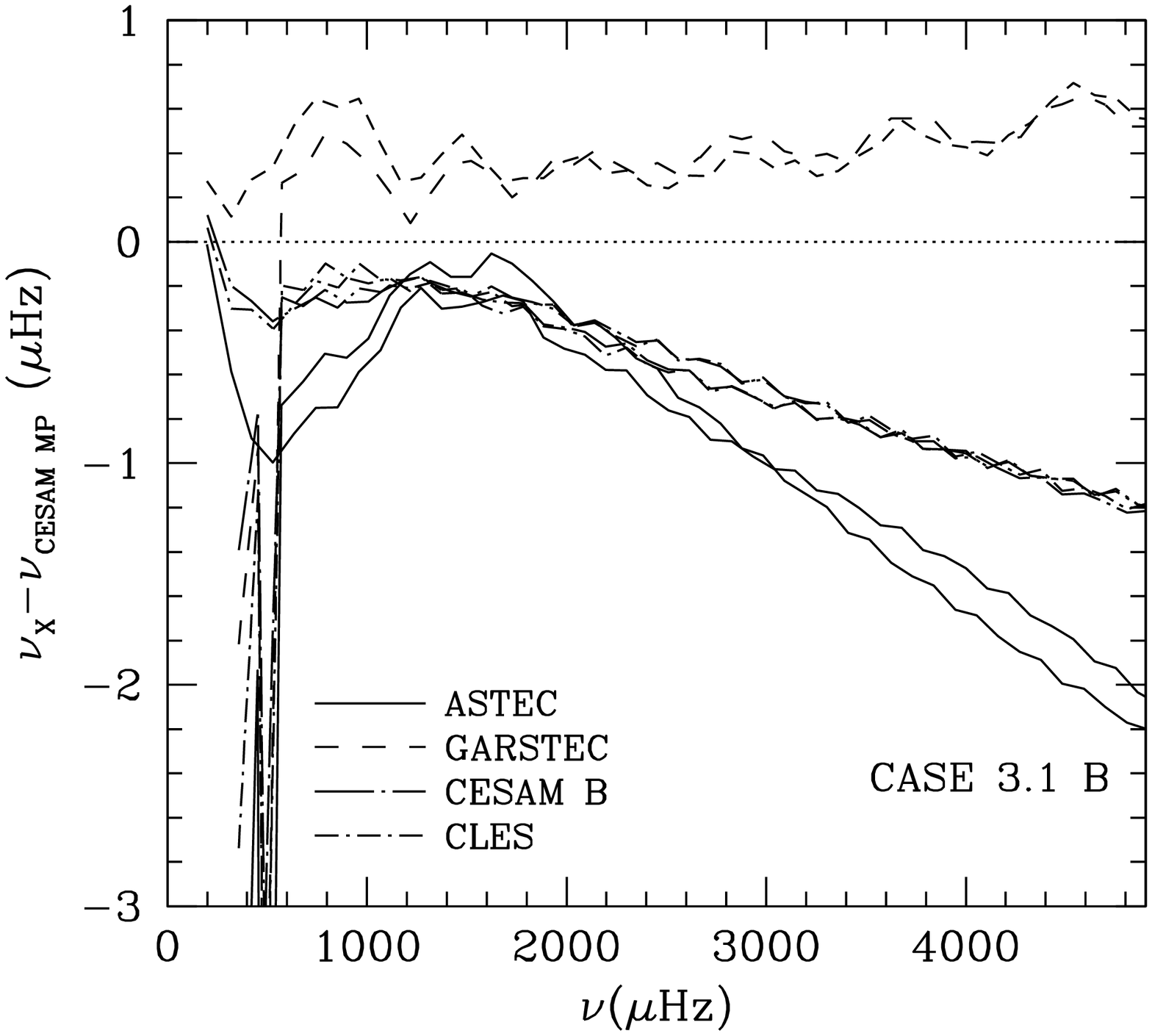}\hspace*{1.8cm}\includegraphics*{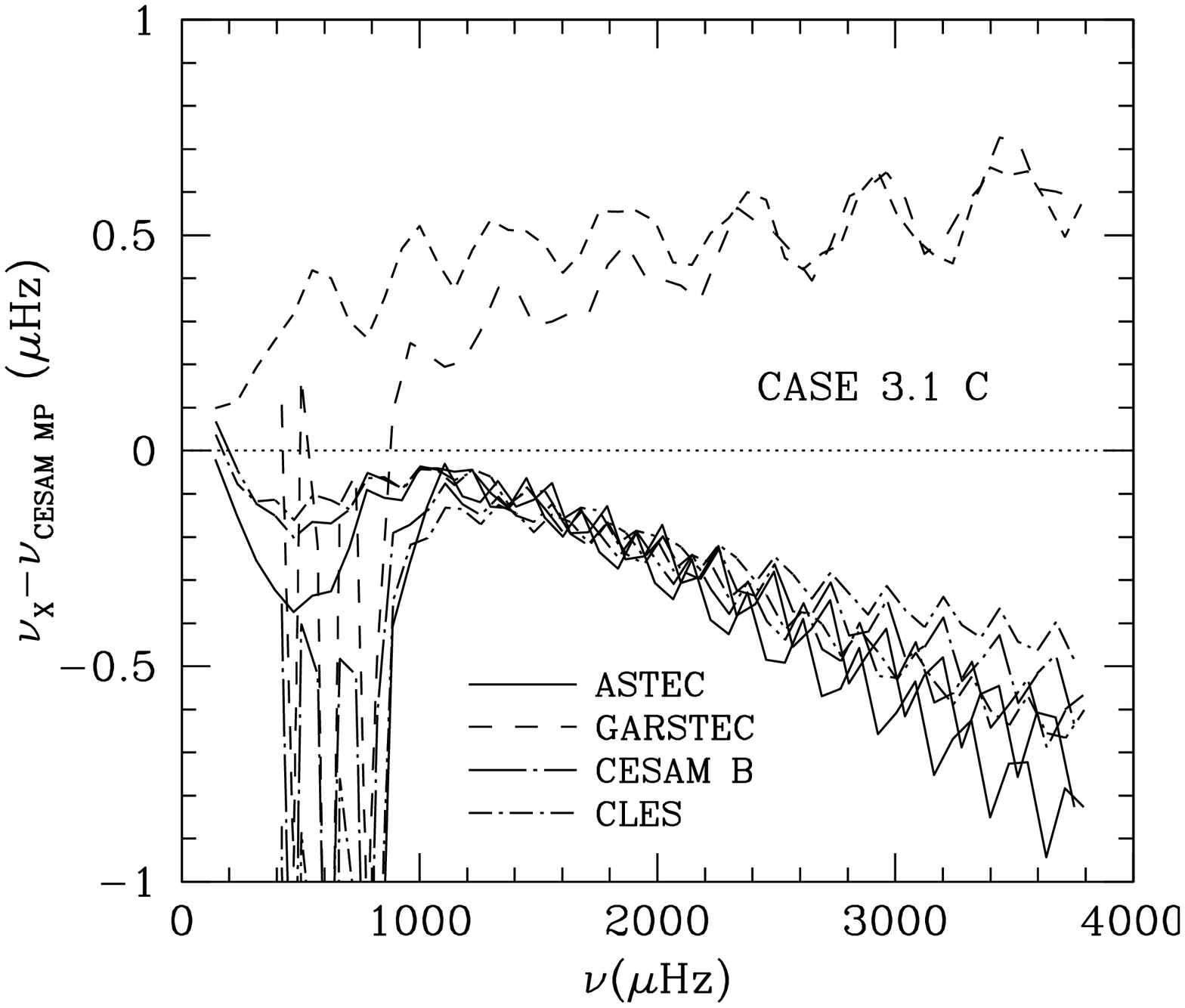}}
\resizebox*{\hsize}{!}{\hspace*{-0.2cm}\includegraphics*{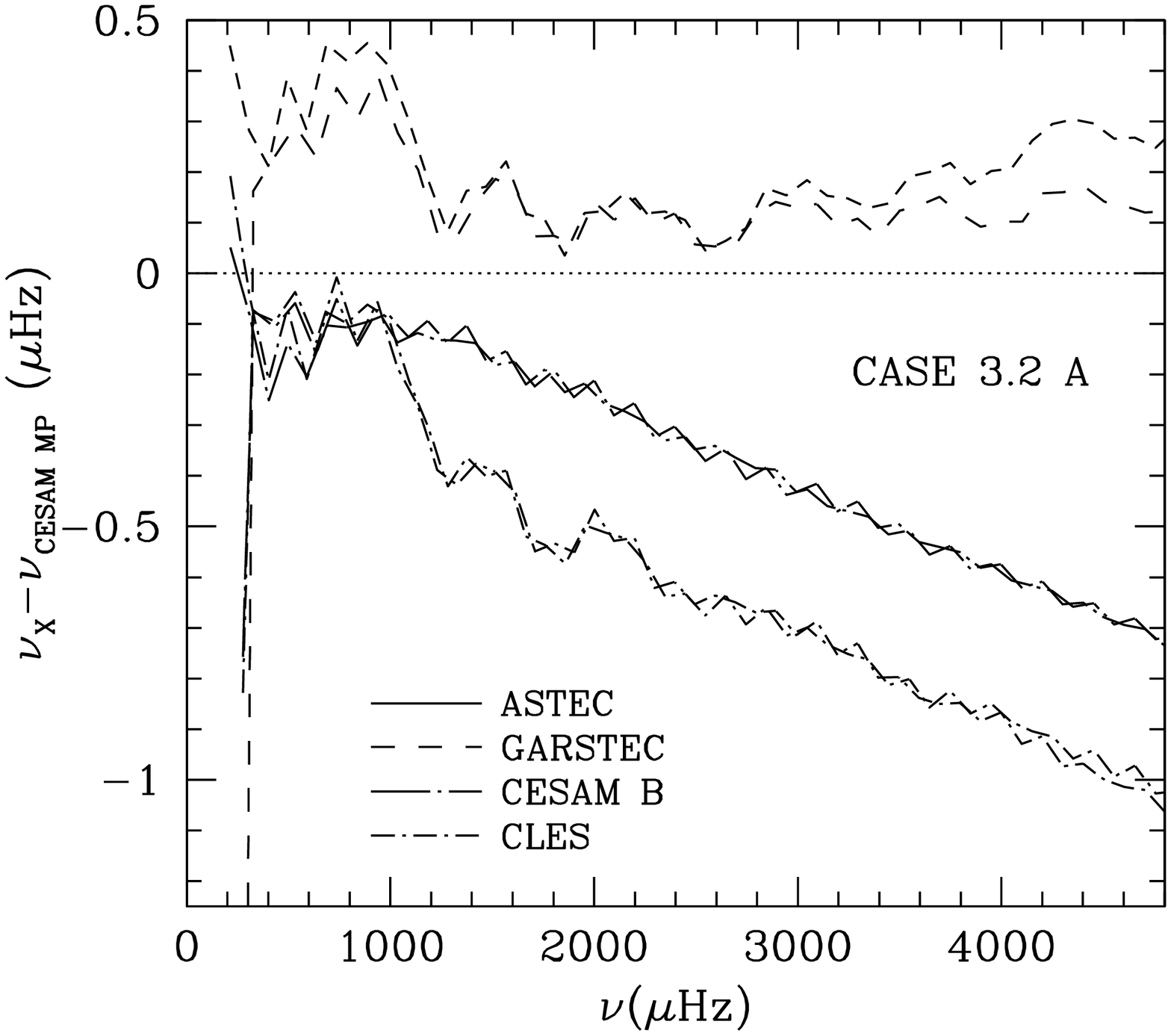}\hspace*{1.5cm}\includegraphics*{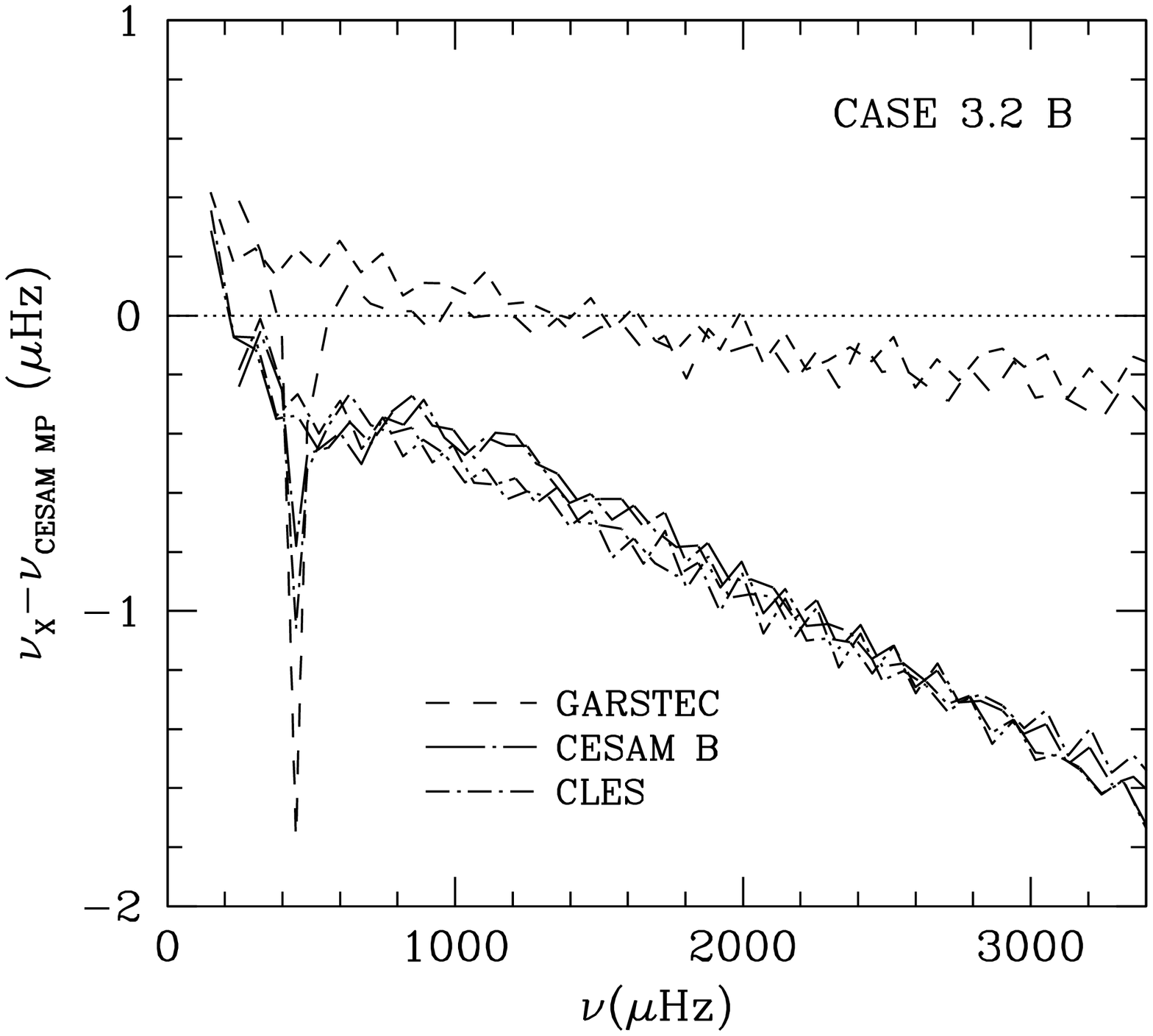}\hspace*{1.8cm}\includegraphics*{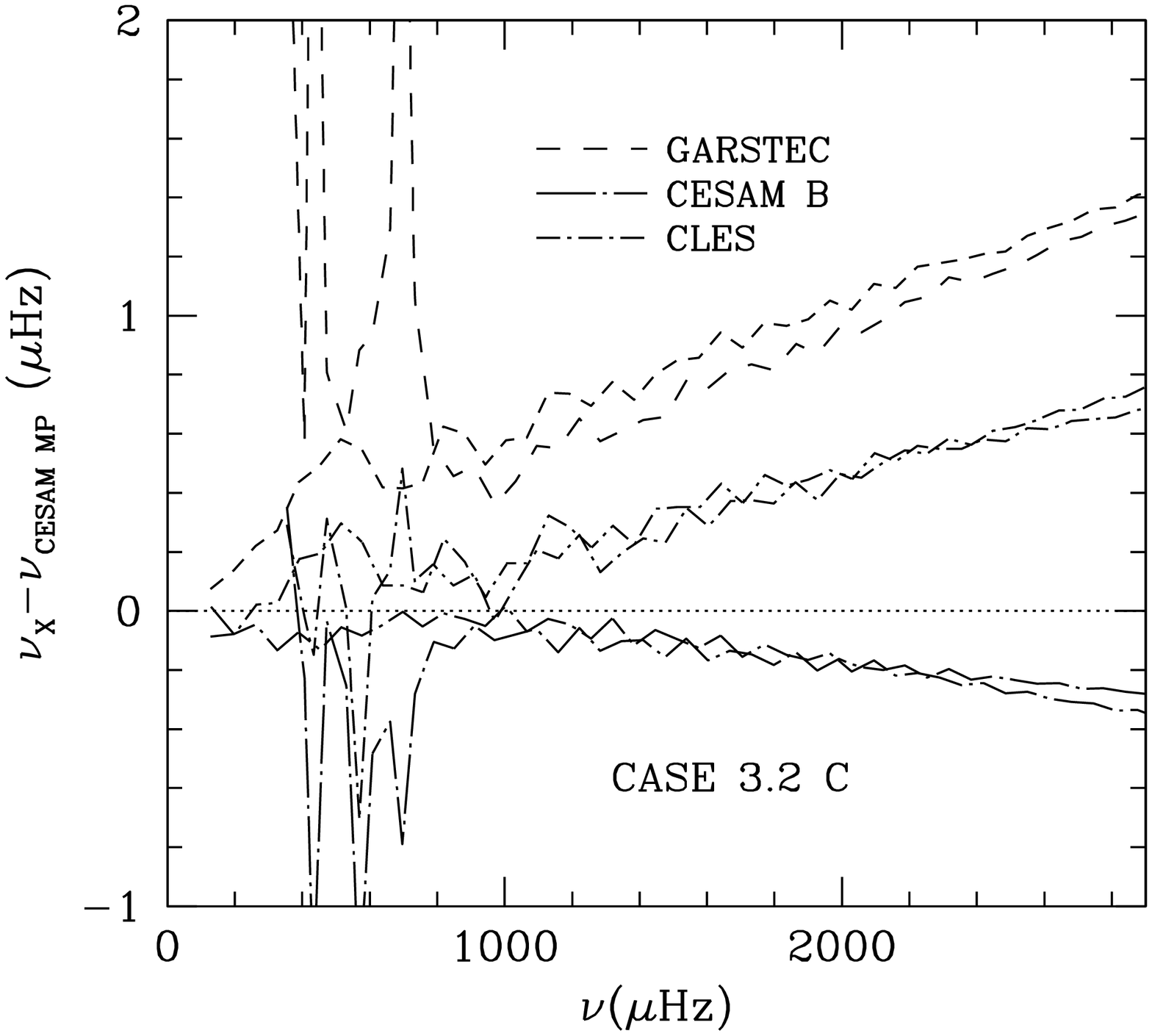}}
\resizebox*{\hsize}{!}{\hspace*{-0.2cm}\includegraphics*{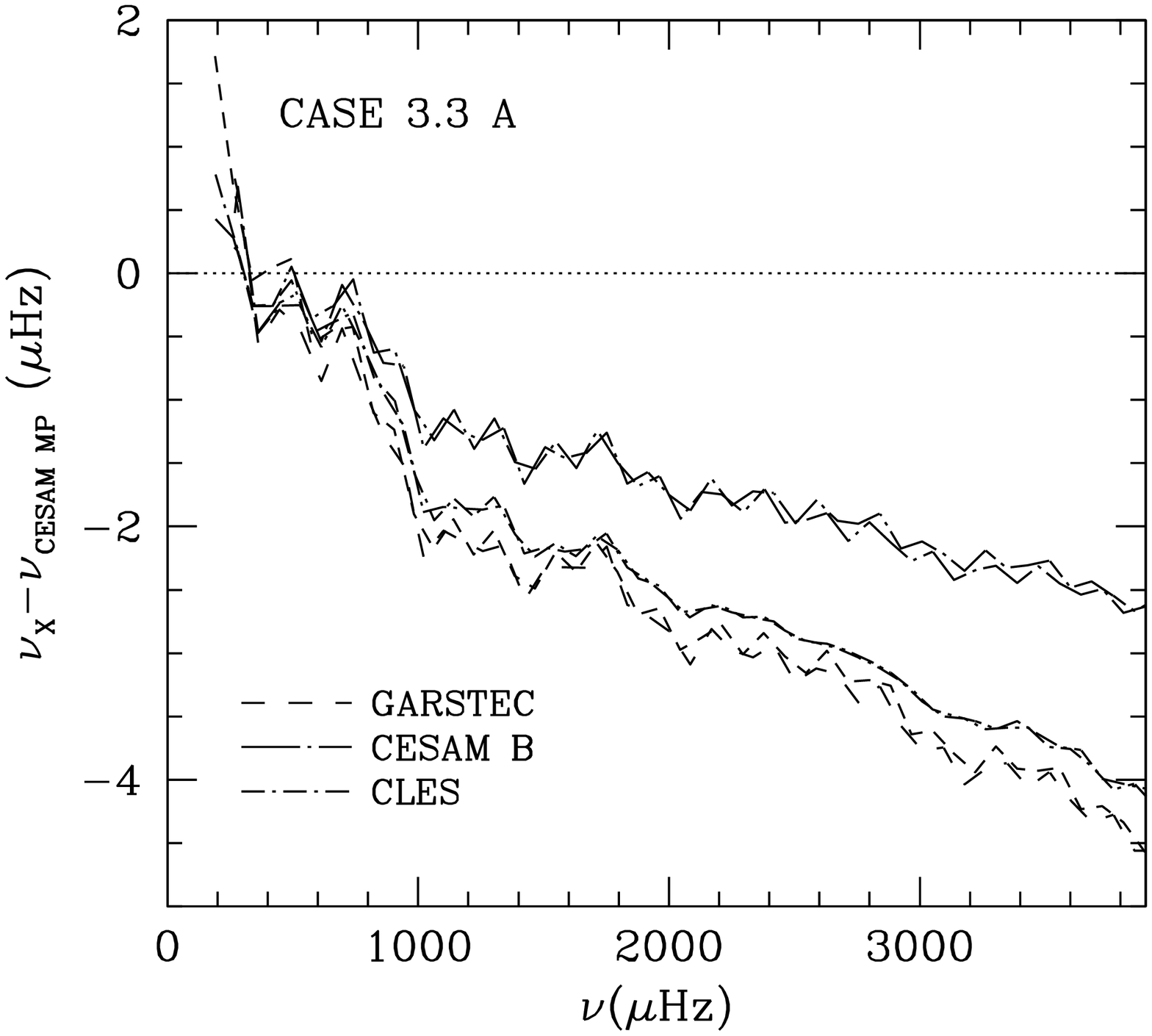}\hspace*{1.5cm}\includegraphics*{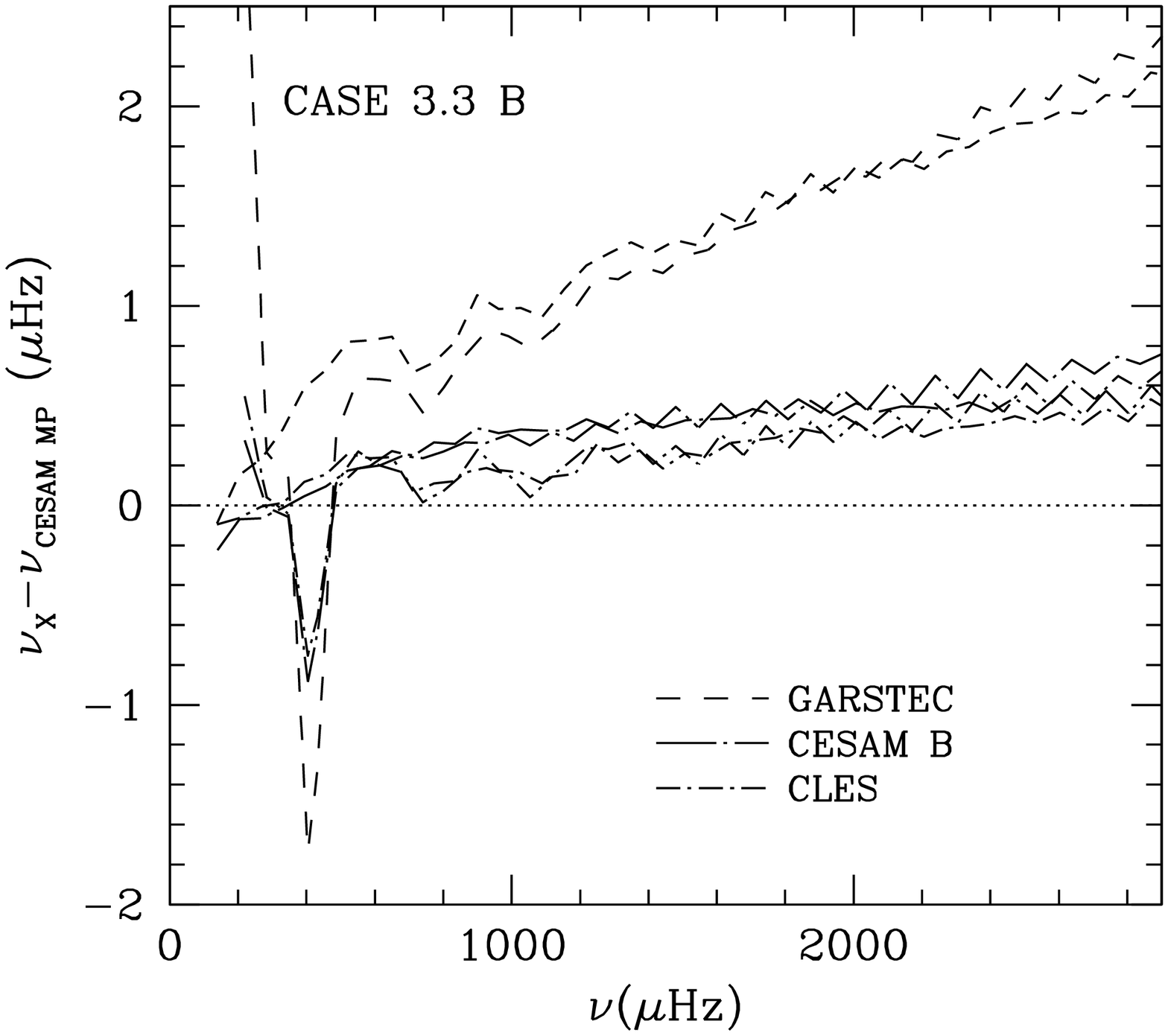}\hspace*{1.8cm}\includegraphics*{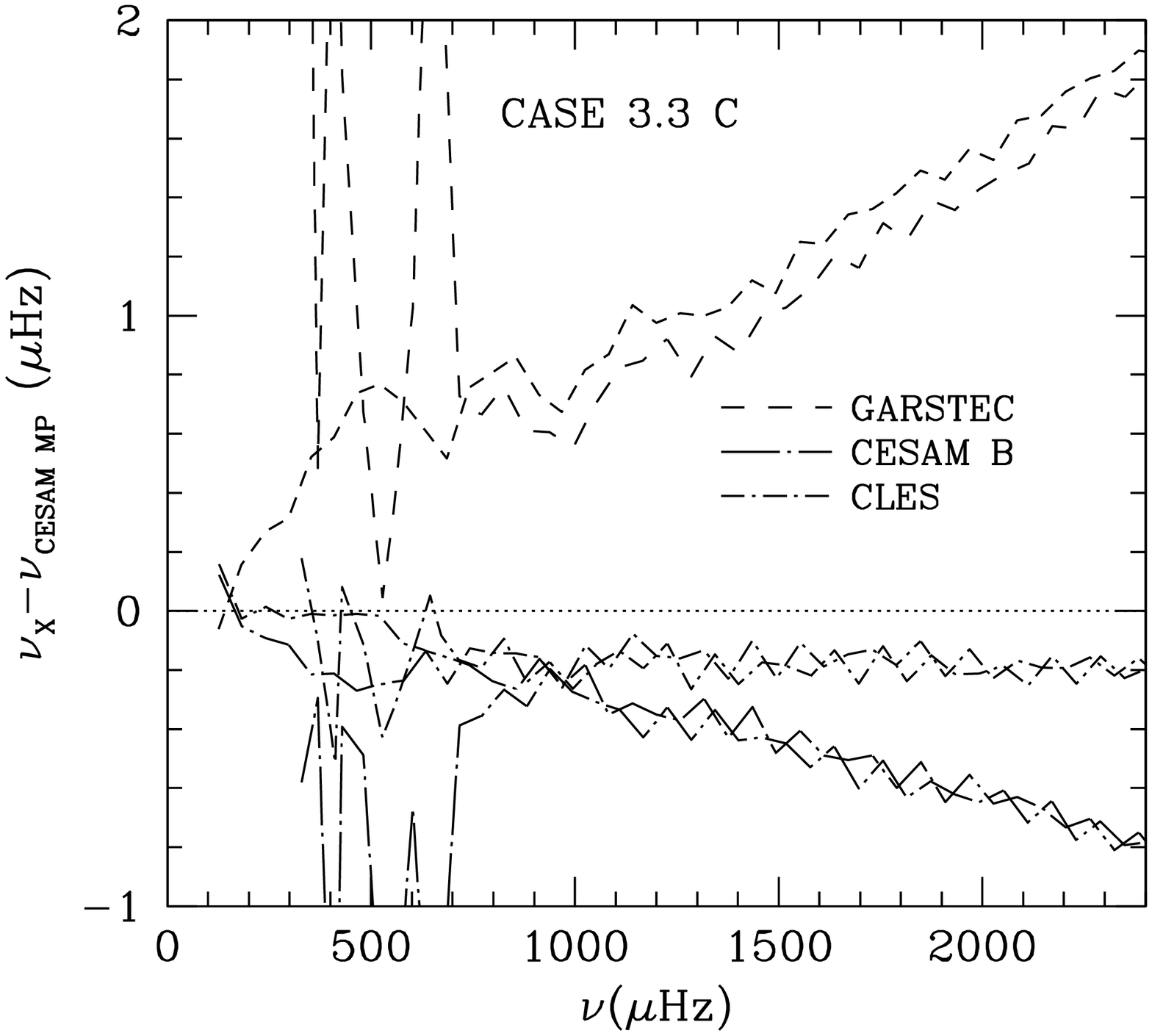}}

\caption{{\task}~3: p-mode frequency differences between models produced by different codes, for Case~3.1 (top row), 3.2 (centre) and 3.3 (bottom). {\cesam} model is taken as reference, and the frequencies have been scaled to remove the effect of different stellar radii. For each code, we plot two curves corresponding to modes with degrees $\ell=0$ and $\ell=1$.}
\label{fig:task3-freq}       % Give a unique label
\end{figure*}

As in Sec.~\ref{sec:task1-osc} we present in Fig.~\ref{fig:task3-freq} the frequency differences $\nu_{\mbox{\scriptsize CODE}}-\nu_{\mbox{\scriptsize CESAM}}$, where again the frequencies have been scaled to correct for differences in the stellar radius. They can be compared to the results obtained in {\task}~1 Cases~1.1, 1.2 and 1.3 (Fig.~\ref{fig:freq12} and \ref{fig:freq3}). 

Differences increase as evolution proceeds and as the mass increases. We find that the trend of the differences found in {\MS} models (phase A) for the 3 cases (3.1, 3.2, 3.3) is very similar to what has been found for Case~1.1 and 1.2 {\MS} models. Again the similar behaviour of curves with different degree indicates that the frequency differences are due to near-surface effects. Differences between curves corresponding to modes of degree $\ell=0$ and $1$, which reflect differences in the interior structure, remain small, below $0.1$--$0.2\mu$Hz (a bit larger for {\astec}). The magnitude of the differences is, on the average, higher in models including microscopic diffusion, due to larger differences in the sound speed in particular in the central regions (or border of the convective core) and at the base of the convective envelope.
Two different oscillatory components with a periodicity of $\sim 2000$~s and $\sim 4000$~s appear in the frequency differences.
The first one which is mainly visible in the  \garstec\ models  is due to differences in the adiabatic exponent,
and its amplitude is related to different helium abundances in the convective envelope. 
The second one makes the ``saw-tooth'' profile, and is due to differences at the border of convective envelope.
 
For {\TAMS} models (phase B), in addition to differences observed for {\MS} models, peaks become clearly visible at low frequencies for $\ell=1$ modes. As in Case~1.3 they can be attributed to differences in the Brunt-V\"ais\"al\"a frequency in the interior and to the mixed character of the corresponding modes. Any difference in the $\mu$ gradient in the region just above the border of the helium core is indeed expected to be seen in the frequency differences. This effect is even larger in {\SGB} models (phase C).

\section{Summary and conclusions}

We have presented detailed comparisons of the internal structures and seismic properties of stellar models in a range of stellar parameters -- mass, chemical composition and evolutionary stage -- covering those of the {\corot} targets.  The models were calculated by 5 codes ({\astec}, {\cesam}, {\cles}, {\starox}, {\garstec}) which have followed rather closely the specifications for the stellar models (input physics, physical and astronomical constants) that were defined by the {\esta} group, although some differences remain, sometimes not fully identified. The oscillation frequencies were calculated by the {\small LOSC} code (see Sect.~\ref{sec:tools}).

%----------------------------------------------------------------------------
%----------------------------------------------------------------------------

%-------------------------------------------------------------------
%-------------------------------------------------------------------

In a first step, we have examined {\esta}-{\task}~1 models, calculated for masses in the range $0.9-5$\msol, with different chemical compositions and evolutionary stages from {\PMS} to {\SGB}. In all these models microscopic diffusion of chemical elements has not been included while one model accounts for overshooting of convective cores. In a second step, we have considered the {\esta}-{\task}~3 models, in the mass range $1.0-1.3$\msol, solar composition, and evolutionary stages from the middle of the {\MS} to the {\SGB}. In all these models, microscopic diffusion has been taken into account.

For both tasks we have discussed the maximum and average differences in the physical quantities from centre to surface (hydrogen abundance $X$, pressure $P$, density $\rho$, luminosity $L_r$, opacity $\kappa$, adiabatic exponent $\Gamma_1$ and gradient $\nabla_{\rm ad}$, specific heat at constant pressure $C_p $ and sound speed $c$). %We have found that the average differences are quite small, generally lower than $0.5$\% but may reach $2-4$\% in some cases for $P$, $\rho$, $L_r$ or $\kappa$.
We have found that the average differences are in general small. Differences in $P$, $T$, $L_r$, $r$, $\rho$ and $\kappa$ are on the percent level while differences in the thermodynamical quantities are often well below $1$\%. Concerning the maximal differences, we have found that they are mostly located in the outer layers and in the zones close to the frontiers of the convective zones. As expected, differences generally increase as the evolution proceeds. They are larger in models with convective cores, in particular in models where the convective core increases during a large part of the {\MS} before receding. They are also higher in models including microscopic diffusion or overshooting of the convective core.

%-------------------------------------------------------------------
\begin{table*}[htbp!]
\caption{{\task}~3: values of the fractional radius and mass at the base of the convective envelope ($r_{\rm cz}/R$, $m_{\rm cz}/M$ ) and border of the convective core ($r_{\rm cc}/R$, $m_{\rm cc}/M$) and hydrogen abundance in the convection zones}
\centering
\label{tab:task3-zc}
\begin{tabular}[h]{llcccccc}
\hline\noalign{\smallskip}
{\bf Code} & {\bf Case} & \boldmath$r_{\rm cz}/R$ & \boldmath$m_{\rm cz}/M$ & \boldmath$X_{\rm cz}$& \boldmath$r_{\rm cc}/R$ & \boldmath$m_{\rm cc}/M$ & \boldmath$X_{\rm cc}$\\[3pt]
\tableheadseprule\noalign{\smallskip} 
 \astec   & 3.1A  &   0.7348 &   0.9835 &   0.7436 &--&--&--\\
 \cesamB  & 3.1A  &   0.7326 &   0.9831 &   0.7446&--&--&-- \\
 \cesamMP & 3.1A  &   0.7321 &   0.9830 &   0.7449 &--&--&--\\
 \cles    & 3.1A  &   0.7315 &   0.9829 &   0.7470&--&--&-- \\
 \garstec & 3.1A  &   0.7357 &   0.9835 &   0.7475&--&--&-- \\
\hline\noalign{\smallskip}
 \astec   & 3.1B  &   0.7202 &   0.9837 &   0.7630&--&--&-- \\
% \cesamB  & C3.1B  &   0.71798 &   0.98358 &   0.76570 \\
% \cesamB  & C3.1B  &   0.71755 &   0.98352 &   0.76570 \\
 \cesamB  & 3.1B  &   0.7153 &   0.9832 &   0.7657 &--&--&--\\
% \cesamMP & C3.1B  &   0.71731 &   0.98345 &   0.76555 \\
% \cesamMP & C3.1B  &   0.71693 &   0.98340 &   0.76555 \\
 \cesamMP & 3.1B  &   0.7143 &   0.9830 &   0.7656&--&--&-- \\
 \cles    & 3.1B  &   0.7151 &   0.9831 &   0.7695&--&--&-- \\
 \garstec & 3.1B  &   0.7159 &   0.9832 &   0.7705 &--&--&--\\
\hline\noalign{\smallskip}
 \astec   & 3.1C  &   0.6676 &   0.9720 &   0.7627&--&--&-- \\
 \cesamB  & 3.1C  &   0.6644 &   0.9716 &   0.7657 &--&--&--\\
% \cesamB  & C3.1C  &   0.65580 &   0.96986 &   0.73908 \\
% \cesamB  & C3.1C  &   0.65379 &   0.96943 &   0.74017 \\
% \cesamB  & C3.1C  &   0.64915 &   0.96843 &   0.73849 \\
% \cesamB  & C3.1C  &   0.64902 &   0.96840 &   0.73847 \\
 \cesamMP & 3.1C  &   0.6631 &   0.9713 &   0.7661 &--&--&--\\
% \cesamMP & C3.1C  &   0.65200 &   0.96899 &   0.74183 \\
% \cesamMP & C3.1C  &   0.65073 &   0.96872 &   0.74194 \\
 \cles    & 3.1C  &   0.6643 &   0.9715 &   0.7684 &--&--&--\\
% \cles    & C3.1C  &   0.65881 &   0.97037 &   0.74842 &--&--&--\\
% \cles    & C3.1C  &   0.65804 &   0.97021 &   0.74539 &--&--&--\\
 \garstec & 3.1C  &   0.6643 &   0.9712 &   0.7691&--&--&-- \\
\hline\noalign{\smallskip}
 \astec   & 3.2A  &   0.8489 &   0.9990 &   0.7754 &  0.0456 &   0.0170 &   0.3500 \\
 \cesamB  & 3.2A  &   0.8451 &   0.9999 &   0.7835  &   0.0451 &   0.0169 &   0.3499 \\
 \cesamMP & 3.2A  &   0.8443 &   0.9990 &   0.7817& 0.0451 &   0.0170 &   0.3500 \\
 \cles    & 3.2A  &   0.8444 &   0.9990 &   0.7888     &   0.0447 &   0.0165 &   0.3504 \\
 \garstec & 3.2A  &   0.8450 &   0.9990 &   0.7883   &  0.0433 &   0.0149 &   0.3499 \\ 
\hline\noalign{\smallskip}
 \cesamB  & 3.2B  &   0.7957 &   0.9969 &   0.7723  &   0.0376 &   0.0325 &   0.0099 \\
 \cesamMP & 3.2B  &   0.7927 &   0.9968 &   0.7723 &  0.0375 &   0.0328 &   0.0100 \\
 \cles    & 3.2B  &   0.7956 &   0.9969 &   0.7796 &0.0373 &   0.0319 &   0.0101 \\
% \garstec & C3.2B  &   1.00000 &   0.99767 &   0.77668 \\
 \garstec & 3.2B  &   0.7953 &   0.9969 &   0.7767 &   0.0372 &   0.0318 &   0.0100 \\
\hline\noalign{\smallskip}
 \cesamB  & 3.2C  &   0.7921 &   0.9972 &   0.7768 &--&--&-- \\
 \cesamMP & 3.2C  &   0.7913 &   0.9972 &   0.7765 &--&--&-- \\
 \cles    & 3.2C  &   0.7903 &   0.9971 &   0.7843 &--&--&-- \\
% \garstec & C3.2C  &   1.00000 &   1.00164 &   0.78141 \\
 \garstec & 3.2C  &   0.7931 &   0.9972 &   0.7814 &--&--&-- \\
 \hline\noalign{\smallskip}
 \cesamB  & 3.3A  &   0.8916 &   0.9999 &   0.8840 &   0.0641 &   0.0575 &   0.3500 \\
 \cesamMP & 3.3A  &   0.8893 &   0.9998 &   0.8631 &   0.0641 &   0.0572 &   0.3490 \\
 \cles    & 3.3A  &   0.8893 &   0.9998 &   0.8947 &   0.0633 &   0.0557 &   0.3503 \\
% \cles    & C3.3A  &   0.88855 &   0.99984 &   0.87975 \\ \cles    & C3.3A  &   0.88505 &   0.99983 &   0.87228 \\ \cles    & C3.3A  &   0.88429 &   0.99983 &   0.86870 \\ \cles    & C3.3A  &   0.88429 &   0.99983 &   0.86870 \\ \cles    & C3.3A  &   0.88253 &   0.99982 &   0.85990 \\ \cles    & C3.3A  &   0.88253 &   0.99982 &   0.85990 \\ \cles    & C3.3A  &   0.88044 &   0.99981 &   0.84697 \\ \cles    & C3.3A  &   0.87990 &   0.99981 &   0.84375 \\
 \garstec & 3.3A  &   0.8920 &   0.9999 &   0.8955 &   0.0631 &   0.0554 &   0.3500 \\
% \garstec & C3.3A  &   0.88374 &   0.99984 &   0.86470 \\% \garstec & C3.3A  &   0.88374 &   0.99984 &   0.86470 \\ 
\hline\noalign{\smallskip}
 \cesamB  & 3.3B  &   0.8336 &   0.9990 &   0.7866 &0.0389 &   0.0404 &   0.0099 \\
 \cesamMP & 3.3B  &   0.8342 &   0.9990 &   0.7886&   0.0385 &   0.0396 &   0.0100 \\
 \cles    & 3.3B  &   0.8335 &   0.9990 &   0.7948&   0.0387 &   0.0399 &   0.0101 \\
 \garstec & 3.3B  &   0.8334 &   0.9989 &   0.7923 &   0.0384 &   0.0391 &   0.0100 \\
\hline\noalign{\smallskip}
 \cesamB  & 3.3C  &   0.8534 &   0.9996 &   0.7902 &--&--&-- \\
 \cesamMP & 3.3C  &   0.8517 &   0.9995 &   0.7926 &--&--&-- \\
 \cles    & 3.3C  &   0.8526 &   0.9996 &   0.7988 &--&--&-- \\
 \garstec & 3.3C  &   0.8514 &   0.9995 &   0.7963 &--&--&-- \\
\hline\noalign{\smallskip}
\end{tabular}
\label{tab:task3-conv}
\end{table*}
%-------------------------------------------------------------------

We have then discussed each case individually and tried to identify the origin of the differences.

The way the codes handle the {\small OPAL-EOS} tables has an impact on the output thermodynamical properties of the models. In particular, the choice of the thermodynamical quantities to be taken from the tables and of those to be recalculated from others by means of thermodynamic relations is critical because it is known that some of the thermodynamical quantities tabulated in the {\small OPAL} tables are inconsistent \citep{2003ApJ...583.1004B}. In particular, it has been shown by one of us (IW) during one of the {\esta} workshops that it is better not to use the tabulated $C_V$-value. Some further detailed comparisons of {\cles} and {\cesam} models by \citet{jm-apss}, have demonstrated that these inconsistencies lead to differences in the stellar models and their oscillation frequencies substantially dominating the uncertainties resulting from the use of different interpolation tools. Similarly, the differences in the opacity derived by the codes do not come from the different interpolation schemes but mainly from the differences in the opacity
tables themselves. Discrepancies depend on stellar mass and for the cases considered in this study the maximum differences in opacities are of the order of 2\% for 2{\msol} models except in a very narrow zone close to $\log T\simeq 4.0$ where there are at a few percents level for all models due to the way the {\small OPAL95} and {\small AF94} tables are combined.

In unevolved models, differences have been found that pertain either to the lack of a detailed calculation of the {\PMS} phase or to the simplifications in the nuclear reactions in the CN cycle. In evolved models we have  identified differences which are due to the method used to solve the set of equations governing the temporal evolution of the chemical composition. The shape and position of the $\mu$ gradient and the numerical handling of the temporal evolution of the border of the convection zones are critical as well. The star keeps the memory of the displacements of the convective core (either growing or receding) through the $\mu$ gradient. Models with microscopic diffusion show differences in the He and metals distributions which result from differences in the diffusion velocities and that affect in turn thermodynamic quantities and therefore the oscillation frequencies. The situation is particularly thorny for models that undergo semiconvection, either below the convective zone or at the border of the convective core (for models with diffusion), because none of the codes treats this phenomenon. 
 
We found that differences in the radius at the bottom of the convective envelope are small, lower than $0.7$\%. The differences in the mass of the convective core are sometimes large as in models that undergo semiconvection or in {\PMS} and low-mass {\ZAMS} models (up to $17-30$\%). In other models, the mass of the convective core differs by $0.5$ to $5$\%. Differences in the surface helium abundance in models including microscopic diffusion are of a few per cent except for the $1.3${\msol} model on the {\MS} where they are in the range $15-30$\% due to the different formalisms used to treat diffusion (see Sect.~\ref{sec:task3}).
 
We have examined the differences in the oscillation frequencies of p and g modes of degrees $\ell= 0, 1, 2, 3$.
For solar-type stars we also calculated and examined the large frequency separation for $\ell=0, 1$ and the frequency separation ratios defined by \cite{1999ASPC..173..257R}. 

For solar-type stars on the {\MS}, the differences in the frequencies calculated by the different codes are in the range $0.05-0.1\mu$Hz (for $\ell=0$, no microscopic diffusion) and $0.1-0.2\mu$Hz (models with diffusion).
For advanced models ({\TAMS} or {\SGB}) differences are larger (up to $1\mu$Hz for $\ell=0$ modes). 
We find that the frequency separation ratios are in excellent agreement which confirms that the differences found in the frequencies and in the large frequency separation have their origin in near-surface effects. Differences are larger in models including microscopic diffusion where the sound speed differences are larger (in the centre and at the borders of convection zones). In addition, in evolved models, at low frequency for $\ell=1$ modes, we found differences of up to $4\mu$Hz that result from differences in the Brunt-V\"ais\"al\"a frequency and from the mixed character of the modes. For stars of $2.0$\msol, frequency differences ($\ell=0, 1$ modes) are lower than $0.5\mu$Hz ({\PMS}) and may reach $1\mu$Hz for the evolved model with overshooting. They are due to structure differences in regions close to the surface, in the region of second He-ionisation and close to the centre.  In the evolved model some modes are mixed modes sensitive to the location of and features in the $\mu$ gradient. Finally, we found that the frequency differences of the massive models ($3$ and $5$\msol) are generally smaller than $0.2\mu$Hz. 

%{\bf Furthermore, in all cases it has been pointed out by \citet{jm-apss} that the relation between the opacities and the frequencies is not evident because a change in opacity modifies the temperature structure and in turn the $\Gamma_1$ value. Any uncertainty in the opacity computation will therefore limit the precision of the theoretical oscillation frequency.}

This thorough comparison work has proven to be very useful in understanding in detail the methods used to handle the calculation of stellar models in different stellar evolution codes. Several bugs and inconsistencies in the codes have been found and corrected. The comparisons have shown that some numerical methods had to be improved and that several simplifications made in the input physics are no longer satisfactory if models of high precision are needed, in particular for asteroseismic applications. We are aware of the weaknesses of the models and therefore of the need for further developments and improvements to bring to them and we are able to give an estimate of the precision they can reach. This gives us confidence on their ability to interpret the asteroseismic observations which are beginning to be delivered by the {\corot} mission where an accuracy of a few $10^{-7}\ {\mathrm{Hz}}$ is expected on the oscillation frequencies \citep{michel06} as well as those that will come from future missions as {\small NASA}'s Kepler mission to be launched in 2009 \citep{2007CoAst.150..350C}.

\begin{acknowledgements} JM thanks A. Noels and A. Miglio for fruitful discussions and 
 acknowledges financial support from the Belgium Science Policy Office (BELSPO) in the frame of the
ESA PRODEX 8 program (contract C90199). YL is grateful to P. Morel and B. Pichon for their kind help on the {\cesam} code. 
The European Helio and Asteroseismology Network (HELAS) is thanked for financial support.
\end{acknowledgements}

%-------------------------------------------------------------------
% BibTeX users please use
%\bibliographystyle{aa}%\bibliographystyle{plainnat}%\bibliographystyle{abbrvnat}\bibliographystyle{spmpsci}
%\bibliographystyle{Spr-mp-nameyear_modif}
%\bibliography{master,preprint}   % name your BibTeX data base

\begin{thebibliography}{299}
%\begin{thebibliography}{}
\ifx \bisbn   \undefined \def \bisbn  #1{ISBN #1}   \fi
\ifx \binits  \undefined \def \binits#1{#1} \fi
\ifx \bauthor  \undefined \def \bauthor#1{#1} \fi
\ifx \batitle  \undefined \def \batitle#1{#1} \fi
\ifx \bjtitle  \undefined \def \bjtitle#1{#1} \fi
\ifx \bvolume  \undefined \def \bvolume#1{#1} \fi
\ifx \byear  \undefined \def \byear#1{#1} \fi
\ifx \bissue  \undefined \def \bissue#1{#1} \fi
\ifx \bfpage  \undefined \def \bfpage#1{#1} \fi
\ifx \blpage  \undefined \def \blpage #1{#1} \fi
\ifx \burl  \undefined \def \burl#1{#1} \fi
\ifx \binterref  \undefined \def \binterref#1{#1} \fi
\ifx \betal  \undefined \def \betal#1{#1} \fi
\ifx \binstitute  \undefined \def \binstitute#1{#1} \fi
\ifx \bctitle  \undefined \def \bctitle#1{#1} \fi
\ifx \beditor  \undefined \def \beditor#1{#1} \fi
\ifx \bpublisher  \undefined \def \bpublisher#1{#1} \fi
\ifx \bbtitle  \undefined \def \bbtitle#1{#1} \fi
\ifx \bedition  \undefined \def \bedition#1{#1} \fi
\ifx \bseriesno  \undefined \def \bseriesno#1{#1} \fi
\ifx \blocation  \undefined \def \blocation#1{#1} \fi
\ifx \bsertitle  \undefined \def \bsertitle#1{#1} \fi
\ifx \bsnm \undefined \def \bsnm#1{#1} \fi
\ifx \bsuffix \undefined \def \bsuffix#1{#1} \fi
\ifx \bparticle \undefined \def \bparticle#1{#1} \fi
\ifx \barticle \undefined \def \barticle#1{#1} \fi
\ifx \botherref \undefined \def \botherref #1{#1} \fi
\ifx \url \undefined \def \url#1{#1} \fi
\ifx \bchapter \undefined \def \bchapter#1{#1} \fi
\ifx \bbook \undefined \def \bbook#1{#1} \fi
\ifx \bcomment \undefined \def \bcomment#1{#1} \fi
\ifx \oauthor \undefined \def \oauthor#1{#1} \fi
\def \endbibitem {}

\bibitem[\protect\citeauthoryear{{Alexander} and {Ferguson}}{1994}]{af94}
  \bauthor{\bsnm{{Alexander}},~\binits{D.R.}},
  \bauthor{\bsnm{{Ferguson}},~\binits{J.W.}}:
\batitle{{Low-temperature Rosseland opacities}}.
\bjtitle{\apj} \bvolume{437},  \bfpage{879} (\byear{1994}).

\bibitem[\protect\citeauthoryear{{Bahcall}, {Pinsonneault}, and
  {Basu}}{2001}]{2001ApJ...555..990B}
\begin{barticle}
\bauthor{\bsnm{{Bahcall}},~\binits{J.N.}},
  \bauthor{\bsnm{{Pinsonneault}},~\binits{M.H.}},
  \bauthor{\bsnm{{Basu}},~\binits{S.}}:
\batitle{{Solar Models: Current Epoch and Time Dependences, Neutrinos, and
  Helioseismological Properties}}.
\bjtitle{\apj} \bvolume{555},  \bfpage{990}--\blpage{1012} (\byear{2001}).
  \binterref{doi:10.1086/321493}
\end{barticle}
\endbibitem

\bibitem[\protect\citeauthoryear{{Boothroyd} and
  {Sackmann}}{2003}]{2003ApJ...583.1004B}
\begin{barticle}
\bauthor{\bsnm{{Boothroyd}},~\binits{A.I.}},
  \bauthor{\bsnm{{Sackmann}},~\binits{I.J.}}:
\batitle{{Our Sun. IV. The Standard Model and Helioseismology: Consequences of
  Uncertainties in Input Physics and in Observed Solar Parameters}}.
\bjtitle{\apj} \bvolume{583},  \bfpage{1004}--\blpage{1023} (\byear{2003}).
  \binterref{doi:10.1086/345407}
\end{barticle}
\endbibitem

\bibitem[\protect\citeauthoryear{{Burgers}}{1969}]{B69}
\begin{botherref}
\oauthor{\bsnm{{Burgers}},~\binits{J.M.}}:
{Flow Equations for Composite Gases}. Flow Equations for Composite Gases, New
  York: Academic Press, 1969 (1969)
\end{botherref}
\endbibitem

\bibitem[\protect\citeauthoryear{{Christensen-Dalsgaard}}{2005}]{jcd-aarhus}
\begin{botherref}
\oauthor{\bsnm{{Christensen-Dalsgaard}},~\binits{J.}}:
{contribution to the CoRoT ESTA Meeting 4, Aarhus, Denmark}. available at {\tt\small $^1$ESTA\_Web\_Meetings/m4/} (2005)
\end{botherref}
\endbibitem

\bibitem[\protect\citeauthoryear{{Christensen-Dalsgaard}}{2007a}]{jcd1-apss}
\begin{botherref}
\oauthor{\bsnm{{Christensen-Dalsgaard}},~\binits{J.}}:
{ASTEC: Aarhus Stellar Evolution Code}.
In: \thisapss. (2007a)
\end{botherref}
\endbibitem

\bibitem[\protect\citeauthoryear{{Christensen-Dalsgaard}}{2007b}]{jcd2-apss}
\begin{botherref}
\oauthor{\bsnm{{Christensen-Dalsgaard}},~\binits{J.}}:
{ADIPLS: Aarhus Adiabatic Pulsation Package}.
In: \thisapss. (2007b)
\end{botherref}
\endbibitem

\bibitem[\protect\citeauthoryear{{Christensen-Dalsgaard}}{2007c}]{jcd-eas}
\begin{botherref}
\oauthor{\bsnm{{Christensen-Dalsgaard}},~\binits{J.}}:
{Comparisons for ESTA-Task3: ASTEC, CESAM and CL{\'E}S}.
In: EAS Publications Series. Engineering and Science,  vol.~26,  pp. 177--185.
  (2007c). doi:10.1051/eas:2007136
\end{botherref}
\endbibitem

\bibitem[\protect\citeauthoryear{{Christensen-Dalsgaard}
  \textit{et~al.}}{2007}]{2007CoAst.150..350C}
\begin{barticle}
\bauthor{\bsnm{{Christensen-Dalsgaard}},~\binits{J.}},
  \bauthor{\bsnm{{Arentoft}},~\binits{T.}},
  \bauthor{\bsnm{{Brown}},~\binits{T.M.}},
  \bauthor{\bsnm{{Gilliland}},~\binits{R.L.}},
  \bauthor{\bsnm{{Kjeldsen}},~\binits{H.}},
  \bauthor{\bsnm{{Borucki}},~\binits{W.J.}},
  \bauthor{\bsnm{{Koch}},~\binits{D.}}:
\batitle{{Asteroseismology with the Kepler mission}}.
\bjtitle{Communications in Asteroseismology} \bvolume{150},
  \bfpage{350}--\blpage{+} (\byear{2007})
\end{barticle}
\endbibitem

\bibitem[\protect\citeauthoryear{{Christensen-Dalsgaard}, {Bedding}, and
  {Kjeldsen}}{1995}]{jcd95}
\begin{barticle}
\bauthor{\bsnm{{Christensen-Dalsgaard}},~\binits{J.}},
  \bauthor{\bsnm{{Bedding}},~\binits{T.R.}},
  \bauthor{\bsnm{{Kjeldsen}},~\binits{H.}}:
\batitle{{Modeling solar-like oscillations in eta Bootis}}.
\bjtitle{\apjl} \bvolume{443},  \bfpage{L29}--\blpage{L32} (\byear{1995}).
  \binterref{doi:10.1086/187828}
\end{barticle}
\endbibitem

\bibitem[\protect\citeauthoryear{{Crowe} and {Matalas}}{1982}]{Crowe82}
\begin{barticle}
\bauthor{\bsnm{{Crowe}},~\binits{R.A.}},
  \bauthor{\bsnm{{Matalas}},~\binits{R.}}:
\batitle{{Semiconvection in Low-Mass Main Sequence Stars}}.
\bjtitle{\aap} \bvolume{108},  \bfpage{55}--\blpage{+} (\byear{1982})
\end{barticle}
\endbibitem

\bibitem[\protect\citeauthoryear{{Dziembowski}, {Moskalik}, and
  {Pamyatnykh}}{1993}]{1993MNRAS.265..588D}
\begin{barticle}
\bauthor{\bsnm{{Dziembowski}},~\binits{W.A.}},
  \bauthor{\bsnm{{Moskalik}},~\binits{P.}},
  \bauthor{\bsnm{{Pamyatnykh}},~\binits{A.A.}}:
\batitle{{The Opacity Mechanism in B-Type Stars - Part Two - Excitation of
  High-Order G-Modes in Main Sequence Stars}}.
\bjtitle{\mnras} \bvolume{265},  \bfpage{588}--\blpage{+} (\byear{1993})
\end{barticle}
\endbibitem

\bibitem[\protect\citeauthoryear{{Floranes}, {Christensen-Dalsgaard}, and
  {Thompson}}{2005}]{2005MNRAS.356..671F}
\begin{barticle}
\bauthor{\bsnm{{Floranes}},~\binits{H.O.}},
  \bauthor{\bsnm{{Christensen-Dalsgaard}},~\binits{J.}},
  \bauthor{\bsnm{{Thompson}},~\binits{M.J.}}:
\batitle{{The use of frequency-separation ratios for asteroseismology}}.
\bjtitle{\mnras} \bvolume{356},  \bfpage{671}--\blpage{679} (\byear{2005}).
  \binterref{doi:10.1111/j.1365-2966.2004.08487.x}
\end{barticle}
\endbibitem

\bibitem[\protect\citeauthoryear{{Gabriel} and {Noels}}{1977}]{Gabriel77}
\begin{barticle}
\bauthor{\bsnm{{Gabriel}},~\binits{M.}}, \bauthor{\bsnm{{Noels}},~\binits{A.}}:
\batitle{{Semiconvection in stars of about 1 solar mass}}.
\bjtitle{\aap} \bvolume{54},  \bfpage{631}--\blpage{634} (\byear{1977})
\end{barticle}
\endbibitem

\bibitem[\protect\citeauthoryear{{Gough}}{1990}]{gough90}
\begin{barticle}
\bauthor{\bsnm{{Gough}},~\binits{D.}}:
\batitle{{Helioseismology - Shaky Clues to Solar Activity}}.
\bjtitle{\nat} \bvolume{345},  \bfpage{768}--\blpage{+} (\byear{1990}).
  \binterref{doi:10.1038/345768a0}
\end{barticle}
\endbibitem

\bibitem[\protect\citeauthoryear{{Iglesias} and {Rogers}}{1996}]{ir96}
  \bauthor{\bsnm{{Iglesias}},~\binits{C.A.}},
  \bauthor{\bsnm{{Rogers}},~\binits{F.J.}}:
\batitle{{Updated Opal Opacities}}.
\bjtitle{\apj} \bvolume{464},  943 (\byear{1996})

\bibitem[\protect\citeauthoryear{{Kjeldsen} and
  {Bedding}}{1995}]{1995A&A...293...87K}
\begin{barticle}
\bauthor{\bsnm{{Kjeldsen}},~\binits{H.}},
  \bauthor{\bsnm{{Bedding}},~\binits{T.R.}}:
\batitle{{Amplitudes of stellar oscillations: the implications for
  asteroseismology.}}
\bjtitle{\aap} \bvolume{293},  \bfpage{87}--\blpage{106} (\byear{1995})
\end{barticle}
\endbibitem

\bibitem[\protect\citeauthoryear{{Lebreton} \textit{et~al.}}{2007a}]{yl1-apss}
\begin{botherref}
\oauthor{\bsnm{{Lebreton}},~\binits{Y.}},
  \oauthor{\bsnm{{Monteiro}},~\binits{M.J.P.F.G.}},
  \oauthor{\bsnm{{Montalb\'an}},~\binits{J.}},
  \oauthor{\bsnm{{Moya}},~\binits{A.}},
  \oauthor{\bsnm{{Baglin}},~\binits{A.}},
  \oauthor{\bsnm{{Christensen-Dalsgaard}},~\binits{J.}},
  \oauthor{\bsnm{{Goupil}},~\binits{M.-J.}},
  \oauthor{\bsnm{{Michel}},~\binits{E.}},
  \oauthor{\bsnm{{Provost}},~\binits{J.}},
  \oauthor{\bsnm{{Roxburgh}},~\binits{I.W.}},
  \oauthor{\bsnm{{Scuflaire}},~\binits{R.}},
  \oauthor{\bsnm{{and the ESTA Team}}}
  :
{The CoRoT evolution and seismic tool activity}.
In: \thisapss. (2007a)
\end{botherref}
\endbibitem
	
\bibitem[\protect\citeauthoryear{{Lebreton} \textit{et~al.}}{2007b}]{yl-eas}
\begin{botherref}
\oauthor{\bsnm{{Lebreton}},~\binits{Y.}},
  \oauthor{\bsnm{{Montalb{\'a}n}},~\binits{J.}},
  \oauthor{\bsnm{{Christensen-Dalsgaard}},~\binits{J.}},
  \oauthor{\bsnm{{Th{\'e}ado}},~\binits{S.}},
  \oauthor{\bsnm{{Hui-Bon-Hoa}},~\binits{A.}},
  \oauthor{\bsnm{{Monteiro}},~\binits{M.J.P.F.G.}},
  \oauthor{\bsnm{{Degl'Innocenti}},~\binits{S.}},
  \oauthor{\bsnm{{Marconi}},~\binits{M.}},
  \oauthor{\bsnm{{Morel}},~\binits{P.}}, \oauthor{\bsnm{{Prada
  Moroni}},~\binits{P.G.}}, \oauthor{\bsnm{{Weiss}},~\binits{A.}}:
{Microscopic Diffusion in Stellar Evolution Codes: First Comparisons Results of
  ESTA-Task 3}.
In: EAS Publications Series. Engineering and Science,  vol.~26,  pp. 155--165.
  (2007b). doi:10.1051/eas:2007134
\end{botherref}
\endbibitem

\bibitem[\protect\citeauthoryear{{Michaud} and {Proffitt}}{1993}]{MP93}
\begin{botherref}
\oauthor{\bsnm{{Michaud}},~\binits{G.}},
  \oauthor{\bsnm{{Proffitt}},~\binits{C.R.}}:
{Particle transport processes}.
In: {Weiss}, W.W., {Baglin}, A. (eds.) IAU Colloq. 137: Inside the Stars.
  Astronomical Society of the Pacific Conference Series,  vol.~40,  pp.
  246--259. (January 1993)
\end{botherref}
\endbibitem

\bibitem[\protect\citeauthoryear{{Michel} \textit{et~al.}}{2006}]{michel06}
\begin{botherref}
\oauthor{\bsnm{{Michel}},~\binits{E.}}, \oauthor{\bsnm{{Baglin}},~\binits{A.}},
  \oauthor{\bsnm{{Auvergne}},~\binits{M.}},
  \oauthor{\bsnm{{Catala}},~\binits{C.}},
  \oauthor{\bsnm{{Aerts}},~\binits{C.}},
  \oauthor{\bsnm{{Alecian}},~\binits{G.}},
  \oauthor{\bsnm{{Amado}},~\binits{P.}},
  \oauthor{\bsnm{{Appourchaux}},~\binits{T.}},
  \oauthor{\bsnm{{Ausseloos}},~\binits{M.}},
  \oauthor{\bsnm{{Ballot}},~\binits{J.}},
  \oauthor{\bsnm{{Barban}},~\binits{C.}},
  \oauthor{\bsnm{{Baudin}},~\binits{F.}},
  \oauthor{\bsnm{{Berthomieu}},~\binits{G.}},
  \oauthor{\bsnm{{Boumier}},~\binits{P.}},
  \oauthor{\bsnm{{Bohm}},~\binits{T.}},
  \oauthor{\bsnm{{Briquet}},~\binits{M.}},
  \oauthor{\bsnm{{Charpinet}},~\binits{S.}},
  \oauthor{\bsnm{{Cunha}},~\binits{M.S.}}, \oauthor{\bsnm{{De
  Cat}},~\binits{P.}}, \oauthor{\bsnm{{Dupret}},~\binits{M.A.}},
  \oauthor{\bsnm{{Fabregat}},~\binits{J.}},
  \oauthor{\bsnm{{Floquet}},~\binits{M.}},
  \oauthor{\bsnm{{Fremat}},~\binits{Y.}},
  \oauthor{\bsnm{{Garrido}},~\binits{R.}},
  \oauthor{\bsnm{{Garcia}},~\binits{R.A.}},
  \oauthor{\bsnm{{Goupil}},~\binits{M.J.}},
  \oauthor{\bsnm{{Handler}},~\binits{G.}},
  \oauthor{\bsnm{{Hubert}},~\binits{A.M.}},
  \oauthor{\bsnm{{Janot-Pacheco}},~\binits{E.}},
  \oauthor{\bsnm{{Lambert}},~\binits{P.}},
  \oauthor{\bsnm{{Lebreton}},~\binits{Y.}},
  \oauthor{\bsnm{{Lignieres}},~\binits{F.}},
  \oauthor{\bsnm{{Lochard}},~\binits{J.}},
  \oauthor{\bsnm{{Martin-Ruiz}},~\binits{S.}},
  \oauthor{\bsnm{{Mathias}},~\binits{P.}},
  \oauthor{\bsnm{{Mazumdar}},~\binits{A.}},
  \oauthor{\bsnm{{Mittermayer}},~\binits{P.}},
  \oauthor{\bsnm{{Montalb\'an}},~\binits{J.}},
  \oauthor{\bsnm{{Monteiro}},~\binits{M.J.P.F.G.}},
  \oauthor{\bsnm{{Morel}},~\binits{P.}},
  \oauthor{\bsnm{{Mosser}},~\binits{B.}}, \oauthor{\bsnm{{Moya}},~\binits{A.}},
  \oauthor{\bsnm{{Neiner}},~\binits{C.}},
  \oauthor{\bsnm{{Nghiem}},~\binits{P.}},
  \oauthor{\bsnm{{Noels}},~\binits{A.}},
  \oauthor{\bsnm{{Oehlinger}},~\binits{J.}},
  \oauthor{\bsnm{{Poretti}},~\binits{E.}},
  \oauthor{\bsnm{{Provost}},~\binits{J.}}, \oauthor{\bsnm{{Renan de
  Medeiros}},~\binits{J.}}, \oauthor{\bsnm{{de Ridder}},~\binits{J.}},
  \oauthor{\bsnm{{Rieutord}},~\binits{M.}},
  \oauthor{\bsnm{{Roca-Cortes}},~\binits{T.}},
  \oauthor{\bsnm{{Roxburgh}},~\binits{I.}},
  \oauthor{\bsnm{{Samadi}},~\binits{R.}},
  \oauthor{\bsnm{{Scuflaire}},~\binits{R.}},
  \oauthor{\bsnm{{Suarez}},~\binits{J.C.}},
  \oauthor{\bsnm{{Theado}},~\binits{S.}},
  \oauthor{\bsnm{{Thoul}},~\binits{A.}},
  \oauthor{\bsnm{{Toutain}},~\binits{T.}},
  \oauthor{\bsnm{{Turck-Chieze}},~\binits{S.}},
  \oauthor{\bsnm{{Uytterhoeven}},~\binits{K.}},
  \oauthor{\bsnm{{Vauclair}},~\binits{G.}},
  \oauthor{\bsnm{{Vauclair}},~\binits{S.}},
  \oauthor{\bsnm{{Weiss}},~\binits{W.W.}},
  \oauthor{\bsnm{{Zwintz}},~\binits{K.}}:
{The Seismology programme of CoRoT}.
In: ''The CoRoT Mission'', (Eds) M. Fridlund, A. Baglin, J. Lochard {\&} L.
  Conroy, ESA Publications Division, ESA Spec.Publ. 1306 (2006).  pp. 39--50.
  (November 2006)
\end{botherref}
\endbibitem

\bibitem[\protect\citeauthoryear{{Miglio}, and {Montalb\'an}}{2005}]{am-aarhus}
\begin{botherref}
\oauthor{\bsnm{{Miglio}},~\binits{A.}},
  \oauthor{\bsnm{{Montalb\'an}},~\binits{J.}}:
{contribution to the CoRoT ESTA Meeting 4, Aarhus, Denmark}. available at {\tt\small $^1$ESTA\_Web\_Meetings/m4/} (2005)
\end{botherref}
\endbibitem

\bibitem[\protect\citeauthoryear{{Miglio}, {Montalb\'an}, and
  {Noels}}{2006}]{migliomg}
\begin{barticle}
\bauthor{\bsnm{{Miglio}},~\binits{A.}},
  \bauthor{\bsnm{{Montalb\'an}},~\binits{J.}},
  \bauthor{\bsnm{{Noels}},~\binits{A.}}:
\batitle{{Effects of extra-mixing processes on the periods of high-order
  gravity modes in main-sequence stars}}.
\bjtitle{Communications in Asteroseismology} \bvolume{147},
  \bfpage{89}--\blpage{92} (\byear{2006})
\end{barticle}
\endbibitem

\bibitem[\protect\citeauthoryear{{Montalb\'an} \textit{et~al.}}{2007a}]{jm-apss}
\begin{botherref}
\oauthor{\bsnm{{Montalb\'an}},~\binits{J.}},
  \oauthor{\bsnm{{Lebreton}},~\binits{Y.}},
  \oauthor{\bsnm{{Miglio}},~\binits{A.}},
  \oauthor{\bsnm{{Scuflaire}},~\binits{R.}},
  \oauthor{\bsnm{{Morel}},~\binits{P.}}, \oauthor{\bsnm{{Noels}},~\binits{A.}}:
{}.
In: \thisapss. (2007a)
\end{botherref}
\endbibitem

\bibitem[\protect\citeauthoryear{{Montalb{\'a}n}, {Th{\'e}ado}, and
  {Lebreton}}{2007b}]{jm-eas}
\begin{botherref}
\oauthor{\bsnm{{Montalb{\'a}n}},~\binits{J.}},
  \oauthor{\bsnm{{Th{\'e}ado}},~\binits{S.}},
  \oauthor{\bsnm{{Lebreton}},~\binits{Y.}}:
{Comparisons for ESTA-Task3: CLES and CESAM}.
In: EAS Publications Series. Engineering and Science,  vol.~26,  pp. 167--176.
  (2007b). doi:10.1051/eas:2007135
\end{botherref}
\endbibitem

\bibitem[\protect\citeauthoryear{{Monteiro}
  \textit{et~al.}}{2006}]{2006corm.conf..363M}
\begin{botherref}
\oauthor{\bsnm{{Monteiro}},~\binits{M.J.P.F.G.}},
  \oauthor{\bsnm{{Lebreton}},~\binits{Y.}},
  \oauthor{\bsnm{{Montalb\'an}},~\binits{J.}},
  \oauthor{\bsnm{{Christensen-Dalsgaard}},~\binits{J.}},
  \oauthor{\bsnm{{Castro}},~\binits{M.}},
  \oauthor{\bsnm{{Degl'Innocenti}},~\binits{S.}},
  \oauthor{\bsnm{{Moya}},~\binits{A.}},
  \oauthor{\bsnm{{Roxburgh}},~\binits{I.W.}},
  \oauthor{\bsnm{{Scuflaire}},~\binits{R.}},
  \oauthor{\bsnm{{Baglin}},~\binits{A.}},
  \oauthor{\bsnm{{Cunha}},~\binits{M.S.}},
  \oauthor{\bsnm{{Eggenberger}},~\binits{P.}},
  \oauthor{\bsnm{{Fernandes}},~\binits{J.}},
  \oauthor{\bsnm{{Goupil}},~\binits{M.J.}},
  \oauthor{\bsnm{{Hui-Bon-Hoa}},~\binits{A.}},
  \oauthor{\bsnm{{Marconi}},~\binits{M.}},
  \oauthor{\bsnm{{Marques}},~\binits{J.P.}},
  \oauthor{\bsnm{{Michel}},~\binits{E.}},
  \oauthor{\bsnm{{Miglio}},~\binits{A.}},
  \oauthor{\bsnm{{Morel}},~\binits{P.}},
  \oauthor{\bsnm{{Pichon}},~\binits{B.}}, \oauthor{\bsnm{{Prada
  Moroni}},~\binits{P.G.}}, \oauthor{\bsnm{{Provost}},~\binits{J.}},
  \oauthor{\bsnm{{Ruoppo}},~\binits{A.}},
  \oauthor{\bsnm{{Suarez}},~\binits{J.C.}},
  \oauthor{\bsnm{{Suran}},~\binits{M.}},
  \oauthor{\bsnm{{Teixeira}},~\binits{T.C.}}:
{Report on the CoRoT Evolution and Seismic Tools Activity}.
In: ''The CoRoT Mission'', (Eds) M. Fridlund, A. Baglin, J. Lochard {\&} L.
  Conroy, ESA Publications Division, ESA Spec.Publ. 1306 (2006) 363.  pp.
  363--+. (December 2006)
\end{botherref}
\endbibitem

\bibitem[\protect\citeauthoryear{{Monteiro} \textit{et~al.}}{2007}]{mm2-apss}
\begin{botherref}
\oauthor{\bsnm{{Monteiro}},~\binits{M.J.P.F.G.}},
  \oauthor{\bparticle{et~}\bsnm{al.},~}:
{}.
In: \thisapss. (2007)
\end{botherref}
\endbibitem

\bibitem[\protect\citeauthoryear{{Morel} and {Lebreton}}{2007}]{pm-apss}
\begin{botherref}
\oauthor{\bsnm{{Morel}},~\binits{P.}},
  \oauthor{\bsnm{{Lebreton}},~\binits{Y.}}:
{CESAM: Code d'Evolution Stellaire Adaptatif et Modulaire}.
In: \thisapss. (2007)
\end{botherref}
\endbibitem

\bibitem[\protect\citeauthoryear{{Paquette}
  \textit{et~al.}}{1986}]{1986ApJS...61..177P}
\begin{barticle}
\bauthor{\bsnm{{Paquette}},~\binits{C.}},
  \bauthor{\bsnm{{Pelletier}},~\binits{C.}},
  \bauthor{\bsnm{{Fontaine}},~\binits{G.}},
  \bauthor{\bsnm{{Michaud}},~\binits{G.}}:
\batitle{{Diffusion coefficients for stellar plasmas}}.
\bjtitle{\apjs} \bvolume{61},  \bfpage{177}--\blpage{195} (\byear{1986}).
  \binterref{doi:10.1086/191111}
\end{barticle}
\endbibitem

\bibitem[\protect\citeauthoryear{{Richard}, {Michaud}, and
  {Richer}}{2001}]{Richard01}
\begin{barticle}
\bauthor{\bsnm{{Richard}},~\binits{O.}},
  \bauthor{\bsnm{{Michaud}},~\binits{G.}},
  \bauthor{\bsnm{{Richer}},~\binits{J.}}:
\batitle{{Iron Convection Zones in B, A, and F Stars}}.
\bjtitle{\apj} \bvolume{558},  \bfpage{377}--\blpage{391} (\byear{2001}).
  \binterref{doi:10.1086/322264}
\end{barticle}
\endbibitem

\bibitem[\protect\citeauthoryear{{Rogers} and
  {Nayfonov}}{2002}]{2002ApJ...576.1064R}
  \bauthor{\bsnm{{Rogers}},~\binits{F.J.}},
  \bauthor{\bsnm{{Nayfonov}},~\binits{A.}}:
\batitle{{Updated and Expanded OPAL Equation-of-State Tables: Implications for Helioseismology}}.
\bjtitle{\apj} \bvolume{576},  \bfpage{1064} (\byear{2002})


\bibitem[\protect\citeauthoryear{{Roxburgh} and
  {Vorontsov}}{1999}]{1999ASPC..173..257R}
\begin{botherref}
\oauthor{\bsnm{{Roxburgh}},~\binits{I.W.}},
  \oauthor{\bsnm{{Vorontsov}},~\binits{S.V.}}:
{Asteroseismological Constraints on Stellar Convective Cores}.
In: ASP Conf. Ser. 173: Stellar Structure: Theory and Test of Connective Energy
  Transport.  pp. 257--+. (1999)
\end{botherref}
\endbibitem

\bibitem[\protect\citeauthoryear{{Roxburgh} and {Vorontsov}}{2003}]{roxvoron03}
\begin{barticle}
\bauthor{\bsnm{{Roxburgh}},~\binits{I.W.}},
  \bauthor{\bsnm{{Vorontsov}},~\binits{S.V.}}:
\batitle{{The ratio of small to large separations of acoustic oscillations as a
  diagnostic of the interior of solar-like stars}}.
\bjtitle{\aap} \bvolume{411},  \bfpage{215}--\blpage{220} (\byear{2003}).
  \binterref{doi:10.1051/0004-6361:20031318}
\end{barticle}
\endbibitem

\bibitem[\protect\citeauthoryear{{Roxburgh}}{2007}]{ir1-apss}
\begin{botherref}
\oauthor{\bsnm{{Roxburgh}},~\binits{I.}}:
{STAROX: Roxburgh's Evolution Code}.
In: \thisapss. (2007)
\end{botherref}
\endbibitem

\bibitem[\protect\citeauthoryear{{Scuflaire} \textit{et~al.}}{2007a}]{rs1-apss}
\begin{botherref}
\oauthor{\bsnm{{Scuflaire}},~\binits{R.}},
  \oauthor{\bsnm{{Th\'eado}},~\binits{S.}},
  \oauthor{\bsnm{{Montalb\'an}},~\binits{J.}},
  \oauthor{\bsnm{{Miglio}},~\binits{A.}},
  \oauthor{\bsnm{{Bourge}},~\binits{P.O.}},
  \oauthor{\bsnm{{Godart}},~\binits{M.}},
  \oauthor{\bsnm{{Thoul}},~\binits{A.}}, \oauthor{\bsnm{{Noels}},~\binits{A.}}:
{CLES: Code Liegeois d'Evolution Stellaire}.
In: \thisapss. (2007a)
\end{botherref}
\endbibitem

\bibitem[\protect\citeauthoryear{{Scuflaire} \textit{et~al.}}{2007b}]{rs2-apss}
\begin{botherref}
\oauthor{\bsnm{{Scuflaire}},~\binits{R.}},
  \oauthor{\bsnm{{Montalb\'an}},~\binits{J.}},
  \oauthor{\bsnm{{Th\'eado}},~\binits{S.}},
  \oauthor{\bsnm{{Bourge}},~\binits{P.O.}},
  \oauthor{\bsnm{{Miglio}},~\binits{A.}},
  \oauthor{\bsnm{{Godart}},~\binits{M.}},
  \oauthor{\bsnm{{Thoul}},~\binits{A.}}, \oauthor{\bsnm{{Noels}},~\binits{A.}}:
{LOSC: Liege Oscillations Code}.
In: \thisapss. (2007b)
\end{botherref}
\endbibitem

\bibitem[\protect\citeauthoryear{{Thoul} and {Montalb{\'a}n}}{2007}]{tm07}
\begin{botherref}
\oauthor{\bsnm{{Thoul}},~\binits{A.}},
  \oauthor{\bsnm{{Montalb{\'a}n}},~\binits{J.}}:
{Microscopic Diffusion in Stellar Plasmas}.
In: EAS Publications Series. Engineering and Science,  vol.~26,  pp. 25--36.
  (2007). doi:10.1051/eas:2007123
\end{botherref}
\endbibitem

\bibitem[\protect\citeauthoryear{{Thoul}, {Bahcall}, and {Loeb}}{1994}]{TBL94}
\begin{barticle}
\bauthor{\bsnm{{Thoul}},~\binits{A.A.}},
  \bauthor{\bsnm{{Bahcall}},~\binits{J.N.}},
  \bauthor{\bsnm{{Loeb}},~\binits{A.}}:
\batitle{{Element diffusion in the solar interior}}.
\bjtitle{\apj} \bvolume{421},  \bfpage{828}--\blpage{842} (\byear{1994}).
  \binterref{doi:10.1086/173695}
\end{barticle}
\endbibitem

\bibitem[\protect\citeauthoryear{{Weiss} and {Schlattl}}{2007}]{aw-apss}
\begin{botherref}
\oauthor{\bsnm{{Weiss}},~\binits{A.}},
  \oauthor{\bsnm{{Schlattl}},~\binits{H.}}:
{GARSTEC: the Garching Stellar Evolution Code}.
In: \thisapss. (2007)
\end{botherref}
\endbibitem

%\end{thebibliography}
%\input{ApSS4.2_vF.bbl}
\end{thebibliography}
%
\end{document}